\documentclass[11pt]{article}
\usepackage[utf8]{inputenc}
\pdfoutput=1
\usepackage{amssymb,amsmath,mathrsfs,enumerate}
\usepackage{graphicx,rotate,multicol}
\usepackage{float}
\usepackage{tocloft}
\usepackage{subfig}
\usepackage[margin=10pt,labelfont=bf]{caption}
\usepackage{cite}
\usepackage{soul}
\usepackage[colorlinks=true,
            linkcolor=blue,
            urlcolor=blue,
            citecolor=teal]{hyperref}
\usepackage{multirow}
\usepackage{placeins}

\makeatletter
\let\Hy@linktoc\Hy@linktoc@page
\makeatother
\usepackage{color}
\definecolor{ourcolor}{rgb}{0.7, 0.25, 0.05}
\usepackage{tikz,braket}
\long\def\rpl#1!!#2!!{\textcolor{red}{#1} \textcolor{blue}{#2}}

\let\bar=\overline

\def \order(#1){{\mathcal O} \left(#1 \right)}

\evensidemargin=\oddsidemargin
\allowdisplaybreaks

\title{\color{black}{\bf Halo uncertainties in electron recoil events at  direct detection experiments}}

\author {\bf Tarak Nath Maity,$^{a,}$\footnote{tarak.maity.physics@gmail.com} 
\hspace{4pt}  Tirtha Sankar Ray,$^{a,b,}$\footnote{tirthasankar.ray@gmail.com} \hspace{4pt}  Sambo Sarkar$^{a,b,}$\footnote{sambosarkar92@gmail.com}
\\[10pt]
\small\em $^a$Department of Physics, Indian Institute of Technology Kharagpur, Kharagpur 721302, India\\
\small\em $^b$Centre for Theoretical Studies, Indian Institute of Technology Kharagpur, Kharagpur 721302, India
}

\date{}

\begin{document}

\maketitle
\begin{abstract}

The dark matter direct detection rates are highly correlated with the phase space distribution of dark matter particles in our galactic neighbourhood. In this paper we make a systematic study of the impact of astrophysical uncertainties on electron recoil events at the direct detection experiments with Xenon and semiconductor detectors. We find that within the standard halo model there can be up to $ \sim 50\%$ deviation from the fiducial choice in the exclusion bounds from these observational uncertainties. For non-standard halo models we report a similar  deviation from the fiducial standard halo model when fitted with recent cosmological $N$-body simulations while even larger deviations are obtained in case of the observational uncertainties.

\end{abstract}

\section{Introduction}
\label{sec:intro}
In the last few decades particulate dark matter (DM) has been probed by its possible scattering with the Standard Model (SM) particles \cite{Goodman:1984dc, Drukier:1986tm, Bertone:2010zza, Profumo:2017hqp, Arcadi:2017kky}. The typical direct detection experiments measure the nuclear recoil of a target material through scattering of ambient DM wind on the surface of the earth \cite{Goodman:1984dc, Drukier:1986tm, Lewin:1995rx, Akerib:2016lao, Cui:2017nnn, Aprile:2018dbl}. While nuclear target experiments are suitable to probe a non-relativistic DM mass at $\mathcal{O}(100)$ GeV, however for  DM masses in sub-GeV range these looses its sensitivity. This is due to the fact that energy deposited by a sub-GeV non-relativistic DM remains below the threshold of these experiments. 

An alternate and novel strategy to search for such light DM is through the DM-electron scattering \cite{Essig:2011nj, Essig:2012yx, Lee:2015qva, Essig:2015cda,  Essig:2017kqs}. For an atomic target  (e.g Xenon) if DM scatters off the electron on the atomic shell then this may lead to the ionization of electrons. Whereas for a semiconductor target material (e.g. Si, Ge etc.) scattering of DM with electron may transfer an electron from valance band to conduction band. These ionization signals could provide a handle in the search for sub-GeV DM.
The boundedness of electron in the target material makes these electron scattering events inelastic in nature. This essentially suggests that the incoming DM particles has to have a sufficient energy to excite these bound electrons. For a semiconductor material the typical energy gap between valance and conduction band is of the order $1 ~{\rm eV}$  whereas for Xe targets the minimum binding energy of a shell is around 10 times larger than that. Thus for a Xe target materials a relatively light DM can excite an electron if the DM is moving fast. These fast moving DM can only be found near the tail of the galactic DM distribution. Subsequently the DM-electron event rate would be suppressed for such Xe targets as compared to semiconductors targets, implying sensitivity of semiconductor detector below the MeV scale.

The DM-electron scattering rate can be divided into three parts viz. particle physics, atomic physics and astrophysics. The particle physics input depends on the particular model under consideration and determines the hard scattering cross section between the DM and the electron. In this paper we take a model independent approach to estimate the cross sections. The electron ionization form factor constitutes the atomic physics part and depends on the wave function of the scattered electron. For this we have used the result of \textsf{QEdark} \cite{QEdark}. In \textsf{QEdark}, the form factor for Xe targets has been calculated using Hartee-Fock method while for semiconductor materials density functional theory has been utilized.

The local distribution of the ambient DM constitutes the astrophysical part. The Standard Halo Model (SHM) with Maxwell-Boltzmann (MB) distribution truncated at galactic escape velocity is usually assumed for the distribution of DM in the galaxy. The dispersion of the MB distribution is determined by the Sun's circular velocity ($v_0$).  The typical choice of these parameters are $v_0=220 ~ {\rm km/s}$ and $v_{\rm esc}=544 ~ {\rm km/s}$, following \cite{Smith:2006ym}. However there is still a considerable uncertainty in the measurement of these astrophysical inputs \cite{Green:2017odb}. In this paper we make a systematic study of the impact of these astrophysical uncertainties in DM electron scattering rate.

Cosmological $N$-body simulations generate a patch of our local universe containing mostly DM particles and in some cases stars and gas to study and compare our local universe including the Milky Way (MW) galaxy and its halo to the present day observations. They usually include the effect of baryons utilizing hydrodynamic simulation. We will refer these collectively as cosmological simulations. These simulations indicate that the SHM may not give an accurate description of a Milky Way-like halo. Modifications of the SHM framework have been introduced to reconcile the astrophysical observations and cosmological simulations \cite{Kavanagh:2013eya, Lee:2014cpa,Kelso:2016qqj, Sloane:2016kyi}. These include the King velocity distribution determines the cut off in the distribution through a self consistent manner \cite{2008gady.book.....B}. The Double Power Law which can nicely explain the high velocity dependence of double power density profiles like that of NFW \cite{Lisanti:2010qx}. The Tsallis is a theoretical distribution based on Gibbs entropy motivated by Tsallis statistics \cite{Tsallis:1987eu}. The Mao et. al. \cite{Mao:2013nda} suggests another distribution showing a strong correlation of particle velocities to their position and characteristic radius of the simulated halo \cite{Mao:2013nda}. In this work we have only considered the impact of isotropic velocity distributions on electron scattering events, while leaving the possible impact of anisotropic distributions \cite{Evans:2018bqy} for a future work.

The paper is organized as follows. In section \ref{sec:RevwDM-e} we briefly review the methodology to calculate rate of the DM-electron scattering for atomic and semiconductor target material. In section \ref{sec:UncertSHM} we present the variations of DM-electron exclusion limit due to the uncertainty in the SHM parameters. Going beyond, in section \ref{sec:UncertBSHM} we explore the effect of non-standard velocity distribution. In section \ref{sec:Com-NB} we present a detailed comparison between the deviations obtained from  cosmological simulations and recent astrophysical observations, with reference to the fiducial SHM. Finally,  we conclude in section \ref{sec:conclusion}.

\section{DM-electron scattering}
\label{sec:RevwDM-e}

In this section we will briefly review the scattering of DM particles with electrons that are bound inside the detector material. We will pair it down to the particle physics effects, the atomic physics effects and the astrophysical factors. 
Let us consider a DM particle $(\chi)$ of mass $m_{\chi}$, and initial velocity $v$ scatters of an electron within the target material. Then in the non-relativistic limit the energy conservation of the system  implies \cite{Essig:2011nj}
\begin{equation}
\label{eq:EConservarion}
\Delta E_e + \frac{|m_{\chi} \mathbf{v}-\mathbf{q}|^2}{2 m_{\chi}} = \frac{1}{2} m_{\chi} v^2,
\end{equation} 
where $q$ is the momentum transfer by DM and $\Delta E_e$ is the energy transferred to electron. Note that for a DM mass in sub-GeV scale, the relevant momentum transfer is small compared to the mass of the nucleus, therefore the nuclear recoil piece has been neglected in equation \eqref{eq:EConservarion}\footnote{The typical momentum transfer to electron is of the order of few keV and the nuclear mass is of the order GeV. Therefore the recoil nuclear energy remains below eV.}. Following equation \eqref{eq:EConservarion}, the minimum DM speed $v_{\rm min}$ required to transfer an energy $\Delta E_e$ for a fixed $q$ is given by
\begin{equation}
\label{eq:vmin}
v_{\rm min}=\frac{q}{2 m_{\chi}}+\frac{\Delta E_e}{q},
\end{equation}
For an atomic target (e.g. Xenon) the differential DM-electron scattering rate is \cite{Essig:2012yx}
\begin{equation}
\label{eq:rateXe}
\frac{dR_{\rm ion}}{d\,ln\, E_e}=N_T\frac{\rho_{\chi}}{m_{\chi}}\,\sum_{nl} \frac{\bar{\sigma}_e}{8\mu_{\chi e}^2} \int q dq \,F_{\rm DM}(q)^2\, |f_{\rm ion}^{n,l}(k^{\prime},q)|^2 \,\eta\left(v_{\rm min}(k^{\prime},q)\right),
\end{equation}
where $N_T$ denotes the number of atoms in the target. $\rho_{\chi}$ is the local DM density. DM-electron reduced mass is denoted by $\mu_{\chi e}$. $\bar{\sigma}_e$ stands for DM-electron cross section for a particular momentum transfer $q=\alpha m_e$ and $F_{\rm DM}$ is the DM form factor. The ionization form factor and  average inverse velocity are represented by $f_{\rm ion}^{\rm n,l}$ and $\eta$ respectively.

Whereas, the differential rate for a semiconductor target (e.g. Ge) can be written as \cite{Essig:2015cda}
\begin{equation}
\label{eq:rateSemi}
\frac{dR_{\rm crystal}}{d\,ln\, E_e}=\frac{\rho_{\chi}}{m_{\chi}}N_{\rm cell}\bar{\sigma}_e \alpha \, \frac{m^2_e}{\mu_{\chi e}^2} \int d {\rm ln} q \, \frac{E_e}{q} \,F_{\rm DM}(q)^2\, |f_{\rm crystal}(k^{\prime},q)|^2 \,\eta\left(v_{\rm min}(k^{\prime},q)\right),
\end{equation}
where $N_{\rm cell}$ stands for the number of unit cell in a crystal targets  \cite{Essig:2015cda}. The $f_{\rm crystal}$ denotes the ionization form factor for the crystal. Other parameters have their usual meaning as discussed.

 Note that both in equations \eqref{eq:rateXe} and \eqref{eq:rateSemi} $F_{\rm DM}(q)$ takes care of the momentum dependency in DM-electron interaction. This DM form factor and $\bar{\sigma}_e$ comprise the main particle physics input in equations \eqref{eq:rateXe} and \eqref{eq:rateSemi}.  Remaining agnostic about any particular model here we have considered three types of interactions between DM and electron. These interactions can be quantified by the DM form factor $F_{\rm DM}$. The three choices for the $F_{\rm DM}$ are $1,~\alpha m_e/q, \left(\alpha m_e/q\right)^2$. Note that $F_{\rm DM}= 1$ can be induced by an exchange of heavy mediator between DM and electron \cite{Holdom:1985ag, Izaguirre:2015yja}, $F_{\rm DM}=\alpha m_e/q$ which could arise through electric dipole moment interaction \cite{Sigurdson:2004zp}, and  $F_{\rm DM}=\left(\alpha m_e/q\right)^2$ which may be induced by a light mediator \cite{Izaguirre:2015yja, Essig:2015cda}.

The ionization form factor defines the suppression of the event rate to ionize an electron from its bound state to a continuum state of momentum $k^{\prime}=\sqrt{2 m_e E_e}$ through  $q$. Hence it depends solely on the target material. Throughout our numerical calculation we have utilized the form factor given in \textsf{QEdark} \cite{QEdark}.

Other than these the usual astrophysical inputs are the local DM density $\rho_{\chi}$ and average inverse DM speed
\begin{equation}
\label{eq:eta}
\eta(v_{\rm min})=\int_{v_{\rm min}}^{\infty} \frac{f_{\oplus}(\mathbf{v})}{v} d^3v
\end{equation}
The $f_{\oplus}(\mathbf{v})$ in equation \eqref{eq:eta} is the DM velocity distribution in the detector rest frame. If we assume $f(\mathbf{v})$ as the DM distribution in the galactic frame then the distribution at lab frame can be obtained by 
\begin{equation}
f_{\oplus}(\mathbf{v}) = f(\mathbf{v+v_e}),
\end{equation}
where $\mathbf{v_e}=\mathbf{v_0+v_{\circledast}+v_{\oplus}}$ and  $\mathbf{v_0}$ and $\mathbf{v_{\circledast}}$ are the Sun's circular velocity at local standard rest and Sun's peculiar velocity respectively. The earth velocity in the Solar rest frame is represented by $\mathbf{v_{\oplus}}$. Note that the variation in $\mathbf{v_{\oplus}}$ with time leads to the familiar annual modulation  \cite{Essig:2015cda,Lee:2015qva} in the DM direct detection rate, has been neglected here. The sun's peculiar motion $\mathbf{v_{\circledast}}=(U_{\circledast}, V_{\circledast}, W_{\circledast} )$ = $ (11.1 \pm 1.5, 12.2 \pm 2, 7.3 \pm 1) $ is adapted from \cite{Sch_nrich_2010}. We set the earth's rotational velocity following the reference \cite{Lewin:1995rx}.

The mean inverse speed is essentially regulated by the astrophysical parameters discussed above. Therefore any uncertainty in determination of these parameters will have a direct impact in the exclusion limit. In the rest of this paper we systematically study the impact of these astrophysical uncertainties including departure from the MB distribution on the exclusion bounds of direct detection experiments.
 

The differential rates given in equations \eqref{eq:rateXe} and \eqref{eq:rateSemi} are with respect to the electron recoil energy $E_e$. While the semiconductor detectors are only sensitive to electron and the Xe detector finally detects photo-electron at the PMT's. For the Xe target material this electronic energy is converted into a number of electrons and subsequently to photo-electrons using the prescription of \cite{Essig:2017kqs}. Note that the Xenon10 experiment \cite{Angle:2011th} which has an exposure of $15$ kg-days, sets most stringent limits in most of the region in the parameter space of interest \cite{Essig:2017kqs}, therefore we would only consider the  bound from Xenon10 in this paper. The exclusion limit for Xenon10 is obtained by making a conservative assumption that all the observed events arise from DM. Whereas for semiconductor targets the electron-hole yields are obtained following reference \cite{Essig:2015cda}.  We have considered  the one electron threshold to present the bound for semiconductor detectors. For semiconductor detectors, the upper limit on the relevant parameters are presented at the $90\%$ CL (corresponds to 2.3 DM events), assuming an one electron
threshold with an exposure of $1$ kg-day.

\section{Astrophysical uncertainties within the Standard Halo Model}
\label{sec:UncertSHM}
The most simplified isotropic and isothermal DM  distribution in the MW halo is usually described by a MB distribution, with a cut off at the escape velocity of the DM particles \cite{LyndenBell:1966bi, 1990ApJ...353..486L}. The distribution function has the following form
\begin{equation}
    f(\mathbf{v})=
    \begin{cases}
      \frac{1}{N}\left[\exp{\left(-\frac{|\mathbf{v}|^{2}}{v_{0}^{2}}\right)}\right] &  |\mathbf{v}| \leq v_{\rm esc} \\
      0 & |\mathbf{v}| > v_{\rm esc},
    \end{cases}
    \label{eq:fv}
  \end{equation}
where $N$ denotes the normalization constant of the distribution, $v_0$ is the measure of its velocity dispersion and $v_{\rm esc}$ sets the maximum allowed DM velocity of the distribution.
Keeping ourselves within this SHM, in this section we will present the uncertainties in the determination of astrophysical parameters and their implications on the DM-electron scattering events\footnote{Note that the impact of these astrophysical uncertainties both in the case of nuclear and electron recoil has studied previously in references \cite{McCabe:2010zh, Frandsen:2011gi, Sloane:2016kyi, Laha:2016iom, Benito:2016kyp, Bozorgnia:2017brl, Fowlie:2017ufs, Fowlie:2018svr, Andersson:2020uwc, Hryczuk:2020trm, Buch:2020xyt}.}. A discussion about the main observational uncertainties in the SHM parameters are now in order:
\begin{enumerate}
\item \textbf{Local DM density:} The typical choice for the local density is $0.3~ \rm GeV\, cm^{-3}$ \cite{Zyla:2020zbs}. Recent estimations suggest that it may vary in the range $(0.2-0.6)~ \rm GeV \, cm^{-3}$ \cite{Catena:2009mf, Salucci:2010qr, Pato:2015dua, McKee_2015, Xia_2016, Green:2017odb, Sivertsson:2017rkp, Schutz:2017tfp, Evans:2018bqy,Buch:2018qdr, Guo_2020, Salomon_2020}. While others predict slightly different values \cite{Sivertsson:2017rkp, Schutz:2017tfp, Buch:2018qdr, Hagen_2018, Guo_2020}. However note that the differential rate given in equations \eqref{eq:rateXe} and \eqref{eq:rateSemi} scale linearly with the local DM density. Therefore for a change in $\rho_{\chi}$ one would expect a proportional vertical shift in the exclusion limits for all the experiments. 
Assuming the central value of the local DM density $0.4~ \rm GeV\, cm^{-3}$, we find that there is a maximum 100\% relative change due to the aforementioned variation of $\rho_{\chi}$. This change is independent of the DM mass and the target materials used for the detection. Hence we have fixed this to $0.4~ \rm GeV\, cm^{-3}$ without considering its variational implications on the exclusion bounds.

\item \textbf{Circular velocity of the Sun:}
The local circular velocity of the Sun ($v_0$) with respect to the galactic center is usually assumed to be $220 ~\rm km/s$ \cite{Aprile:2018dbl, Agnese:2014aze, Akerib:2016vxi}. This would be considered as the fiducial choice of the parameter $v_0$ for rest of the paper.  From the orbit of the GD-1 stellar stream, the reference \cite{Koposov:2009hn} constrained $v_0$ in the range $221 \pm 18 ~\rm km/s$. A similar range of $v_0$, namely $225 \pm 29 ~\rm km/s$ is found to be in consonance with the kinematics of maser \cite{McMillan:2009yr}. These estimates seem to have around $10\%$ error in the measurement of $v_0$. However a more precise assessment of $v_0$ can be done using the measurement of apparent proper motion of Sgr A$^*$ relative to a distant quasar \cite{Reid:2004rd, Bland_Hawthorn_2016}. This measurement fixes the total angular velocity of the Sun ($(v_0+V_{\circledast})/R_{\odot}$) in the range $30.24 \pm 0.12 ~\rm km \,s^{-1} kpc^{-1}$. On the other hand recently GRAVITY collaboration has estimated the value of $R_{\odot}$ with quite a high accuracy: $8.122 \pm 0.031$ kpc \cite{Abuter:2018drb}. Also note that the relevant component of the peculiar velocity $V_{\circledast}$ varies in the range $12.24 \pm 2.47$ \cite{Sch_nrich_2010}. Combining all these observations, the circular velocity of the Sun has been found to be $233 \pm 3\, \rm km/s$ \cite{Evans:2018bqy}. Other recent assessments \cite{Gillessen:2009ht, Reid_2014, Eilers_2019, Hogg_2019} would also lead to similar result for $v_0$. We will explore the impact of the deviations of this from the fiducial value on the exclusion limits from direct detection experiments in the electron scattering events.

As can be seen from the equation \eqref{eq:fv}, $v_0$ is related to the standard deviation of the distribution. Thus any increment in $v_0$  would flatten the distribution. Therefore  this would make more DM particles available to interact with electrons in the tail region. This will lead to a relatively stronger bound in DM-electron cross section. The effect will be reversed for a decrement in $v_0$.  Further a change in $v_0$ would also alter the Galilean boost.
\begin{figure*}[t]
\begin{center}
\subfloat[\label{sf:XeSHMa}]{\includegraphics[scale=0.18]{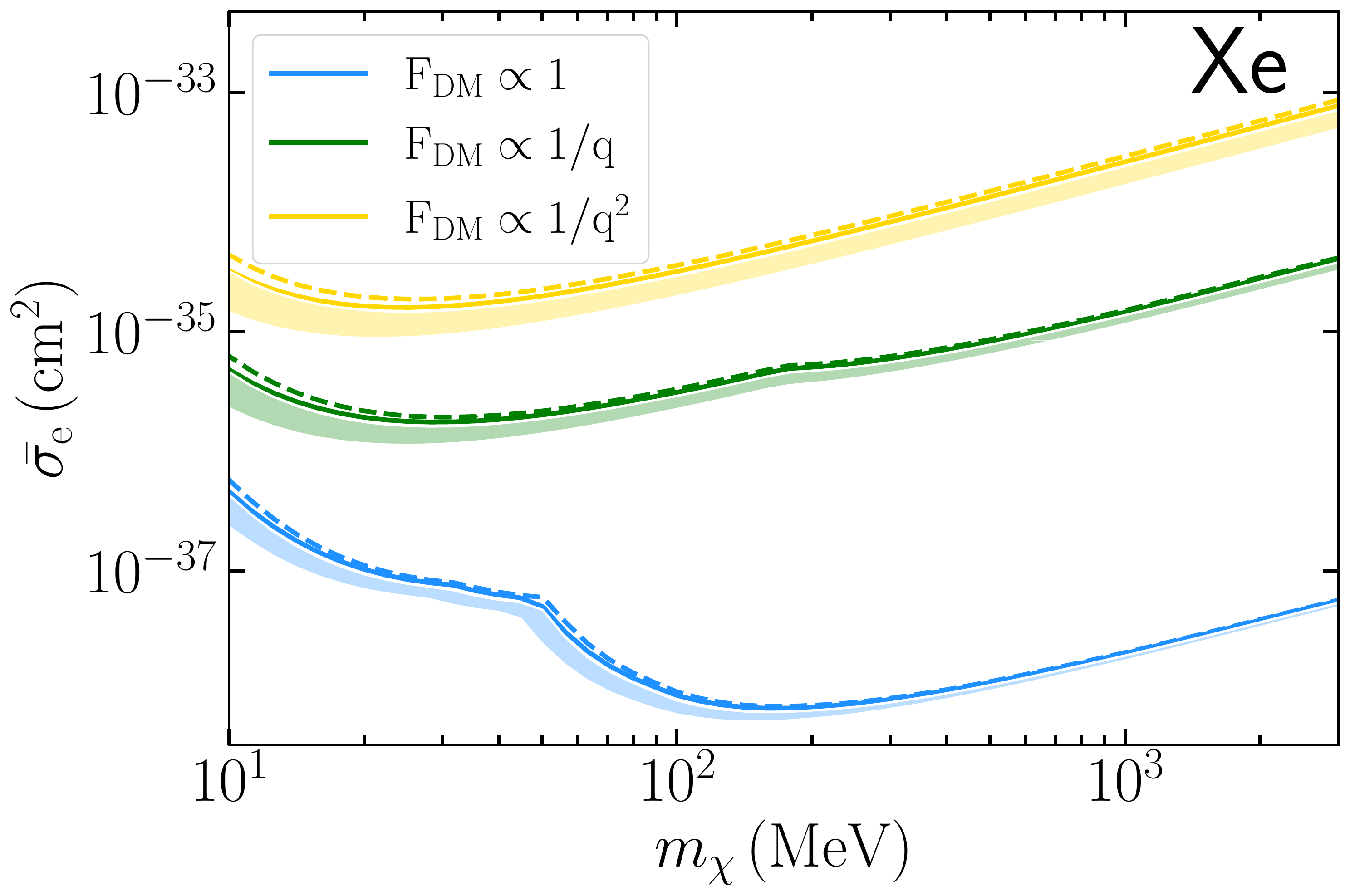}}
\subfloat[\label{sf:GeSHMa}]{\includegraphics[scale=0.18]{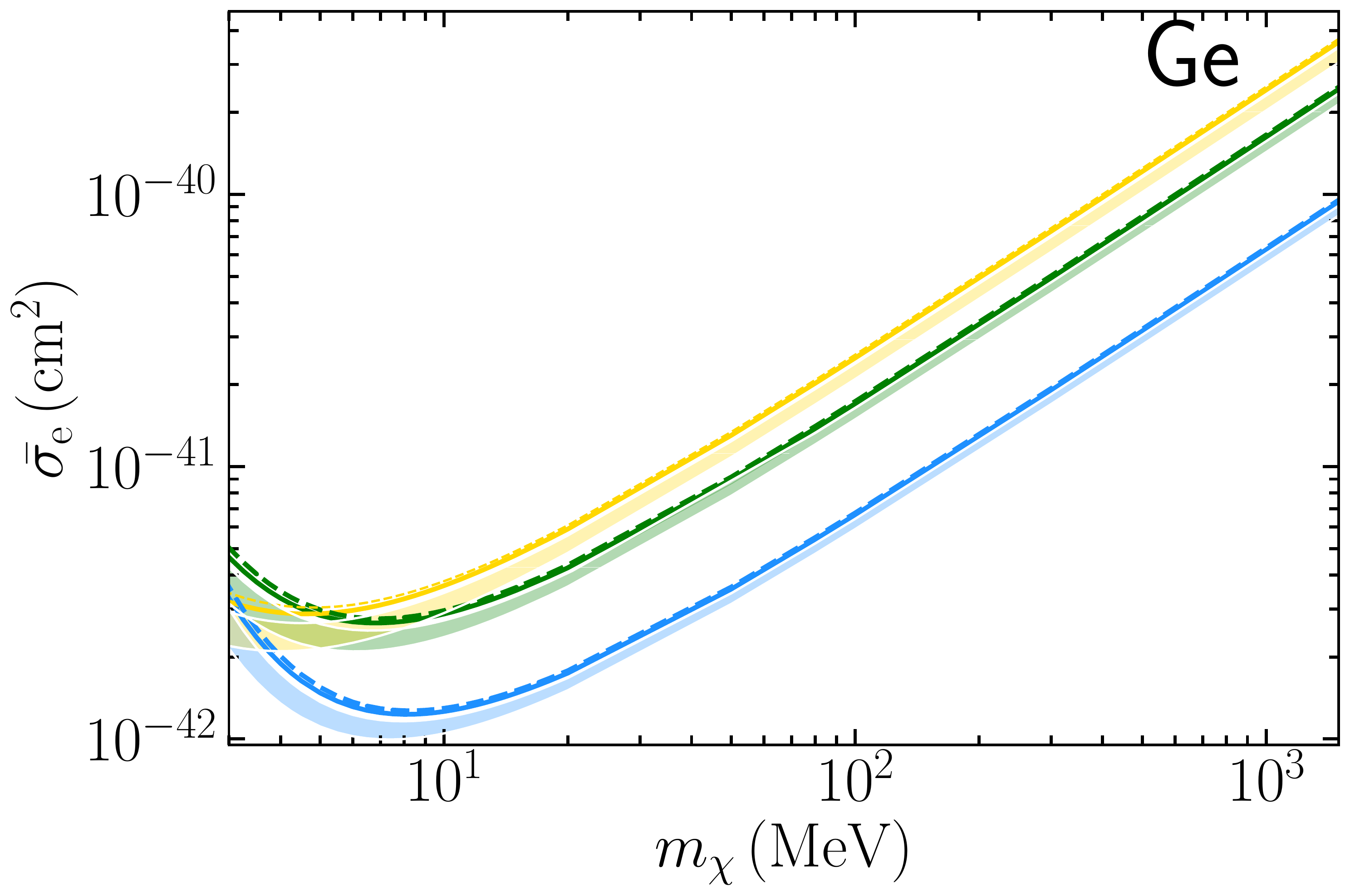}}
\subfloat[\label{sf:SiSHMa}]{\includegraphics[scale=0.18]{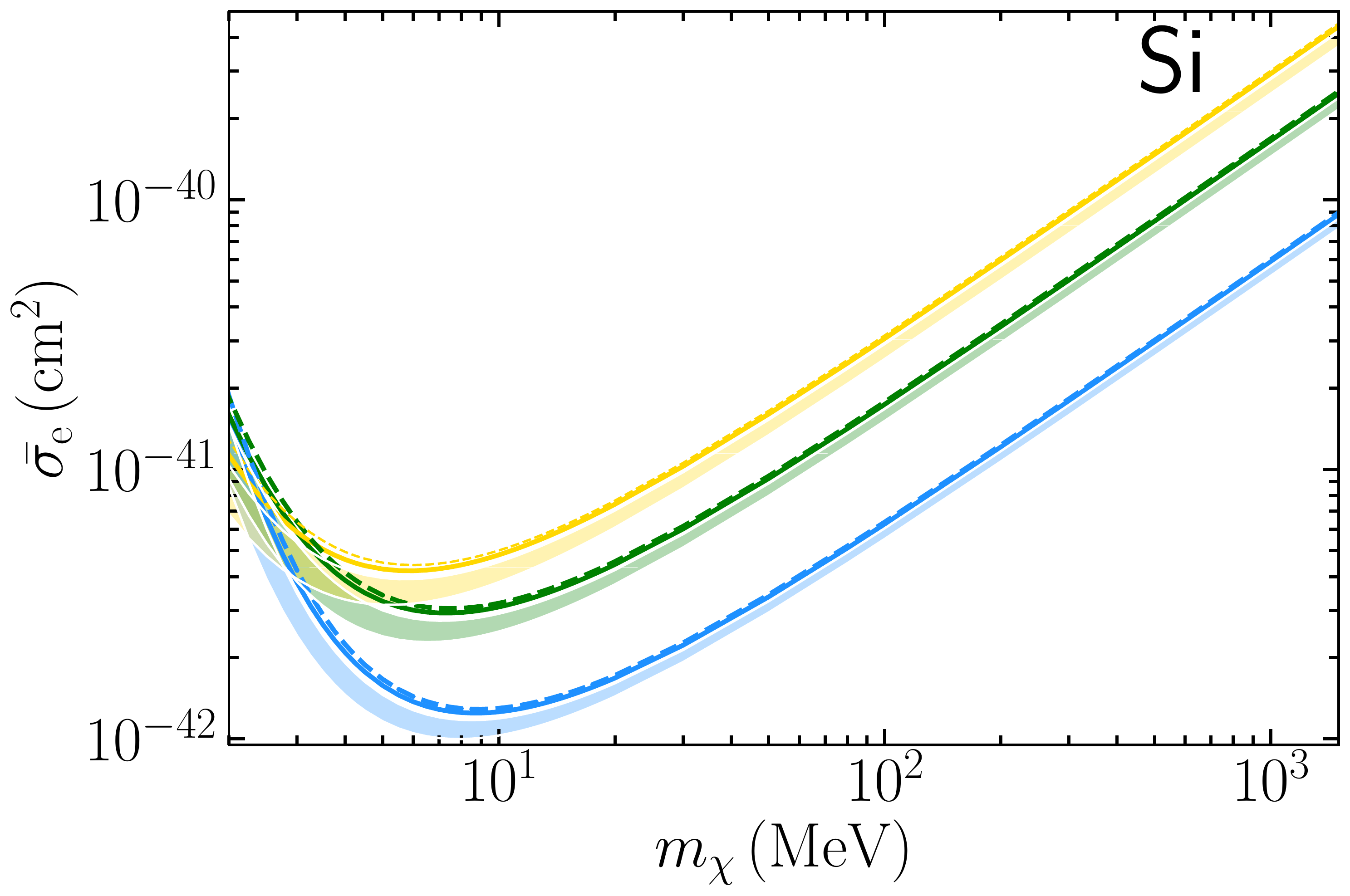}}
\newline
\subfloat[\label{sf:Xeastroeff}]{\includegraphics[scale=0.19]{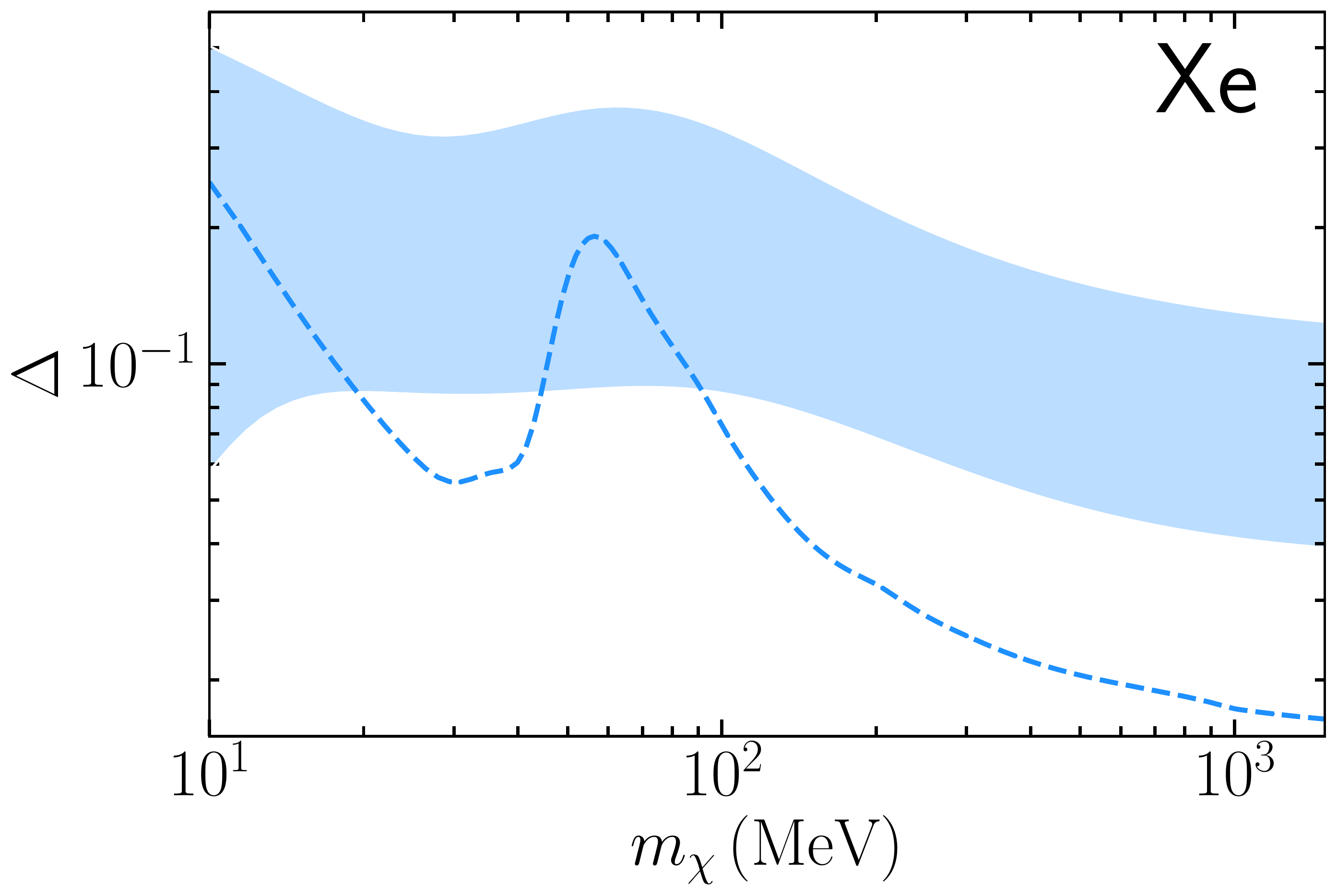}}
\subfloat[\label{sf:Siastroeff}]{\includegraphics[scale=0.195]{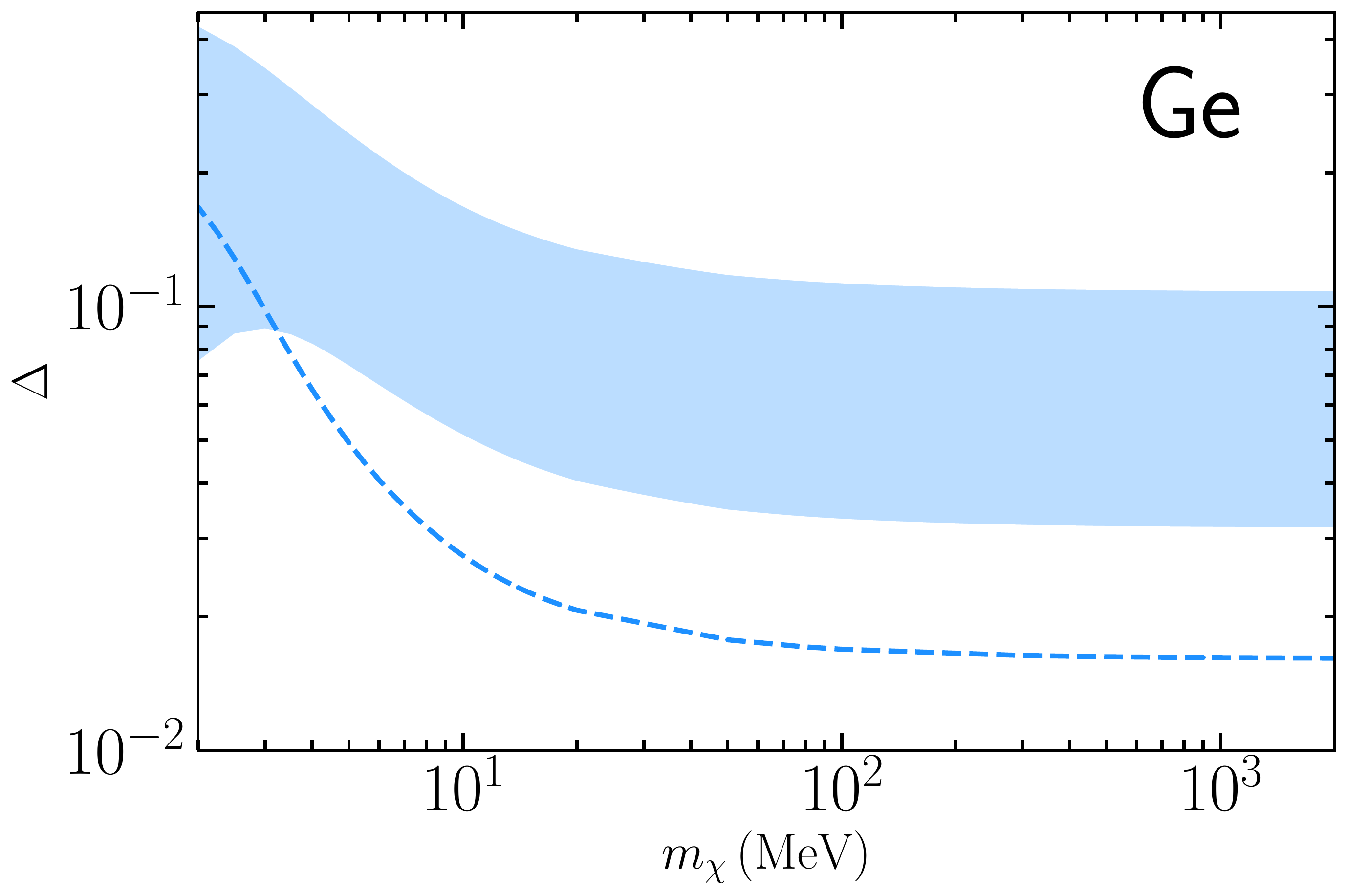}}
\subfloat[\label{sf:Geastroeff}]{\includegraphics[scale=0.19]{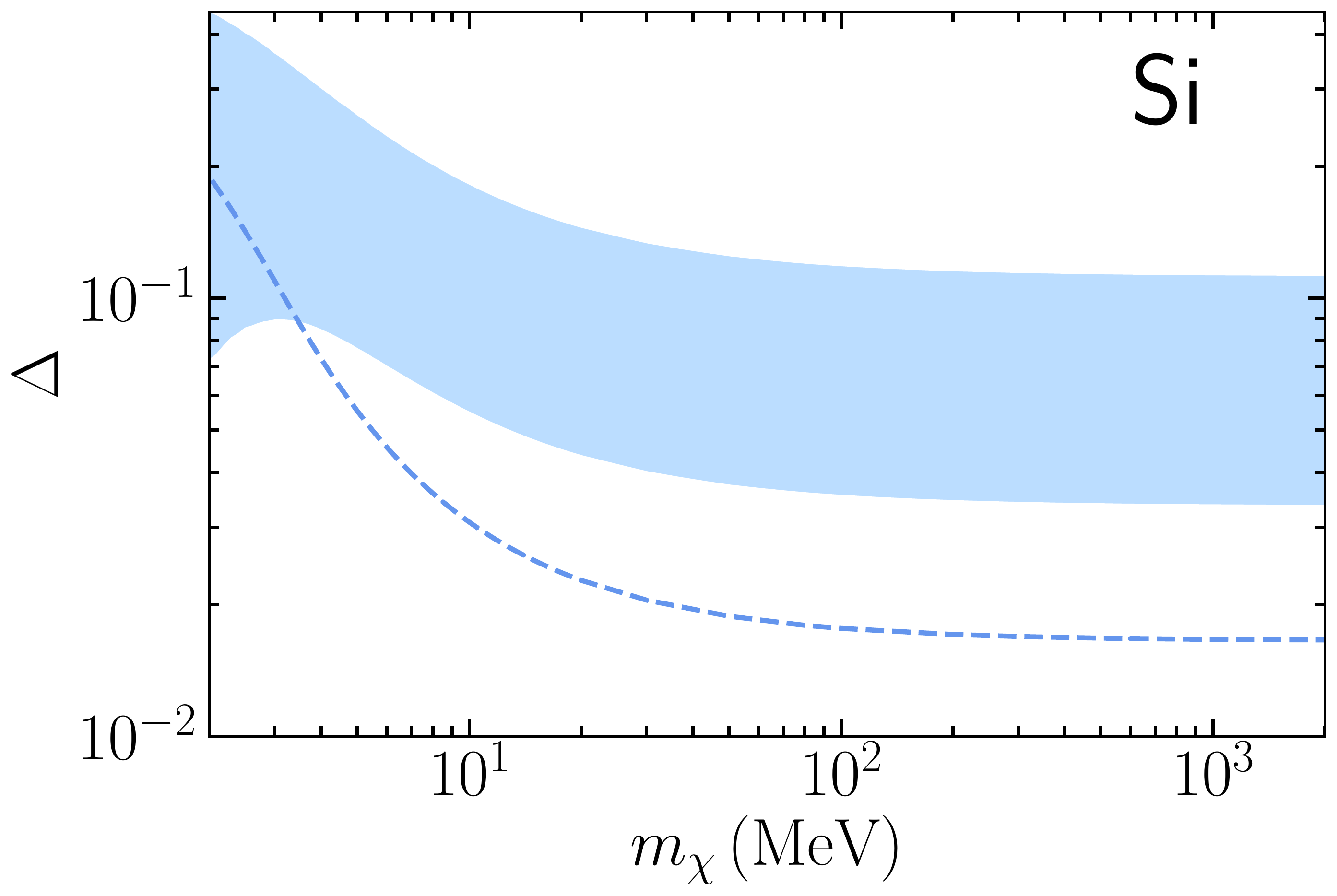}}
\caption{Exclusion bounds in the SHM for fiducial and with recently obtained astrophysical parameters. In the upper panel, the solid lines are for the fiducial values and the dashed lines are the exclusion bounds for a Modified MB distribution (given in equation \eqref{eq:MMB}) with the fiducial values. The light blue, green and yellow shaded regions express the uncertainties in astrophysical observations for $F_{\rm DM}=$  $1,~\alpha m_e/q$ and $\left(\alpha m_e/q\right)^2$ respectively. In the lower panel we show the corresponding fractional changes relative to the fiducial choice in the SHM (defined in equation \eqref{eq:sigma-effective}). The uncertainties associated with the recent astrophysical observations are utilized to form light blue shaded bands. Whereas the blue dashed curves denote the relative fractional changes in the Modified Maxwell-Boltzmann distribution for $F_{\rm DM}=1$.} 
\label{fig:v0-astro}
\end{center}
\end{figure*}

\item \textbf{Galactic escape velocity:}
The escape velocity of a massive body in a galaxy is defined by the velocity above which they will no longer remain bound to its gravitational potential. Measurements from the high velocity stars of the RAVE survey determines the $v_{\rm esc}$ in the range $498-608 $ km/s \cite{Smith:2006ym} with  the median $544$ km/s. In the rest of the paper we would consider this as the fiducial choice for $v_{\rm esc}$.  Based on the recent analysis from the velocities of $2850$ halo stars from the \textit{Gaia} velocity survey Data Release-2 \cite{Monari_2018}, the local escape speed has been revised to $580 \pm 63$ km/s. However it has been argued that this result is sensitive to the prior chosen for describing  high velocity tail of the distribution function. With a prior estimation from simulations, and a more localized sample of $2300$ high velocity counter-rotating stars, the escape speed has been obtained to be $528^{+24}_{-25}$ km/s \cite{Deason_2019}. This is also in consonance with the previous results. We will be using the central value of the latter as the new central value for the escape velocity of the DM particles within the SHM.
\end{enumerate}

To estimate the relative fractional change in the cross-section we will adopt equation \eqref{eq:sigma-effective} throughout this paper, 
\begin{equation}
  \Delta =\lvert \frac{ \bar{\sigma}_e^{\rm i} - \bar{\sigma}_e^{\rm fiducial,SHM}}{ \bar{\sigma}_e^{\rm fiducial,SHM}}\rvert,
	\label{eq:sigma-effective}
	\end{equation}
where $\bar{\sigma}_e^{\rm i}$ denotes the cross section corresponding to non-fiducial values of the distribution in consideration.

With a decrease in the escape velocity there will be less energetic particles in the halo capable of scattering, hence a larger $ \bar{\sigma}_{e}$ is required to produce the same number of events at a given experiment. Due to the exponential suppression of the MB distribution near the tail, increasing $v_{\rm esc}$ has a smaller effect than decreasing it by the same amount. We find that for similar relative change in $v_0$ and $v_{\rm esc}$, the effect due to the change in $v_0$ is more pronounced. This can be ascribed to the fact that any change in $v_0$ causes an overall change in the shape of distribution, whereas the $v_{\rm esc}$ determines the distribution's cut-off near the exponentially suppressed tail.

 The inverse relation between $m_{\chi}$ and $v_{\rm min}$ implies that the required minimum DM velocity is rather close to $v_{\rm esc}$ for light DM. This is because DM particles having lower mass and hence lower kinetic energy can generate the required recoil energy  if their minimum velocity becomes closer to the escape velocity of the distribution. And the tail of the distribution is quite sensitive to the choice of astrophysical parameters. Thus the observed fractional change in the exclusion limit is significantly larger for light DM. This is evident from all figures shown in the paper.

In the upper panel of figure \ref{fig:v0-astro}, the light blue, green and yellow shaded regions represent uncertainties associated with an updated measurements of $v_{\rm esc}=528^{+24}_{-25}$ km/s and $v_0=233 \pm 6$ km/s for the three choices of the DM form factor $F_{\rm DM}=$  $1,~\alpha m_e/q $ and $ \left(\alpha m_e/q\right)^2$ respectively at $ 95\%$ confidence level. In the lower panel we show the deviation from the fiducial SHM for $F_{\rm DM}=1$, the light blue bands arise due the uncertainties related to the measurement of $v_{\rm esc}$ and $v_0$. The corresponding deviations for the other choices of  $F_{\rm DM}$ lie in a similar range, hence have not been shown in the plots to reduce clutter. Like figure \ref{fig:v0-astro}, in rest of the paper the variations of a Xenon target are shown in the left panels, whereas similar variations for Germanium and Silicon target materials are depicted in the middle and right panels respectively. In conclusion our study indicates a variation  in $\Delta$  between $ 2 \% $ to $ 50 \%$, due to the astrophysical uncertainties in the exclusion bounds within the SHM. This is over and above the modifications due to the ambiguity in the measurement of the local DM density mentioned earlier.

\subsection{Cosmological simulations}
\label{subsec:Nbody}

%
\begin{table}
\centering   
\bigskip
\begin{tabular}{|p{5cm}|p{3cm}|p{3cm}|}
    \hline
    \textbf{ Simulation} & \textbf{$v_{\rm esc} ~({\rm km/s})$} & \textbf{$v_{o}~ ({\rm km/s})$ } \\
     
\hline 
\setlength\arraycolsep{10pt} 

APOSTLE DMO \cite{Sawala:2015cdf, Fattahi_2016, Bozorgnia:2016ogo}& $ 646 $ & $ 212.7 $ \\ \hline
APOSTLE DMB \cite{Sawala:2015cdf, Fattahi_2016, Bozorgnia:2016ogo} & $ 646 $ & $ 224.1 $  \\ \hline 
ARTEMIS DMO \cite{Font_2020, Poole-McKenzie:2020dbo}& $ 521.6 $ & $ 161.4 $ \\ \hline
ARTEMIS DMB \cite{Font_2020, Poole-McKenzie:2020dbo}& $ 521.6 $ & $ 184.3 $ \\ \hline   
\end{tabular}
\caption{Best fit parameters for the SHM. DMO refers to the DM only simulation. DMB indicates the simulation which takes into account both DM and baryon smooth particle hydrodynamics.}
\label{tab:SHMbestfit}
\end{table}

An alternate way to estimate the value of $v_0$ and $v_{\rm esc}$ is to fit a model of DM velocity distribution to  cosmological simulations. Here we have utilized the results of the APOSTLE \cite{Sawala:2015cdf, Fattahi_2016, Bozorgnia:2016ogo} and ARTEMIS \cite{Font_2020, Poole-McKenzie:2020dbo} simulations \footnote{We have provided brief details of the simulations used in this paper in appendix \ref{app:Nbody}.} to find the best-fit values of the velocity  distribution parameters. For APOSTLE, two MW-like haloes identified as A1 and A2 in \cite{Poole-McKenzie:2020dbo} are chosen based on the selection criterion: total halo mass ($M_{200}$) in the range $5 \times 10^{11} < M_{200}/M_{\odot} < 2 \times 10^{13}$, having rotation curves similar to the observed MW like rotation curves, stellar mass ($M_{\star}$) within the observed stellar mass of the MW i.e in the range $4.5 \times 10^{10} < M_{\star}/M_{\odot} < 8.3 \times 10^{10}$ and having a substantial stellar disc \cite{Calore:2015oya, Bozorgnia:2016ogo}. In this work we have used the A1 halo for all our fits. For the ARTEMIS simulation \cite{Font_2020}, 42 MW like haloes are chosen by considering galaxies in the mass range  $8 \times 10^{11} < M_{200}/M_{\odot} <2 \times 10^{12}$.  We have utilized the median DM distribution of the 42 MW-like galaxies reported in \cite{Poole-McKenzie:2020dbo}.

\begin{figure*}[t]
\begin{center}
\subfloat[\label{sf:Xerest1}]{\includegraphics[scale=0.18]{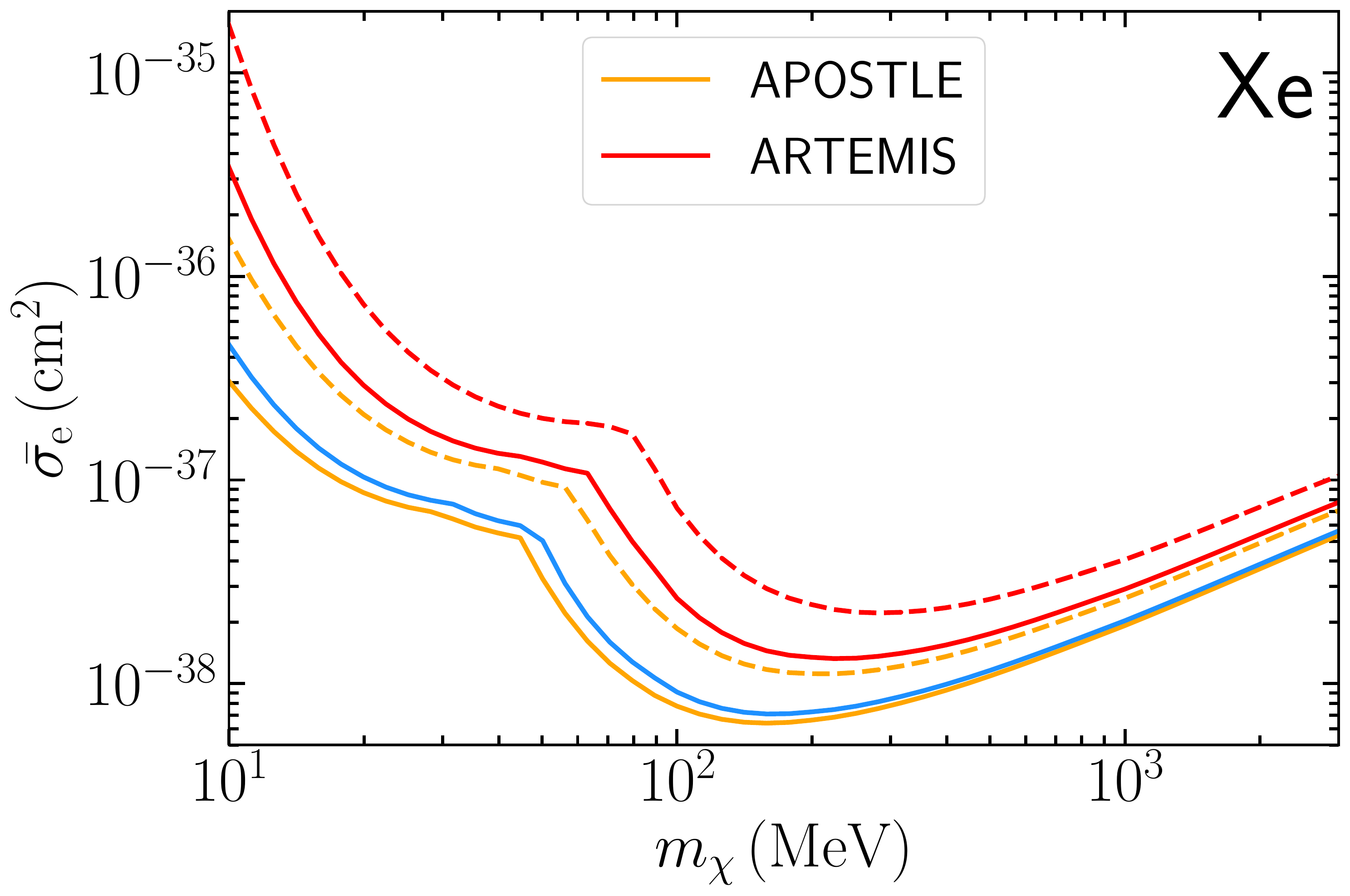}}
\subfloat[\label{sf:Sirest1}]{\includegraphics[scale=0.18]{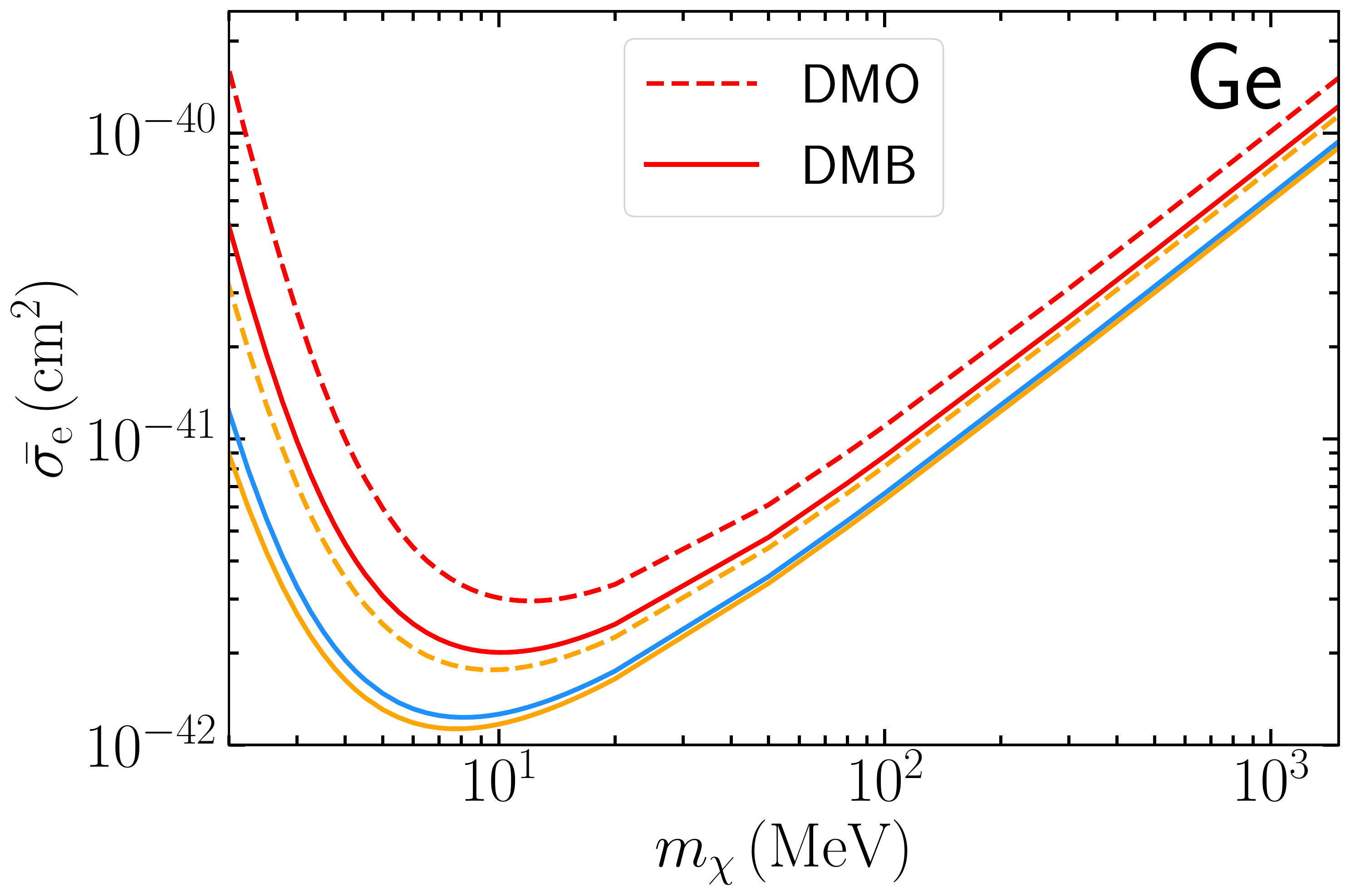}}
\subfloat[\label{sf:Gerest1}]{\includegraphics[scale=0.18]{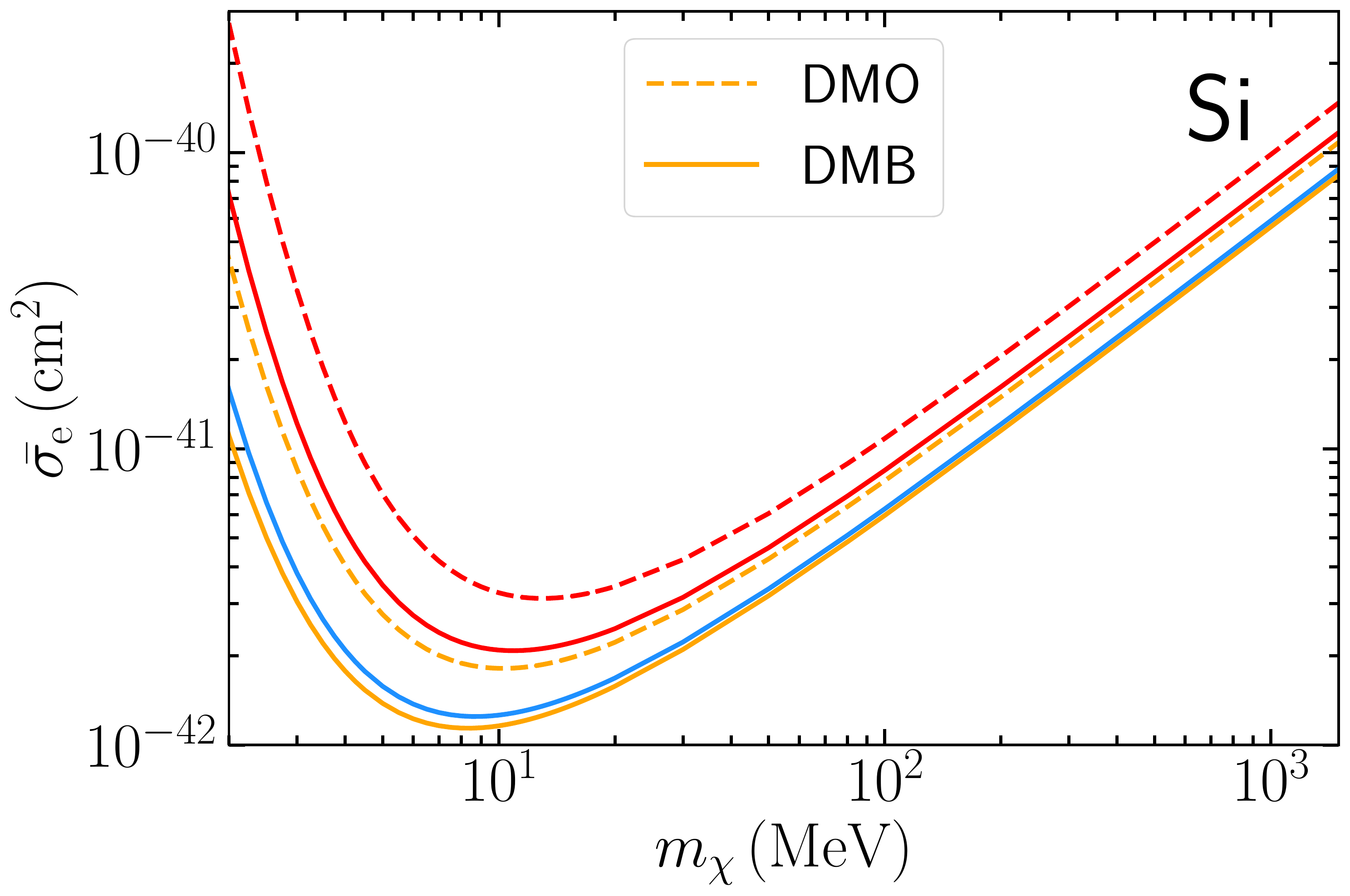}}
\caption{Variation in the SHM for the APOSTLE and ARTEMIS simulations for $F_{\rm DM}=1$. The light blue solid lines denote the fiducial SHM exclusion limit. The dashed lines corresponds to upper limit on DM-electron cross when SHM is fitted with DM only simulation (denoted by DMO). The solid lines represents exclusion limit when SHM is fitted to the simulations which includes both baryon and DM (denoted by DMB). The orange and red lines represent the APOSTLE and ARTEMIS simulations. The other relevant details are mentioned in figure \ref{fig:v0-astro}.}
\label{fig:Nbody-SHM}
\end{center}
\end{figure*}

Note that these two simulations capture the impact of baryons through hydrodynamic corrections and have relatively high force resolution. This may be contrasted with the older DM only simulations like GHALO \cite{Stadel:2008pn} or Via Lactea \cite{Diemand:2008in} which have large deviations from the SHM. Therefore we do not consider them further in this study.
Given these DM distribution of A1 for APOSTLE and the median DM distribution for ARTEMIS, the best fit values of the relevant DM distribution parameters are summarized in table \ref{tab:SHMbestfit}. Throughout the paper, we have utilized an inbuilt python function {\tt scipy} \cite {Virtanen:2019joe} to fit the simulation data with a given distribution. We find that our result matches relatively well with \cite{Lisanti:2010qx,Bozorgnia:2016ogo}. Note that throughout the work DMO implies simulations containing only DM for both APOSTLE and ARTEMIS. Both simulations also have a hydrodynamic version containing both DM and baryons. Simulations that include both DM and baryon effects hereafter will be referred to as DMB. We have employed the values given in table \ref{tab:SHMbestfit}, to present the shift in the exclusion bounds as shown in figure \ref{fig:Nbody-SHM} for $F_{\rm DM}=1$. The light blue curves represent the exclusion bounds for the fiducial choice of $v_0$ and $v_{\rm esc}$. The other solid lines represent exclusion limits with the DMB simulations. The dashed lines represent exclusion limits with DMO simulations.  The orange and red lines corresponds to APOSTLE and ARTEMIS simulations respectively. For the APOSTLE DMB simulation, we observe a maximum change $\Delta \sim \mathcal{O}(30\%)$ in the cross section relative to the fiducial SHM.

\subsection{Modified Maxwell-Boltzmann distribution}
\label{subsec:MMB}
In the passing, we briefly discuss the modified version of Maxwell-Boltzmann velocity distribution in the galactic frame, truncated at the galaxy escape velocity $v_{ \rm esc}$. This can be written in the form
\begin{equation}
    f(\mathbf{v})=
    \begin{cases}
      \frac{1}{N}\left[\exp\left(-\frac{|\mathbf{v}|^{2}}{v_0^2}\right)-\beta \exp\left(-\frac{v_{\rm esc}^{2}}{v_0^2}\right)\right] &  |\mathbf{v}| \leq v_{ \rm esc} \\
      0 &  |\mathbf{v}| > v_{\rm esc},
    \end{cases}
    \label{eq:MMB}
  \end{equation}
where the symbols have their usual meaning.  
In this work $\beta=1$ has been chosen with the desire for an exponential cut-off. The exclusion bounds for this distribution of DM are shown by the dashed lines in figure \ref{fig:v0-astro}. We use the fiducial values to generate the exclusion curves. We do not observe any significant changes in the exclusion bounds as compared to the standard MB distribution.

\section{Beyond the Standard Halo Model}
\label{sec:UncertBSHM}

Several high resolution cosmological simulations suggest that the DM velocity distribution may depart from standard MB distribution, particularly in the high velocity tail \cite{Vogelsberger:2008qb, Kuhlen:2009vh,Lisanti:2010qx}. Due to the sharp cut off at the escape velocity, the SHM over predicts the number of high energetic DM  particles that are available for scattering. A possibility in this regard is to look for a non-SHM distribution of DM motivated by  cosmological simulations.  In what follows, we have considered some of these distributions and their implications on the  DM-electron scattering rates in the direct detection experiments.

\subsection{King Model}
\label{subsec:king}
For a physical system of finite size, the truncated MB distribution, given in equation \eqref{eq:fv} may not be a natural solution of collision-less Boltzmann equation. The King distribution  \cite{King:1966fn} is an alternate model which can be formulated self-consistently for a finite size DM halo. In this model instead of the escape velocity, the maximum DM particle velocity $ v_{\rm max}$ determines the cut-off criterion ($ v_{\rm max} < v_{\rm esc}$). It is predicted on the assumption that if a DM particle moves with $v_{\rm max}$ at any position of the halo then it can reach the halo boundary  where by construction the density vanishes \cite{Chaudhury:2010hj}. This boundary is often called the truncated radius which represents the physical size of a halo. Such a finite size halo provides a more realistic  description of galaxies as compared to the isothermal sphere. These so called lowered isothermal model is also preferred by simulations \cite{Ling:2009eh}.
The distribution function can be written as 
\begin{figure*}[t!]
\begin{center}
\subfloat[\label{sf:XeK1}]{\includegraphics[scale=0.18]{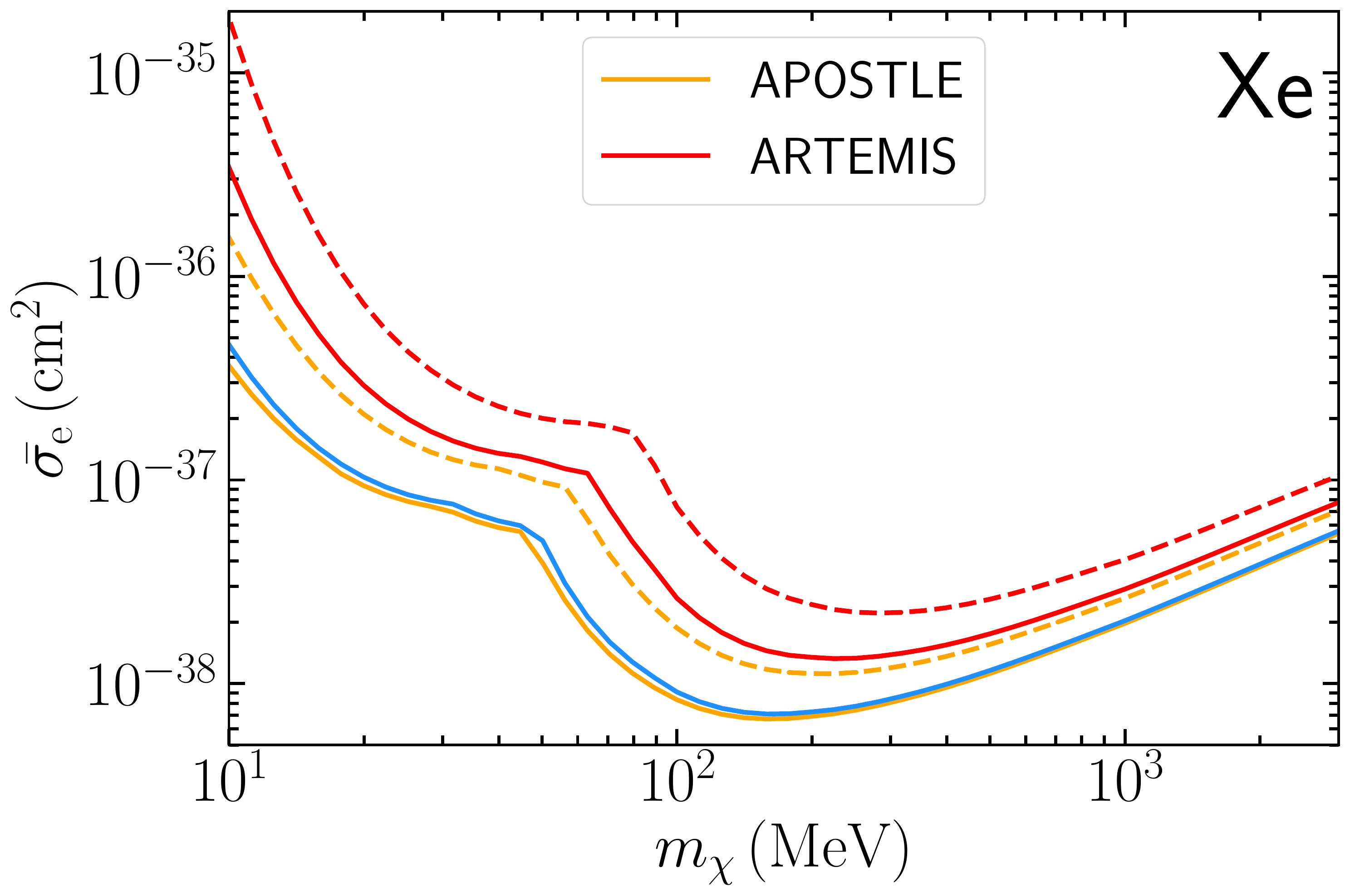}}
\subfloat[\label{sf:SiK1}]{\includegraphics[scale=0.18]{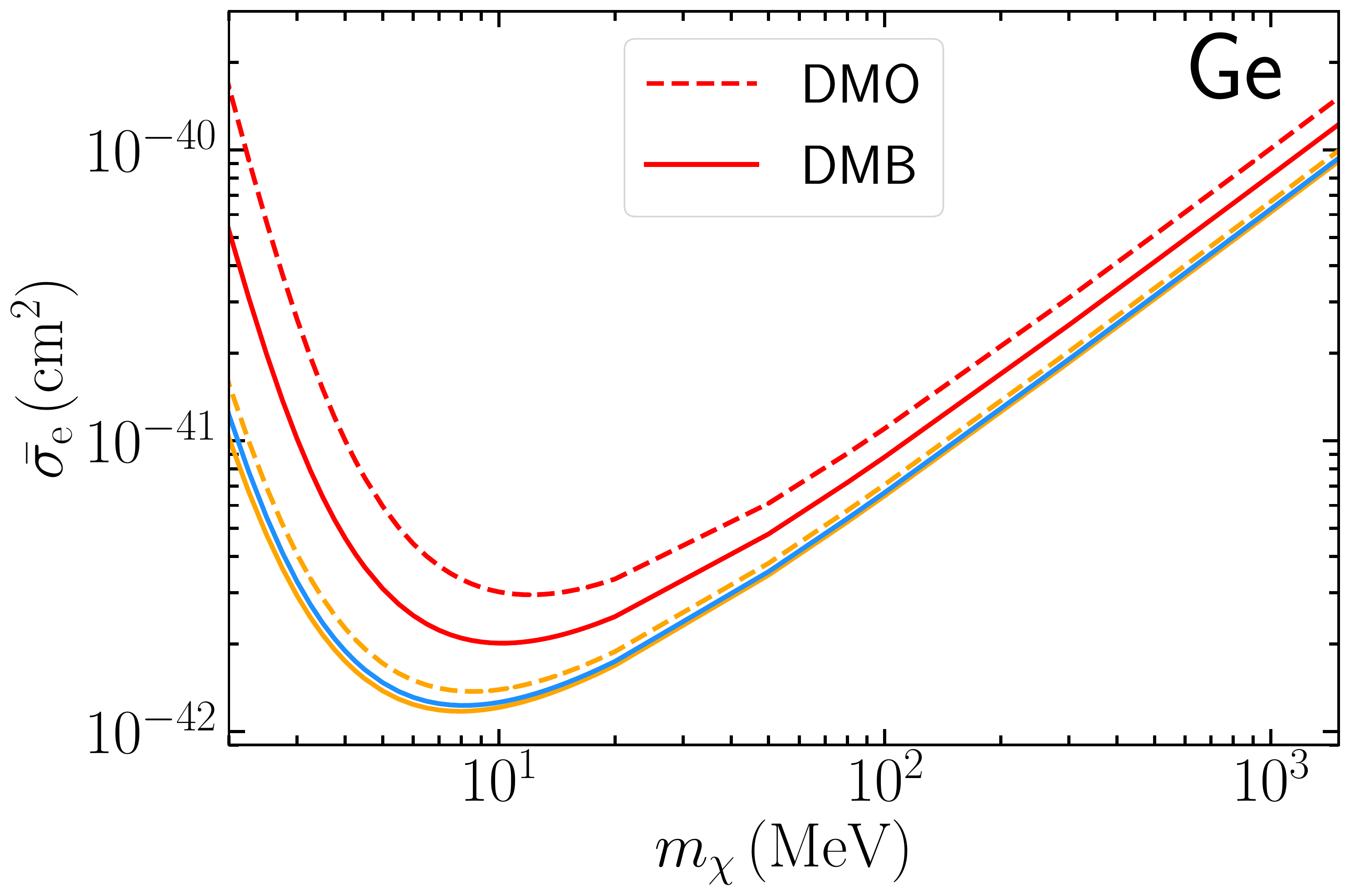}}
\subfloat[\label{sf:GeK1}]{\includegraphics[scale=0.18]{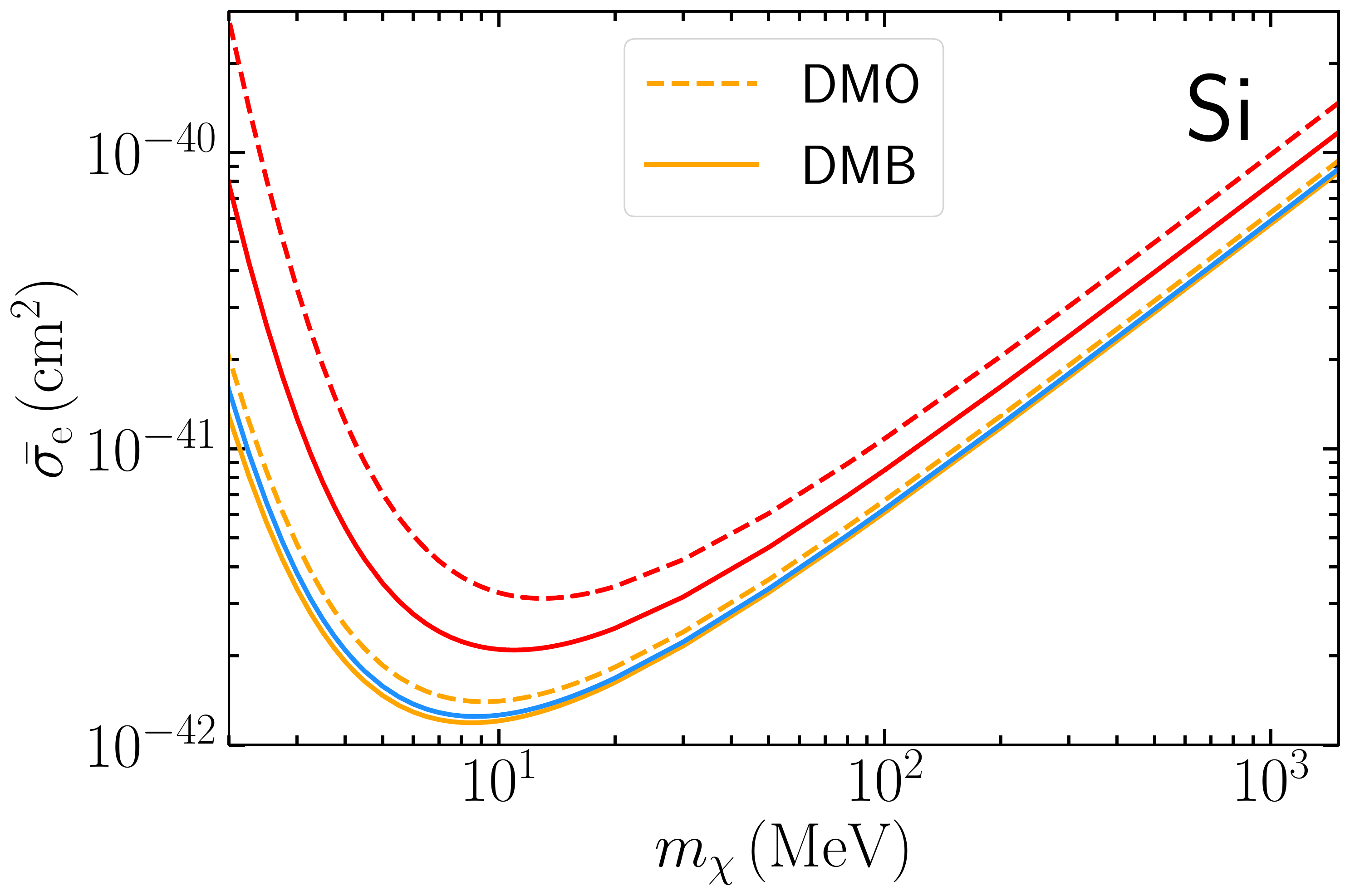}}
\newline
\subfloat[\label{sf:XeKingfa}]{\includegraphics[scale=0.18]{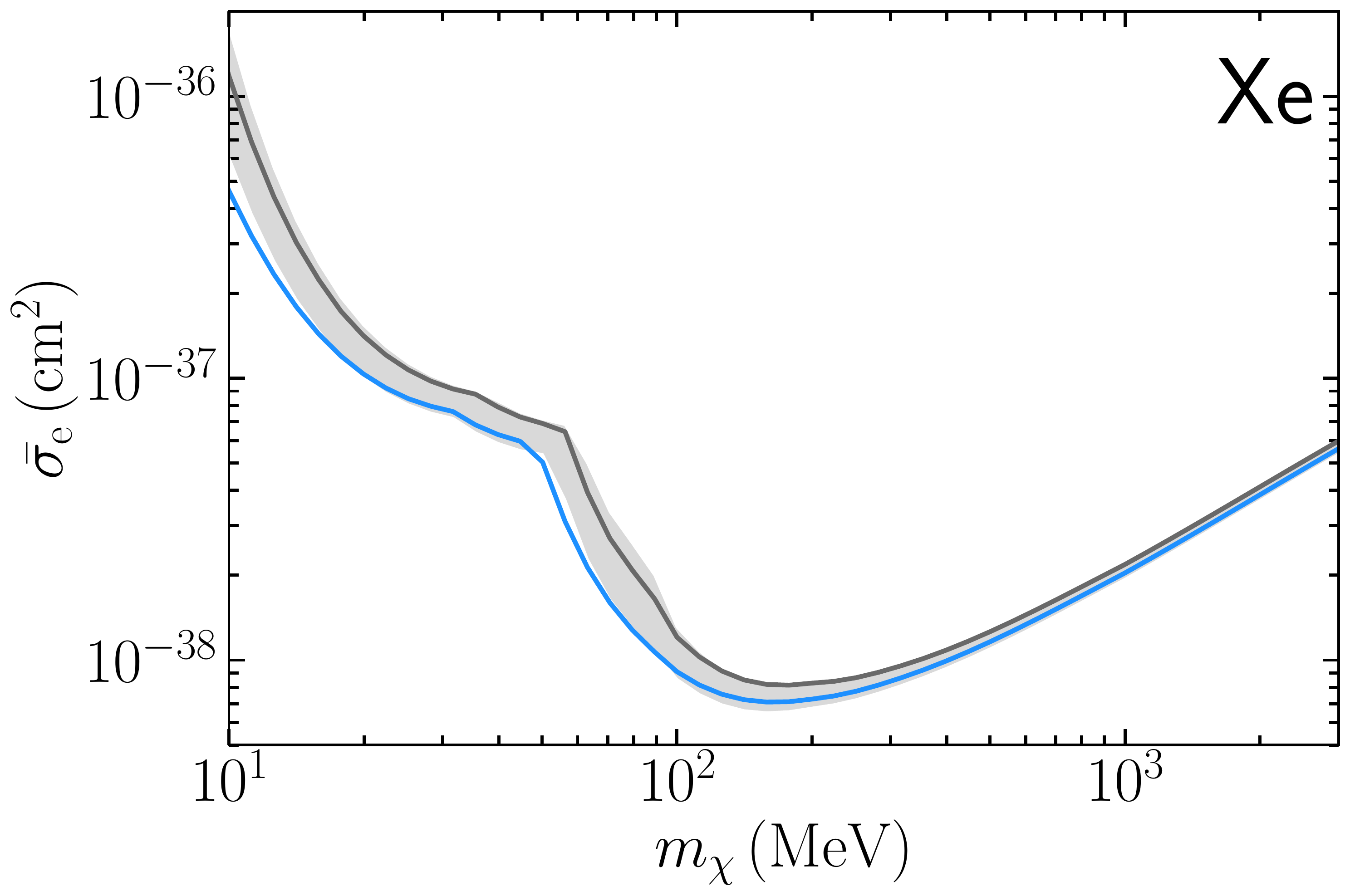}}
\subfloat[\label{sf:SiKingfa}]{\includegraphics[scale=0.18]{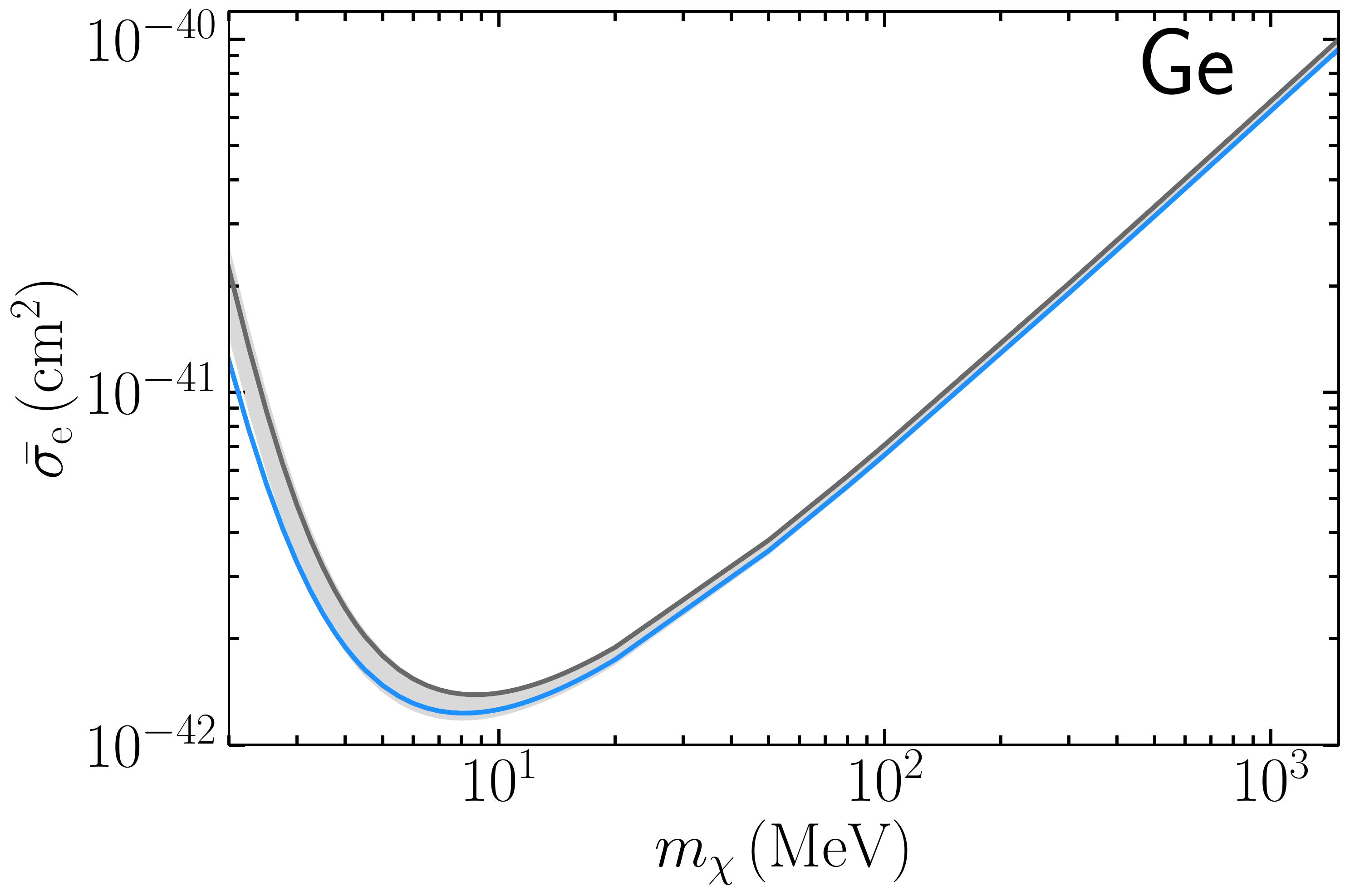}}
\subfloat[\label{sf:GeKingfa}]{\includegraphics[scale=0.18]{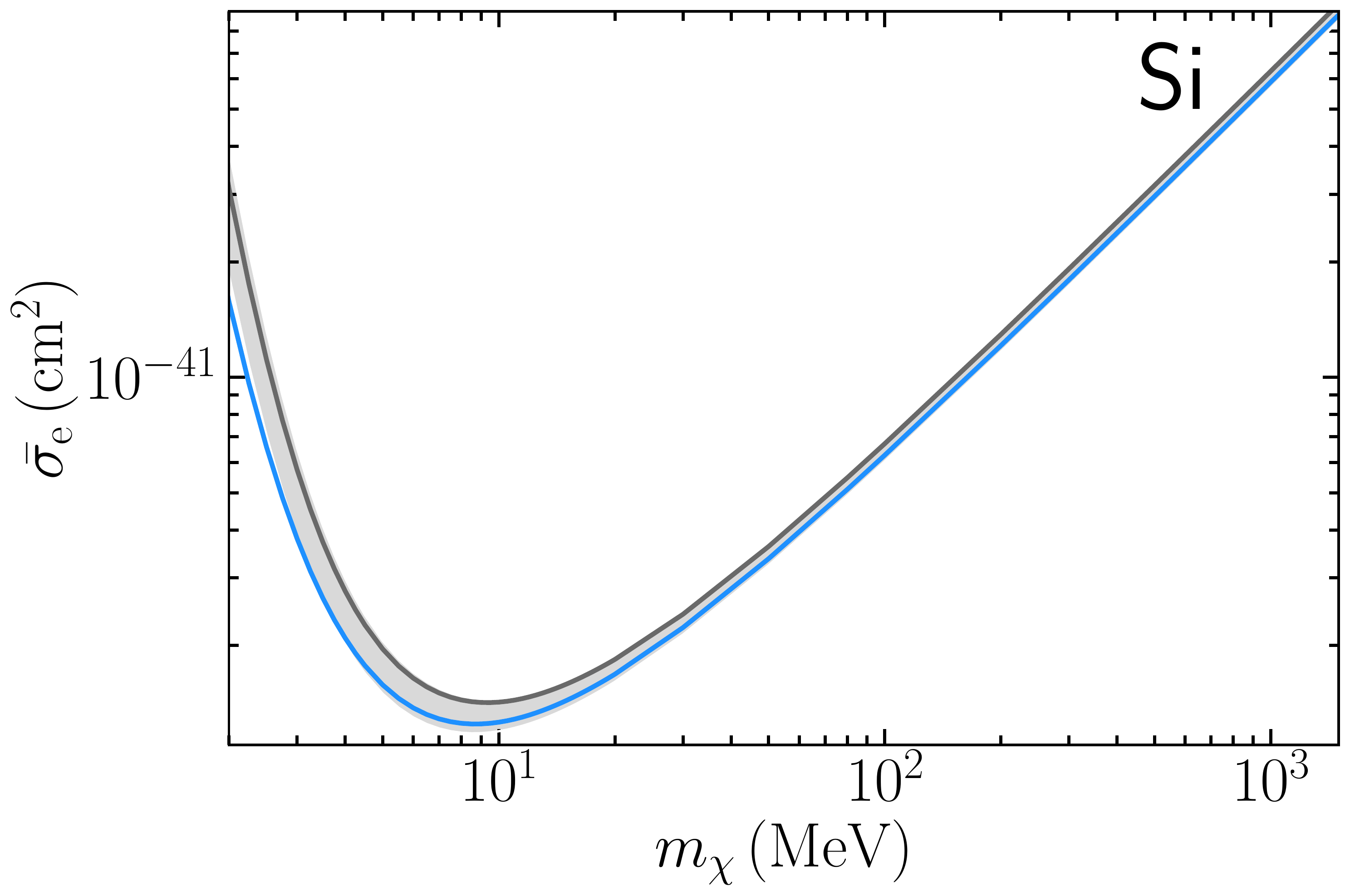}}
\newline
\subfloat[\label{sf:XeK1eff}]{\includegraphics[scale=0.19]{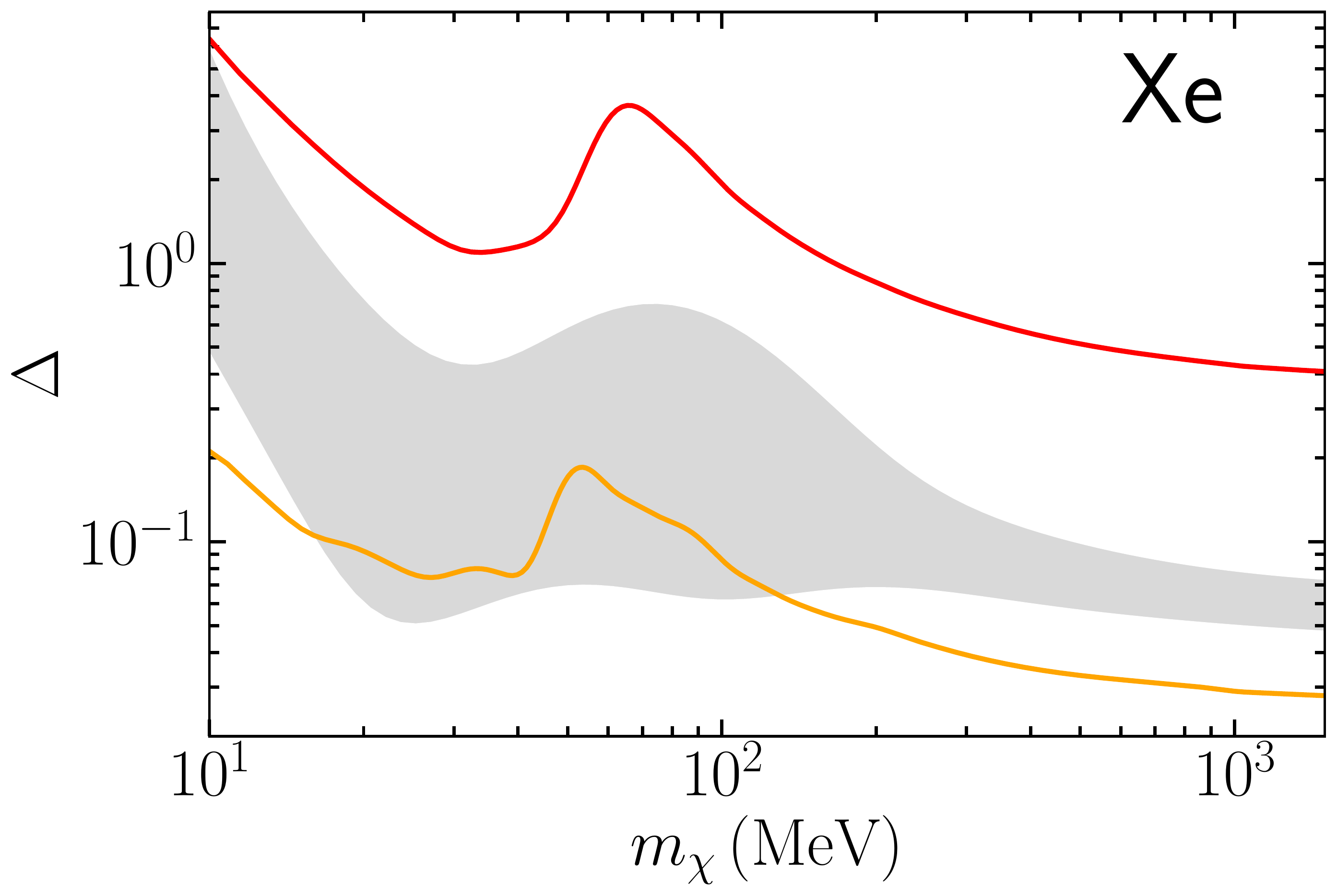}}
\subfloat[\label{sf:SiK1eff}]{\includegraphics[scale=0.19]{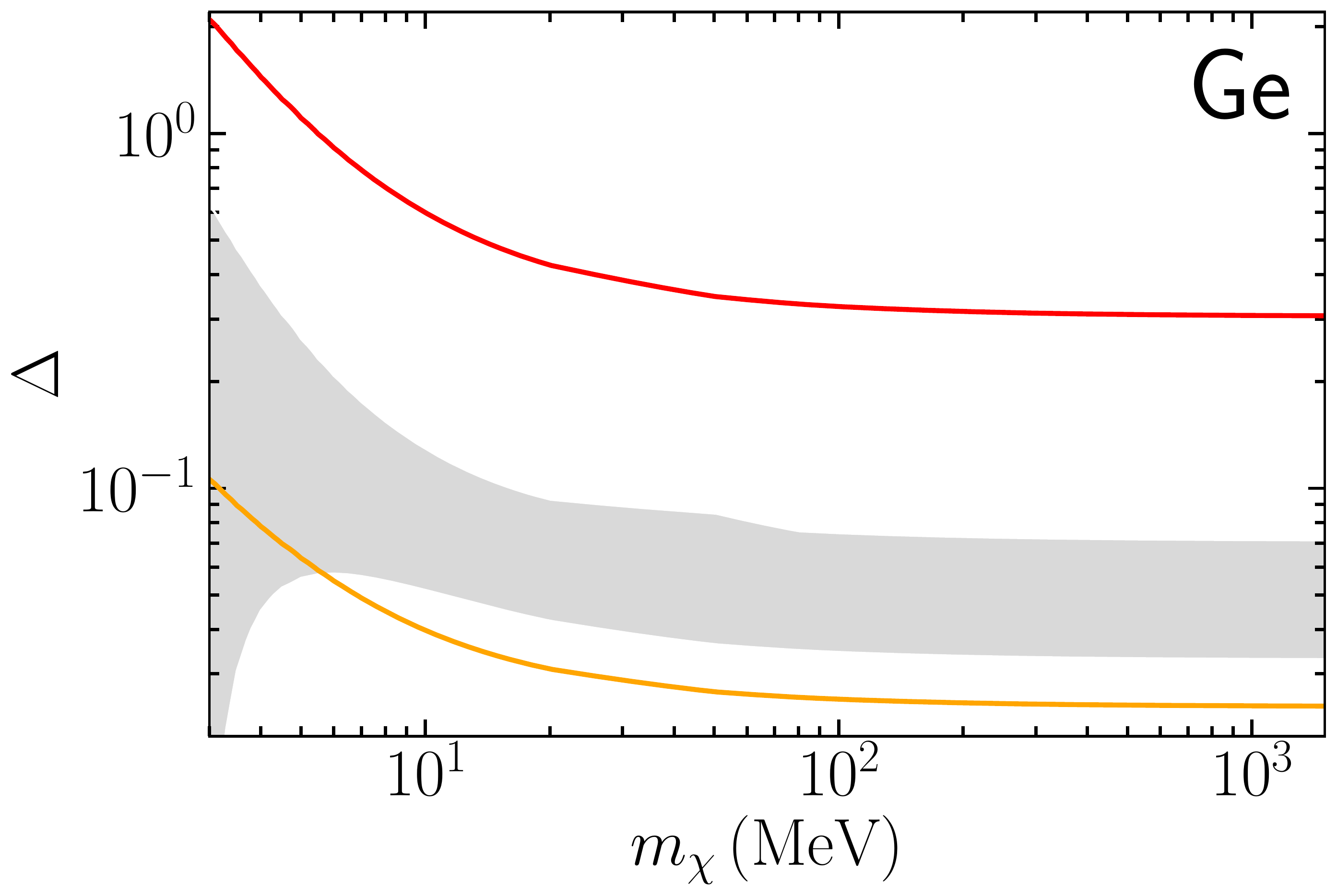}}
\subfloat[\label{sf:GeK1eff}]{\includegraphics[scale=0.19]{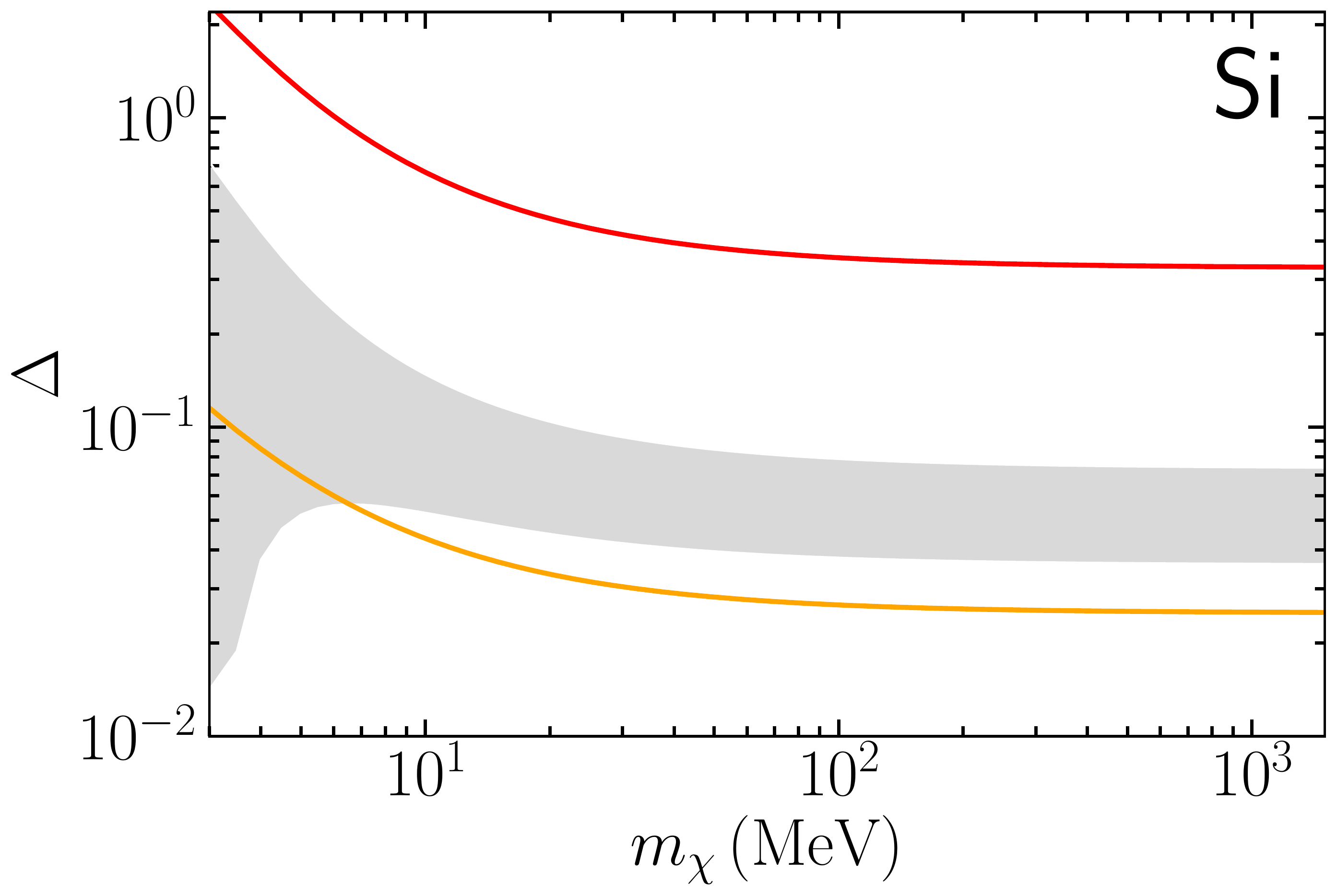}}
\caption{Variations in the exclusion bounds for the King distribution with $F_{\rm DM} =1$. In the upper panel the orange and red exclusion lines correspond to the APOSTLE and ARTEMIS simulations respectively with best-fit values from table \ref{tab:bestfit}. In the middle panel we show the exclusion bounds using King distribution with astrophysical parameters obtained from recent observations. The shaded bands  represent observational error at the $95\%$ confidence level. The solid light blue and grey lines correspond to the exclusion bounds with the fiducial parameters of the SHM and the King distribution respectively. In the lower panel the fractional changes relative to the fiducial choice (defined in equation \eqref{eq:sigma-effective}) for the APOSTLE and ARTEMIS simulations containing both DM and baryon (DMB) are shown by the solid orange and red lines respectively. Whereas the gray bands represent the deviation due to the recent astrophysical observations.}
\label{fig:king}
\end{center}
\end{figure*}
%
%
\begin{equation}
    f(\mathbf{v})=
    \begin{cases}
      \frac{1}{N}\left[\exp{\left(\frac{v_{\rm esc}^{2}-|\mathbf{v}|^{2}}{v_{0}^{2}}\right)-1}\right] &  |\mathbf{v}| \leq v_{\rm max} \\
      0 &  |\mathbf{v}| > v_{\rm max},
    \end{cases}
    \label{eq:king}
\end{equation}
where the symbols have their usual meaning\footnote{The normalisation constant $N$ has been computed numerically for all the non-standard velocity distributions considered in this paper.}. Equating $v_{\rm max}$ with the maximum velocities reported by the  cosmological simulations provides a reasonable estimate. The best fit values of the parameters for  the considered simulations are provided in table \ref{tab:bestfit}. The corresponding shift in the exclusion curves are shown  in the upper panel of figure \ref{fig:king} for contact interaction ($F_{\rm DM}=1$) between DM and electron. In the middle panel the solid grey lines correspond to exclusion limits for the King distribution with fiducial parameters. The grey shaded bands represent the uncertainties associated with the recent astrophysical observations of $v_0=233 \pm 6$ km/s and $v_{\rm max}=443^{+27}_{-30}$ km/s \footnote{Following reference \cite{Posti_2019}, the numerical value of $v_{\rm max}$ is obtained using the radius and mass of a typical milky-way like galaxies.} for the King distribution. We have also shown the fiducial SHM exclusion bounds by the light blue solid lines for reference. The grey bands in the lower panel represent the relative change from fiducial values for the King model, fitted with astrophysical parameters from recent observations. The orange and red lines represent the same deviation for best fitted APOSTLE and ARTEMIS simulations respectively. The variation for the other types of interactions are shown in appendix \ref{app:q&q2}.


\begin{table}[t]
\centering  
\bigskip
\begin{tabular}{|c|c|c|c|c|c|c|c|c|}
    \hline
\multirow{2}*{ Simulation} & $v_{\rm esc}$ &  King 				  & \multicolumn{2}{|c|}{ DPL} & \multicolumn{2}{|c|}{ Mao } & \multicolumn{2}{|c|}{ Tsallis } \\ \cline{3-9} 
			    & 		(km/s)  		  & \begin{tabular}[c]{@{}c@{}}$v_0$ \\ (km/s) \end{tabular} & \begin{tabular}[c]{@{}c@{}}$v_0$ \\ (km/s) \end{tabular} & $k$  & \begin{tabular}[c]{@{}c@{}}$v_0$ \\ (km/s) \end{tabular} & $p$ & \begin{tabular}[c]{@{}c@{}}$v_0$ \\ (km/s) \end{tabular} & $q$ \\ \cline{1-9} 
     
\setlength\arraycolsep{10pt} 

APOSTLE DMO  \cite{Sawala:2015cdf, Fattahi_2016, Bozorgnia:2016ogo} &  646    & 192.5 & 192.4   & 0.5 & 105.2 & 2.36 & 210.3   & 0.894 \\ \hline
APOSTLE DMB  \cite{Sawala:2015cdf, Fattahi_2016, Bozorgnia:2016ogo} &  646    & 223	 & 212.7 & 0.1 & 165   & 2.2 & 257 & 0.84 \\ 
\hline
ARTEMIS DMO  \cite{Font_2020, Poole-McKenzie:2020dbo} & 521.6  & 161.5 & 161.5 & 0.5 & 120   & 3.6 & 177.4 & 0.884  \\ \hline 
ARTEMIS DMB  \cite{Font_2020, Poole-McKenzie:2020dbo} &  521.6  & 184.6 & 184.3 & 0.67& 174.7 & 3.4 & 209.4 & 0.839 \\ \hline
\end{tabular}
\caption{Best fit values used to derive the exclusion limit for King, DPL, Mao, and Tsallis  model.}
\label{tab:bestfit}
\end{table}

The special feature observed around the DM mass of 50 MeV on the fiducial SHM curve in figures \ref{sf:XeK1} and \ref{sf:XeKingfa} is originating from the atomic structure of the Xenon atom. This is because around the aforesaid mass the maximum accessible DM energy crosses a threshold to ionize electrons from inner shell. This leads to an increment in the event rate and subsequently the bound becomes tighter. Further, as can be seen from figures \ref{sf:XeK1} and \ref{sf:XeKingfa}, compared to the fiducial values this special feature shift towards higher mass for a lower choice of $v_0$. Note that as $v_0$ decreases the DM population in the high velocity region decreases substantially. Due to this unavailability of high velocity DM the required DM mass shift towards higher mass to overcome a certain threshold. While for a semiconductor target, recoil electrons need to overcome only one energy barrier i.e. the energy gap between valance band and  conduction band. Hence we do not observe such features in figures \ref{sf:SiK1} and \ref{sf:GeK1}. The apparent spike in the lower left panel of figure \ref{sf:XeK1eff} is a manifestation of the offset between the two kinks of the SHM and the King distribution.

As indicated in table \ref{tab:bestfit}, the best fit values of $v_0$ for DM only simulations is smaller than the fiducial values.  Due to this in the high velocity region, less number of DM particles are available to interact with the electrons. This leads to weaker bounds  in DM electron cross section\footnote{For APOSTLE simulations $v_{\rm esc}$ is larger than the fiducial value however as indicated earlier effect of  $v_0$ will be more pronounced.}. This has been shown by the dashed lines in figure \ref{fig:king}. The best fit values of $v_0$ with the DMB simulations lie close to fiducial value, implying exclusion limits close to the fiducial exclusion curve. This has been displayed by the orange and red solid lines for APOSTLE and ARTEMIS respectively.

We note that the deviation for King's model lies between $2\%$ to $20\%$ for Xe, Ge and Si target material for the best fit values of the APOSTLE DMB simulation. In contrast to that the best fit values of APOSTLE DMO and ARTEMIS DMO simulations seem to produce larger deviation from the SHM. The corresponding deviation induced by the recent astrophysical observations lie between $5\%$ to $500\%$ for Xe  detector and $1\%$ to $70\%$ for both Ge and Si semiconductor detectors.

\subsection{Double Power Law}
\label{subsec:DPL}

The Double Power Law (DPL) distribution is an isotropic distribution of DM which has been obtained empirically \cite{Lisanti:2010qx}. The DPL very well describes the empirical matter distributions such as NFW, Hernquist, etc \cite{Lisanti:2010qx}. The matter distribution of a generic empirical double power law has the form
\begin{equation}
   \rho(r) = \frac{\rho_{s}}{(\frac{r}{r_{s}})^{\alpha}(1+\frac{r}{r_{s}})^{\gamma-\alpha}},
\label{eq:dpden}
  \end{equation}
where $r_s$ is the characteristic radius, $\rho_s$ is the characteristic density, $\alpha$ and $\gamma$ determines the slope of the density profile at small and large radii respectively. For $(\alpha, \gamma)$ = (1, 3) this reduces to the NFW profile \cite{Navarro:1996gj} while for $(\alpha, \gamma)$ = (1, 4) this reproduces Hernquist profile \cite{Hernquist:1990be}. The DPL velocity distribution is expressed by the form
\begin{equation}
    f(\mathbf{v})=
    \begin{cases}
      \frac{1}{N}\left[\exp{\left(\frac{v_{\rm esc}^{2}-|\mathbf{v}|^{2}}{kv_{0}^{2}}\right)}-1\right]^{k} &  |\mathbf{v}| \leq v_{\rm esc} \\
      0 &  |\mathbf{v}| > v_{\rm esc},
    \end{cases}
\label{eq:DPLfv}
  \end{equation}  
where the symbols have their usual meaning.  
\begin{figure*}[t!]
\begin{center}
\subfloat[\label{sf:XeDPL1}]{\includegraphics[scale=0.18]{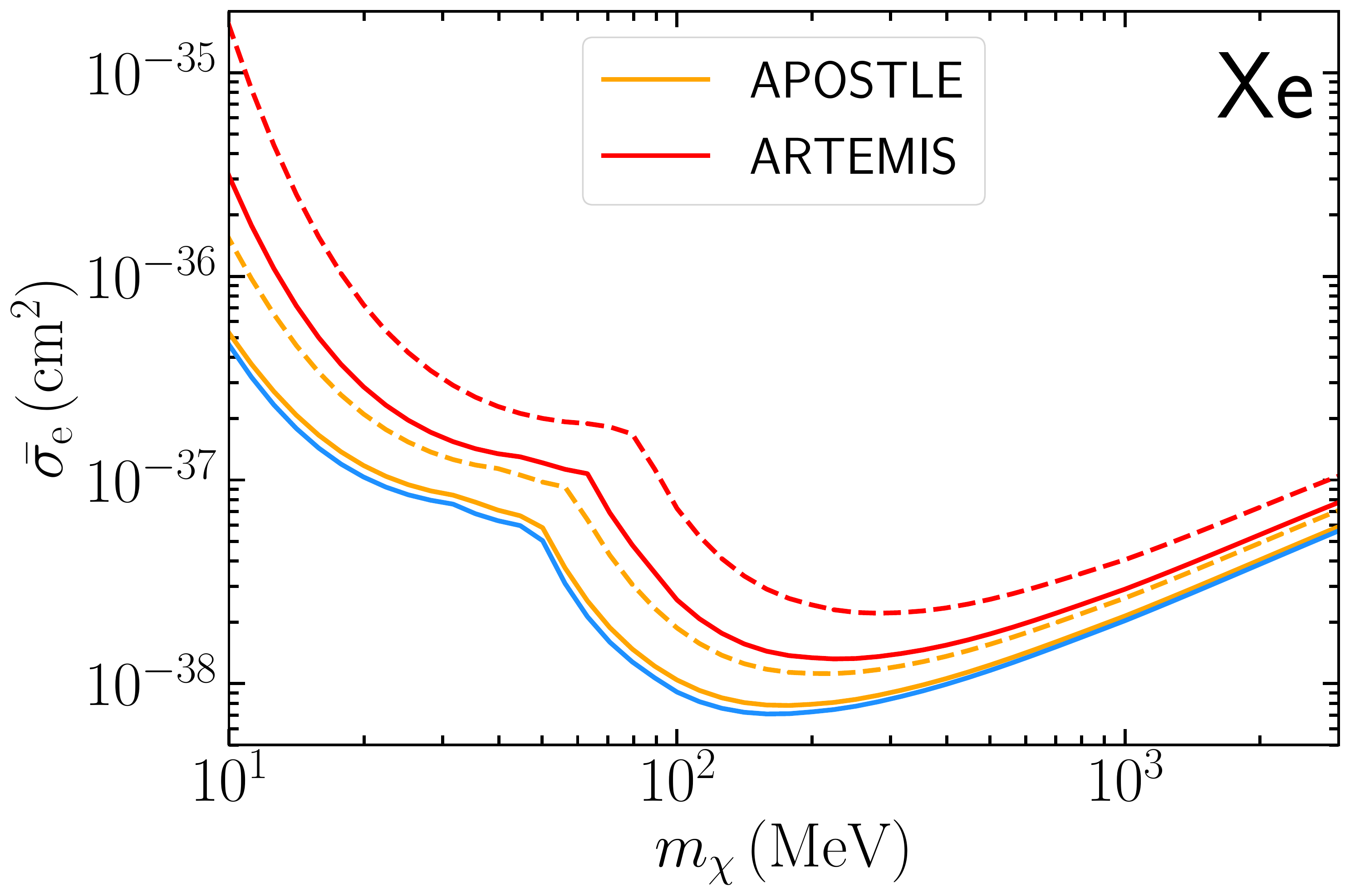}}
\subfloat[\label{sf:SiDPL1}]{\includegraphics[scale=0.18]{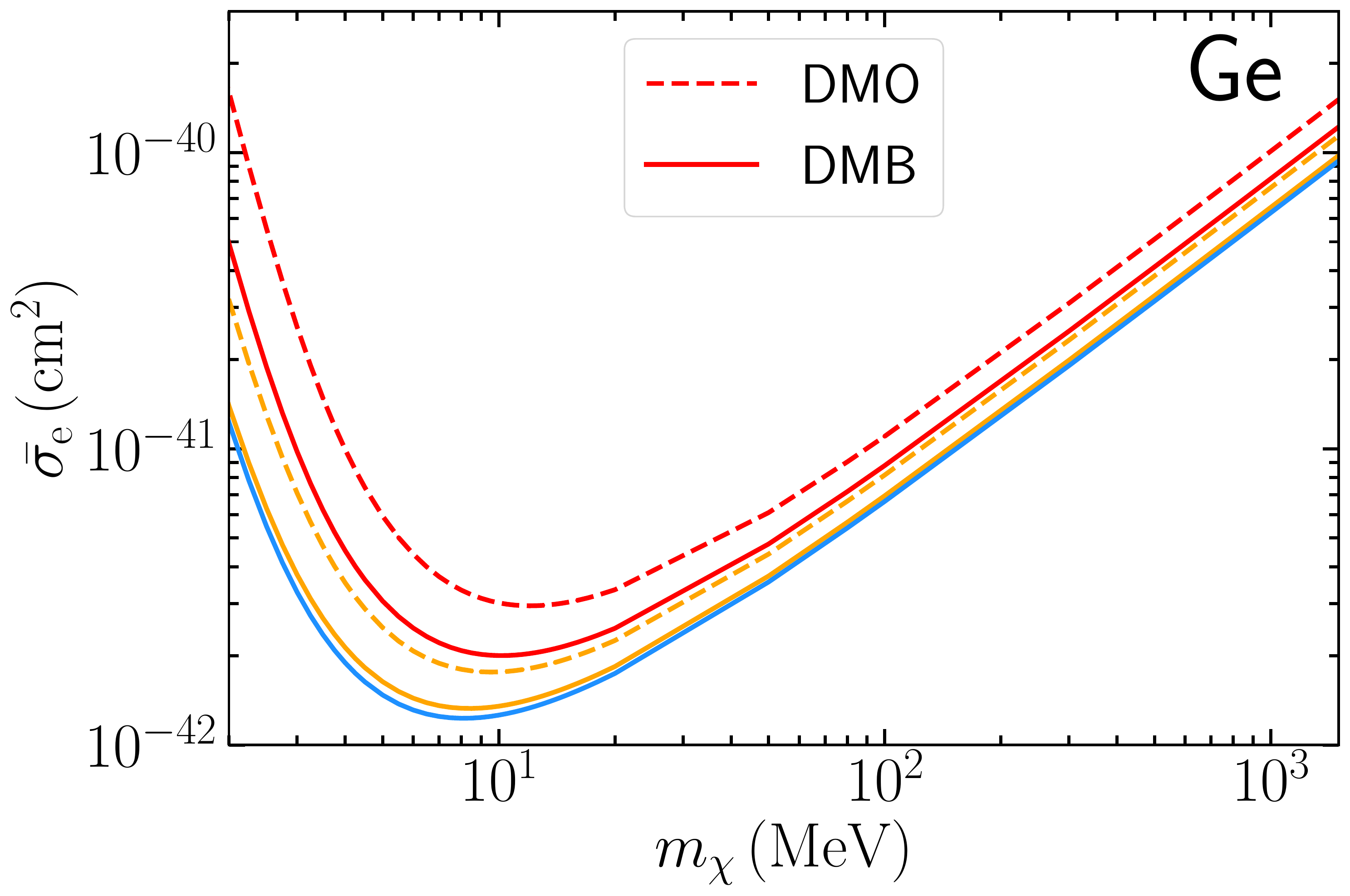}}
\subfloat[\label{sf:GeDPL1}]{\includegraphics[scale=0.18]{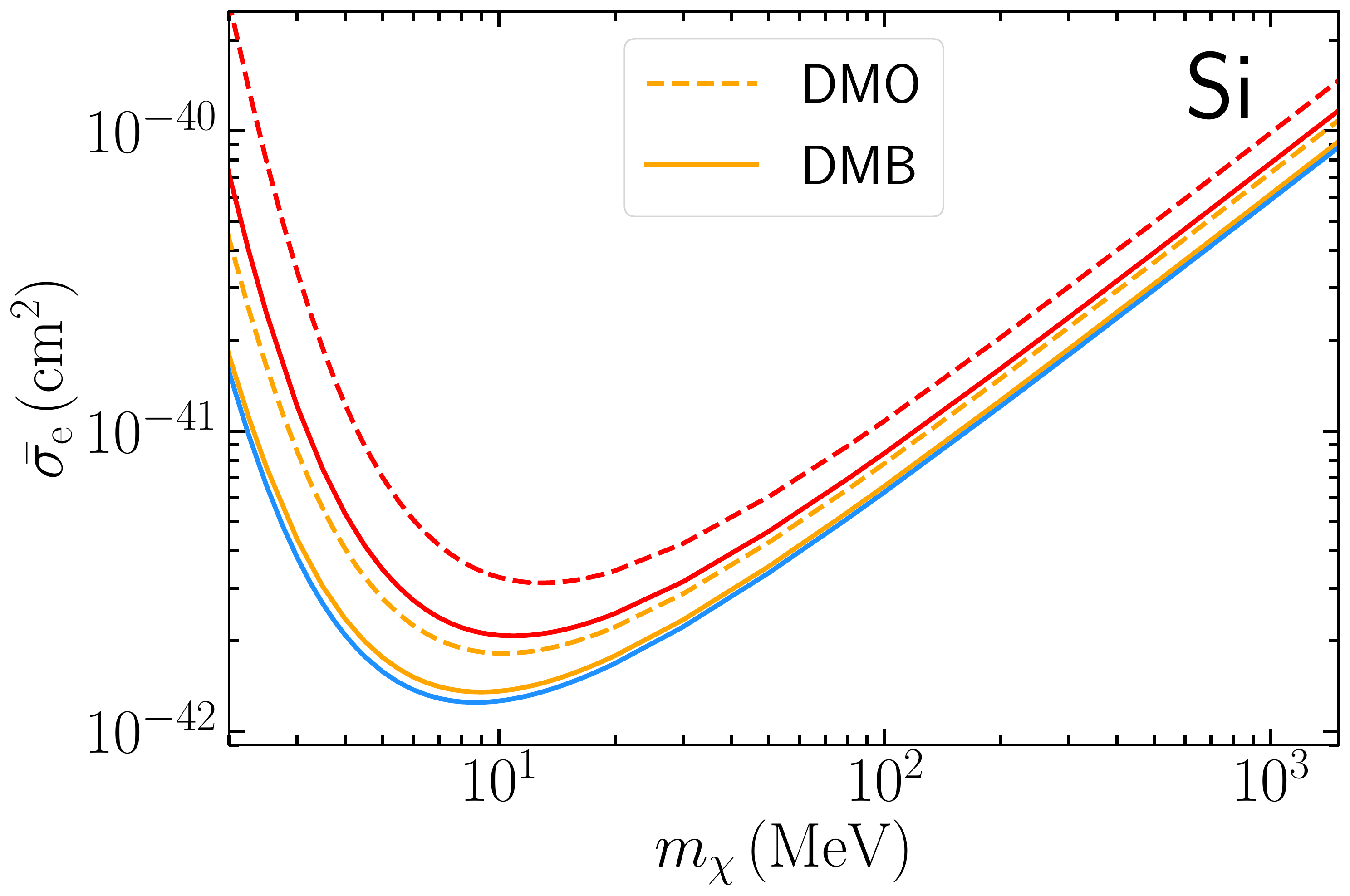}}
\newline
\subfloat[\label{sf:XeDPLfa}]{\includegraphics[scale=0.18]{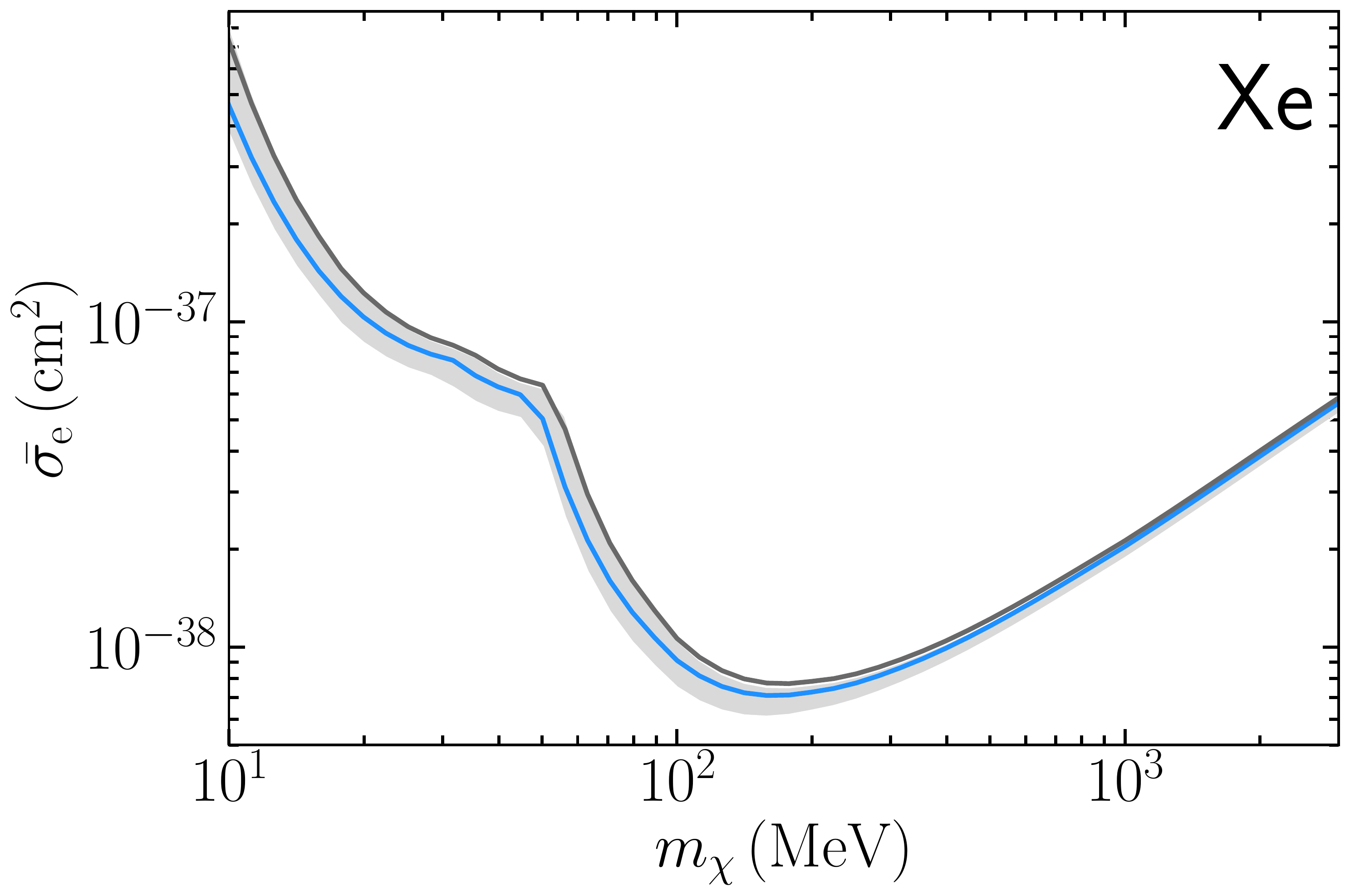}}
\subfloat[\label{sf:SiDPLfa}]{\includegraphics[scale=0.18]{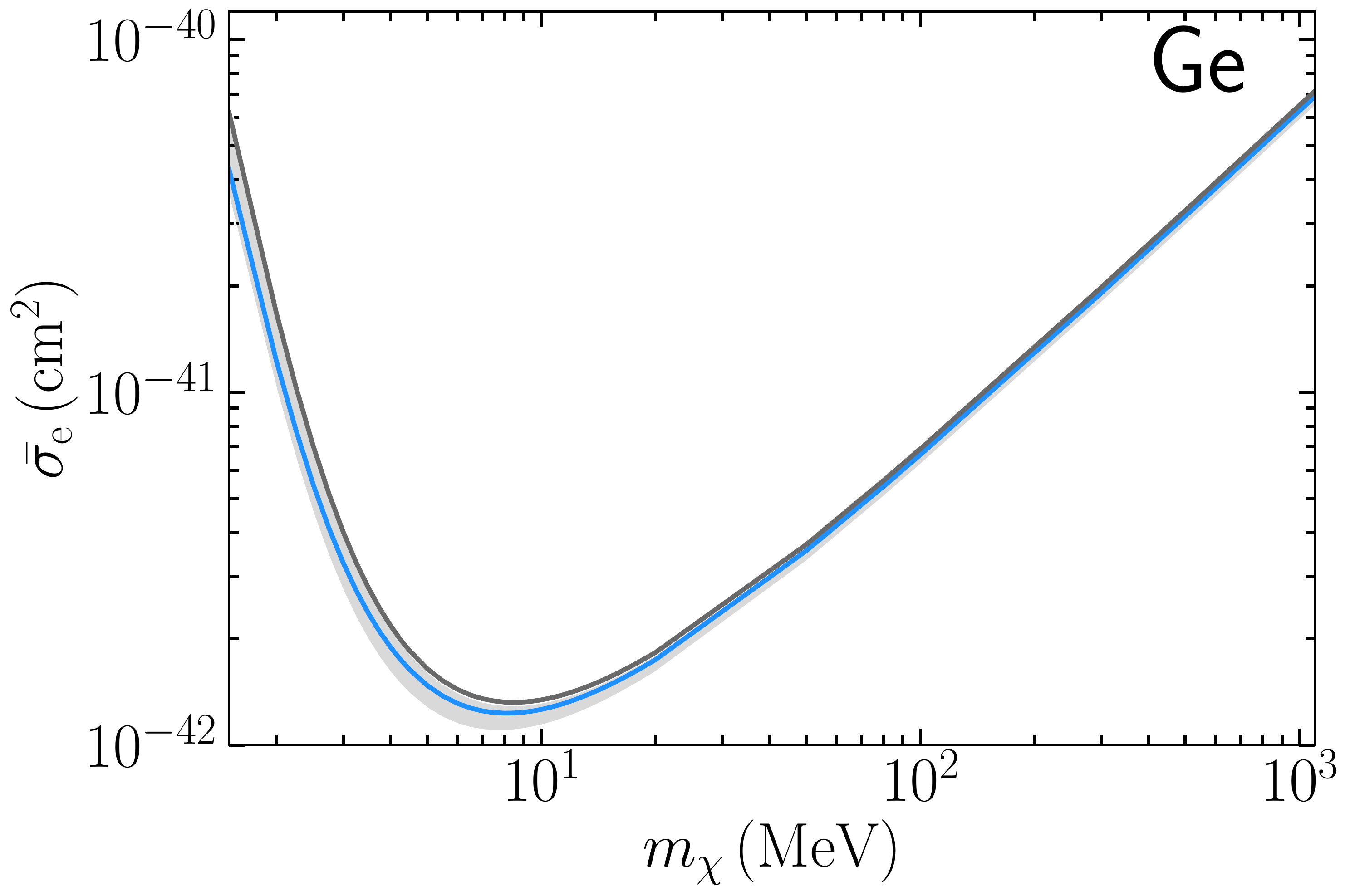}}
\subfloat[\label{sf:GeDPLtfa}]{\includegraphics[scale=0.18]{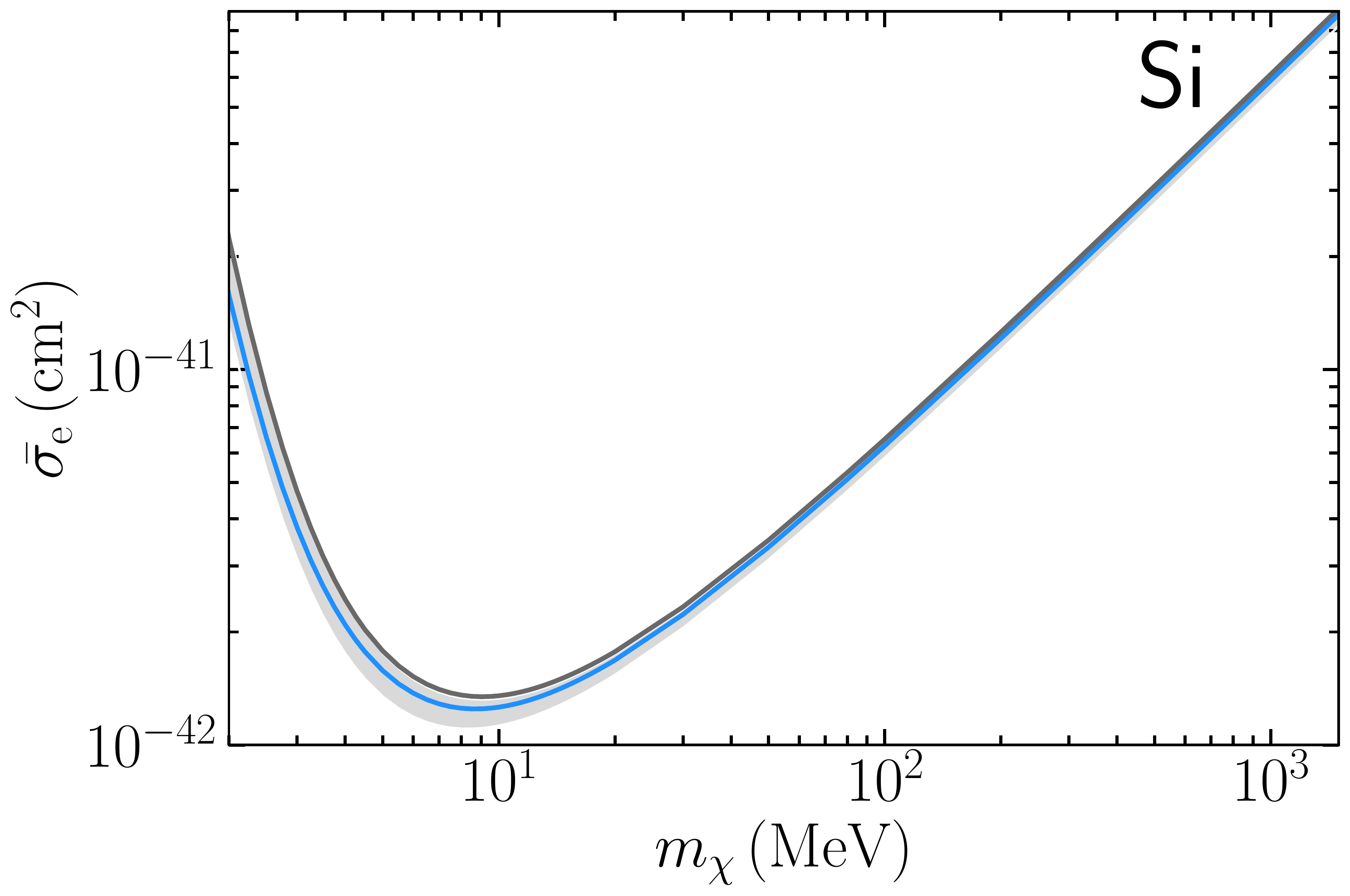}}
\newline
\subfloat[\label{sf:XeDPL1eff}]{\includegraphics[scale=0.19]{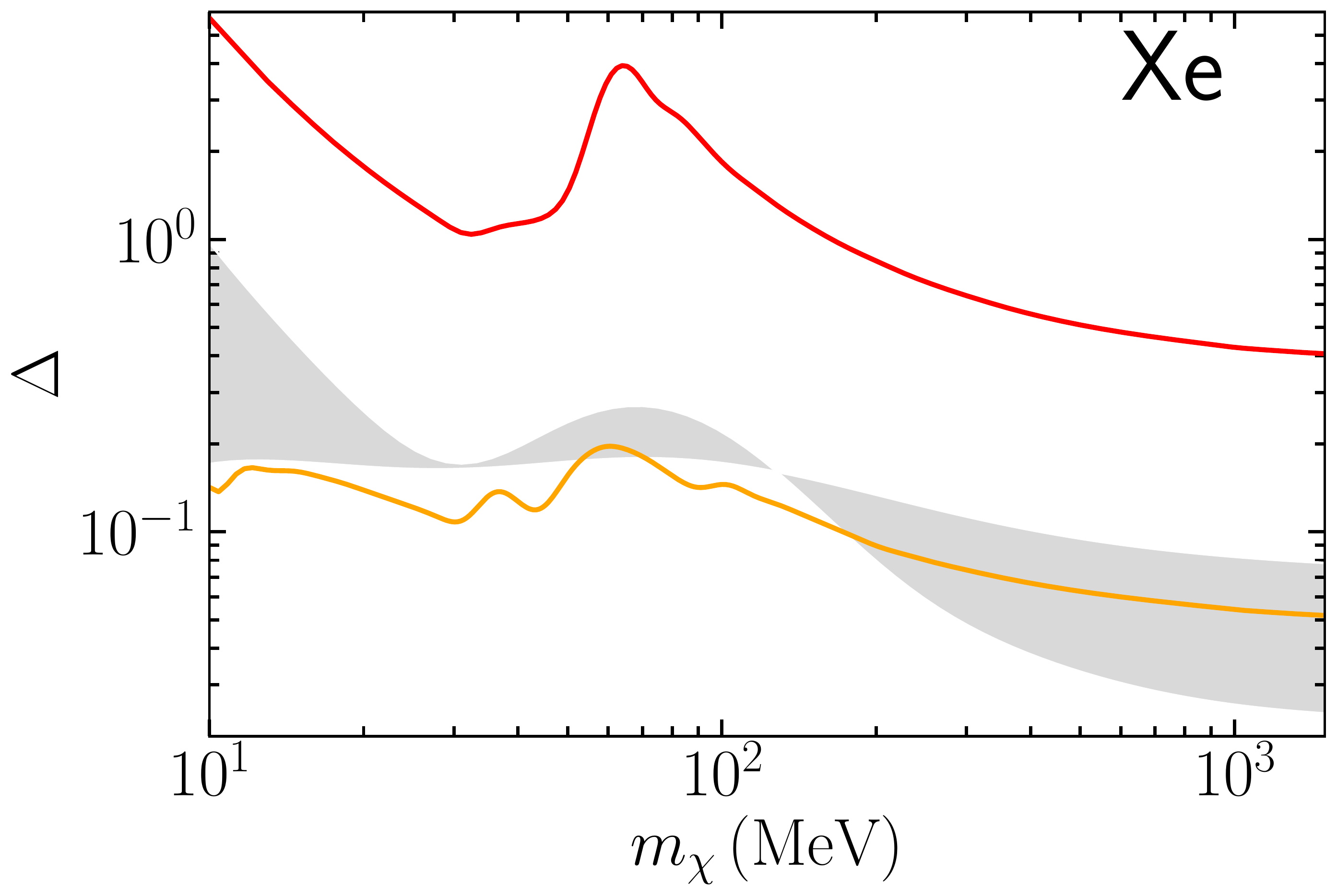}}
\subfloat[\label{sf:SiDPL1eff}]{\includegraphics[scale=0.19]{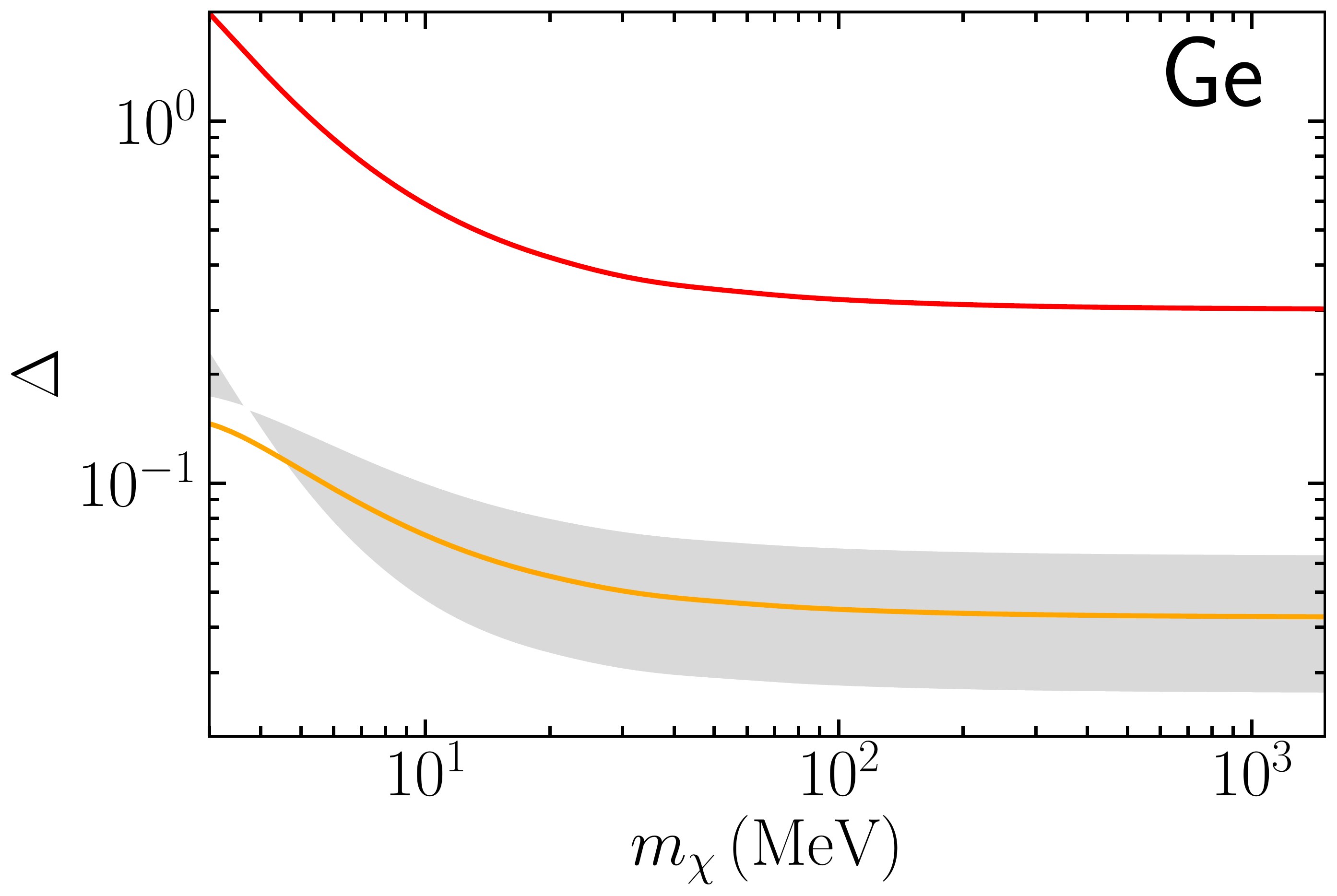}}
\subfloat[\label{sf:GeDPL1eff}]{\includegraphics[scale=0.19]{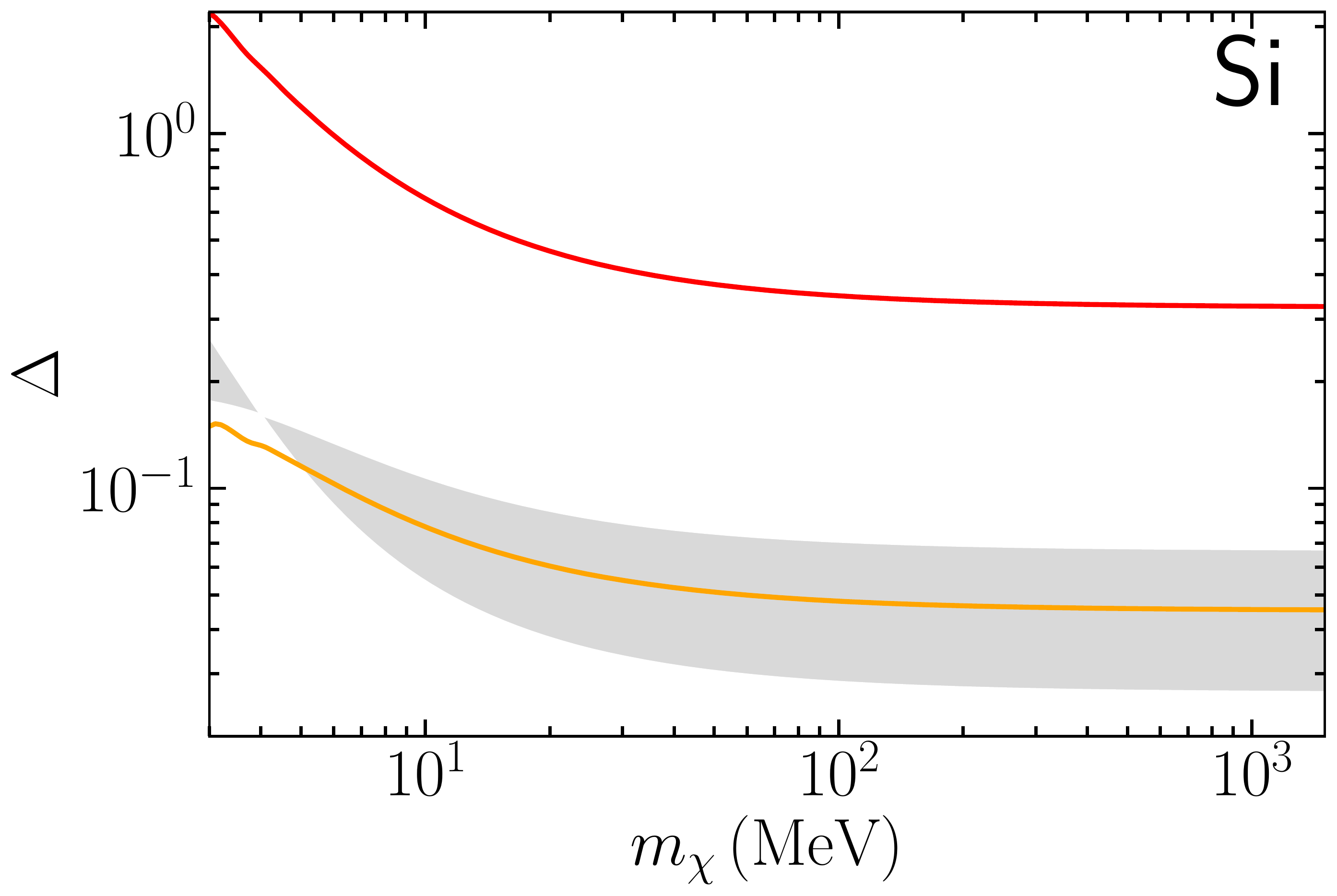}}
\caption{Variations in the exclusion bounds for the Double Power Law with $F_{\rm DM} =1$.  In both the middle and lower panels $k=1.5$ is represented by the gray region. The other relevant details are same as of figure \ref{fig:king}.}
\label{fig:DPL}
\end{center}
\end{figure*}
Cosmological simulation which take into account the quasi-static equilibrium nature of the virialised objects \cite {LyndenBell:1966bi} and it's formation history attributed from  hierarchical merging, smooth accretion and violent relaxations \cite{Wang:2008un} favour such distribution \cite{Lisanti:2010qx}. Unlike the SHM, the velocity distribution in DPL smoothly goes to zero at the escape velocity. Thus it differs in the high velocity tail region from  the SHM and predicts lower number of DM particles near the tail of the velocity distribution.
For $ k \rightarrow 0 $ it reduces to the SHM and for $k=1$ it tends to the King distribution. The best fit parameters from   cosmological simulations, that have been used in this work are given in table \ref{tab:bestfit}. We note that our best fit values is matched with \cite{Lisanti:2010qx}.

Note that the DPL distribution can be viewed as a generalization of the King model, described in section \ref{subsec:king}. Therefore the impact on electron recoil event rates here would be similar to  what has been discussed in section \ref{subsec:king}. The difference here is in the power index $k$.  Any change in the numerical value of $k$ would proportionally change the event rate and subsequently the direct detection limits. This can also be understood by comparing the relative fractional change in cross section depicted in lower panel of the figures \ref{fig:king} and \ref{fig:DPL}. Like figure \ref{fig:king}, in the upper panel of figure \ref{fig:DPL} the orange and red coloured lines correspond to the APOSTLE and ARTEMIS simulations respectively. Whereas in the middle panel, the solid grey lines correspond to the exclusion bounds for a representative $k=1.5$ with the fiducial parameters. The shaded bands represent the uncertainties associated with the recent astrophysical observations of $v_0=233 \pm 6$ km/s and $v_{\rm esc}=528^{+24}_{-25}$ km/s for $k=1.5$. We have also shown the fiducial SHM exclusion bounds by the light blue solid lines for reference.

We find that for APOSTLE DMB simulation with DPL distribution, the deviation obtained ranges between $ 5\%$ to $20\%$ for Xe and $ 4\%$ to $ 14\%$ for both Ge and Si semiconductor detectors. This deviation is fairly flat in the DM mass range of interest. The corresponding deviation induced by the recent astrophysical observation ranges between $2\%$ to $100\%$ for Xe  detector, between $1\%$ to $21\%$ for Ge and $1\%$ to $26\%$ for Si semiconductor detectors.

\subsection{Tsallis}
\label{subsec:Tsallis}

%
\begin{figure*}[t]
\begin{center}
\subfloat[\label{sf:XeTsa1}]{\includegraphics[scale=0.18]{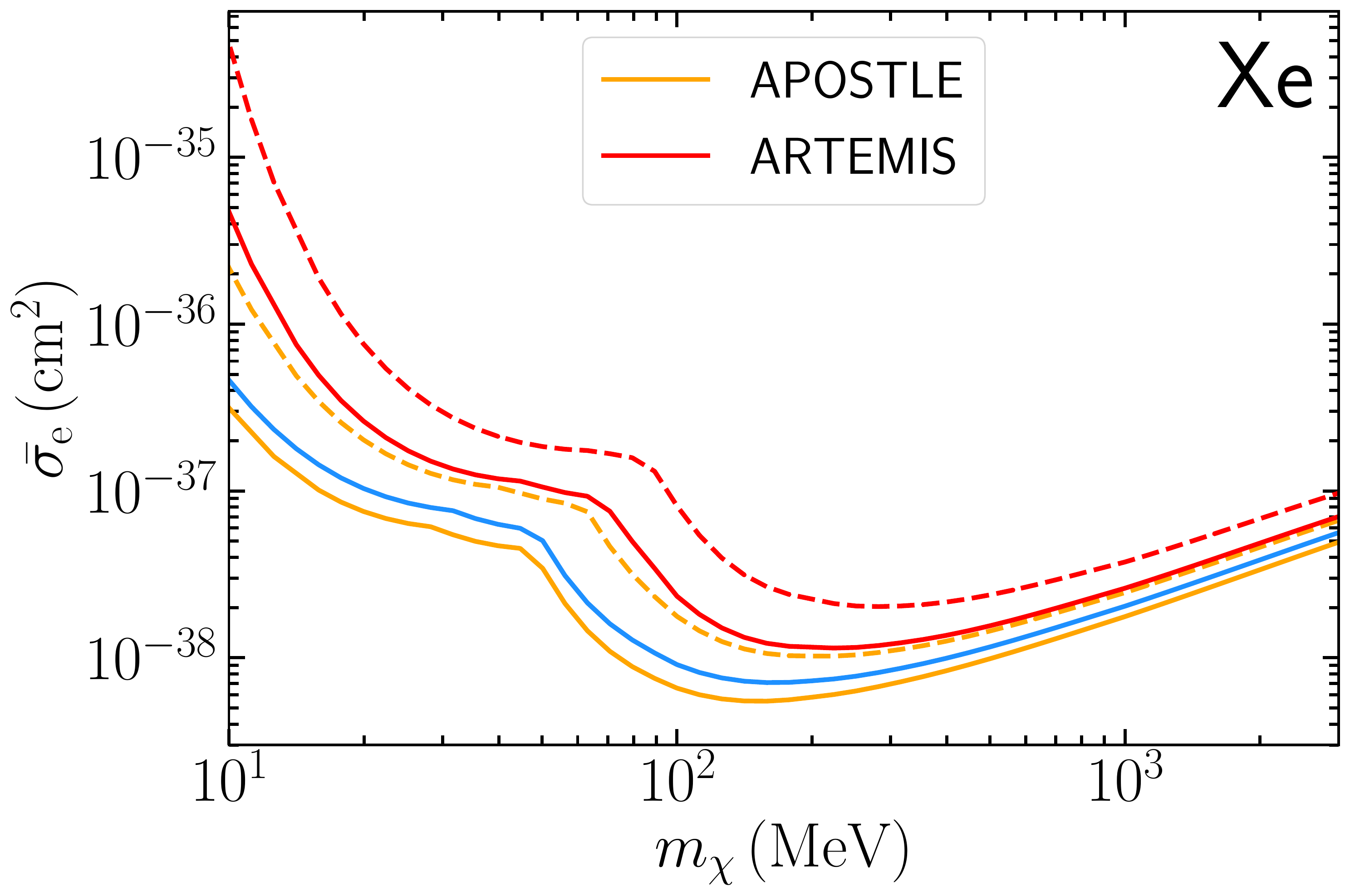}}
\subfloat[\label{sf:SiTsa1}]{\includegraphics[scale=0.18]{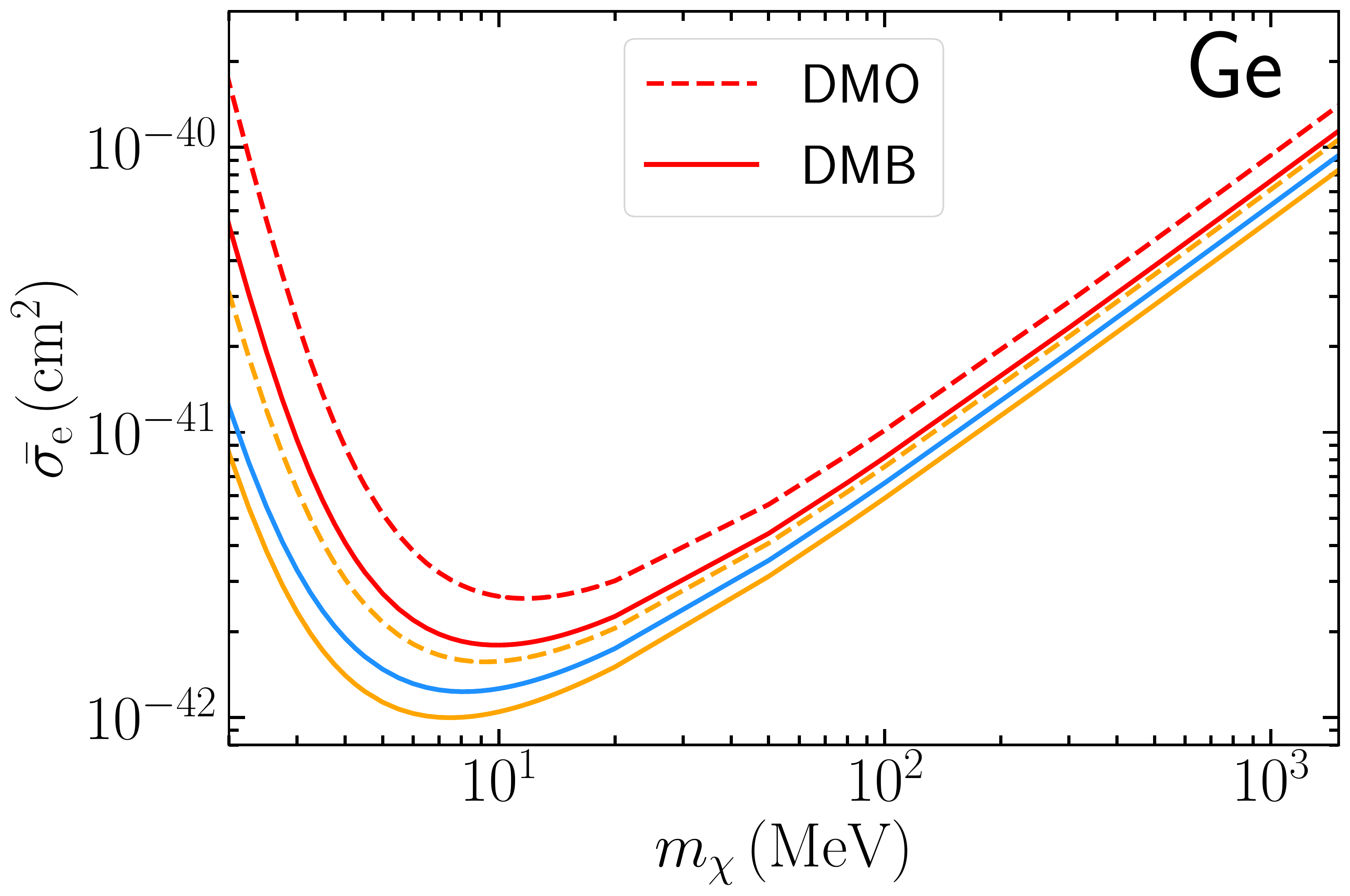}}
\subfloat[\label{sf:GeTsa1}]{\includegraphics[scale=0.18]{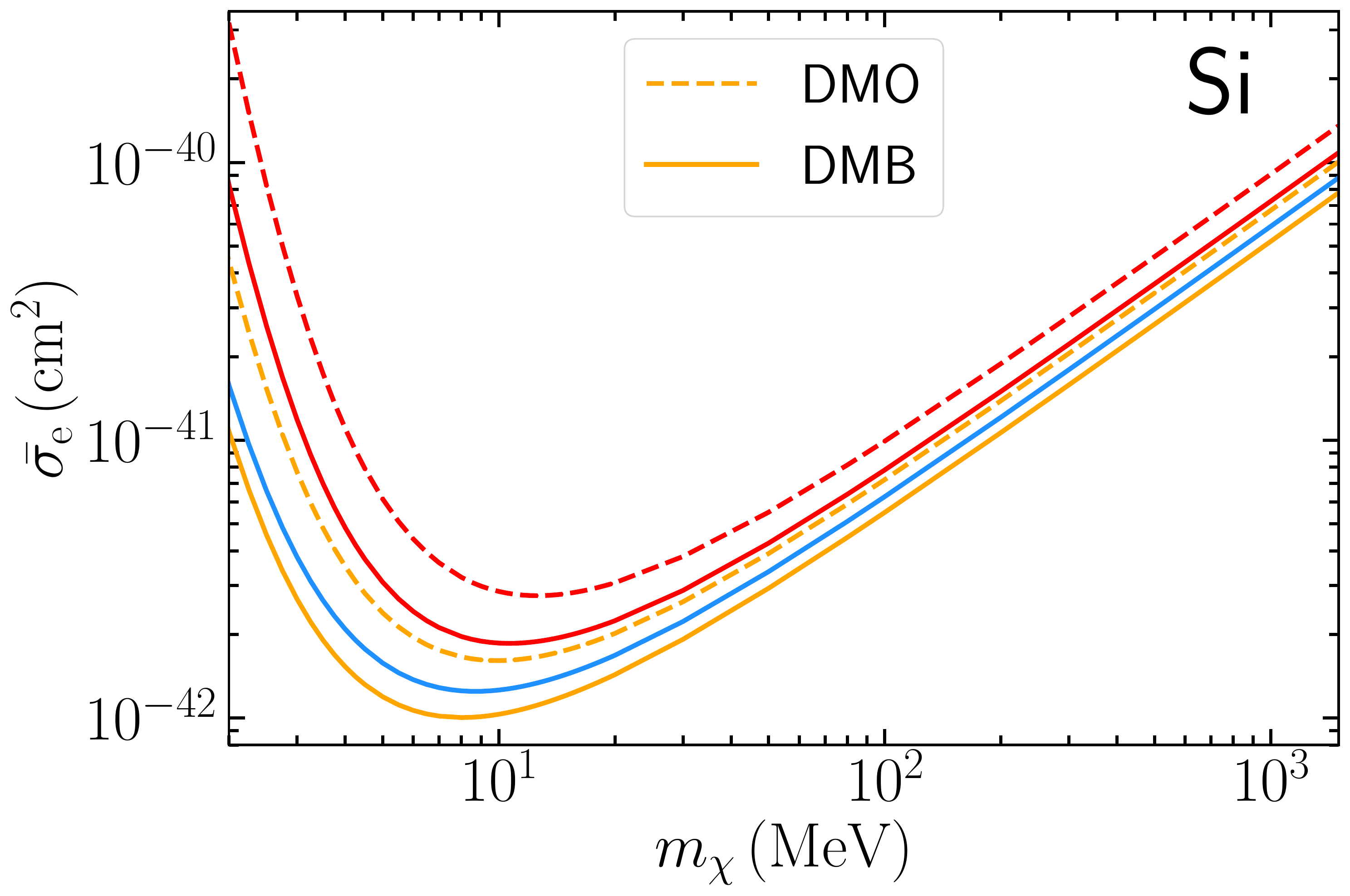}}
\newline
\subfloat[\label{sf:XeTslfa}]{\includegraphics[scale=0.18]{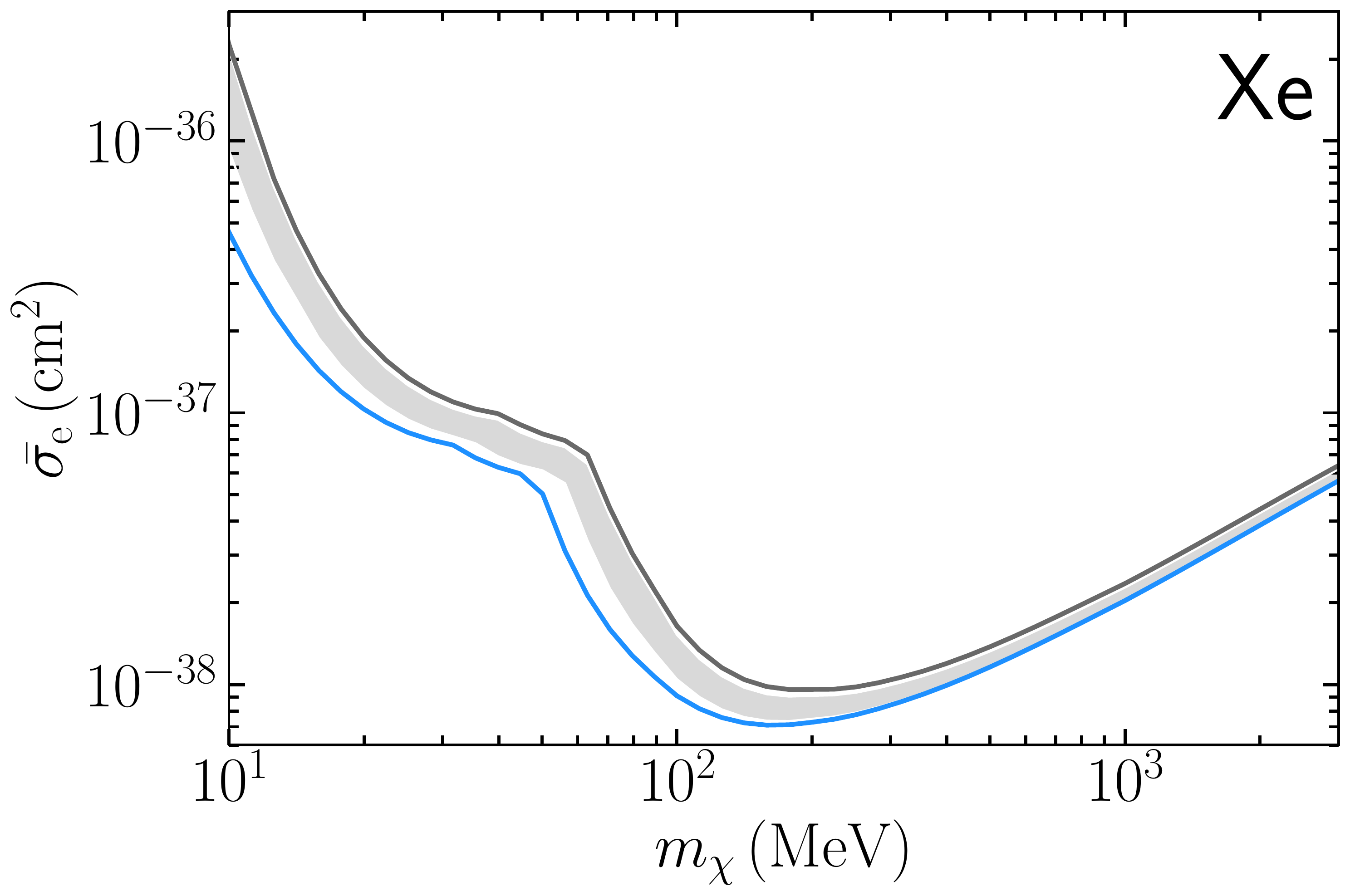}}
\subfloat[\label{sf:SiTslfa}]{\includegraphics[scale=0.18]{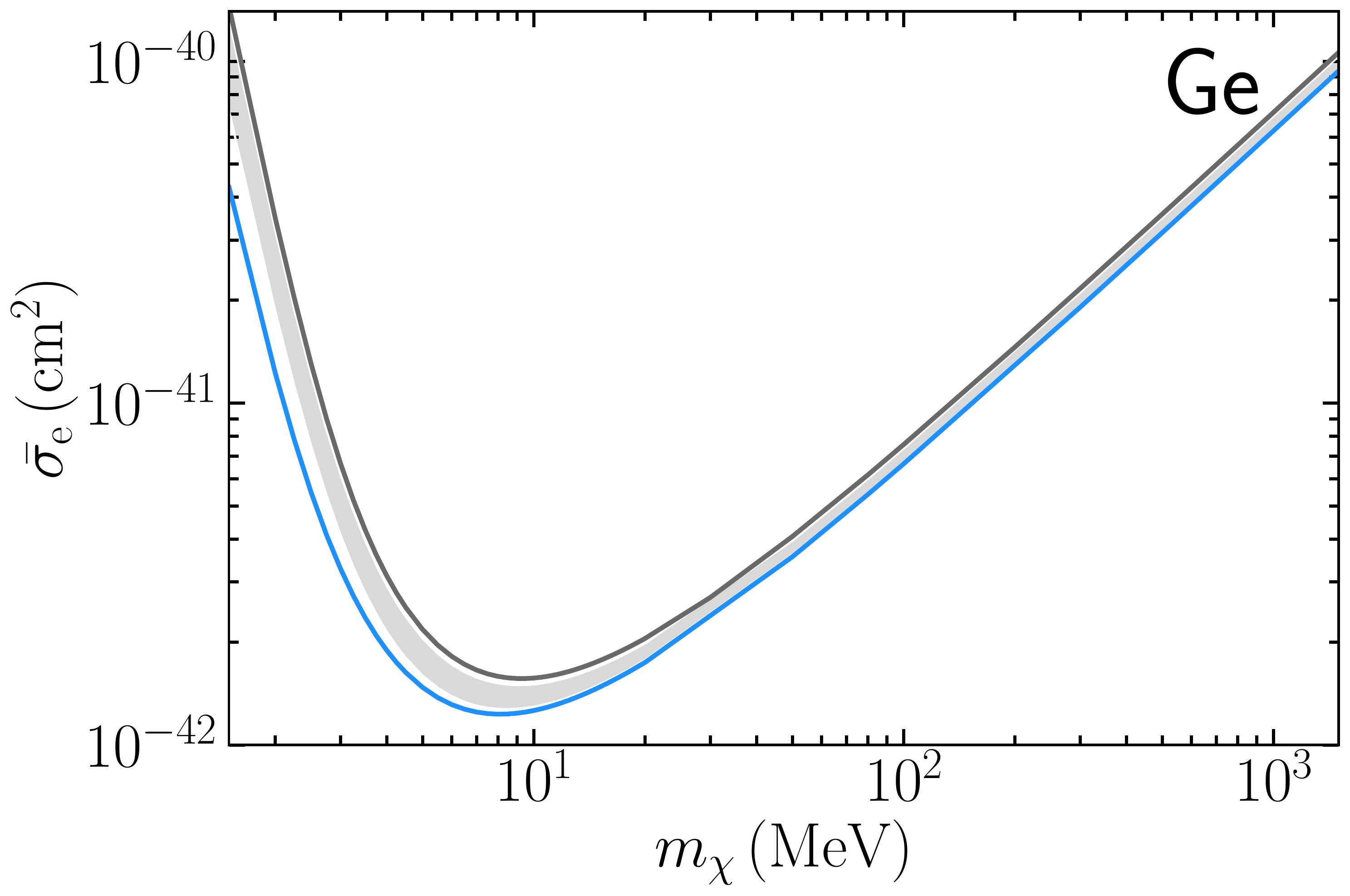}}
\subfloat[\label{sf:GeTslfa}]{\includegraphics[scale=0.18]{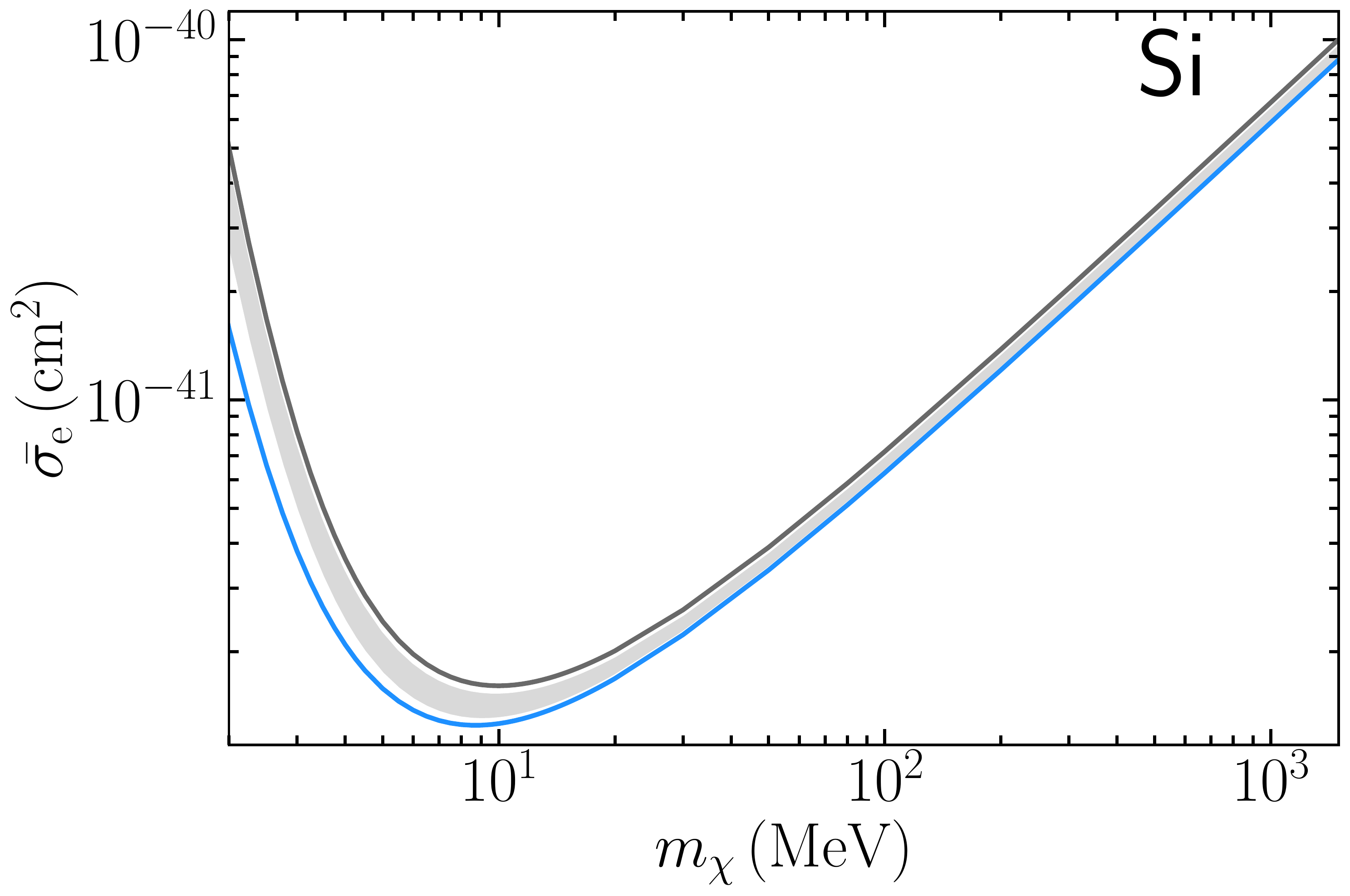}}
\newline
\subfloat[\label{sf:XeTsaeff}]{\includegraphics[scale=0.20]{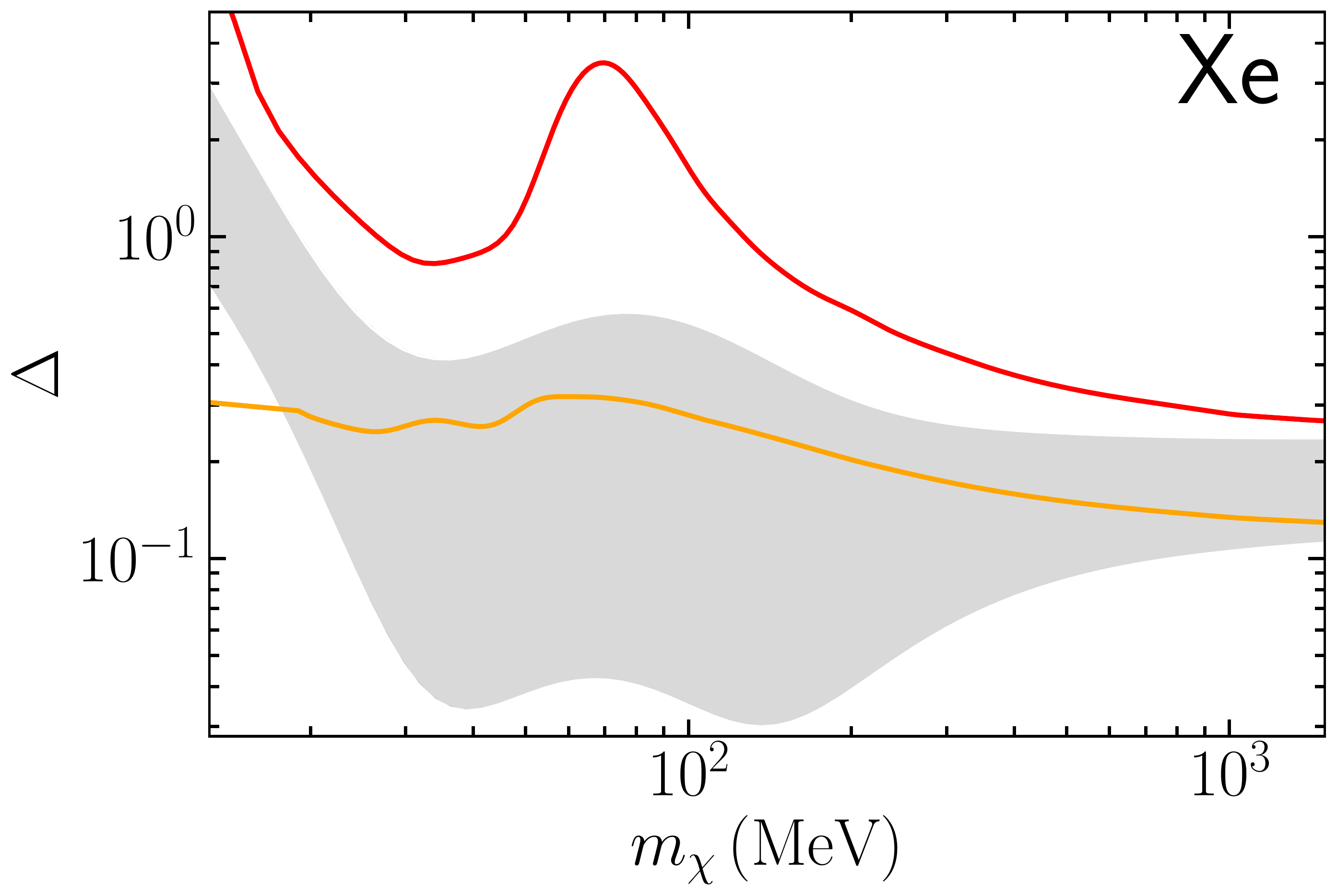}}
\subfloat[\label{sf:SiTsaeff}]{\includegraphics[scale=0.20]{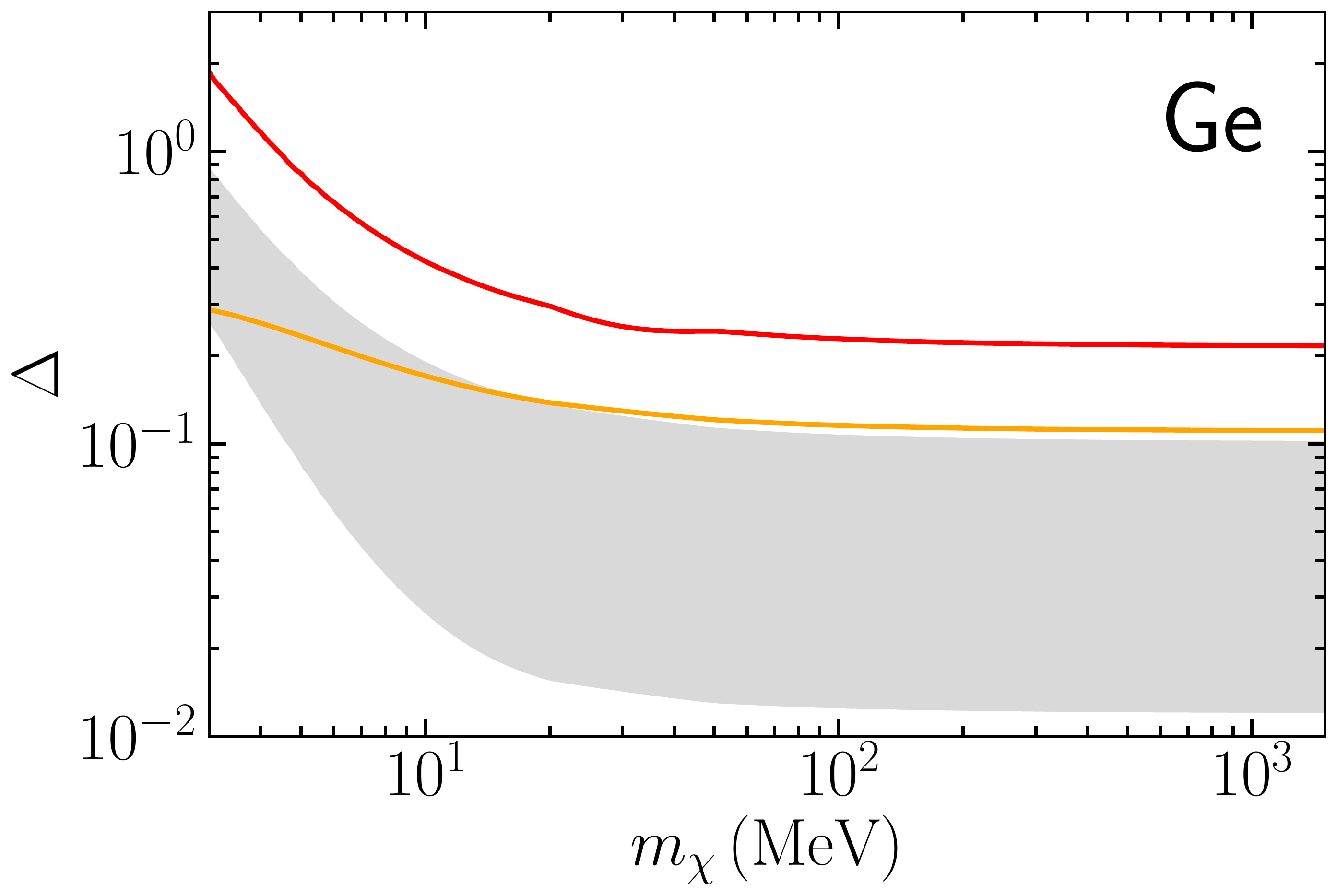}}
\subfloat[\label{sf:GeTsaeff}]{\includegraphics[scale=0.20]{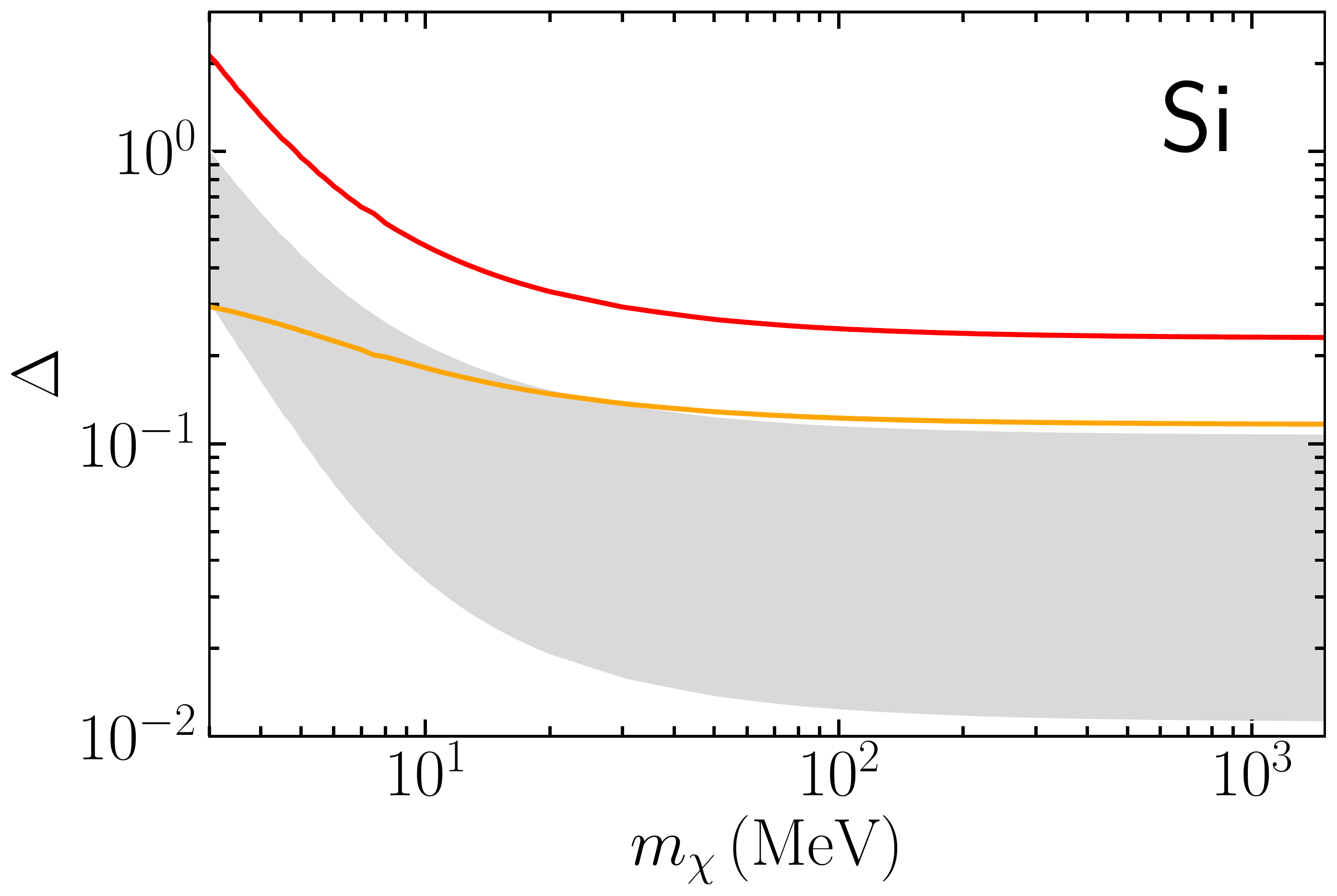}}
\caption{Variations in the exclusion bounds for Tsallis with $F_{\rm DM} =1$.  All the relevant details are same as of figure \ref{fig:king}.}
\label{fig:Tsallis}
\end{center}
\end{figure*}

The Tsallis distribution is explicitly derived through a factorization approximation of the Tsallis statistics \cite{Tsallis:1987eu} which is a generalisation of Boltzmann-Gibbs entropy. The distribution is widely used in high energy collisions \cite{Cleymans:2015lxa}, Bose-Einstein condensation \cite{MILLER2006357}, black-body radiation, neutron star \cite{Menezes:2014wqa}, early universe cosmology \cite{szczniak2018nonparametric} and  superconductivity \cite{Parvan:2019hqf}. The velocity distribution function goes by the form
\begin{equation}
    f(\mathbf{v})=
    \begin{cases}
      \frac{1}{N}\left[1-\left(1-q\right)\frac{|\mathbf{v}|^{2}}{v_{0}^{2}}\right]^{\frac{1}{1-q}} &  |\mathbf{v}| \leq v_{\rm esc} \\
      0 &  |\mathbf{v}| > v_{\rm esc},
    \end{cases}
    \label{eq:Tsallis}
  \end{equation}
where the symbols have their usual meaning. For this distribution with $q<1$ the escape velocity is determined by the relation $v_{\rm esc}^2=v_0^2/(1-q)$.  This inherent cut off criterion makes this distribution appealing as compared to the SHM. While for $q>1$ escape velocity still remains a somewhat arbitrary parameter. In $q \to 1$ limit the Tsallis distribution reduces to the Gaussian form of the SHM. Further from equation \eqref{eq:Tsallis} it is evident that this distribution predicts a continuous and smooth fall near the tail, favoured by  cosmological simulations \cite{Vogelsberger:2008qb, Kuhlen:2009vh, Ling:2009eh}. In particular, it has been argued in reference \cite{Ling:2009eh} that the  Tsallis distribution seems to fit better with Milky Way like simulations including Baryonic physics.

With the best fit values provided in table \ref{tab:bestfit}, the bound on $\bar{\sigma}_e$ considering Tsallis as the distribution for DM is presented  in the upper panel of figure \ref{fig:Tsallis}.  In the middle panel, the solid grey curve corresponds to the Tsallis distribution when fitted with the fiducial value. The grey bands correspond to the Tsallis distribution when fitted to the recent astrophysical observations of $v_0=233 \pm 6$ km/s and $v_{\rm esc}=528^{+24}_{-25}$ km/s. The deviation obtained for APOSTLE DMB simulation ranges in between $ 12\%$ to $ 33\%$ for the three set of target materials considered here.  The corresponding deviations induced by the recent astrophysical measurement are between $3\%$ to $290\%$ for Xe targets, $1\%$ to $90\%$ for Ge and $1\%$ to $100\%$ for Si semiconductor detectors.

\begin{figure*}[t]
\begin{center}
\subfloat[\label{sf:XeMao1}]{\includegraphics[scale=0.18]{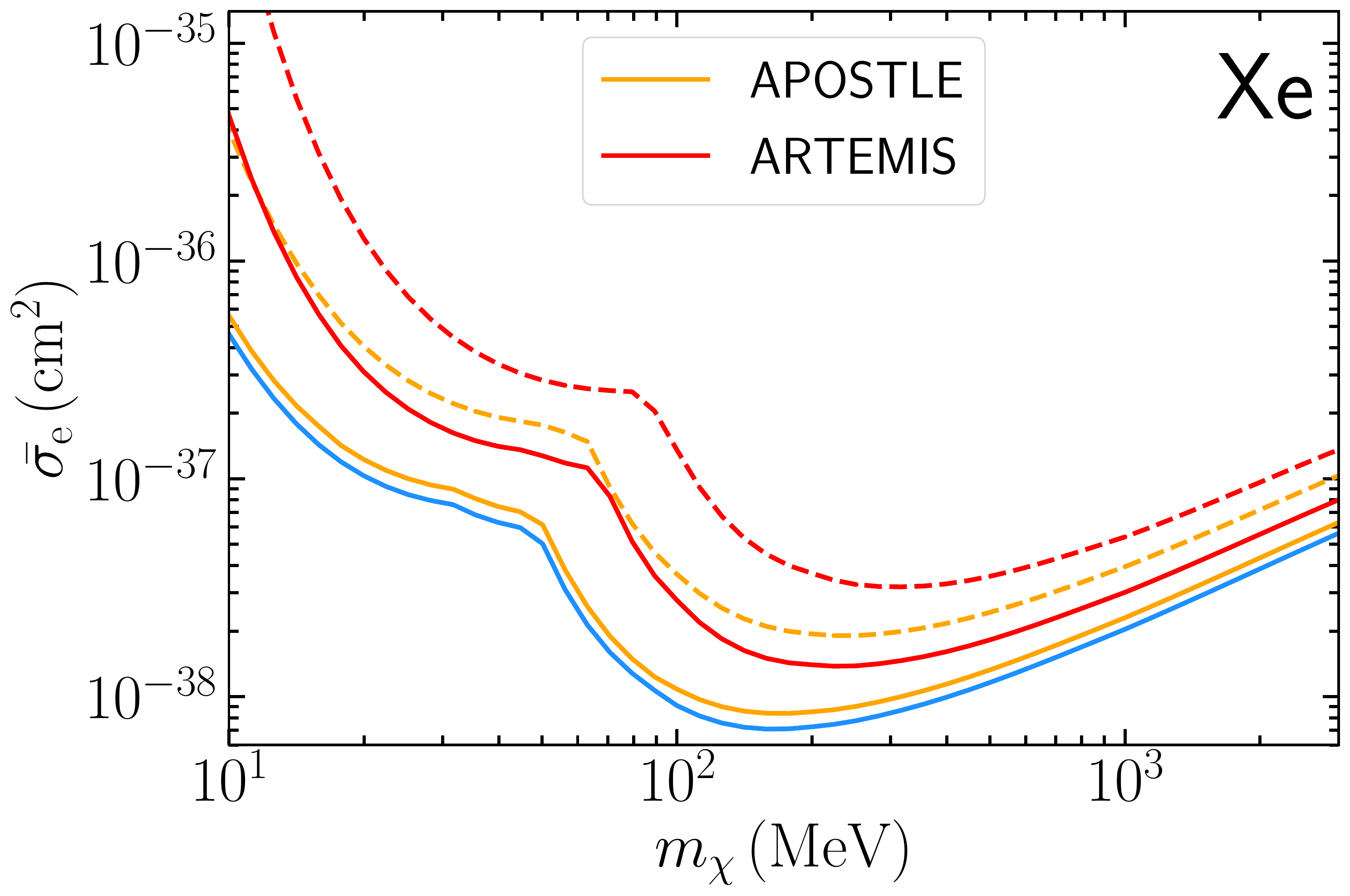}}
\subfloat[\label{sf:SiMao1}]{\includegraphics[scale=0.18]{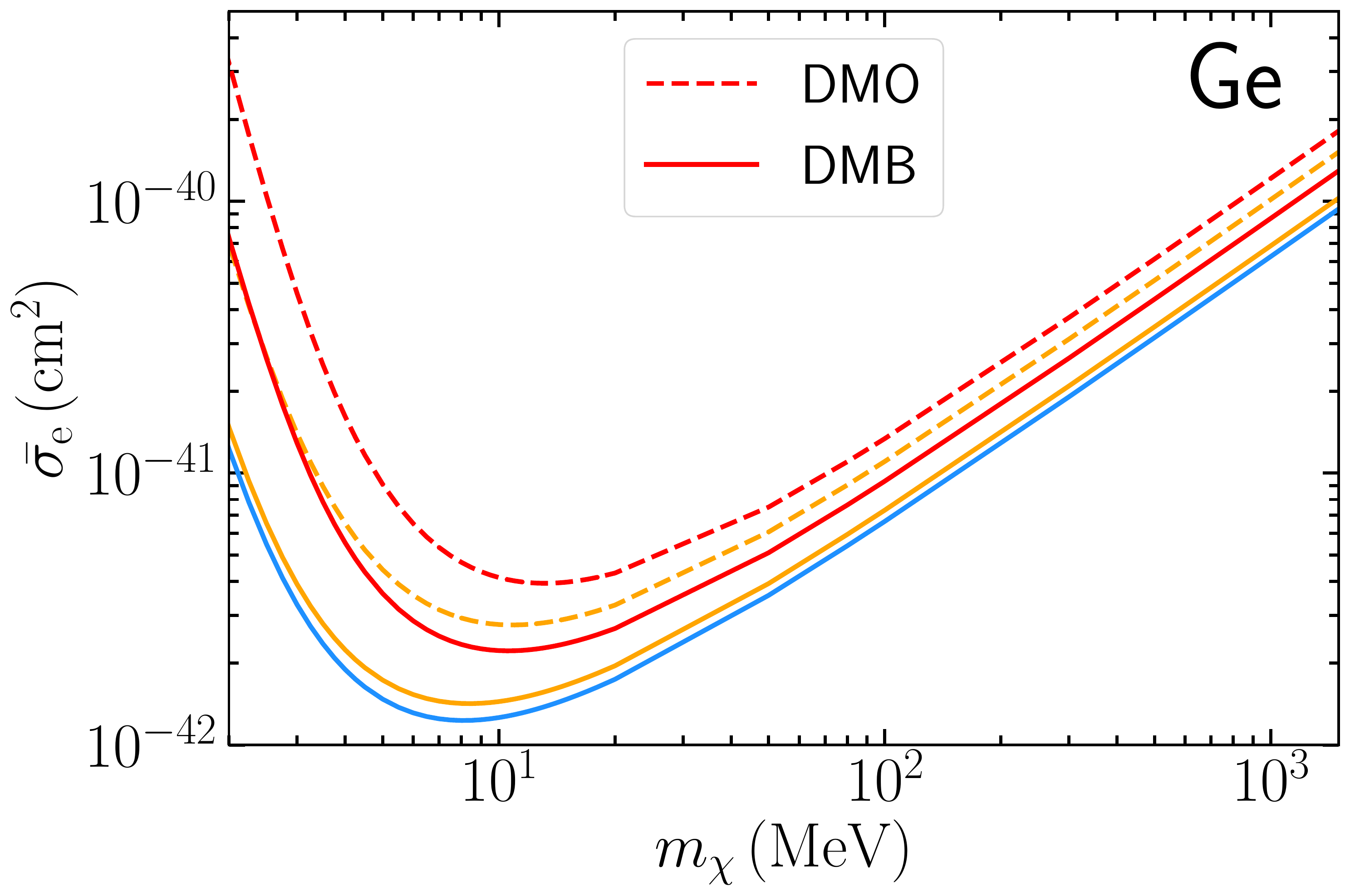}}
\subfloat[\label{sf:GeMao1}]{\includegraphics[scale=0.18]{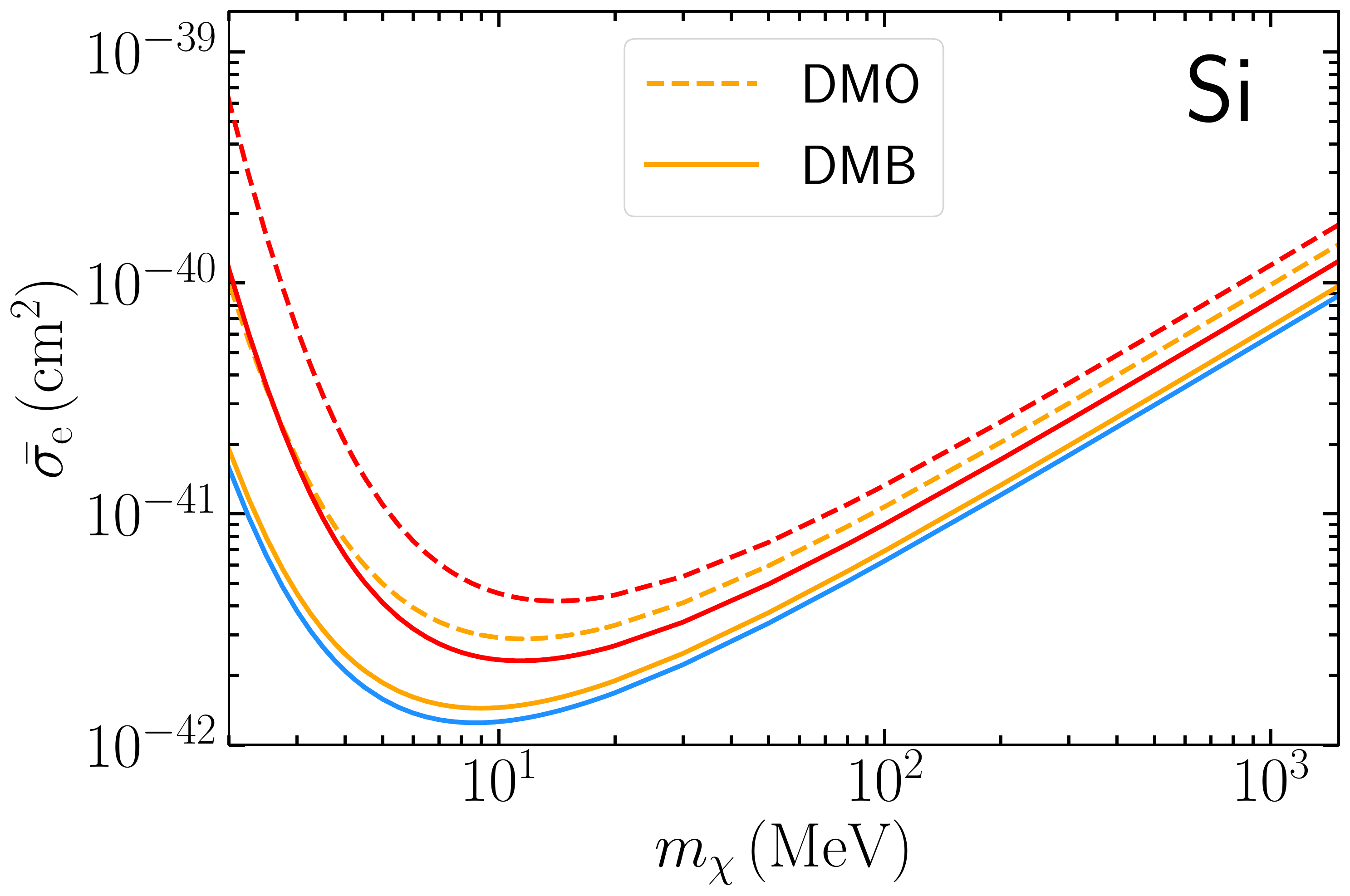}}
\newline
\subfloat[\label{sf:XeMaofa}]{\includegraphics[scale=0.18]{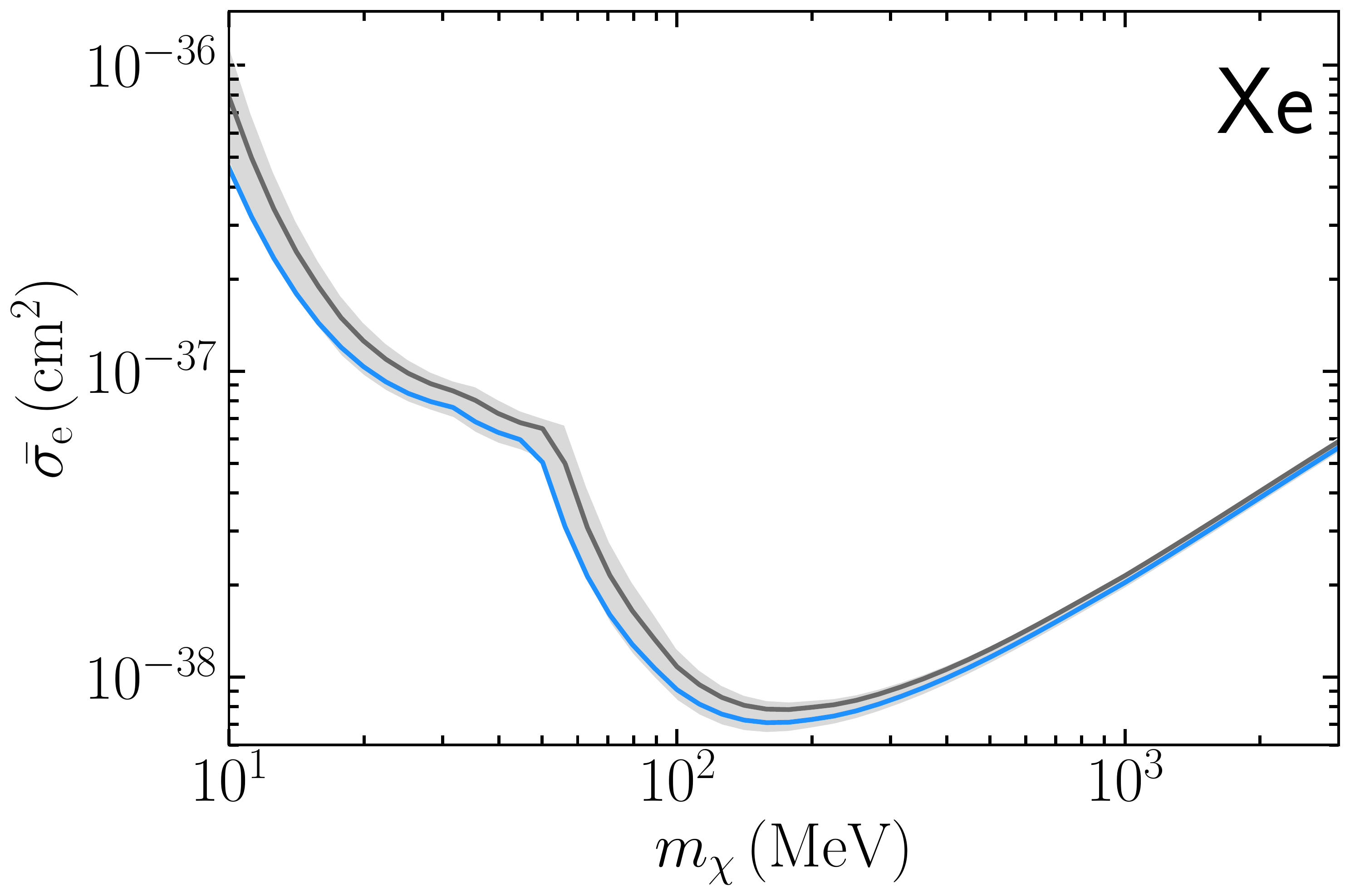}}
\subfloat[\label{sf:SiMaofa}]{\includegraphics[scale=0.18]{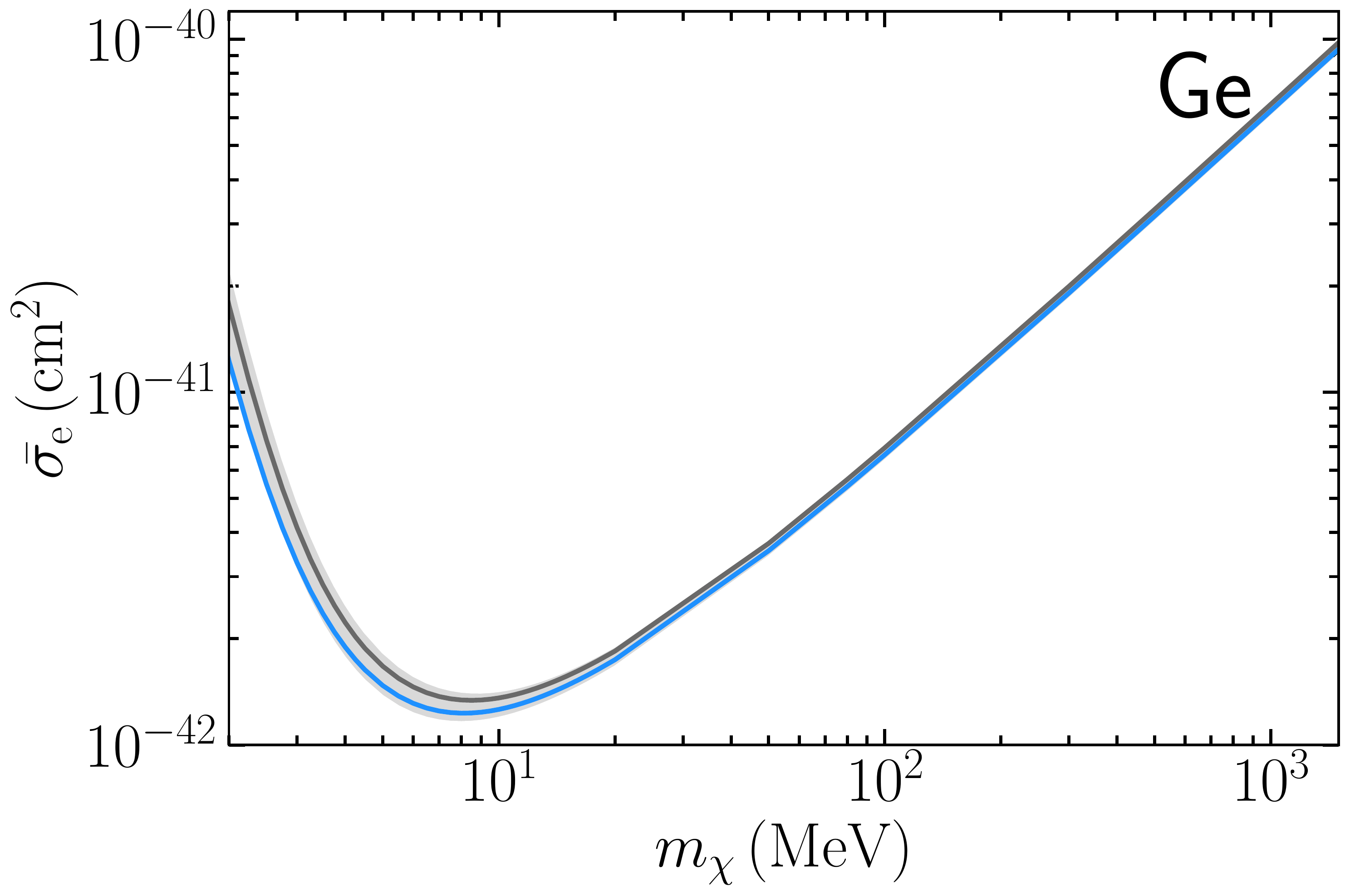}}
\subfloat[\label{sf:GeMaofa}]{\includegraphics[scale=0.18]{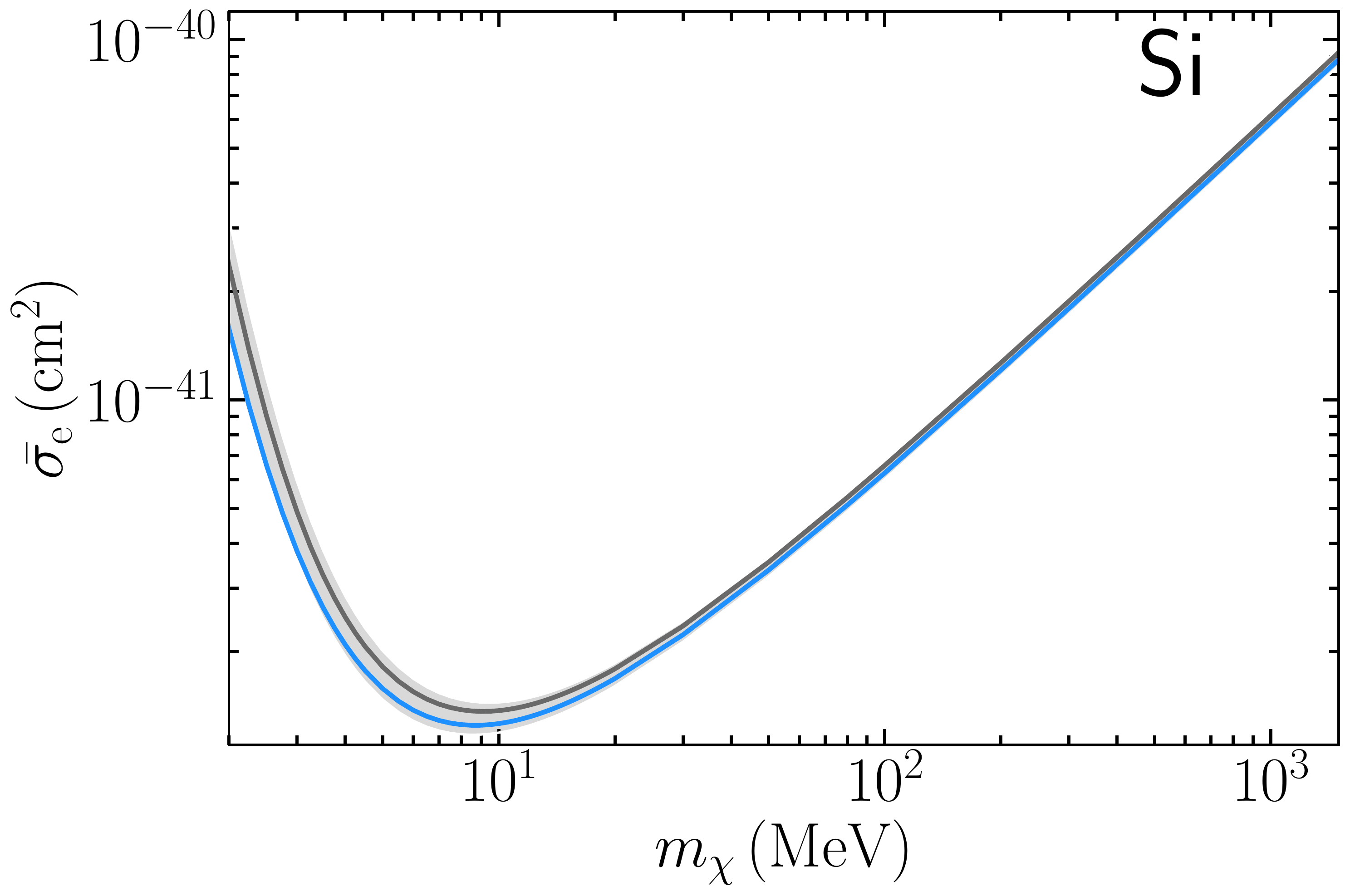}}
\newline
\subfloat[\label{sf:XeMaoeff}]{\includegraphics[scale=0.20]{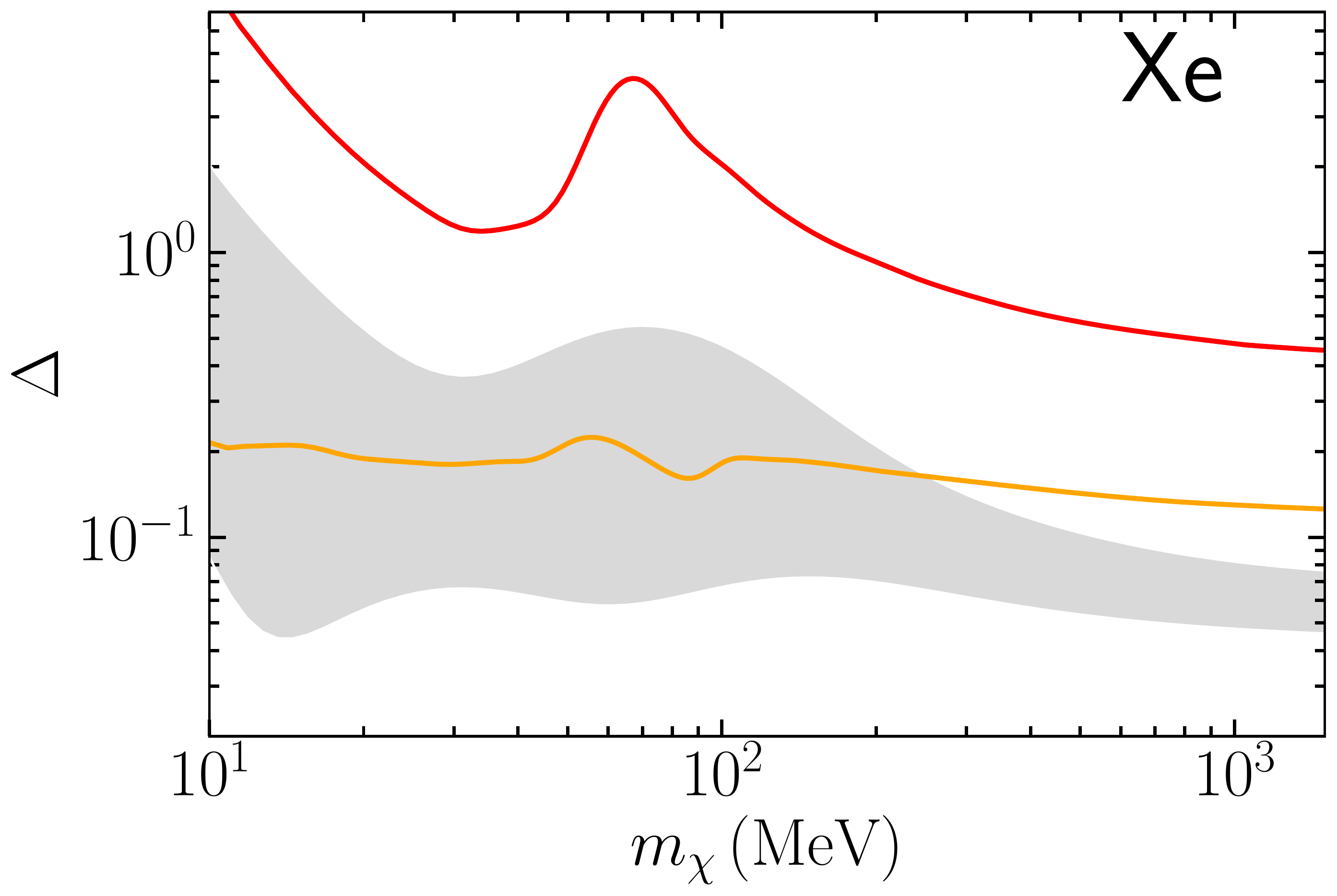}}
\subfloat[\label{sf:SiMaoeff}]{\includegraphics[scale=0.20]{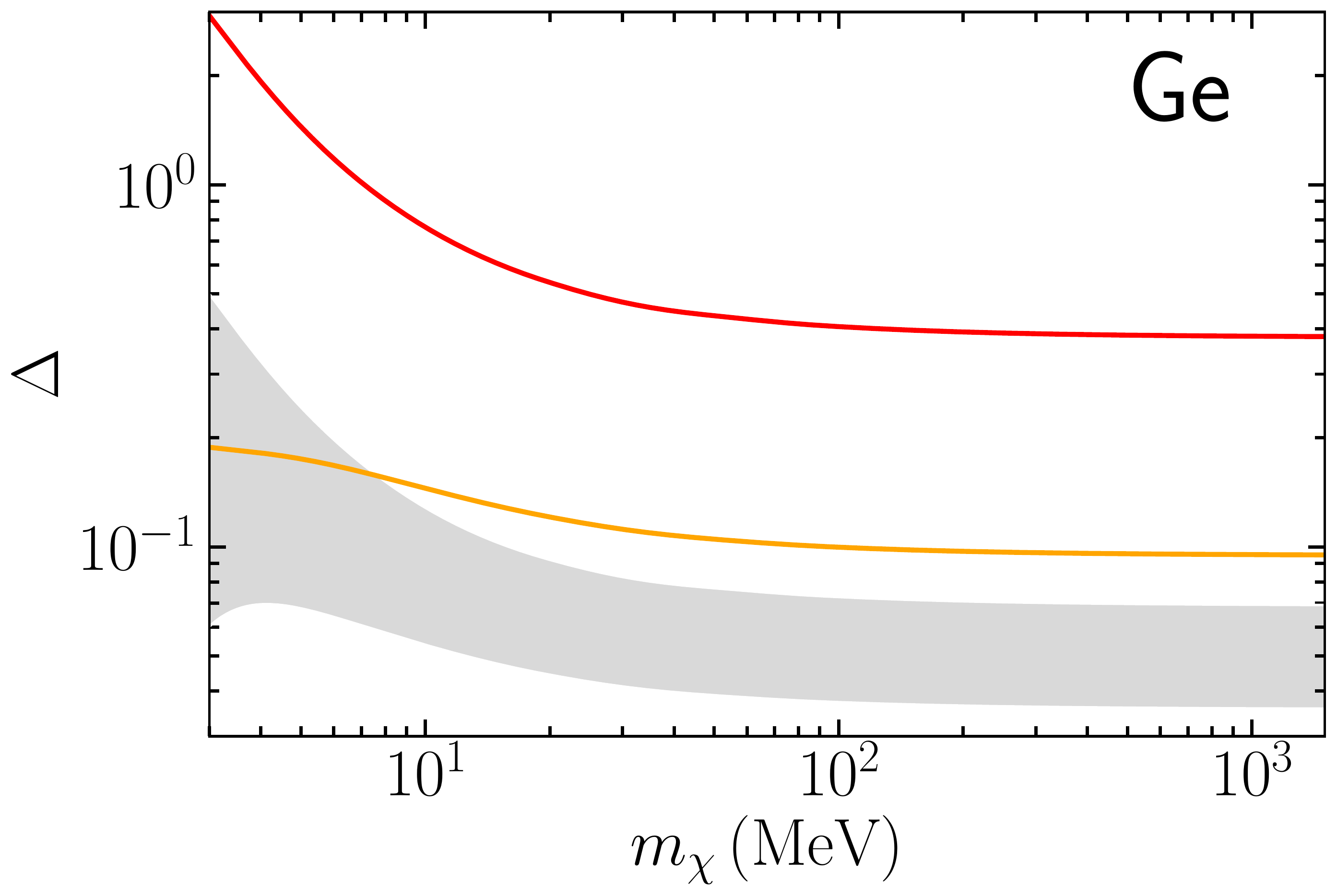}}
\subfloat[\label{sf:GeMaoeff}]{\includegraphics[scale=0.20]{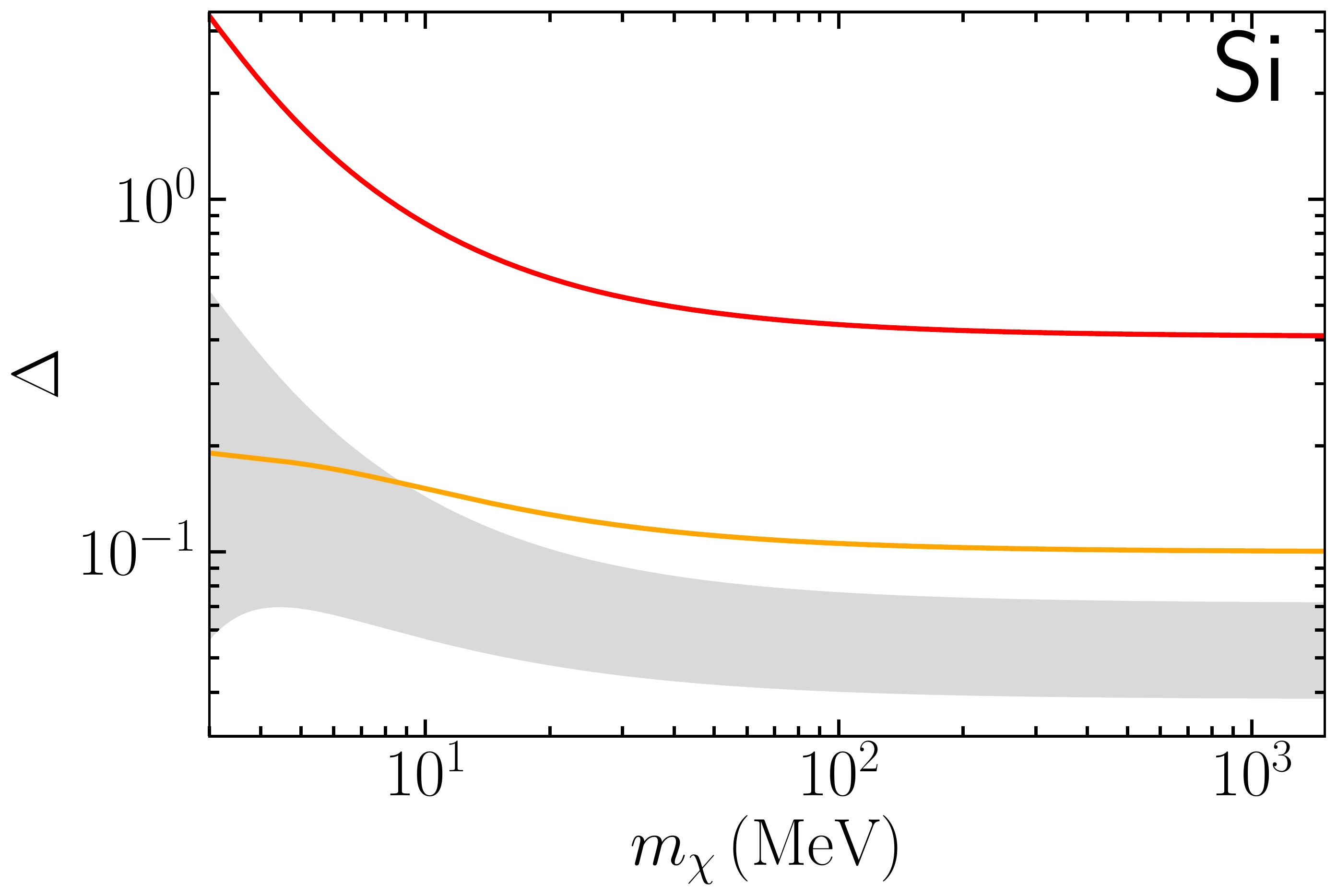}}

\caption{Variations in the exclusion bounds for Mao et. al. with a representative value of $p=2.7$ for $F_{\rm DM} =1$ . All other relevant details are same as of figure \ref{fig:king}.}
\label{fig:Mao}
\end{center}
\end{figure*}

\subsection{Mao et. al.}
\label{subsec:mao}

Mao et. al. \cite{Mao:2012hf} postulates an empirical model for the velocity distribution of DM having a wider peak and a steeper tail in comparison to the MB distribution. The Mao distribution function is given by
\begin{equation}
    f(\mathbf{v})=
    \begin{cases}
      \frac{1}{N}\left[\left( v_{\rm esc}^{2}-|\mathbf{v}|^{2} \right)^{p} e^{-\frac{v}{v_{0}}}  \right] &  |\mathbf{v}| \leq v_{\rm esc} \\
      0 &  |\mathbf{v}| > v_{\rm esc},
    \end{cases}
    \label{eq:Mao}
  \end{equation}
where the symbols have their usual meaning.
  It is favoured by the simulations that have taken into account the sequence of mergers, violent-relaxation and accretion in the simulated halos \cite{Hansen:2005yj}. Unlike other variants of the SHM, this empirical model is not based on a Gaussian but rather on an exponential distribution function having a power law cut-off at the binding energy or the equivalent escape velocity.  We fit the distribution given in \eqref{eq:Mao} with two different simulation namely APOSTLE and ARTEMIS with the corresponding best fit values given in table \ref{tab:bestfit}.
%

%

%

Assuming Mao as the velocity distribution for DM, in the upper panel of figure \ref{fig:Mao} we have presented the shifts in the exclusion limits for the best fit value given in table \ref{tab:bestfit}. In the  middle panel the grey curves correspond to the exclusion bounds for a representative value of $p=2.7$, while keeping $v_0$ and $v_{\rm esc}$ at the fiducial choice. In the lower panel, with $p=2.7$, the grey shaded band emerge due to the uncertainties associated with the recent astrophysical observations of $v_0=233 \pm 6$ km/s and $v_{\rm esc}=528^{+24}_{-25}$ km/s.

From the upper panel of figure \ref{fig:Mao} it is clear that compared to the fiducial SHM, the red curves set weaker bounds on DM-electron scattering cross section. This is mainly due to the large difference between the best fit values of $v_0$ and the fiducial values of the SHM. For the best fit DMO simulations, we observe a significant deviation from the SHM, compared to DMB simulations. Further the best fit  APOSTLE DMB simulation which takes into account the baryonic contribution, shows  a variation between $10\%$ to $20\%$ for all the three considered targets. This signifies the correlation between the exclusion bounds and the nature of the underlying fitted  cosmological simulations.  The corresponding deviation relative to the fiducial SHM induced for the recent astrophysical observations are between $4\%$ to $200\%$ for Xe targets, for Ge $3\%$ to $43\%$ and $3\%$ to $57\%$ for Si semiconductor detectors.
%

%
\section{Comparison of deviation: observational and cosmological simulation} 
\label{sec:Com-NB}

A discussion about the effects of best fit parameters for the various  cosmological simulations on the derived exclusion bounds and its comparison with observational estimation of astrophysical parameters  is now in order.

 Comparing the bounds from various simulations, it can be inferred that DM only simulations show large deviation with respect to SHM fiducial as compared to simulation which takes into account the baryonic effects. Further among the two simulations, considered in the paper the  APOSTLE DMB  closely resemble the fiducial SHM scenario. For instance, the King, DPL, Tsallis, and Mao distributions show a maximum deviation of $\mathcal{O}(10\%)$ relative to the fiducial SHM. We typically observe an  order of magnitude deviation from the fiducial SHM for DM only counterpart of the considered simulations. This indicates that these deviations are correlated with the underlying dynamics of the simulations. Interestingly, for these non-SHM models, the recent astrophysical measurements  indicate a maximum $ \mathcal{O}(100\%)$ and $\mathcal{O}(10\%)$ deviations from fiducial SHM for Xe and semiconductor detectors respectively.

\begin{table}[]
\centering
\resizebox{\textwidth}{!}{%
\begin{tabular}{|c|c|c|c|c|c|c|}
\hline
\multirow{2}{*}{\begin{tabular}[c]{@{}c@{}}Halo\\ Model\end{tabular}} & \multicolumn{2}{c|}{$\Delta$ in percentage for Xe} & \multicolumn{2}{c|}{$\Delta$ in percentage for  Ge } & \multicolumn{2}{c|}{$\Delta$ in percentage for  Si} \\ \cline{2-7} 
 & simulation & \begin{tabular}[c]{@{}c@{}}Recent\\ Measurements\end{tabular} & simulation & \begin{tabular}[c]{@{}c@{}}Recent\\ Measurements\end{tabular} & simulation & \begin{tabular}[c]{@{}c@{}}Recent\\ Measurements\end{tabular} \\ \hline
King 	& 3-20  & 5-500 & 2-12   & 1-60 		    & 3-15 	& 1-74 	\\ \hline
DPL	 	& 5-20  & 2-100 & 4-14   & 1-21 			& 5-14 	& 1-26	\\ \hline
Tsallis & 12-30 & 3-290 & 12-32  & 1-90	        	& 13-33	& 1-100	\\ \hline
Mao		& 10-21 & 4-200 & 10-20  & 3-43 		    & 10-20 & 3-57	\\ \hline
\end{tabular}%
}
\caption{A comparison of the uncertainties between APOSTLE DMB simulations for DPL, King, Tsallis, and Mao distributions and in the observed estimates of $v_0$ and $v_{\rm esc}$.}
\label{tab:PerUnc}
\end{table}

In table \ref{tab:PerUnc}, we have provided a comparison of the percentage deviations from the fiducial SHM for sophisticated  cosmological simulations and for the recent observational values of $v_{\rm esc}$ and $v_0$. As evident from the table, we conclude that for most of the halo distributions it is the updated astrophysical measurements that gives the most significant deviations in the exclusion bounds which are traditionally represented using the fiducial choice. This can imply significant reinterpretation of the conclusion drawn from direct detection experiments.

\section{Conclusions}
\label{sec:conclusion}
The non-observation of DM in the typical nuclear recoil direct detection experiments and the inability of GeV scale cold DM to address certain small scale structure formation issues have increased the interest in sub-GeV scale DM. An elegant avenue to probe such light DM is to consider the scattering of DM with electron in the direct detection experiments. Bounds on DM electron cross section are typically presented assuming the SHM for the DM distribution in our galaxy, with a fiducial choice for $v_0=220~ {\rm km/s}$ and $v_{\rm esc}=544~ {\rm km/s}$. However, recent progress in the measurement of these parameters shows a deviation from these values. In this paper we have systematically investigated the effects of uncertainties associated with the determination of these astrophysical quantities on the exclusion limits of DM electron cross section. We consider the uncertainties within the SHM and empirical models of DM distribution beyond the SHM that have been motivated by recent high resolution  cosmological simulations.

We find that the exclusion bounds are expectedly sensitive to the population of DM particles in the high velocity tails of the distributions. Within the SHM the velocity distribution is assumed to be MB like. The tail shape is controlled by the Sun's circular velocity ($v_0$) and to a lesser extent by the escape velocity ($v_{\rm esc}$).  We find that within SHM for contact interaction between DM and electron, these uncertainties  imply a  $2 \%$ to $ 50\%$ change in the event rates in the three target materials that have been considered. Further, inclusion of uncertainties in the local DM density leads to  additional change in the exclusion bounds.

Going beyond the SHM, we have considered the simulation motivated  King's model, Double Power Law, Mao, and Tsallis distributions. Relative to SHM all these non-SHM models fall smoothly near the high velocity tail, predicting less number particles in the region. This  causes a reduction in the event rate.  Therefore for the same set of astrophysical parameters, the non-SHM models seem to provide weaker bounds as compared to their SHM counterparts. Further for these models, depending on the fitted parameters we find that for most of the region in the parameter space the fractional changes in the cross section could vary substantially. Interestingly the amount of deviations observed from cosmological simulations fit can be traced to their treatment of baryonic content. In these models the deviation from the fiducial choice ranges between $1\%-30\%$ for the APOSTLE hydrodynamic simulation which includes baryons. Whereas for these non-SHM models the uncertainties associated with the recent astrophysical observations lead to a maximum $\mathcal{O}(100\%)$ change in the exclusion limit.

\textbf{Note added}: While this paper was under preparation a related work \cite{Radick:2020qip} appeared in the arXiv. Their treatment of halo uncertainties for semiconductor targets with Tsallis and Mao distribution is complementary with our discussion in section \ref{sec:UncertBSHM}.

\paragraph*{Acknowledgements\,:} 
TNM acknowledges the Department of Science and Technology, Government of India, under the Grant Agreement No. ECR/2018/002192 (Early Career Research Award) for financial assistance. SS acknowledges the University Grants Commission (UGC) of the Government of India for providing financial assistance through Senior Research Fellowship (SRF) with reference ID: 522157.

\appendix
\section{A brief detail of the  cosmological simulations}
\label{app:Nbody}
In this appendix we briefly summarize the cosmological simulations used in this paper. We give a brief and generic outline of the APOSTLE and ARTEMIS simulations along with their relevant resolutions.

\subsection{APOSTLE}
\label{subapp:apostle}

The APOSTLE (A Project Of Simulating The Local Environment) project comprises of a suite of high-resolution cosmological simulation of 12 Milky way like environments, selected from a volumes of the $\Lambda$CDM universe. Each contain a pair of halo with a virial mass in the range $5 \times 10^{11} M_{\odot}$ to $2.5 \times 10^{12}M_{\odot}$ , with median values of $1.4 \times 10^{12}M_{\odot}$ for the more massive and $0.9 \times 10^{12}M_{\odot}$ for the less massive haloes. The project uses a combination of a hydrodynamical SPH implementation named \texttt{Anarchy} superimposed on the Tree-PM SPH \texttt{Gadget} code. The numerical resolution of the hydrodynamical simulation reaches up to a mass of $10^4 \, M_{\odot}$ per gas particle, using the sub-grid developed for the EAGLE project \cite{Schaye:2014tpa}. Each of the main galaxy haloes contain more than $2 \times 10^7$ particles. The high resolution initial conditions were generated using second-order Lagrangian perturbation theory. The Apostle simulation suite has three different resolution levels for the primordial gas (DM) particle masses of approximately $1.0 (5.0)\times 10^4 M_{\odot} \,, 1.2(5.9) \times 10^5 M_{\odot}$ and $1.5(7.5) \times 10^6 M_{\odot}$ respectively. Maximum gravitational softening lengths of 134 pc, 307 pc and 711 pc. In this work we use the DM distribution data of the APOSTLE IR halo marked A1, which fits the selection criteria as discussed in section \ref{subsec:Nbody} having a stellar mass of $ 4.88 \times 10^{10}\, M_{\odot}$ and a virial mass of $1.64 \times 10^{12}\, M_{\odot}$.
 

\subsection{ARTEMIS}
\label{subapp:art}

The ARTEMIS (Assembly of high-ResoluTion Eagle-simulations of MIlky Way-type galaxieS) is a suite of high-resolution cosmological hydrodynamical simulations of Milky Way like haloes \cite{Font_2020}. It uses the 'zoom-in' technique to simulate the halos at a high resolution with DM and by incorporating baryons with DM using SPH. The initial conditions were generated using the \texttt{MUSIC} code \cite{Hahn:2011uy}. Haloes were selected from a base periodic box is $25$ Mpc $h^{-1}$ on one side with $256^3$ particles. The initial conditions were generated at a redshift of $ 127$ using a transfer function computed using the \texttt{CAMB} Boltzmann code cite for a flat $\Lambda$CDM WMAP cosmology was run to $z = 0$ using the Tree-PM SPH \texttt{Gadget-3} code. Haloes were selected with masses in the range $8 \times 10^{11} < M_{200} /M_{\odot} < 2 \times 10^{12}$, where $M_{200}$ is the mass enclosed inside a sphere of radius $R_{200}$, when the mean density is 200 times the critical density at $z = 0$. There were 63 such halos in this mass range. A subset of $42$ high-resolution collision-less simulations (DMO), together with their full hydrodynamical counterparts including baryons. The subset of 42 halos were not selected based on any physical criterion. They are the subset that ran to $z = 0$. The DM particles have a mass around $1.17 \times 10^5 M_{\odot} h^{-1}$ and the baryon particle mass is $2.23 \times 10^4 M_{\odot} h^{-1}$. The force resolution (Plummer-equivalent softening) is $125 \, \rm pc/h^{-1}$. ARTEMIS used the stellar mass estimates  to be in the range $(1.75 - 5.45) \times 10^{10} M_{\odot} $ with a mean value of $2.87 \times 10^{10} M_{\odot} $.

\section{Momentum dependent DM-electron interactions}
\label{app:q&q2}
\begin{figure*}[t]
\begin{center}
\subfloat[\label{sf:XeSq}]{\includegraphics[scale=0.18]{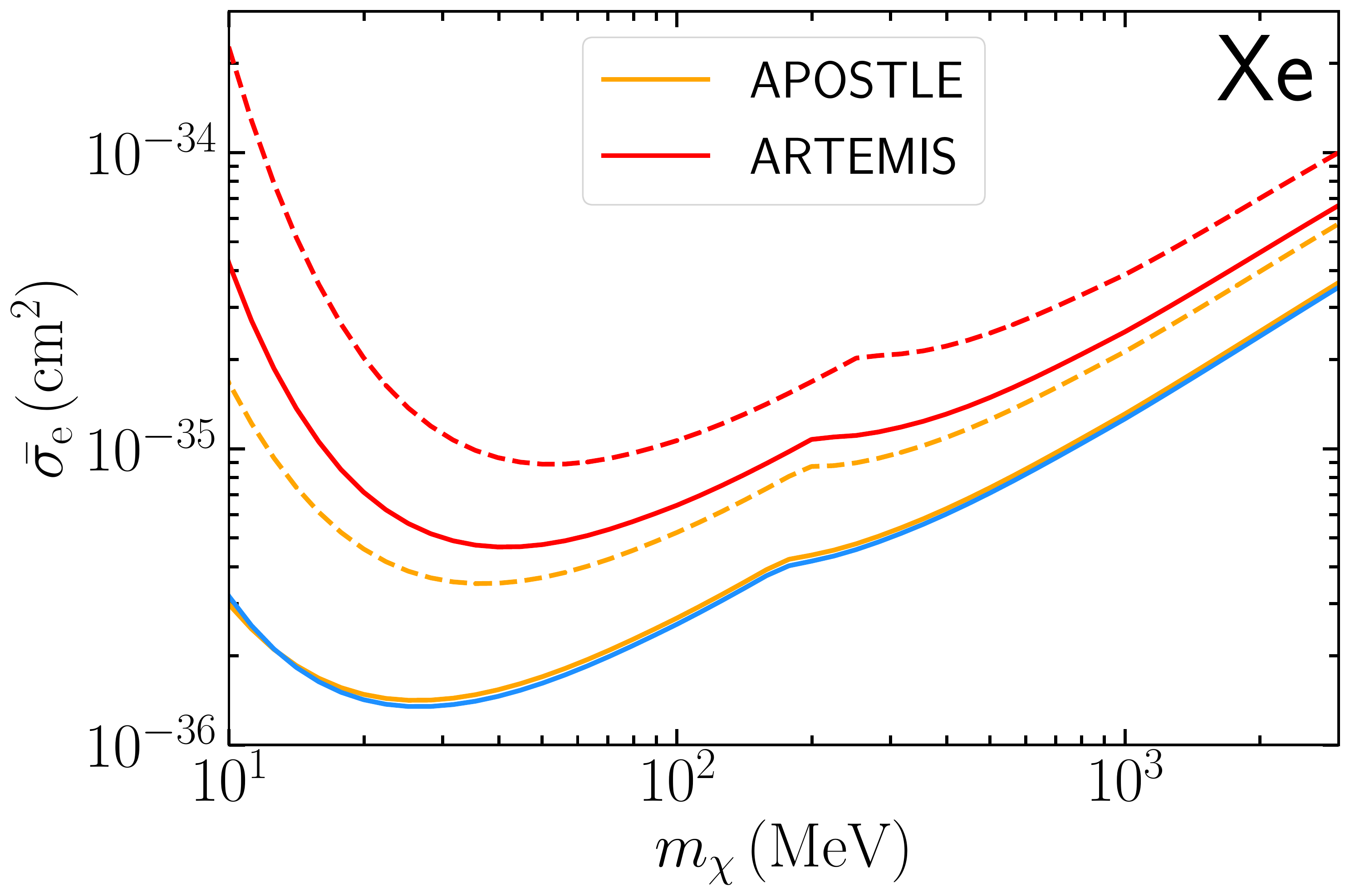}}
\subfloat[\label{sf:SiSq}]{\includegraphics[scale=0.18]{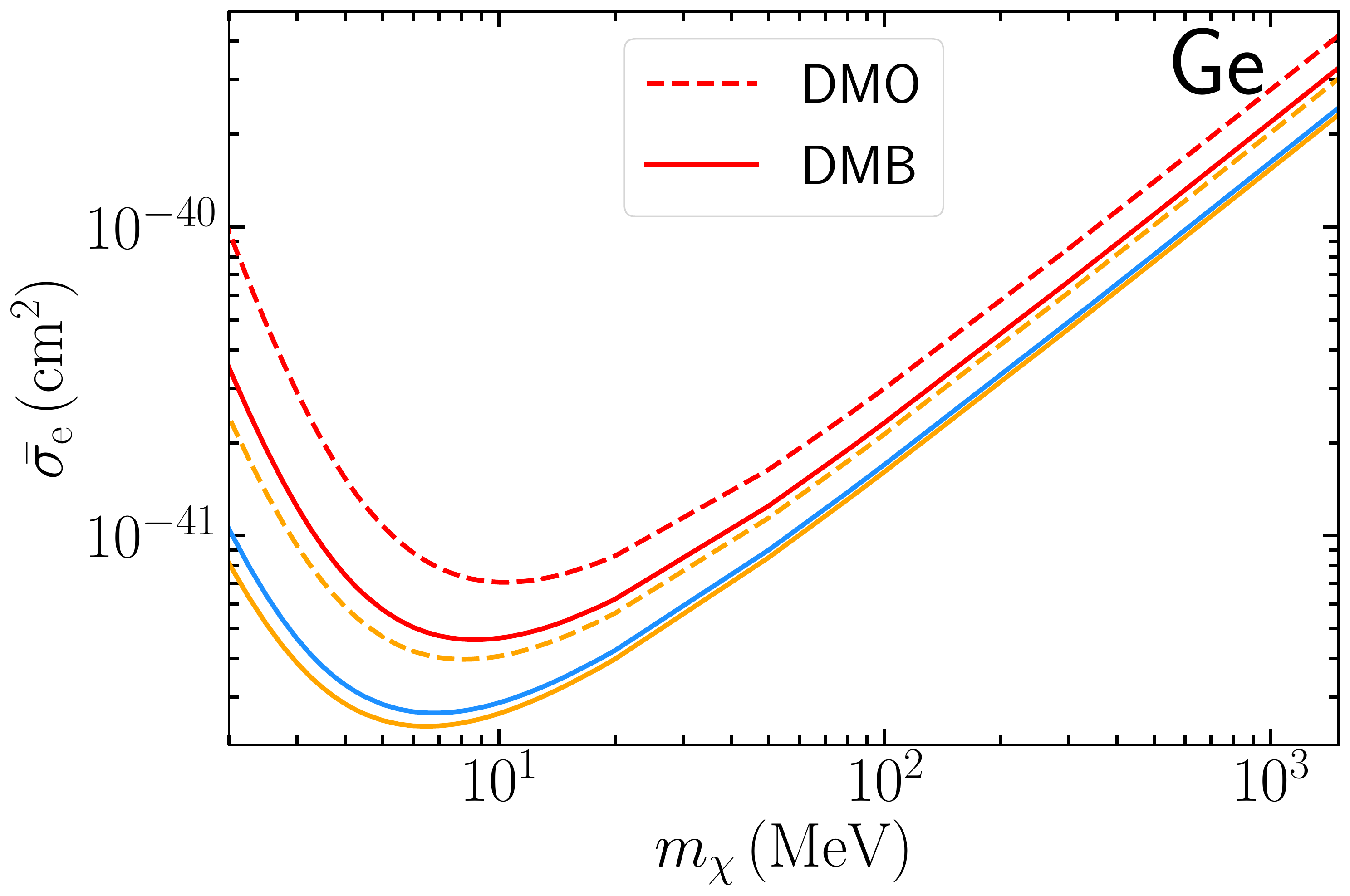}}
\subfloat[\label{sf:GeSq}]{\includegraphics[scale=0.18]{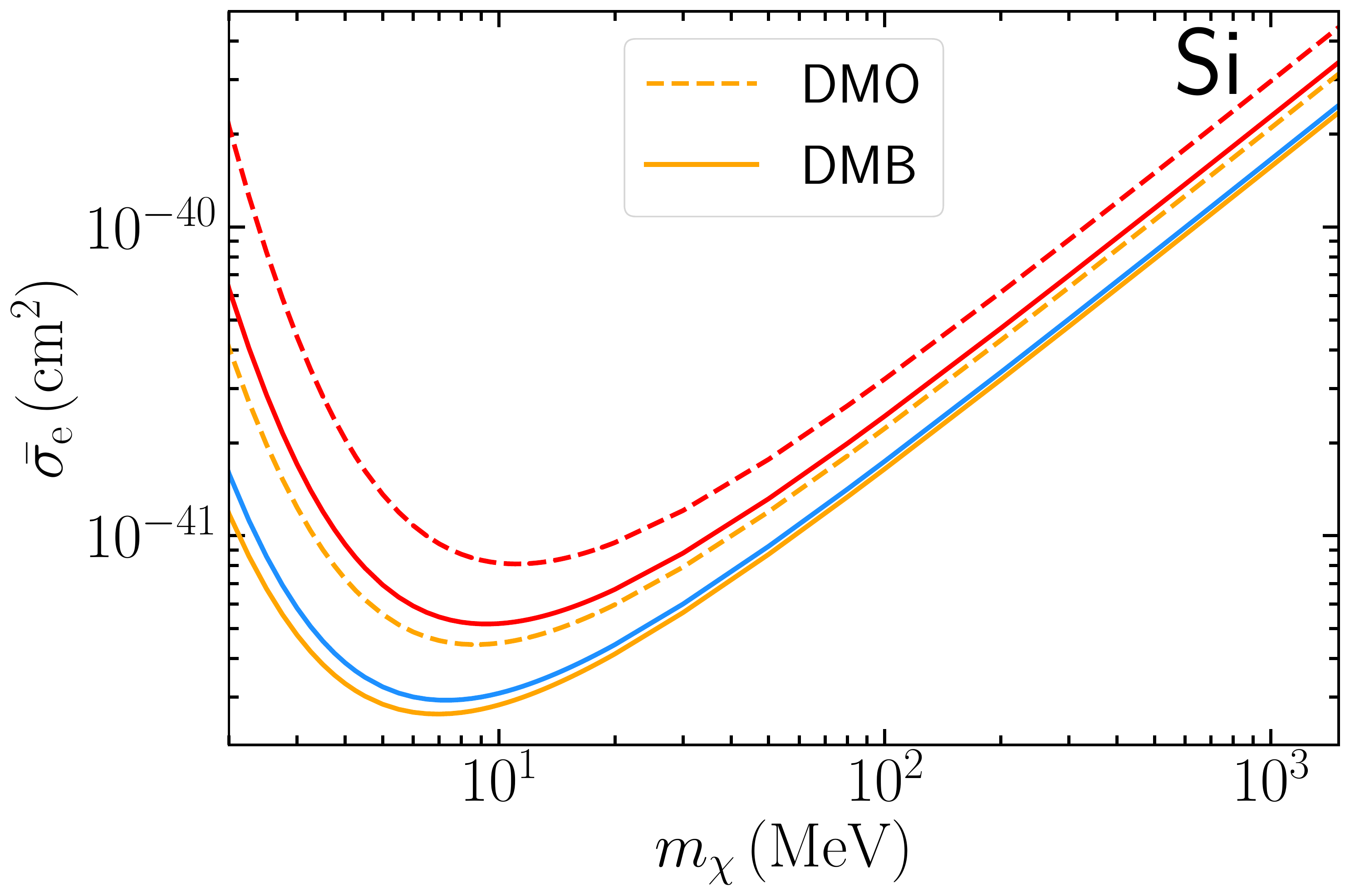}}
\caption{Shifts in the exclusion limits for the standard halo model with $F_{\rm DM}  \propto q^{-1}$ for the best fit values of the different  cosmological simulations. The other relevant details are same as of figures \ref{fig:Nbody-SHM} .
}
\label{fig:SHMappq}
\end{center}
\end{figure*}

\begin{figure*}[t]
\begin{center}
\subfloat[\label{sf:XeSq2}]{\includegraphics[scale=0.178]{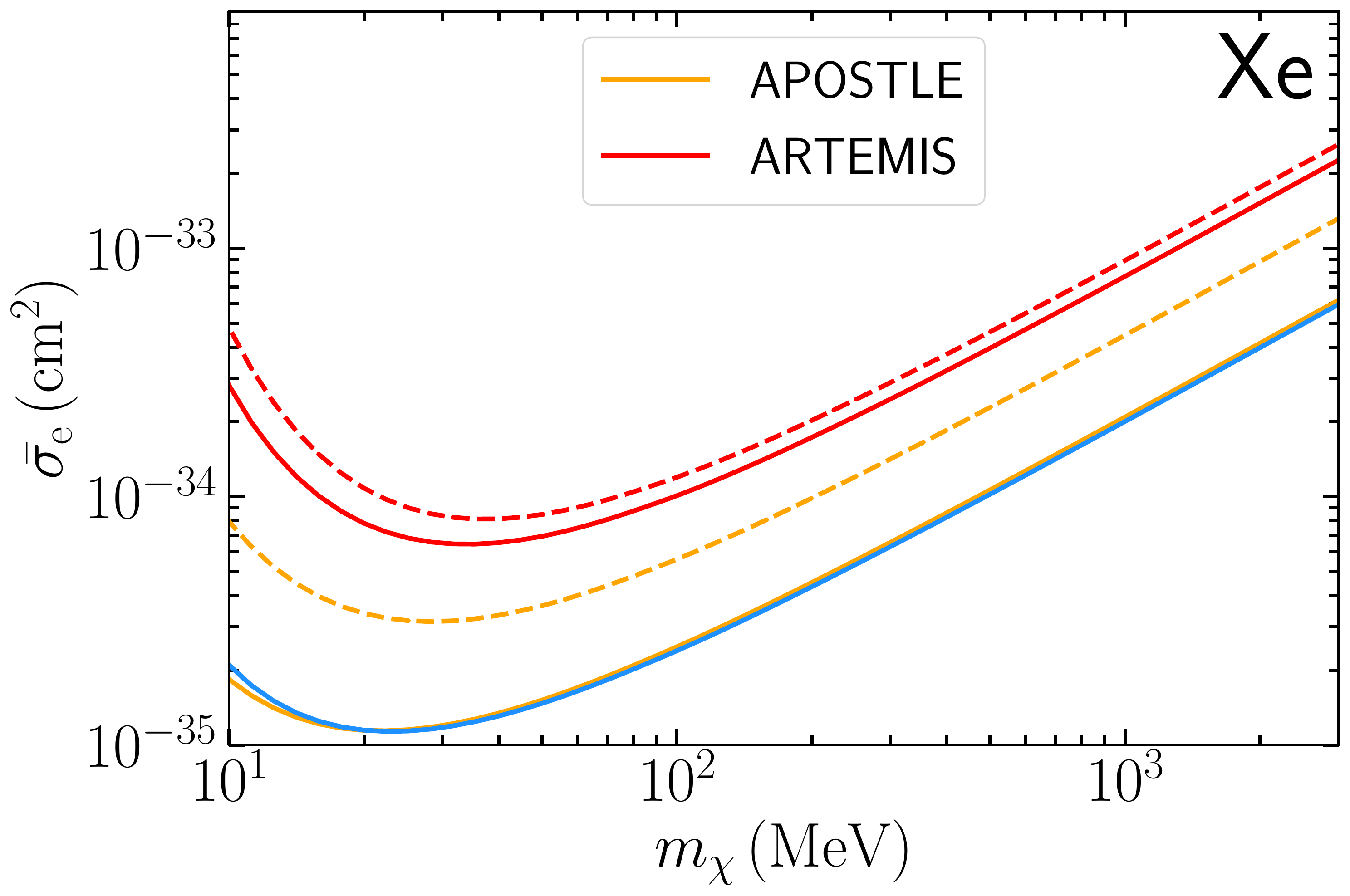}}
\subfloat[\label{sf:SiSq2}]{\includegraphics[scale=0.178]{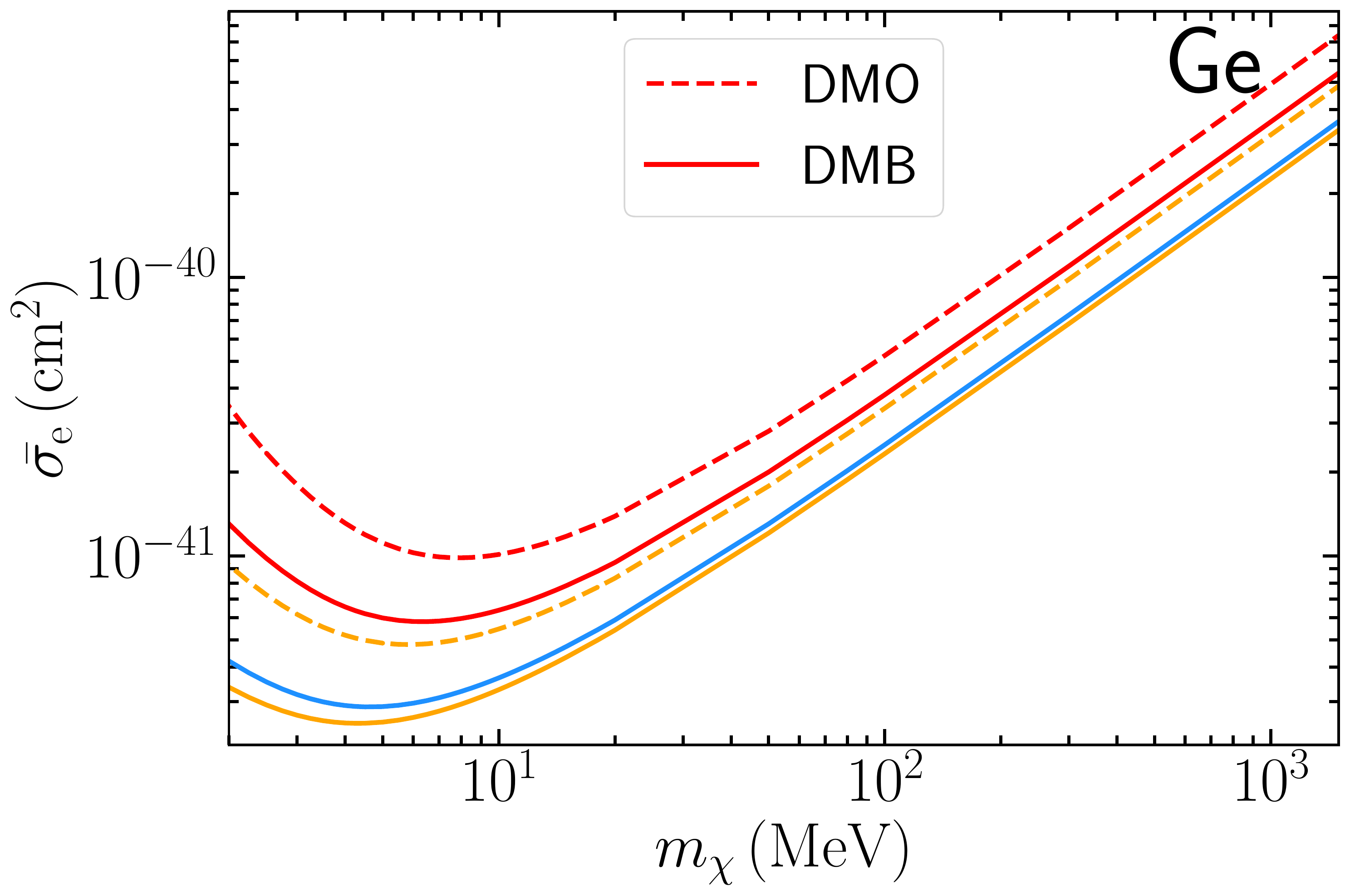}}
\subfloat[\label{sf:GeSq2}]{\includegraphics[scale=0.178]{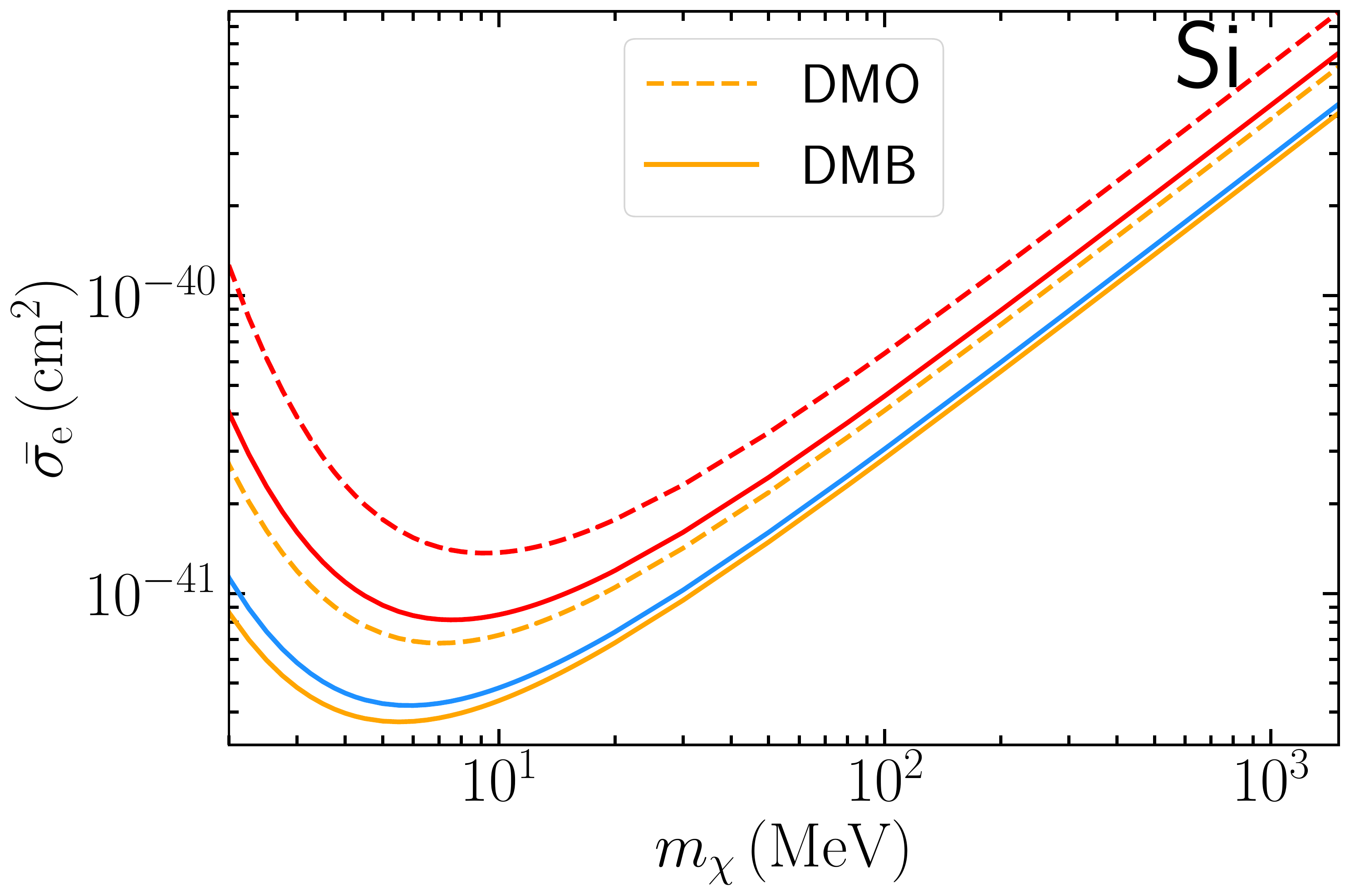}}
\caption{Same as figure \ref{fig:SHMappq} but for $F_{\rm DM}  \propto q^{-2}$.}
\label{fig:SHMappq2}
\end{center}
\end{figure*}

\begin{figure*}[h!]
\begin{center}
\subfloat[\label{sf:XeKq}]{\includegraphics[scale=0.18]{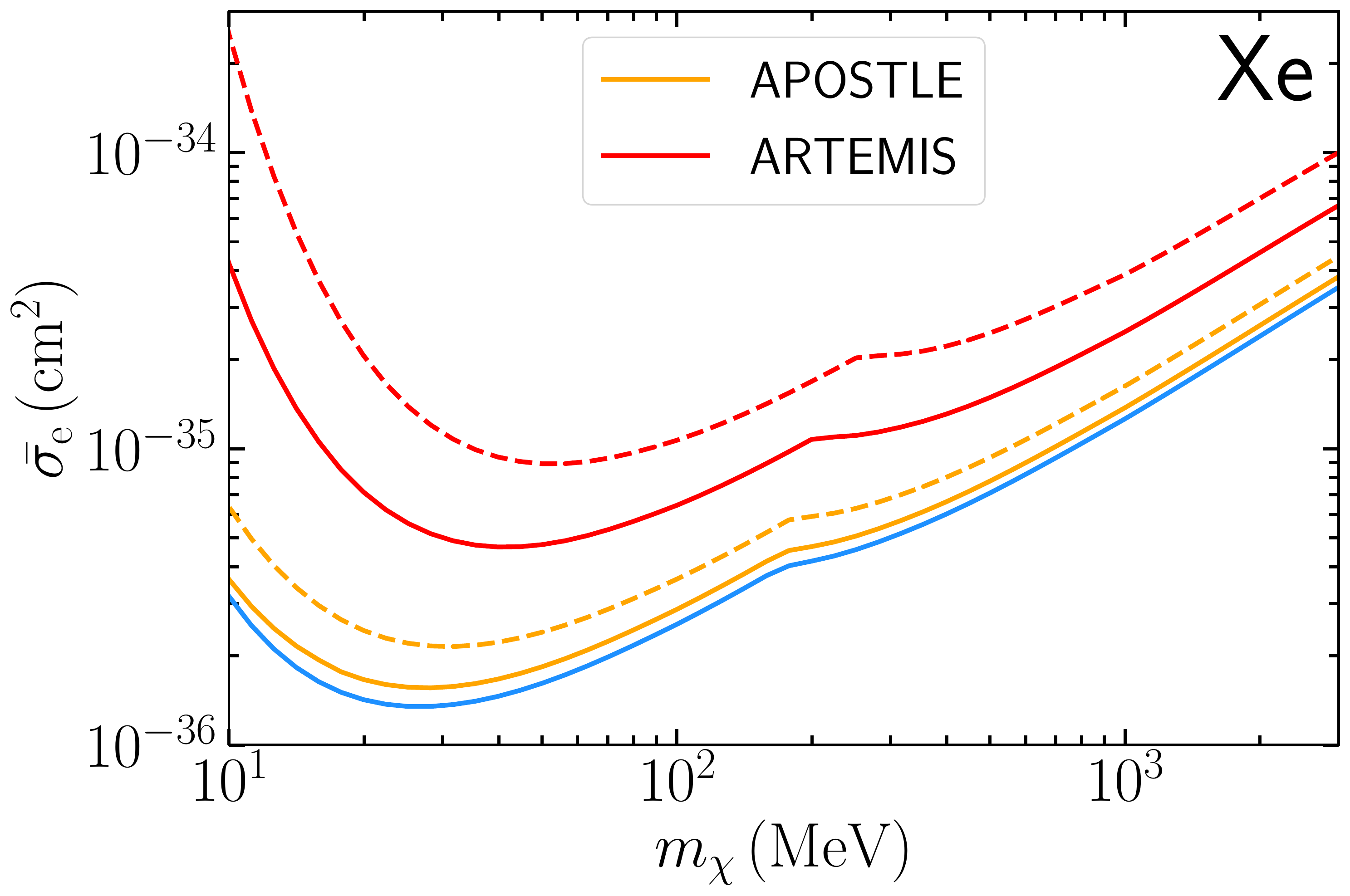}}
\subfloat[\label{sf:SiKq}]{\includegraphics[scale=0.18]{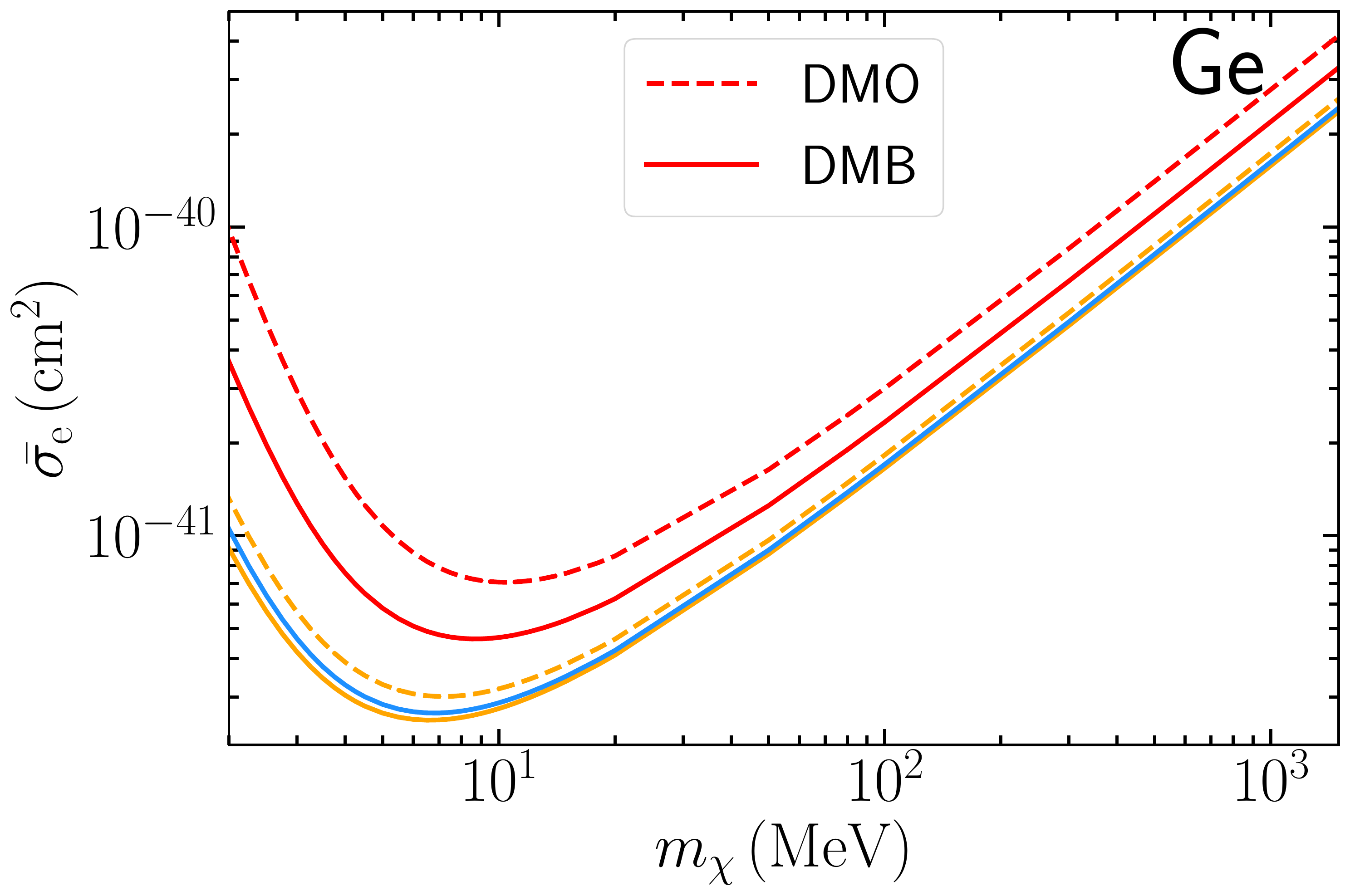}}
\subfloat[\label{sf:GeKq}]{\includegraphics[scale=0.18]{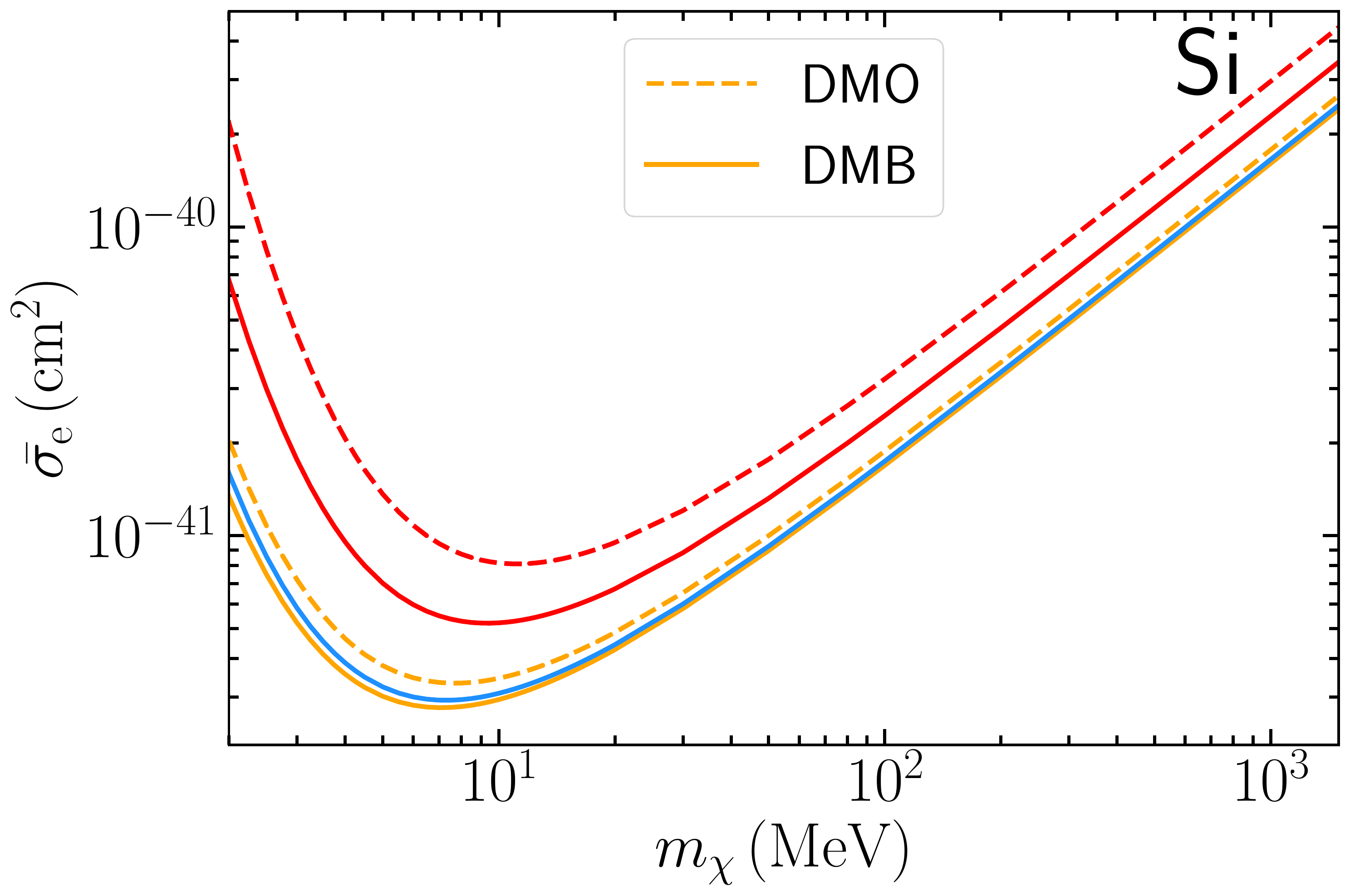}}
\newline
\subfloat[\label{sf:XeKingqfa}]{\includegraphics[scale=0.18]{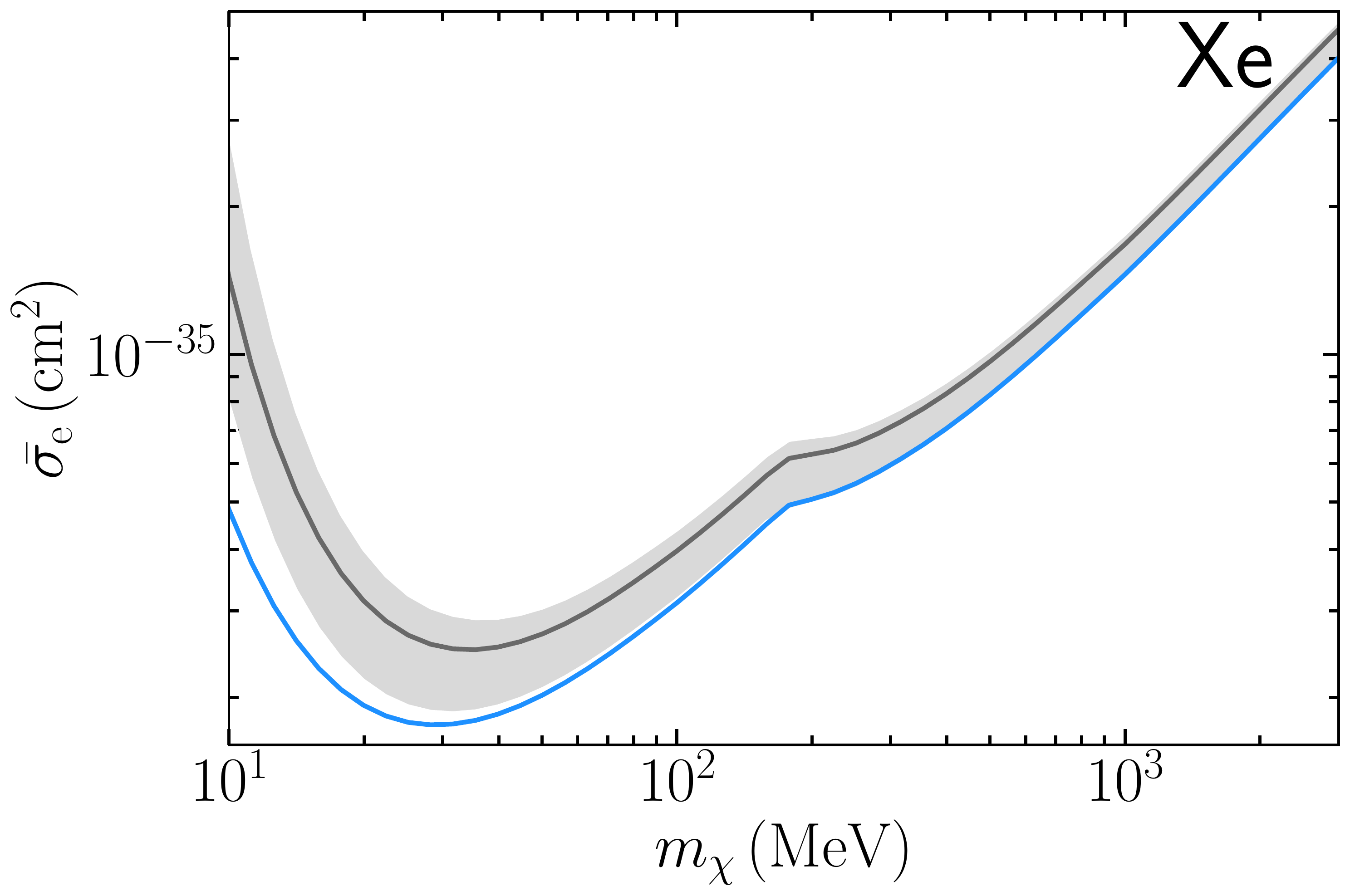}}
\subfloat[\label{sf:SiKingqfa}]{\includegraphics[scale=0.18]{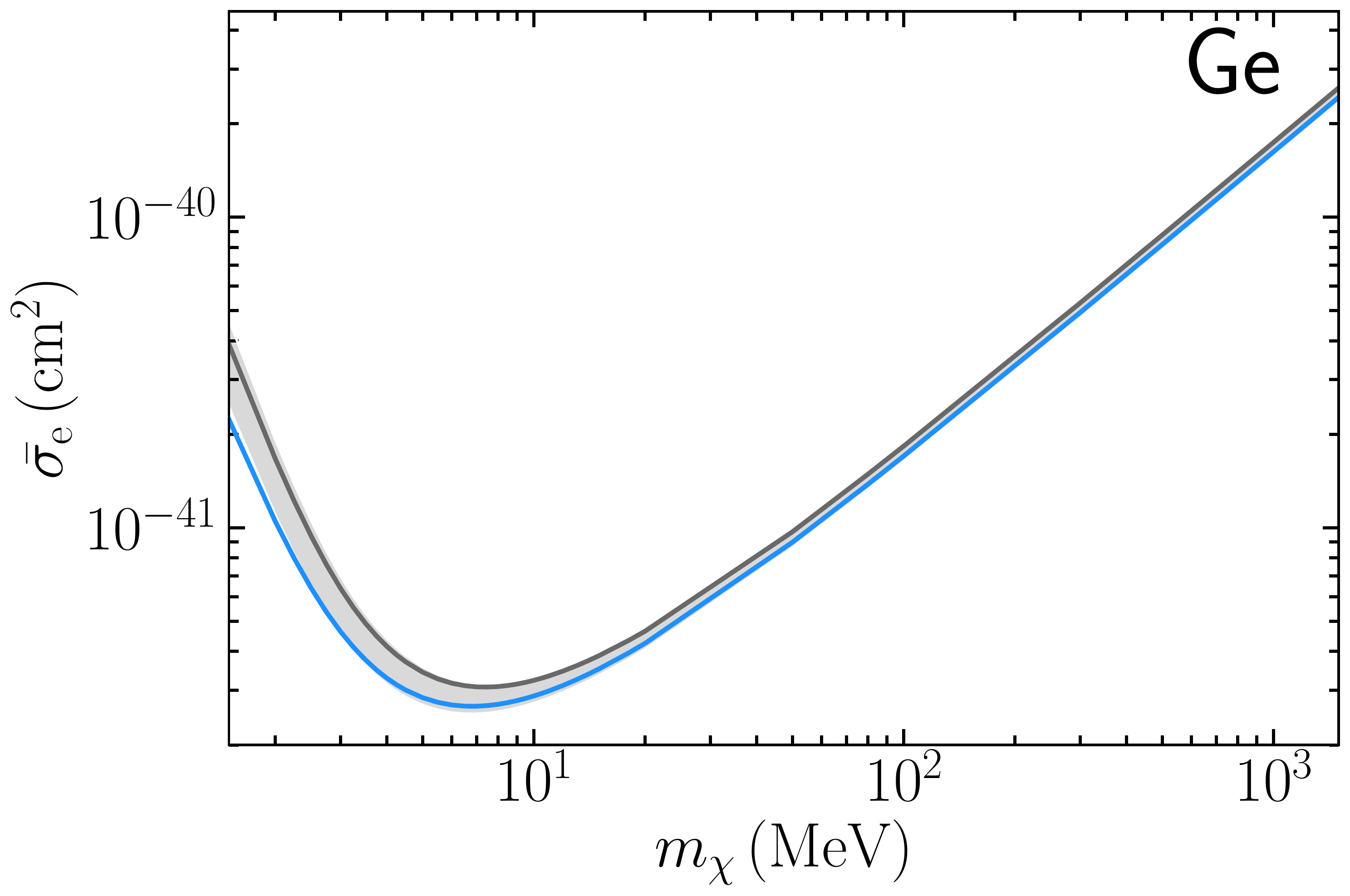}}
\subfloat[\label{sf:GeKingqfa}]{\includegraphics[scale=0.18]{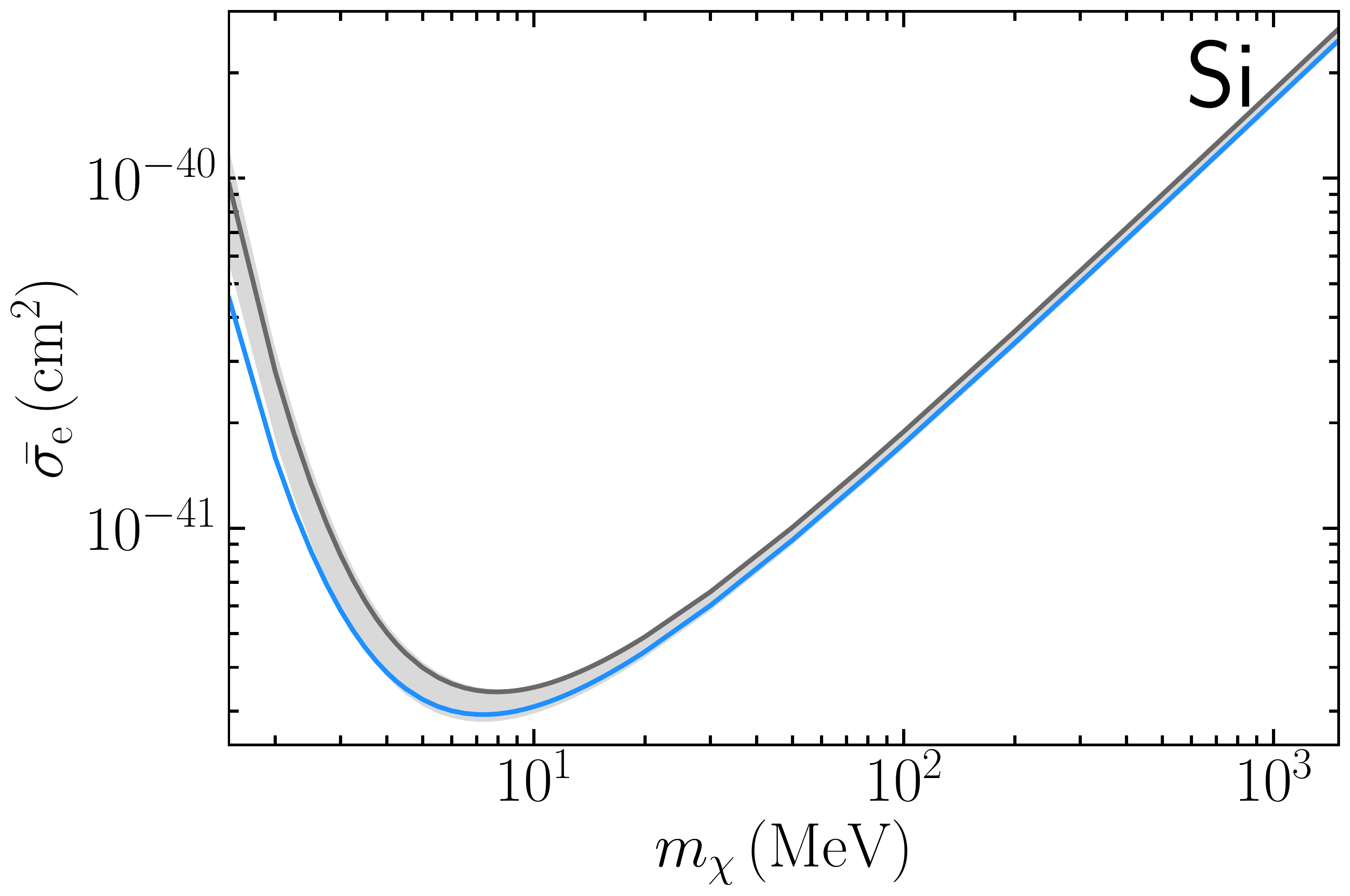}}
\caption{Shifts in the exclusion limits for King distribution with $F_{\rm DM}  \propto q^{-1}$. Upper panel shows the variations for the best fit values of different cosmological simulations. The bands in the lower panel represent uncertainties in the recent astrophysical measurements.  The other relevant details are same as of figures \ref{fig:v0-astro} and \ref{fig:king}.
}
\label{fig:Kingappq}
\end{center}
\end{figure*}

\begin{figure*}[t]
\begin{center}
\subfloat[\label{sf:XeKq2}]{\includegraphics[scale=0.18]{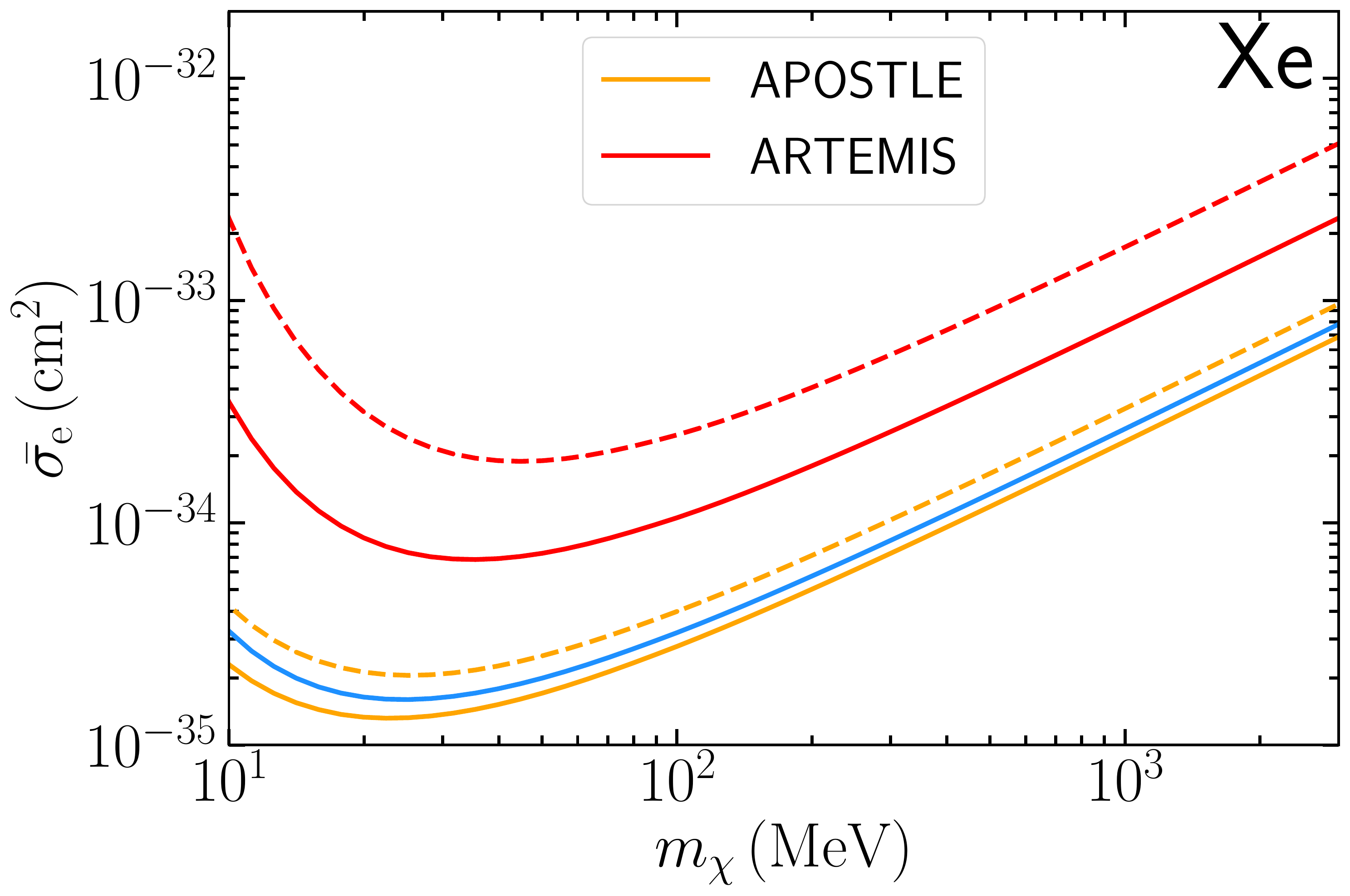}}
\subfloat[\label{sf:SiKq2}]{\includegraphics[scale=0.18]{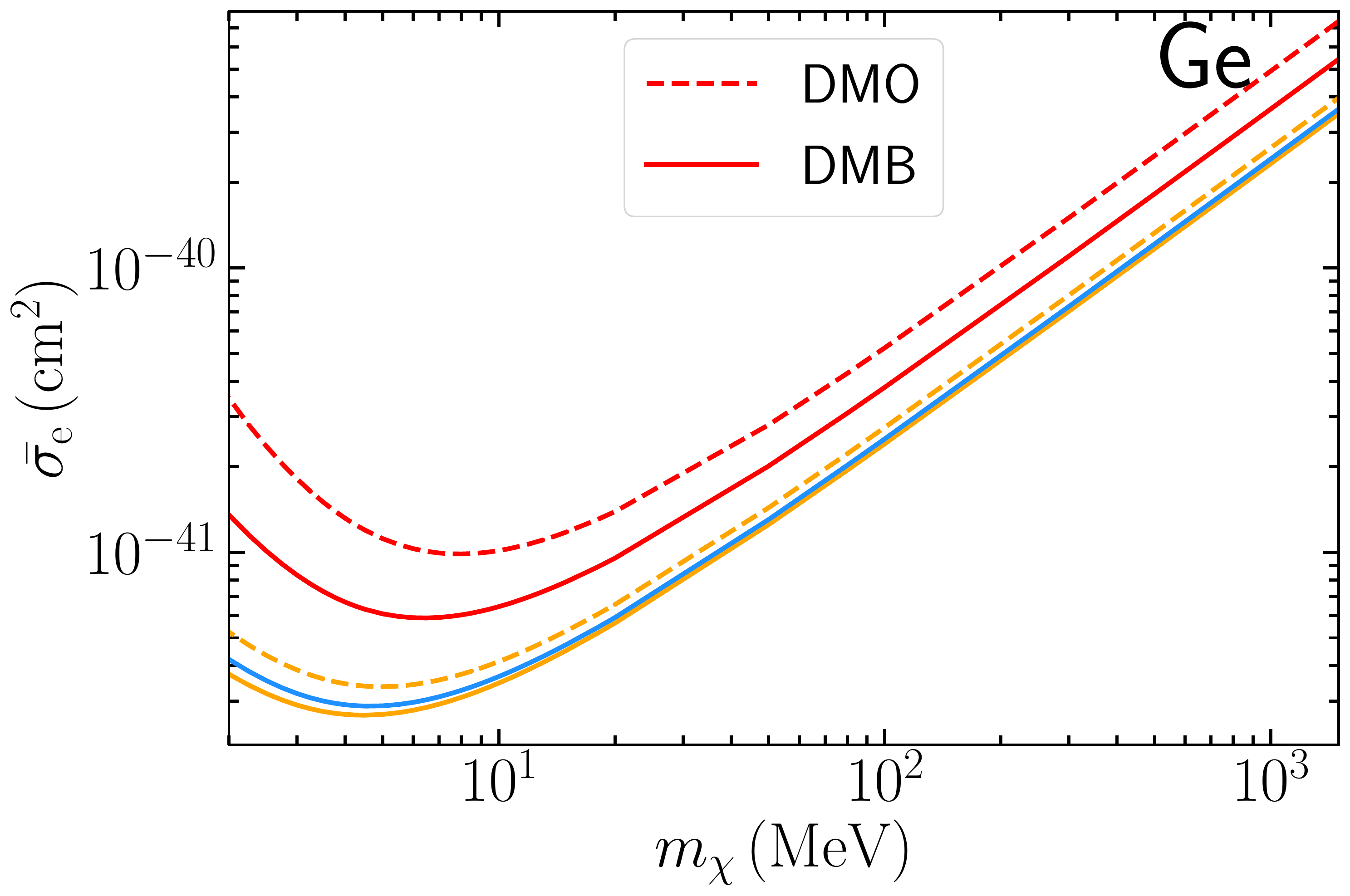}}
\subfloat[\label{sf:GeKq2}]{\includegraphics[scale=0.18]{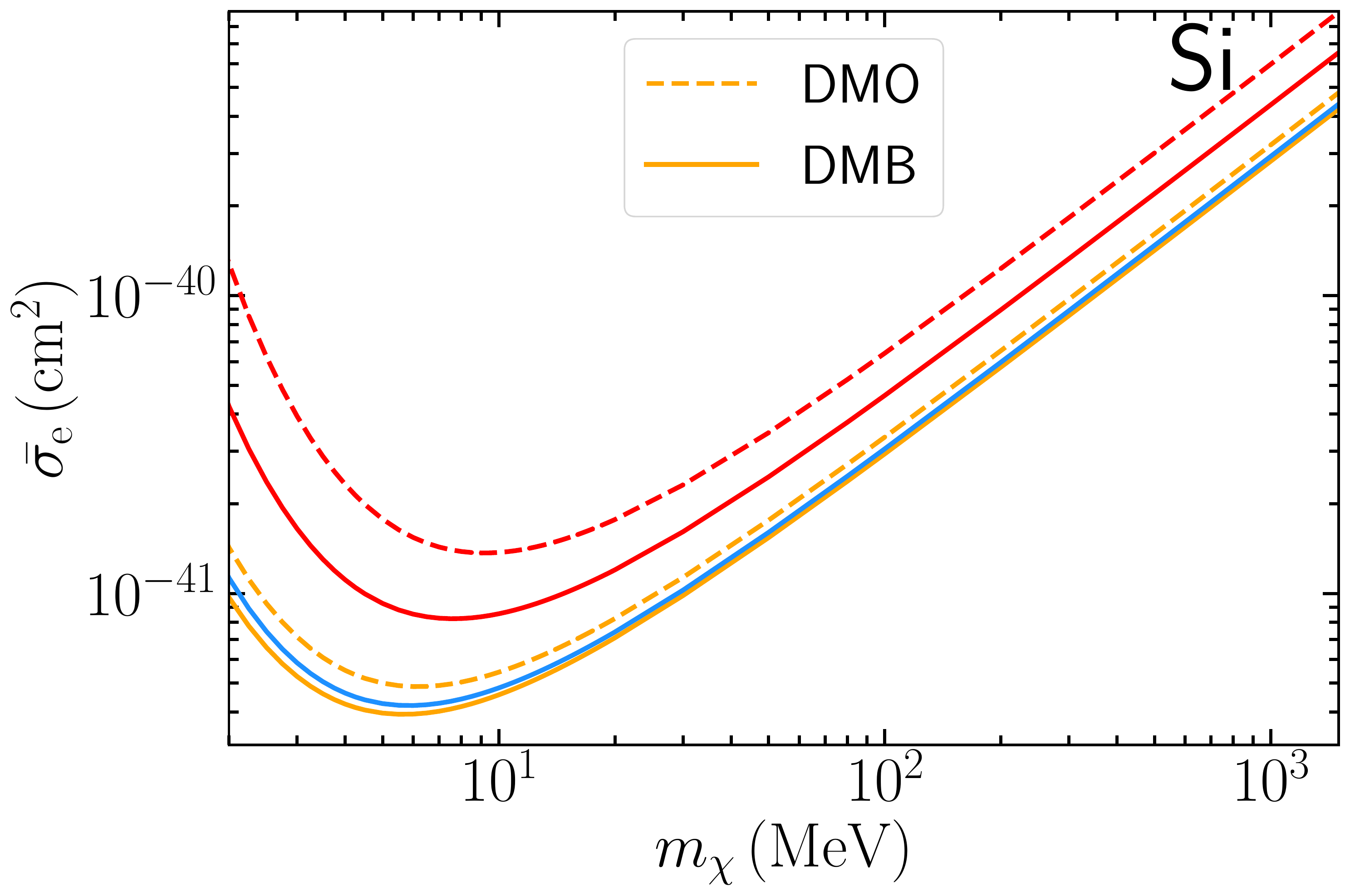}}
\newline
\subfloat[\label{sf:XeKq2fa}]{\includegraphics[scale=0.18]{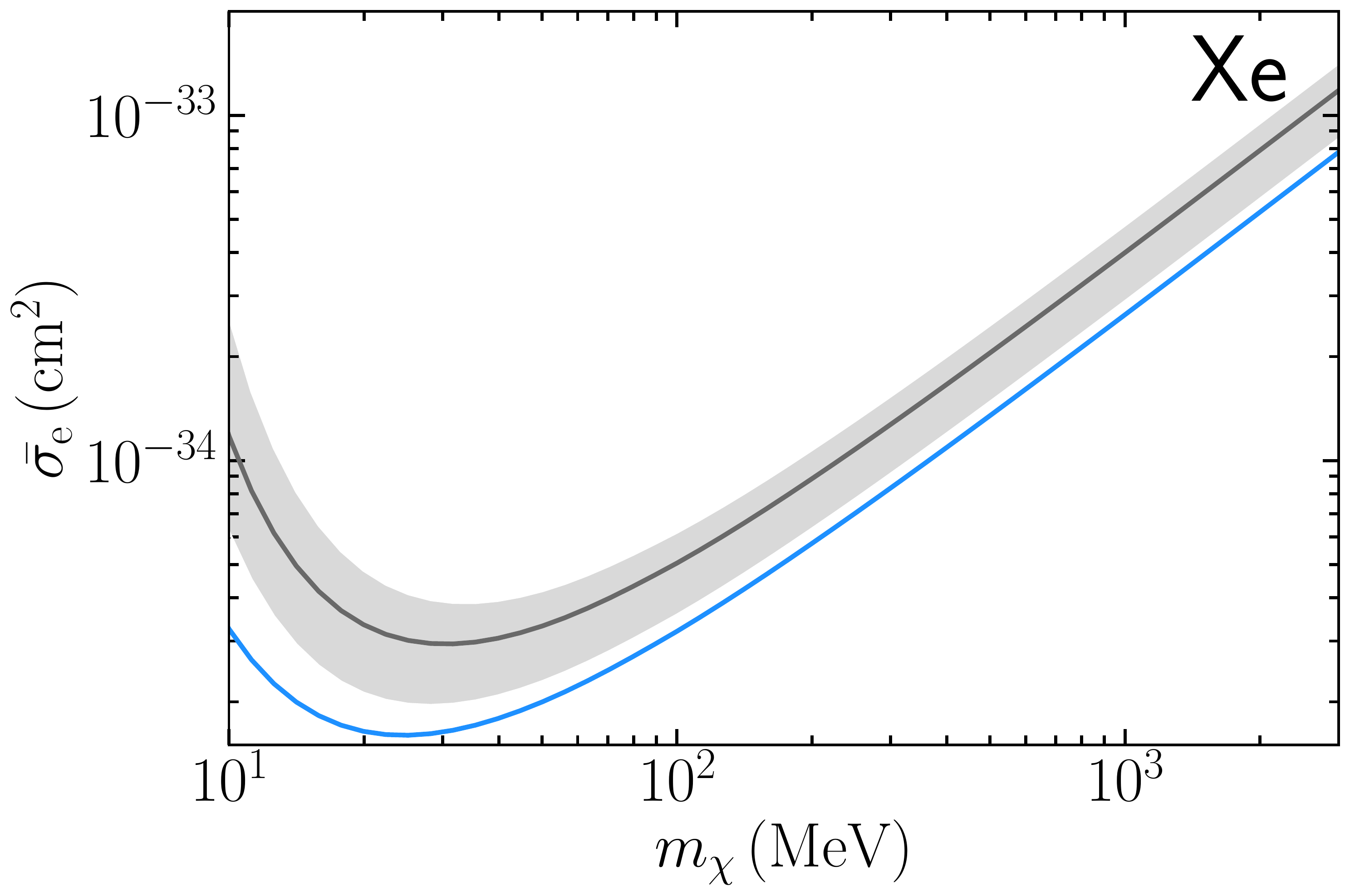}}
\subfloat[\label{sf:SiKq2fa}]{\includegraphics[scale=0.18]{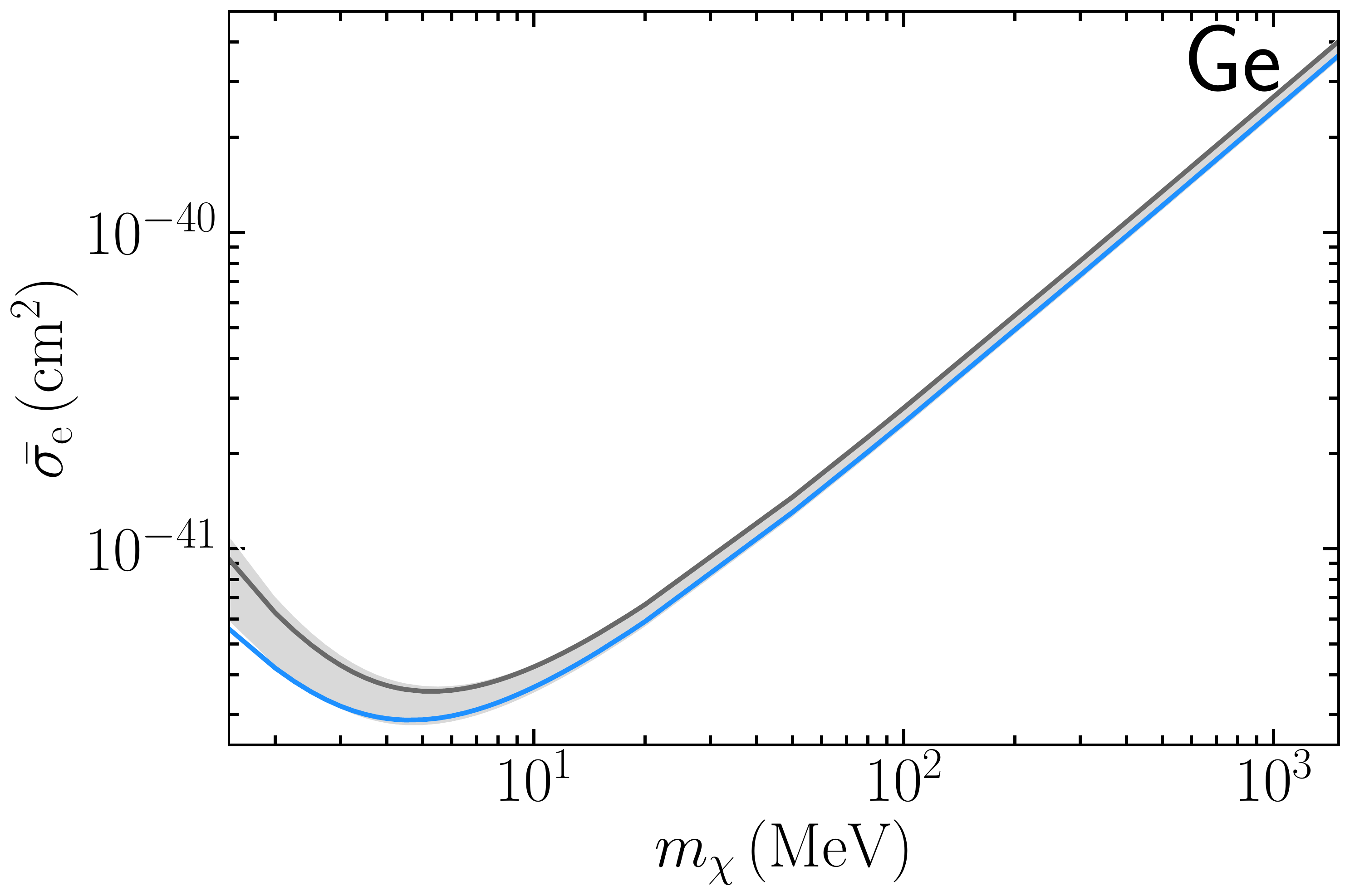}}
\subfloat[\label{sf:GeKq2fa}]{\includegraphics[scale=0.18]{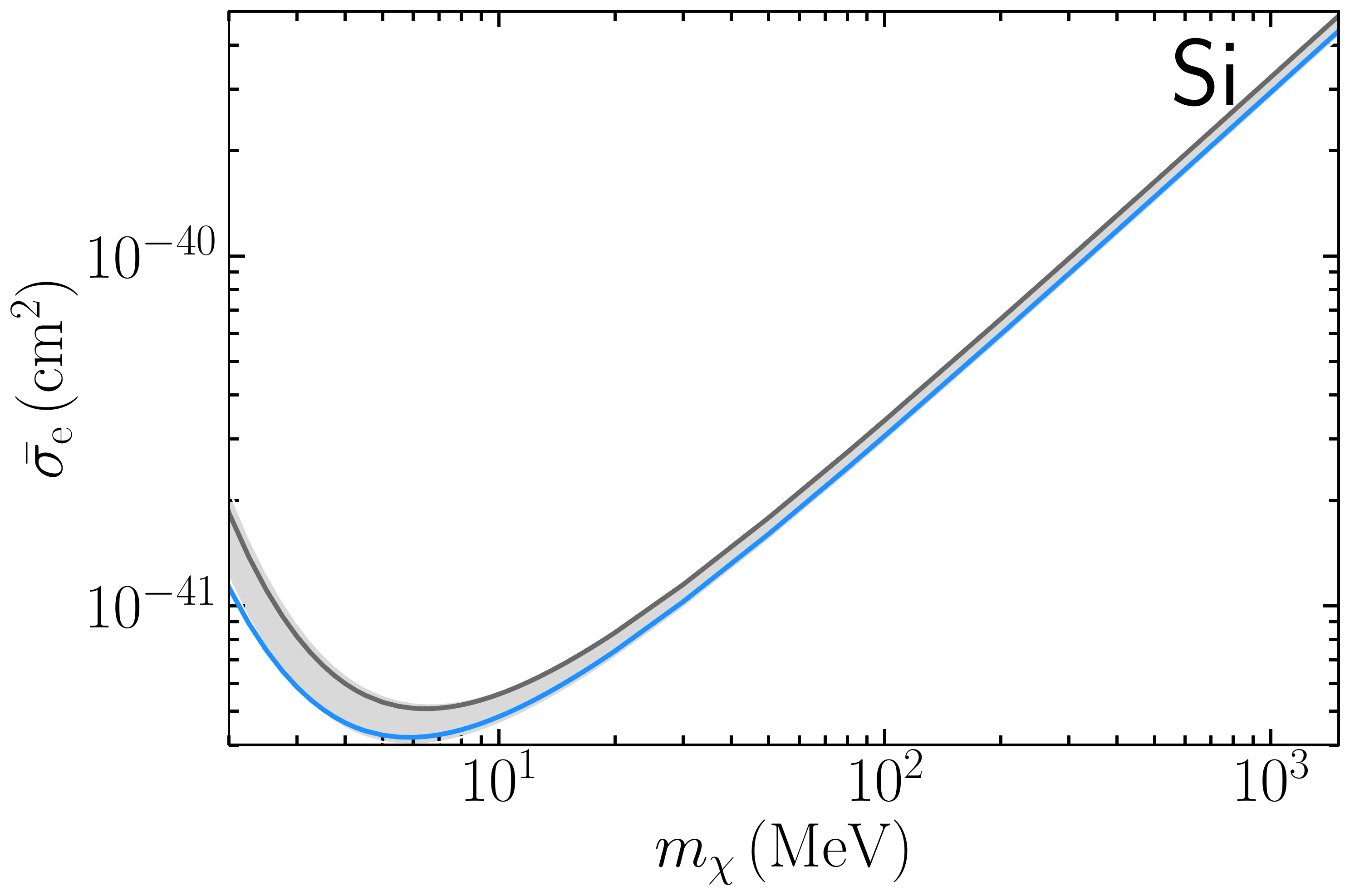}}
\caption{Same as figure \ref{fig:Kingappq} but for $F_{\rm DM}  \propto q^{-2}$.}
\label{fig:Kingappq2}
\end{center}
\end{figure*}

\begin{figure*}[t!]
\begin{center}
\subfloat[\label{sf:XeDPLq}]{\includegraphics[scale=0.18]{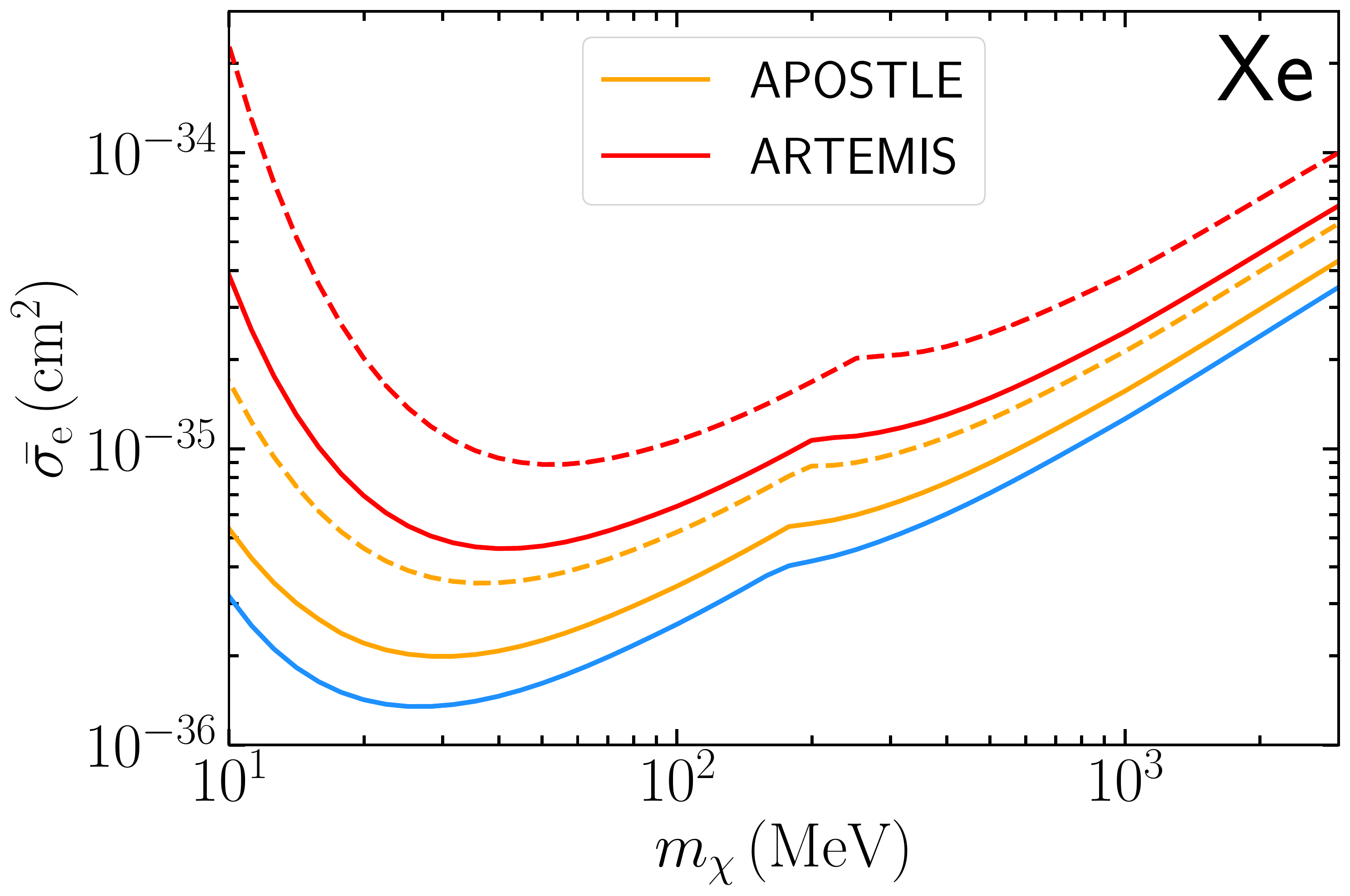}}
\subfloat[\label{sf:SiDPLq}]{\includegraphics[scale=0.18]{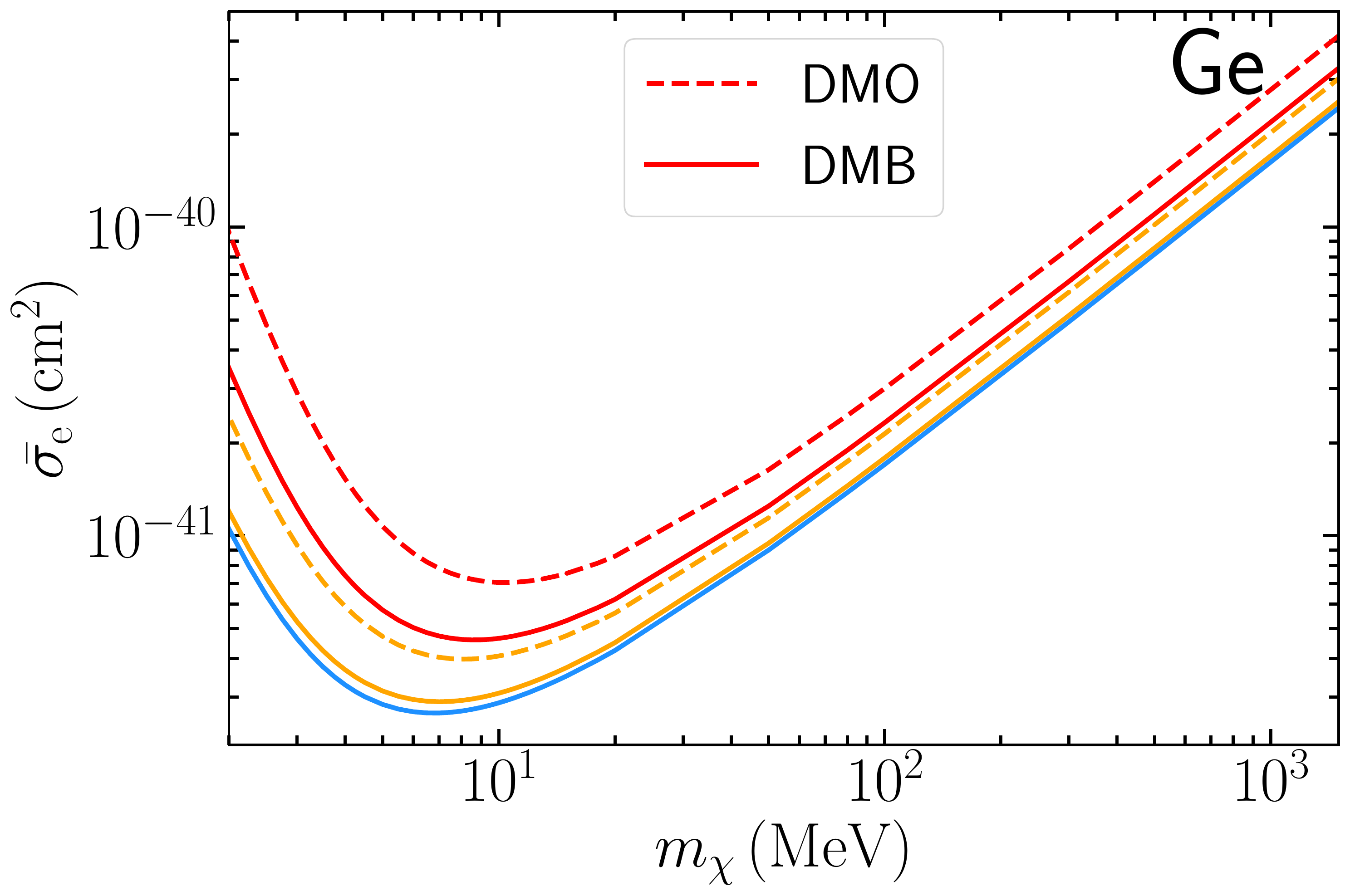}}
\subfloat[\label{sf:GeDPLq}]{\includegraphics[scale=0.18]{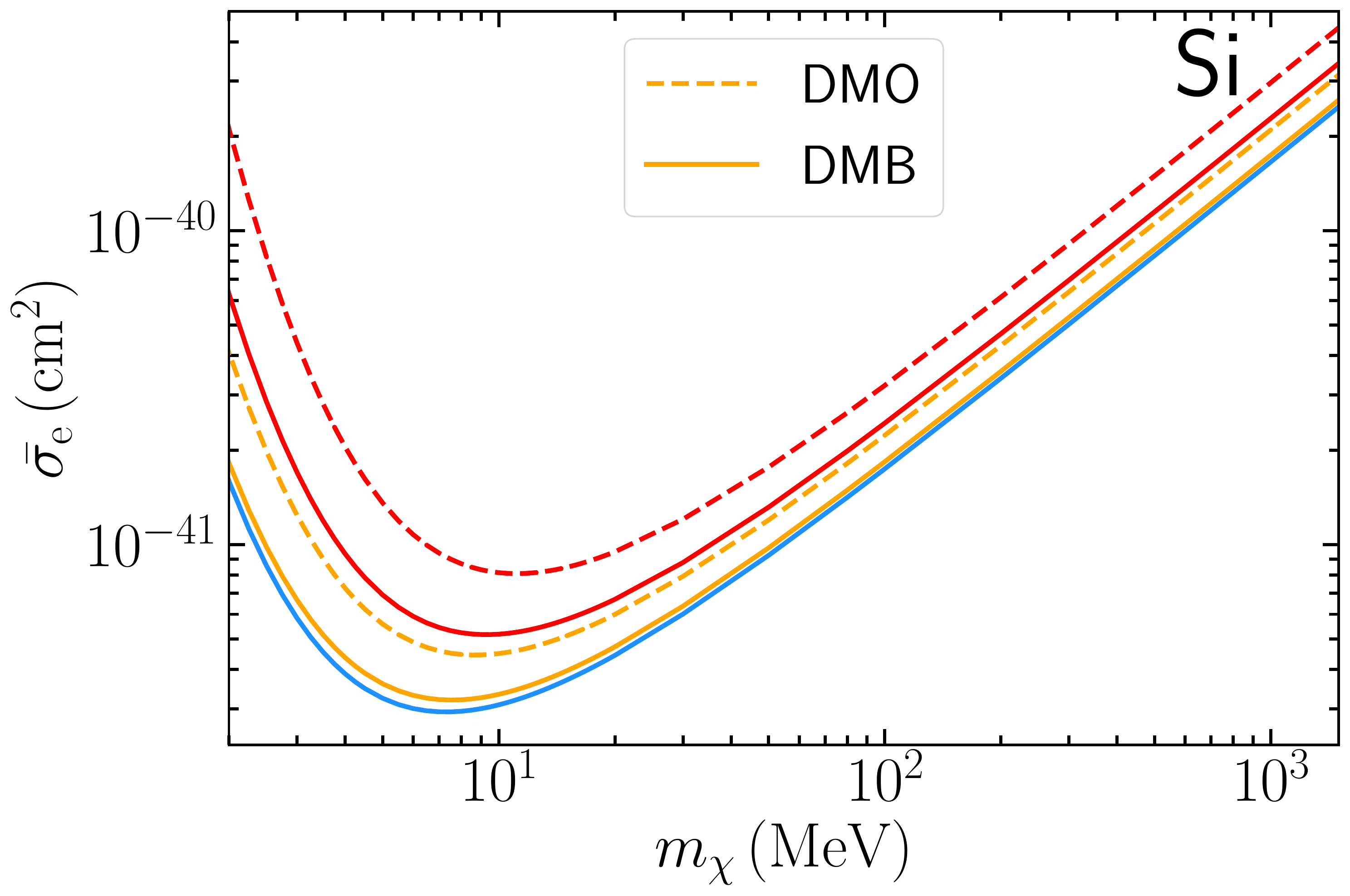}}
\newline
\subfloat[\label{sf:XeDPLqfa}]{\includegraphics[scale=0.18]{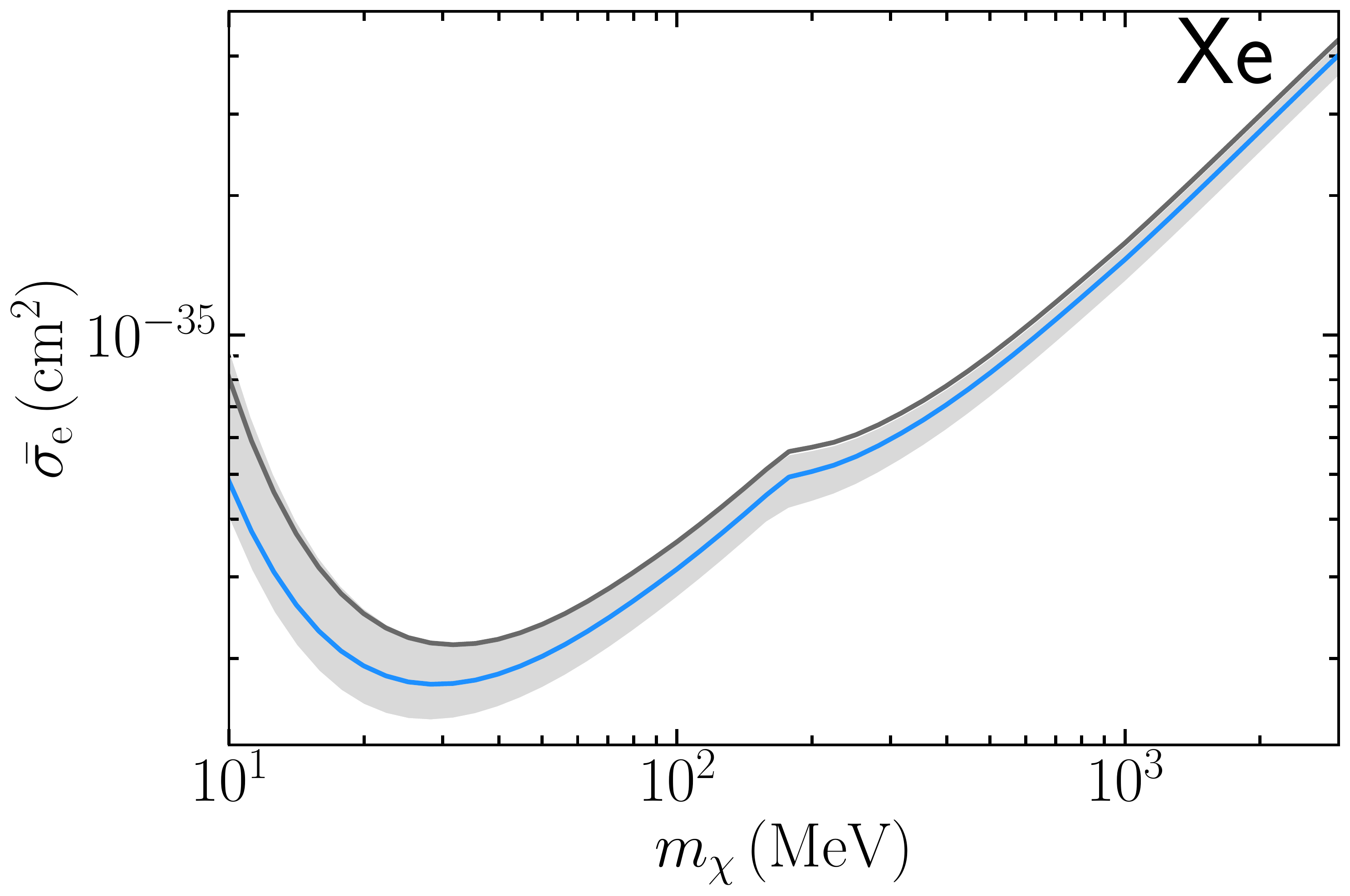}}
\subfloat[\label{sf:SiDPLqfa}]{\includegraphics[scale=0.18]{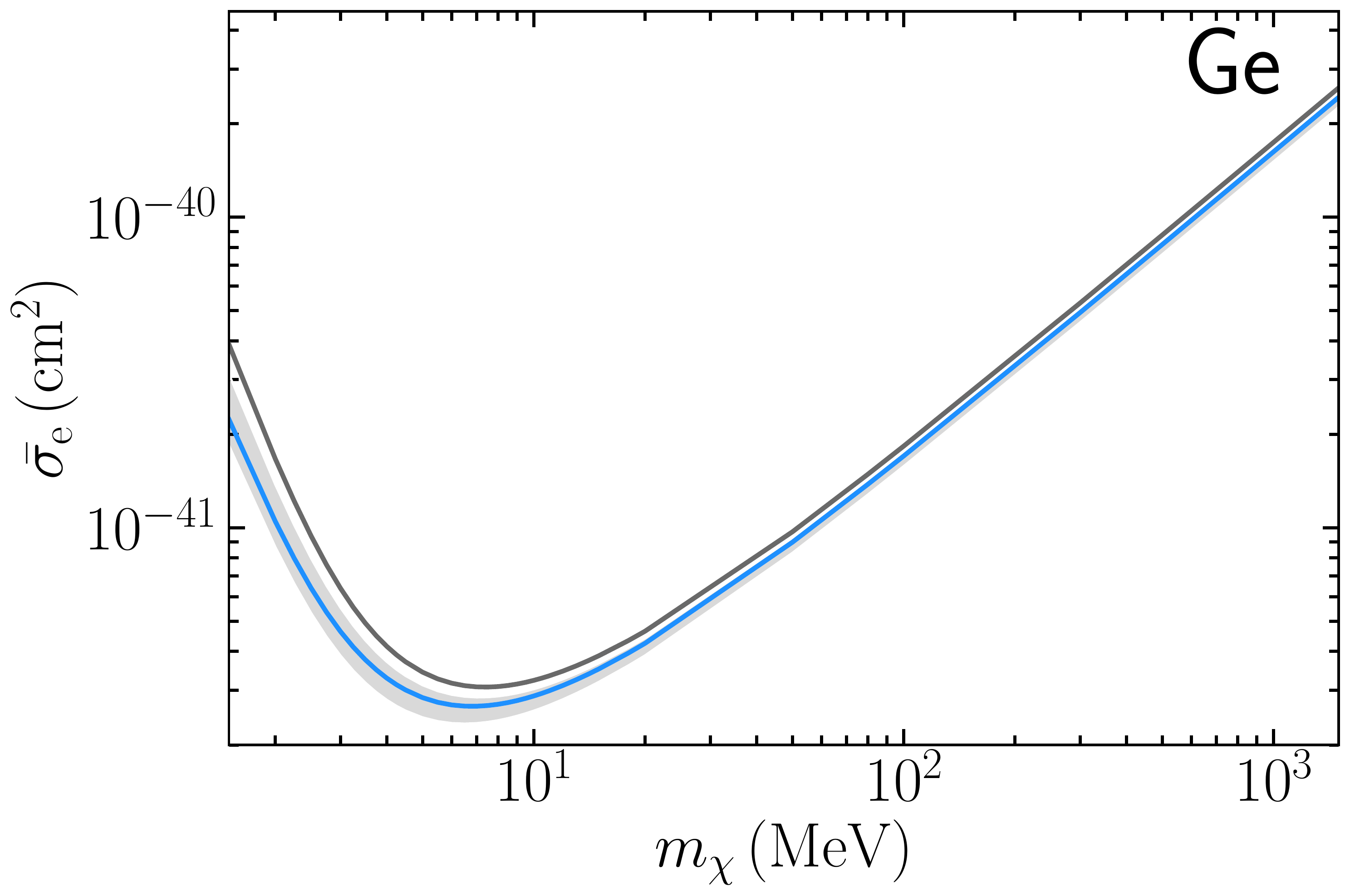}}
\subfloat[\label{sf:GeDPLqfa}]{\includegraphics[scale=0.18]{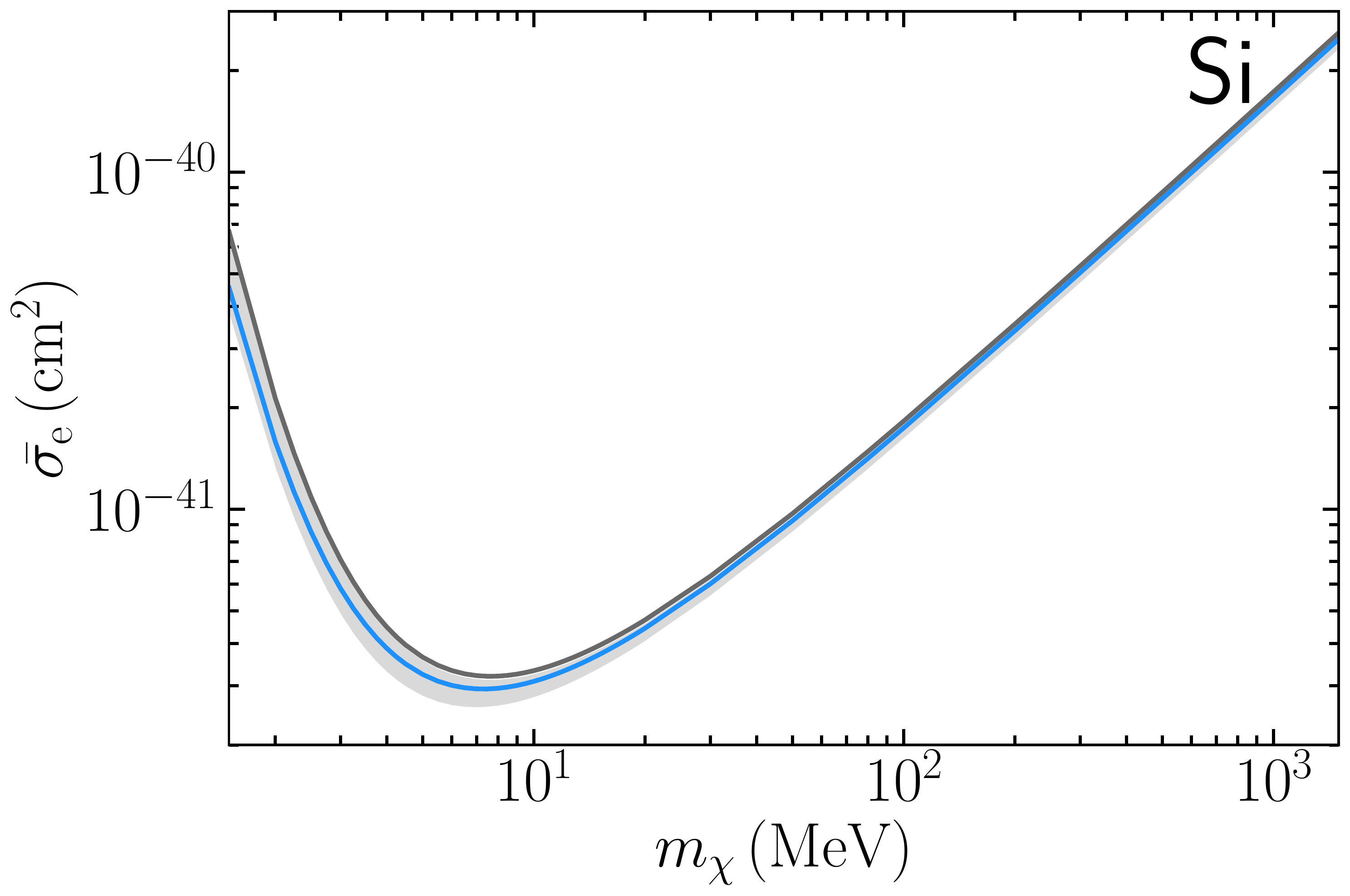}}
\caption{Shifts in the exclusion limits for DPL distribution with $F_{\rm DM}  \propto q^{-1}$. In the upper panel we show the variations for the best fit values of different  cosmological simulations. The bands in the lower panel represents uncertainties in the recent astrophysical measurements.  The other relevant details are same as of figures \ref{fig:v0-astro} and \ref{fig:DPL}.}
\label{fig:DPLappq}
\end{center}
\end{figure*}

\begin{figure*}[t]
\begin{center}
\subfloat[\label{sf:XeDPLq2}]{\includegraphics[scale=0.18]{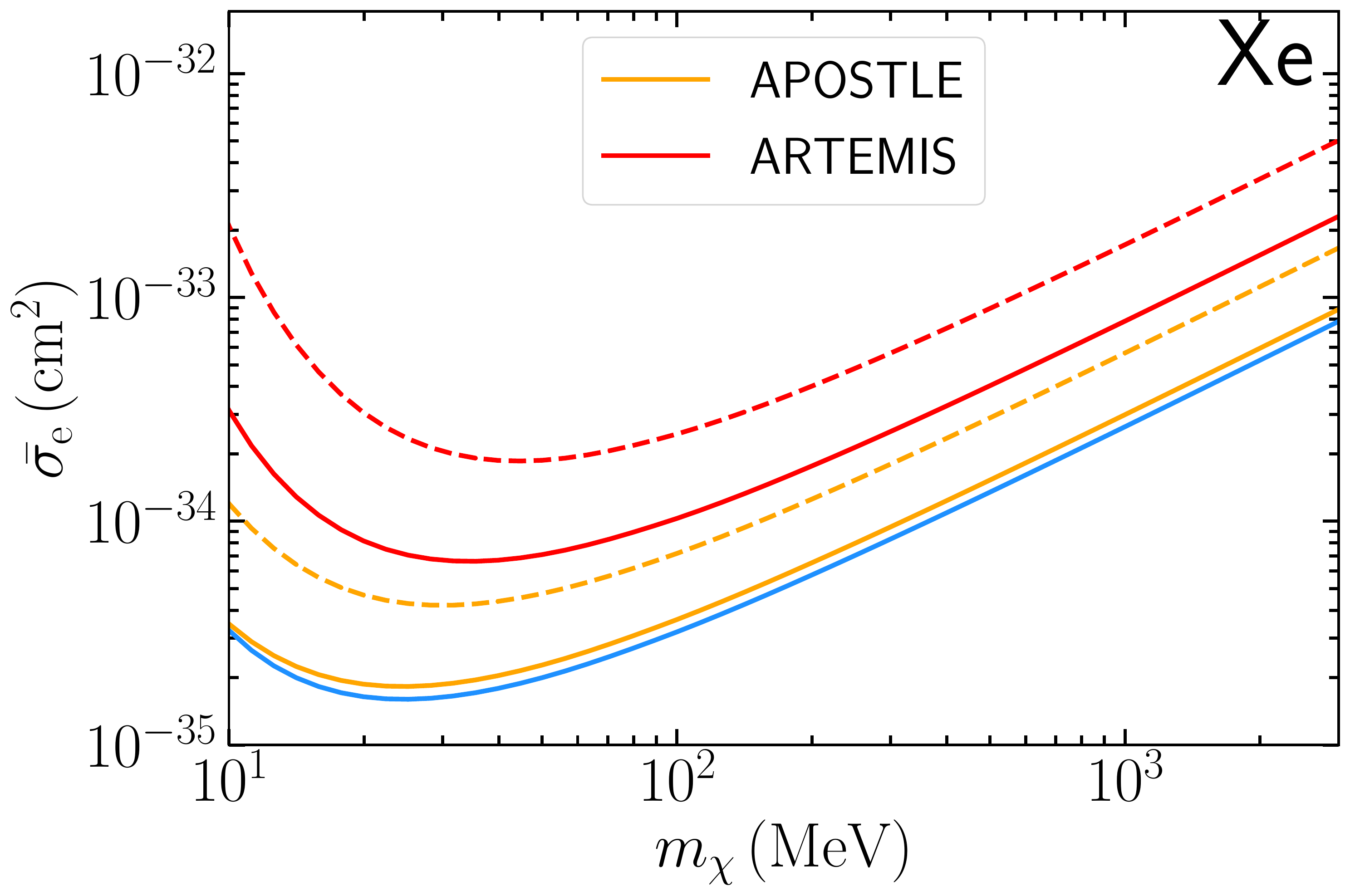}}
\subfloat[\label{sf:SiDPLq2}]{\includegraphics[scale=0.18]{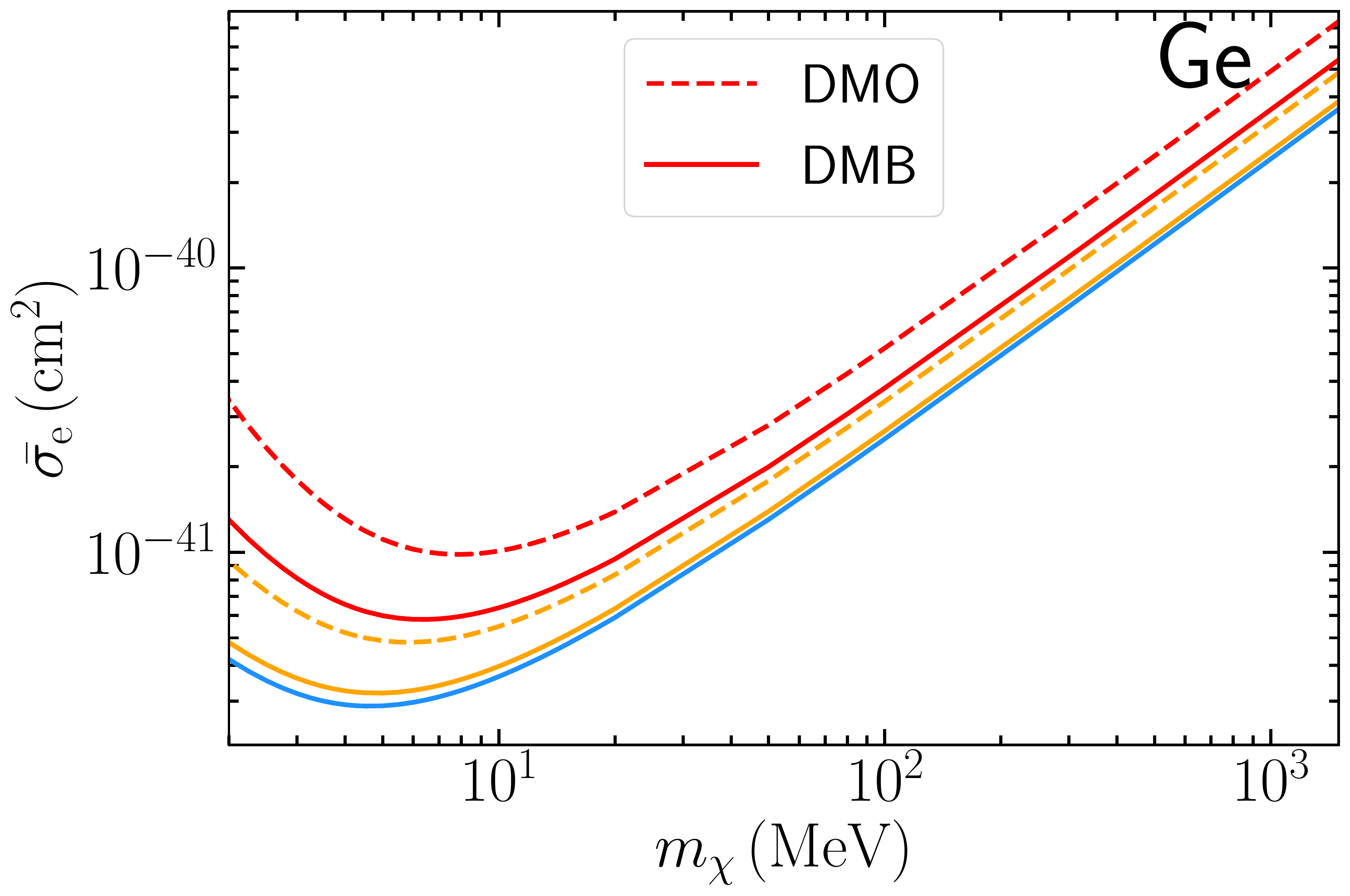}}
\subfloat[\label{sf:GeDPLq2}]{\includegraphics[scale=0.18]{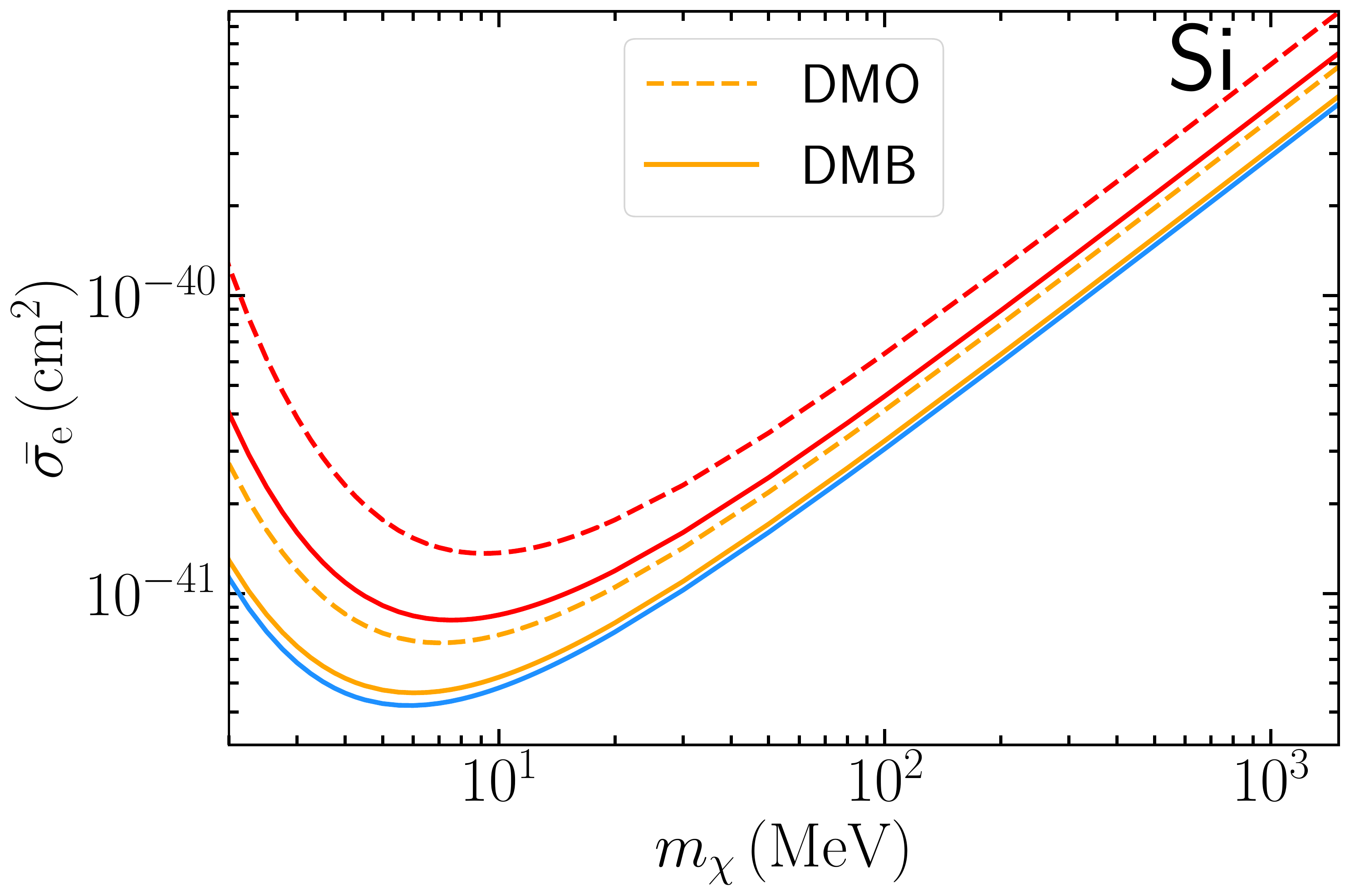}}
\newline
\subfloat[\label{sf:XeDPLq2fa}]{\includegraphics[scale=0.19]{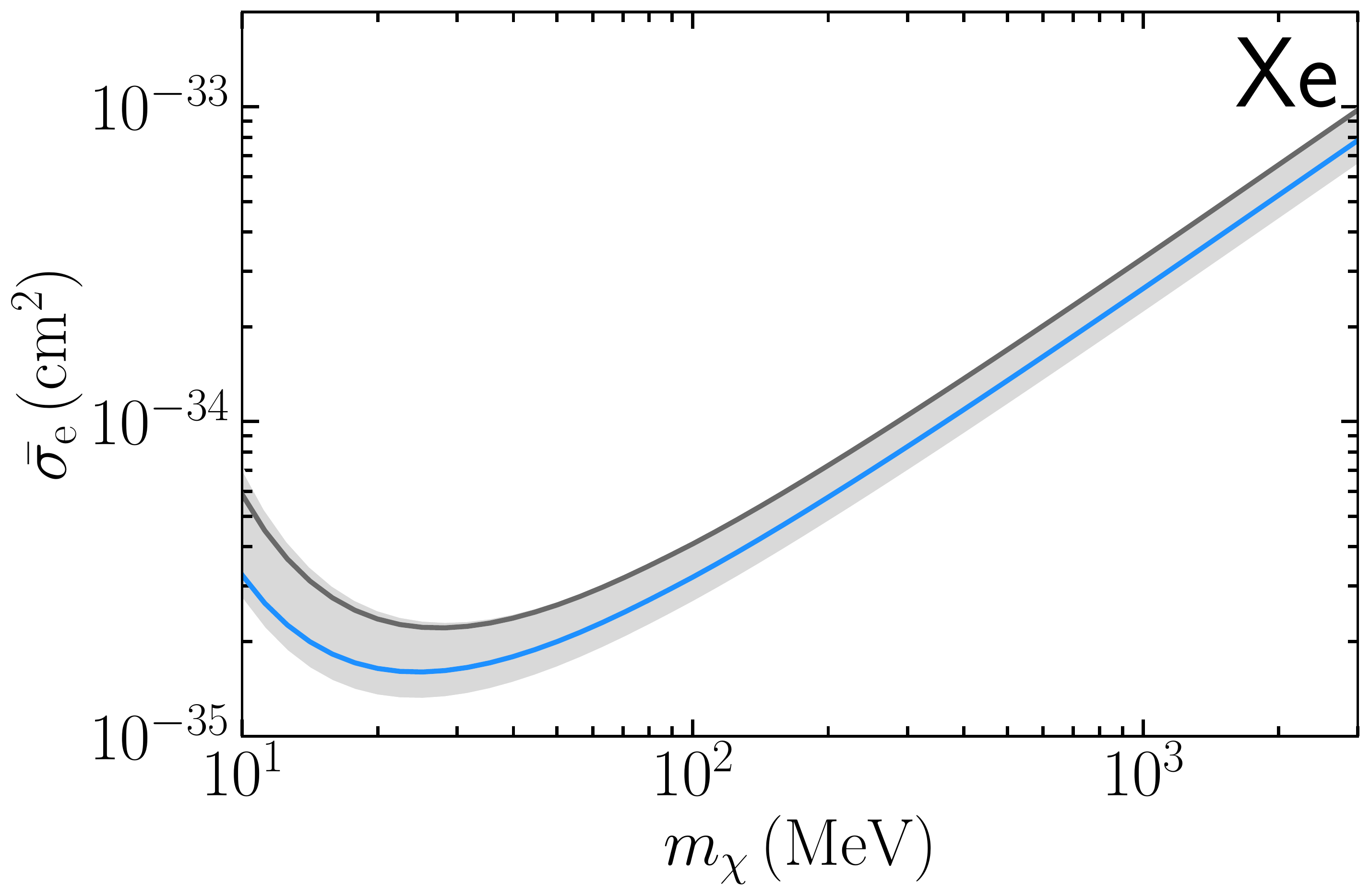}}
\subfloat[\label{sf:SiDPLq2fa}]{\includegraphics[scale=0.18]{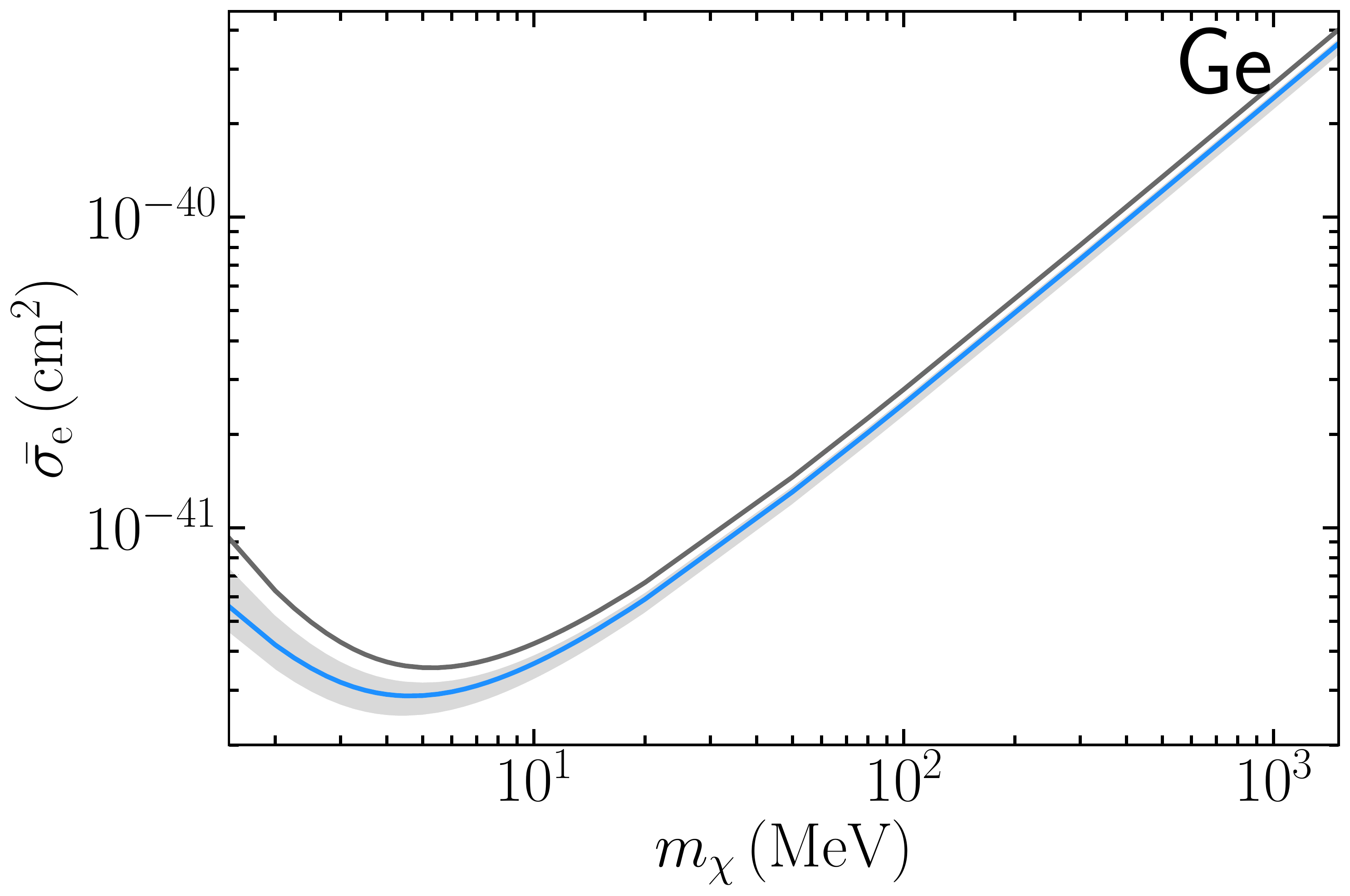}}
\subfloat[\label{sf:GeDPLq2fa}]{\includegraphics[scale=0.18]{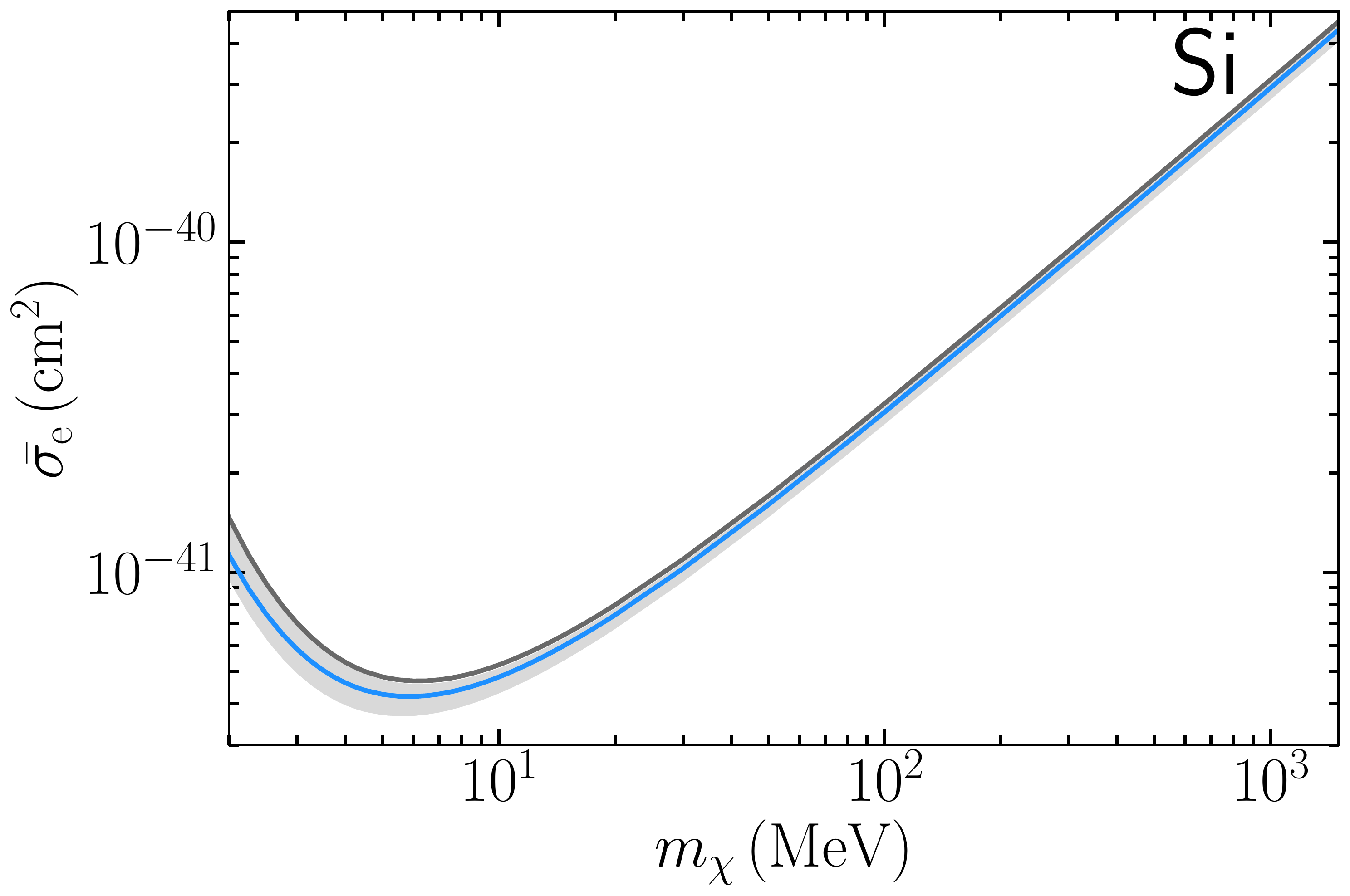}}
\caption{Same as figure \ref{fig:DPLappq} but for $F_{\rm DM}  \propto q^{-2}$.}
\label{fig:DPLappq2}
\end{center}
\end{figure*}

\begin{figure*}[t]
\begin{center}
\subfloat[\label{sf:XeTsaq}]{\includegraphics[scale=0.18]{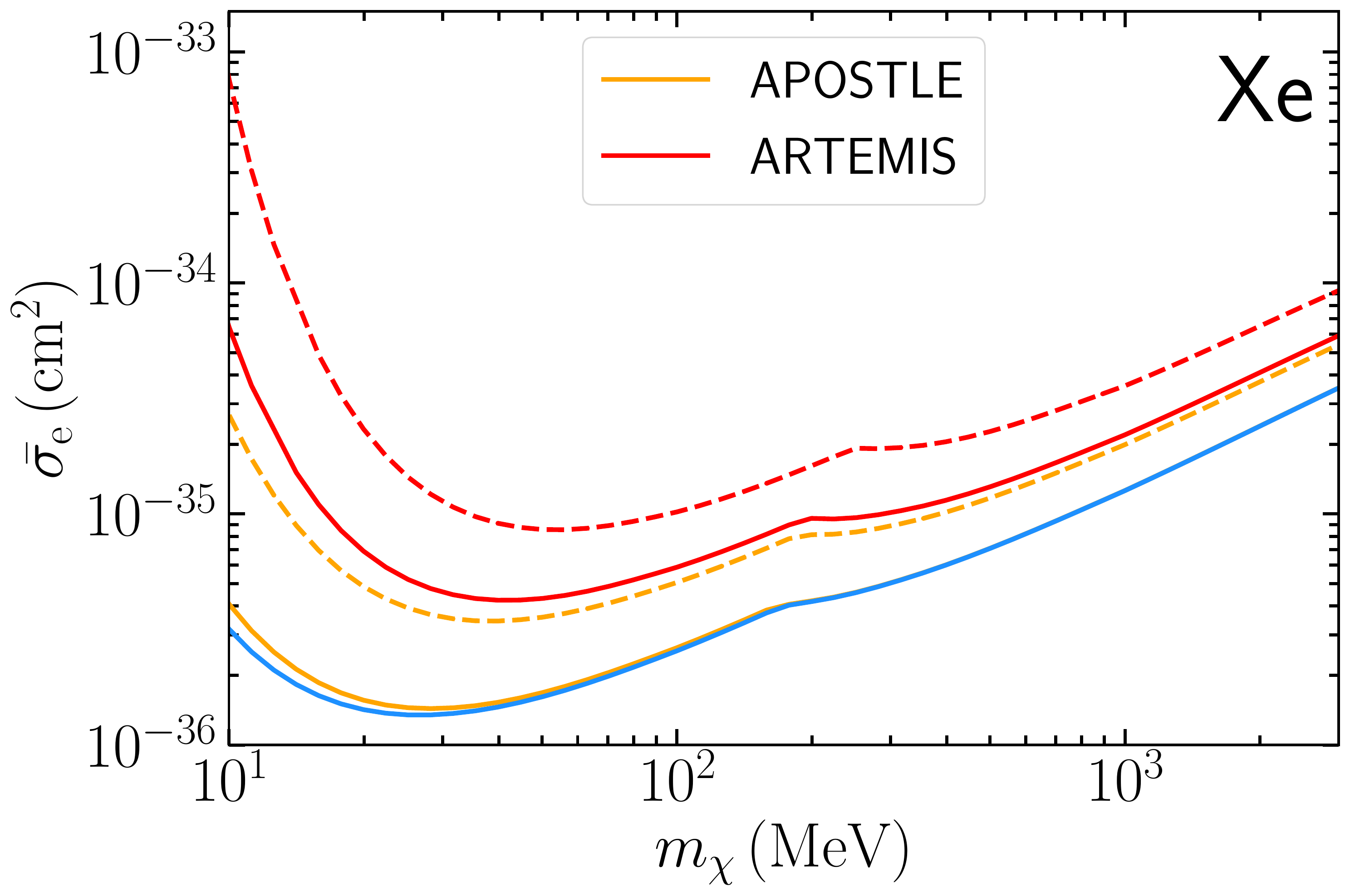}}
\subfloat[\label{sf:SiTsaq}]{\includegraphics[scale=0.18]{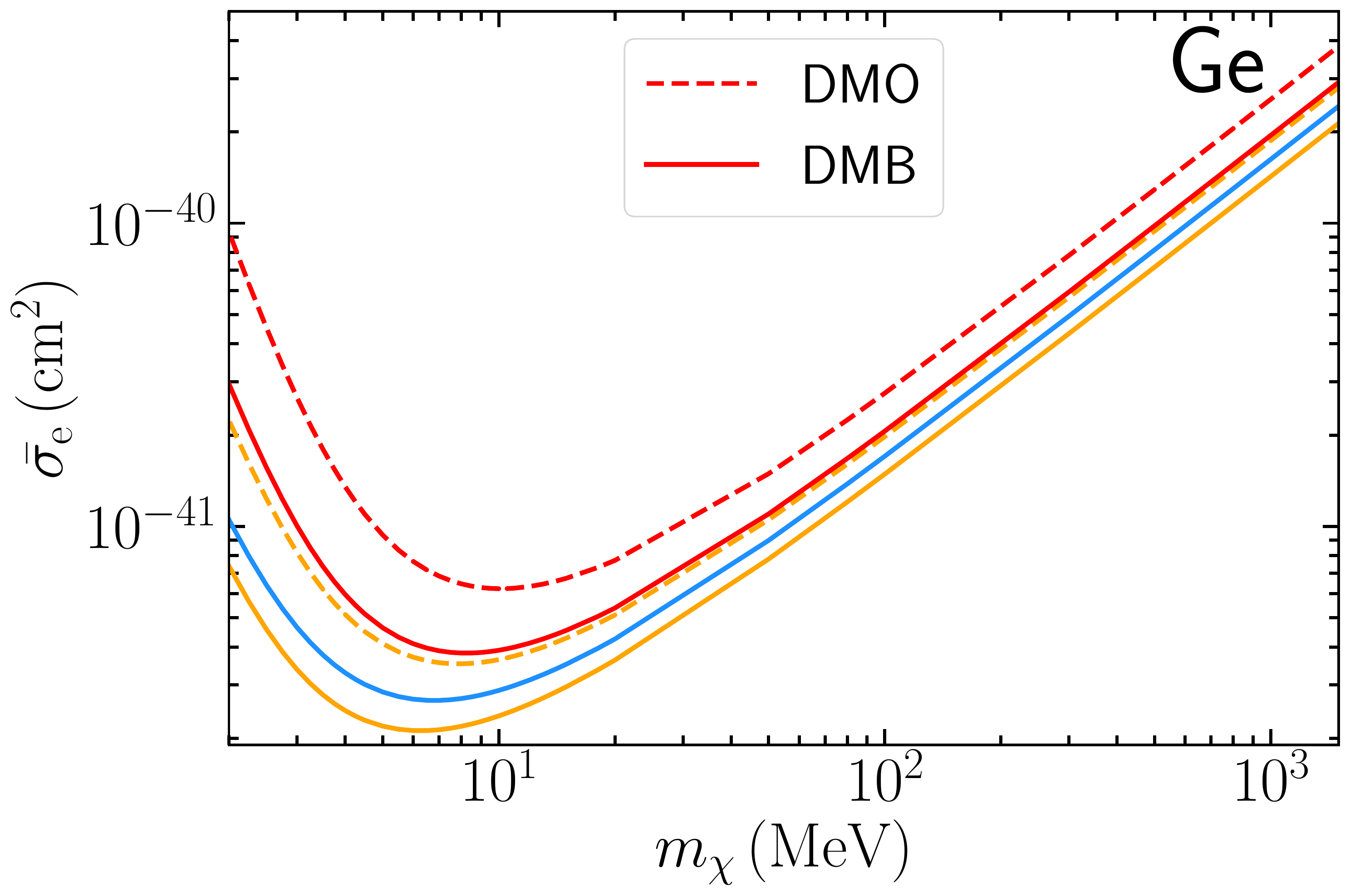}}
\subfloat[\label{sf:GeTsaq}]{\includegraphics[scale=0.18]{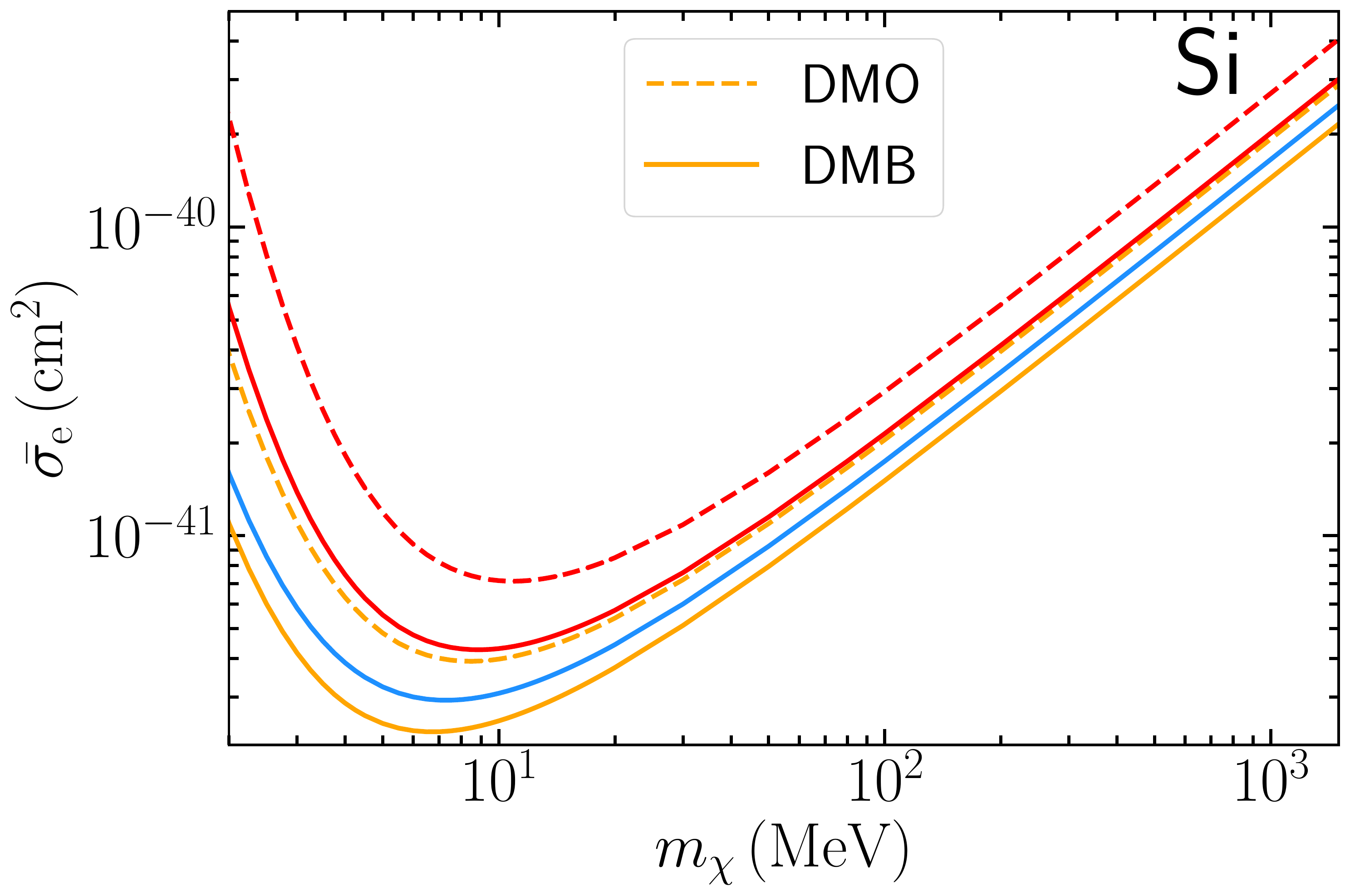}}
\newline
\subfloat[\label{sf:XeTsaqfa}]{\includegraphics[scale=0.18]{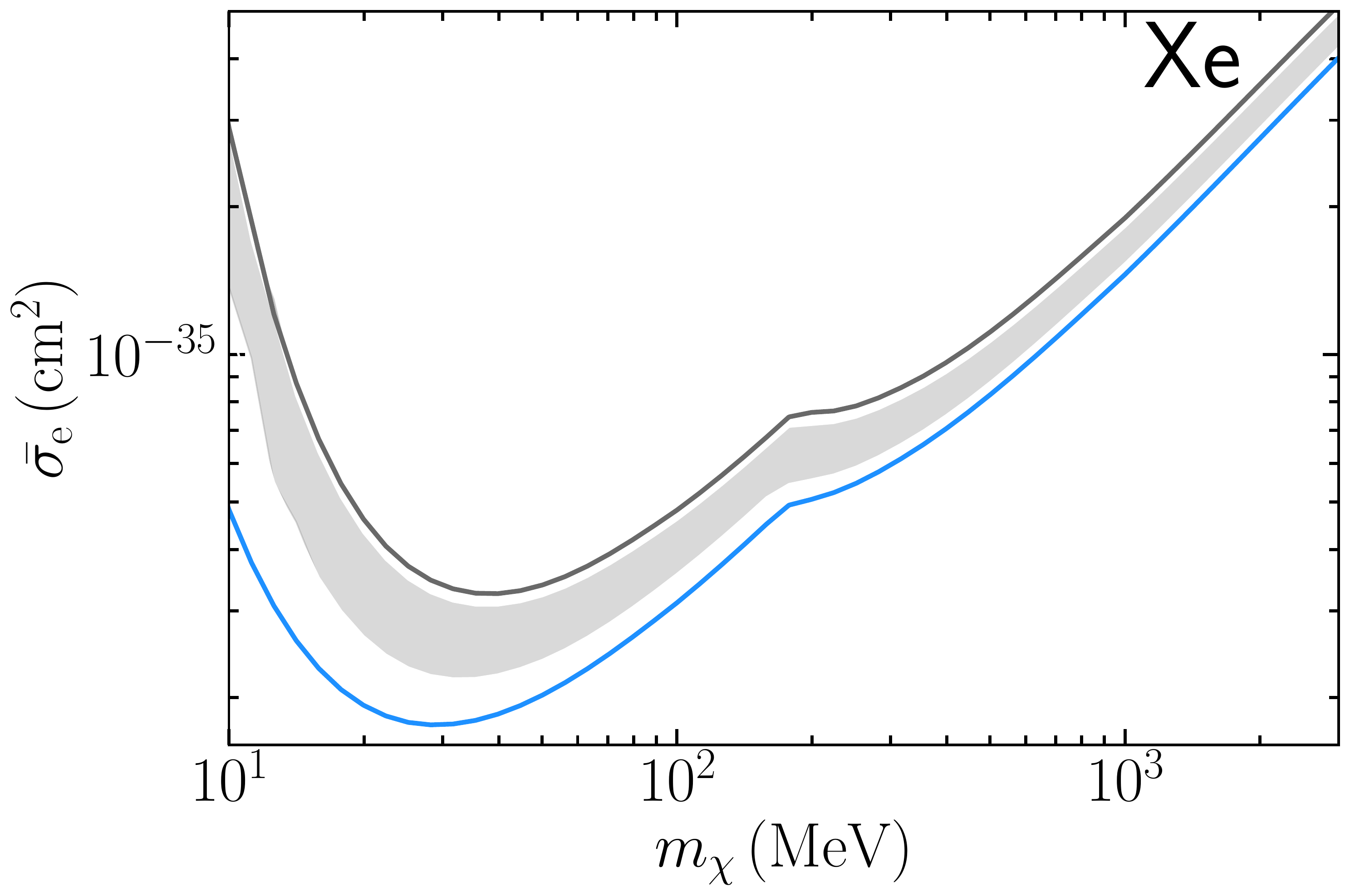}}
\subfloat[\label{sf:SiTsaqfa}]{\includegraphics[scale=0.18]{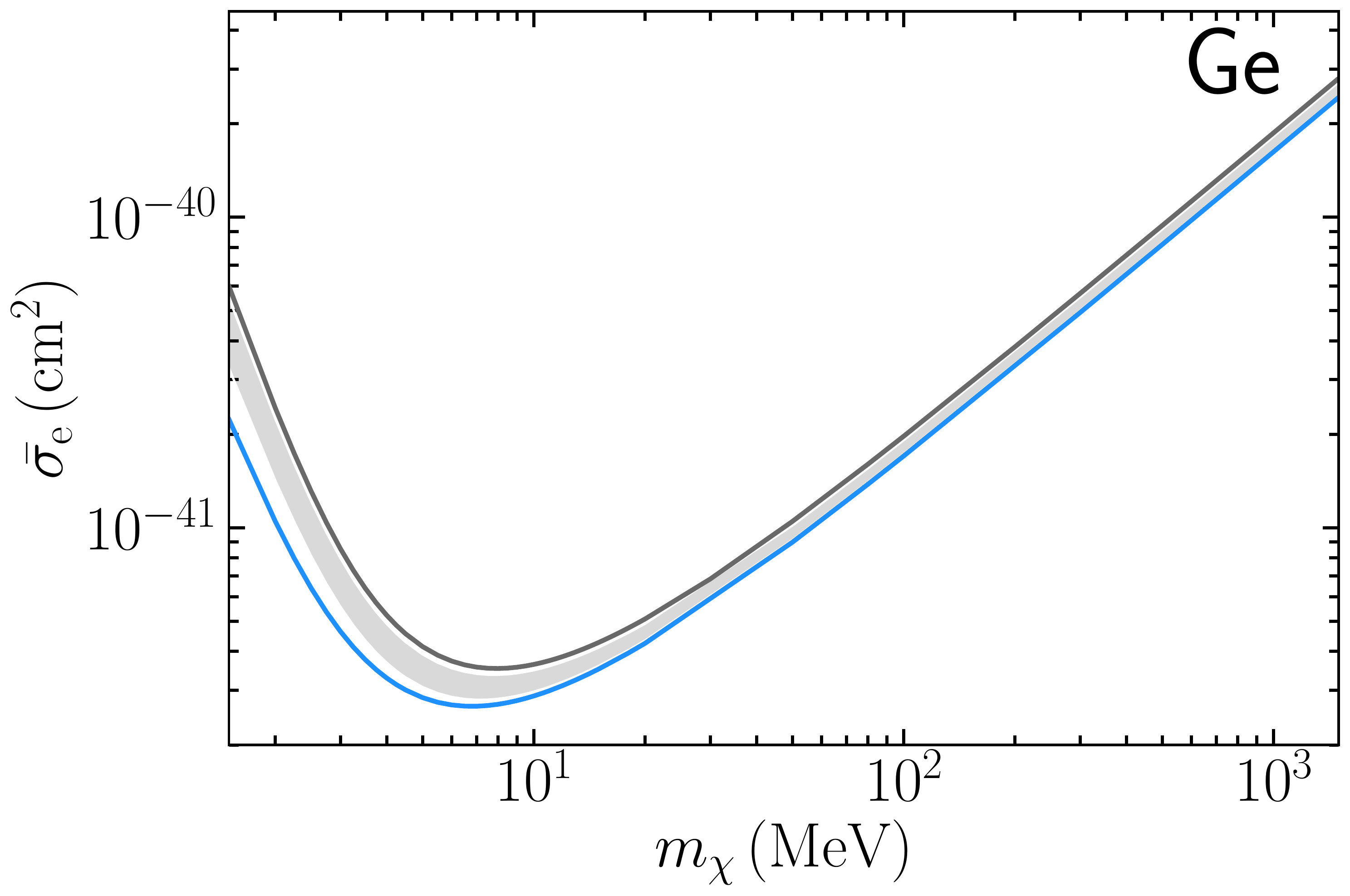}}
\subfloat[\label{sf:GeTsaqfa}]{\includegraphics[scale=0.18]{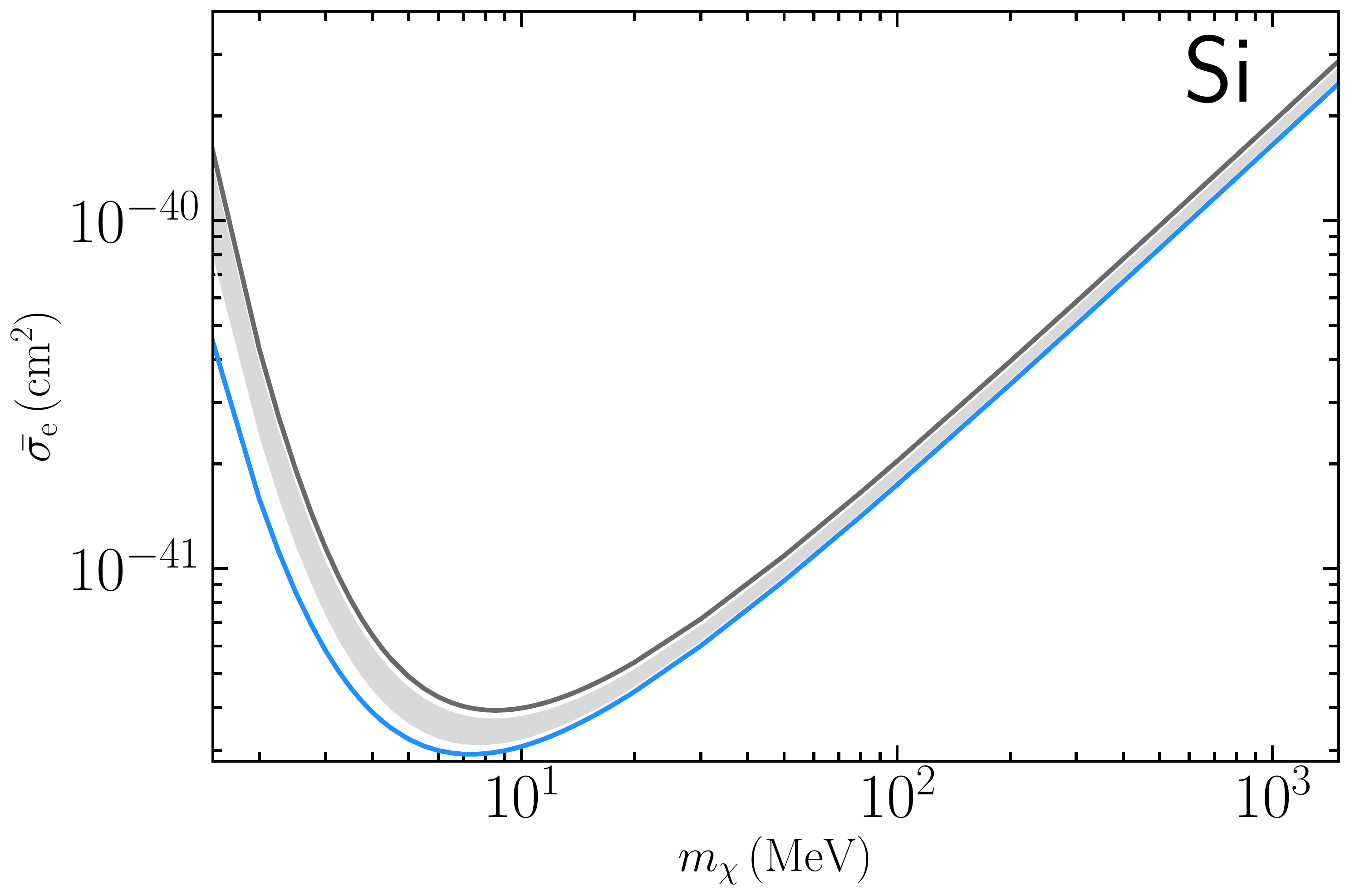}}
\caption{Shifts in the exclusion limits for Tsallis distribution with $F_{\rm DM}  \propto q^{-1}$. In the upper panel we show the variations for the best fit values of different  cosmological simulations. The bands in the lower panel represents uncertainties in the recent astrophysical measurements.  The other relevant details are same as of figures \ref{fig:v0-astro} and \ref{fig:Tsallis}.}
\label{fig:Tsaappq}
\end{center}
\end{figure*}

\begin{figure*}[t]
\begin{center}
\subfloat[\label{sf:XeTsaq2}]{\includegraphics[scale=0.18]{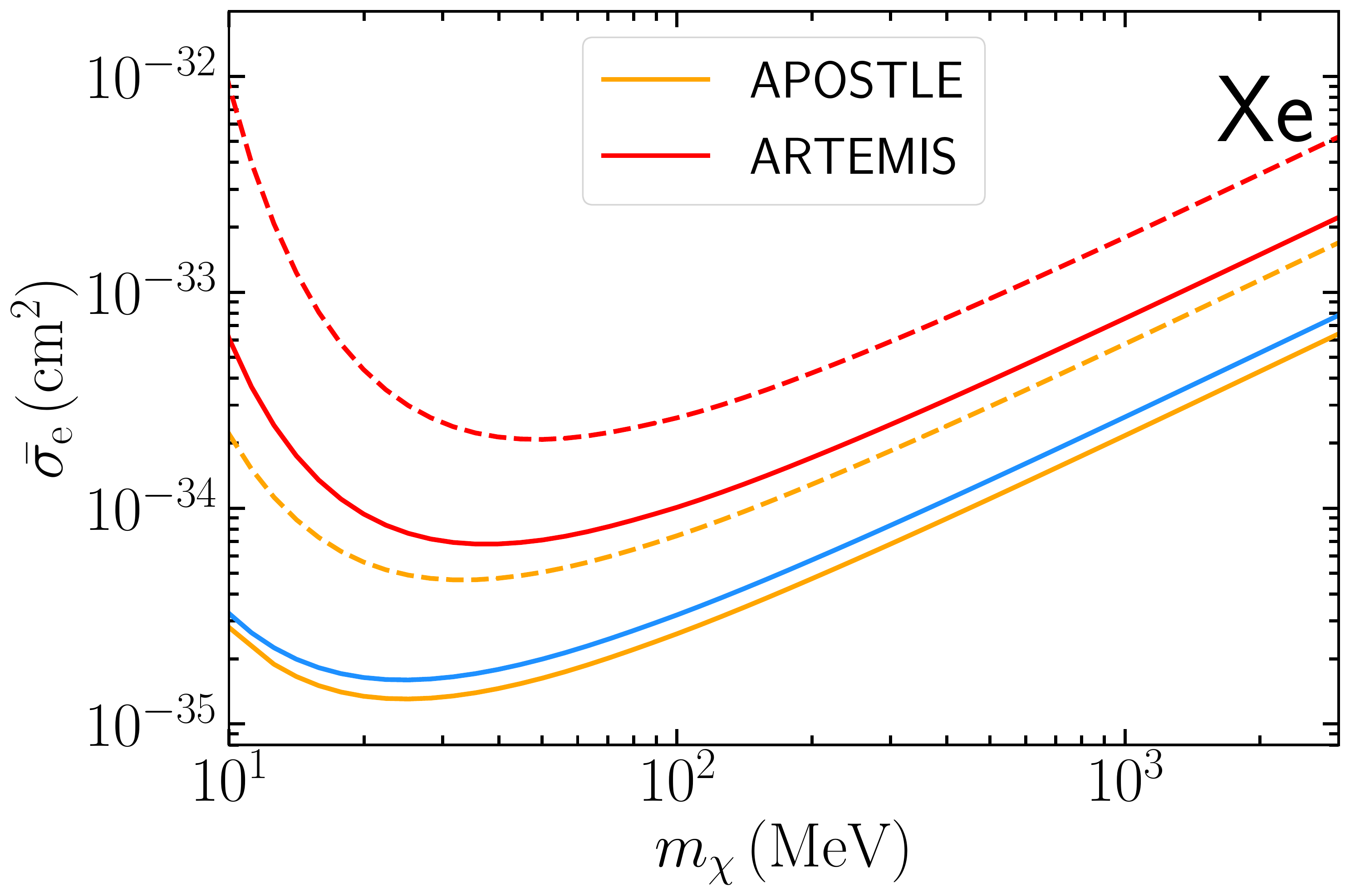}}
\subfloat[\label{sf:SiTsaq2}]{\includegraphics[scale=0.18]{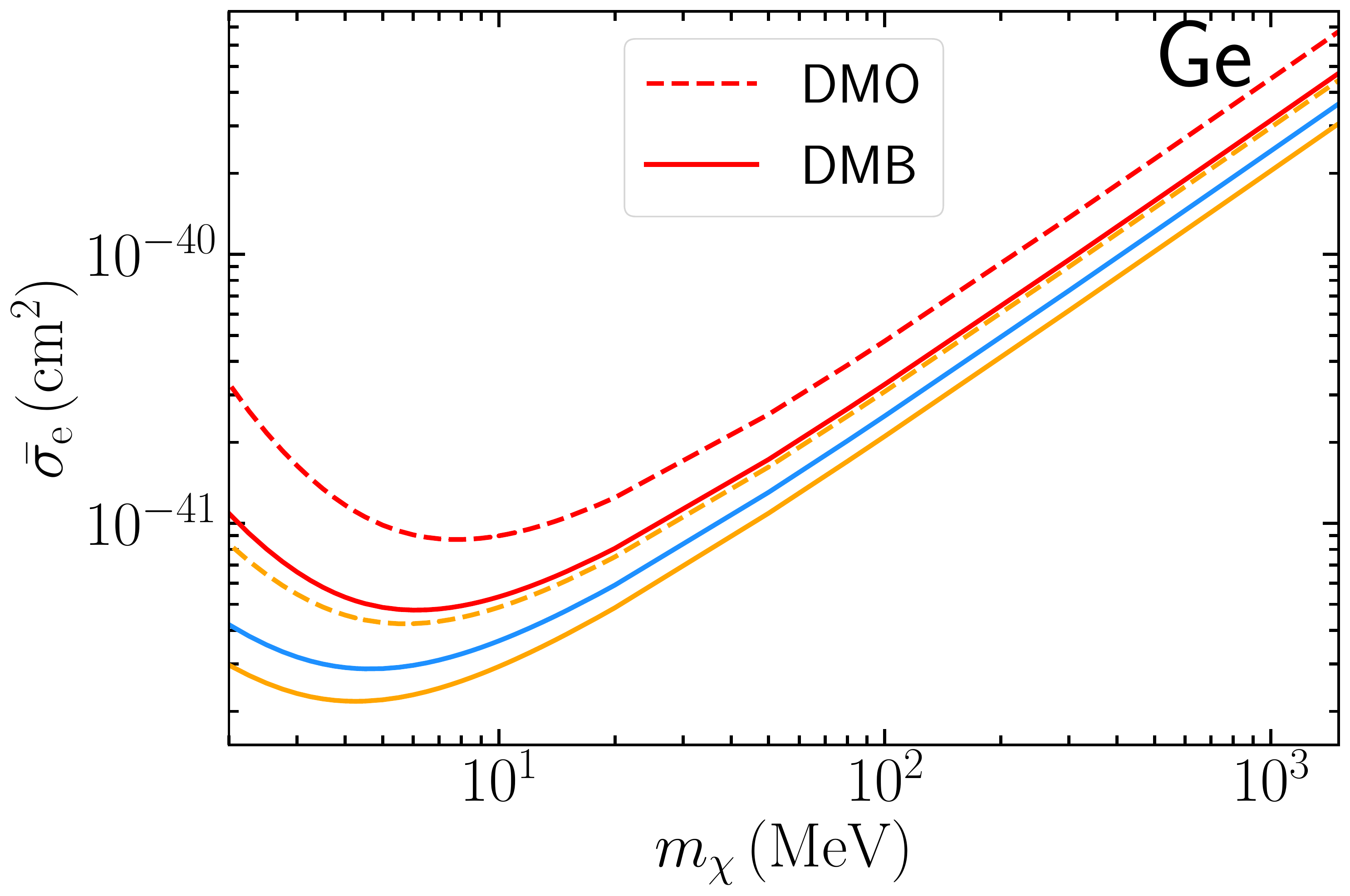}}
\subfloat[\label{sf:GeTsaq2}]{\includegraphics[scale=0.18]{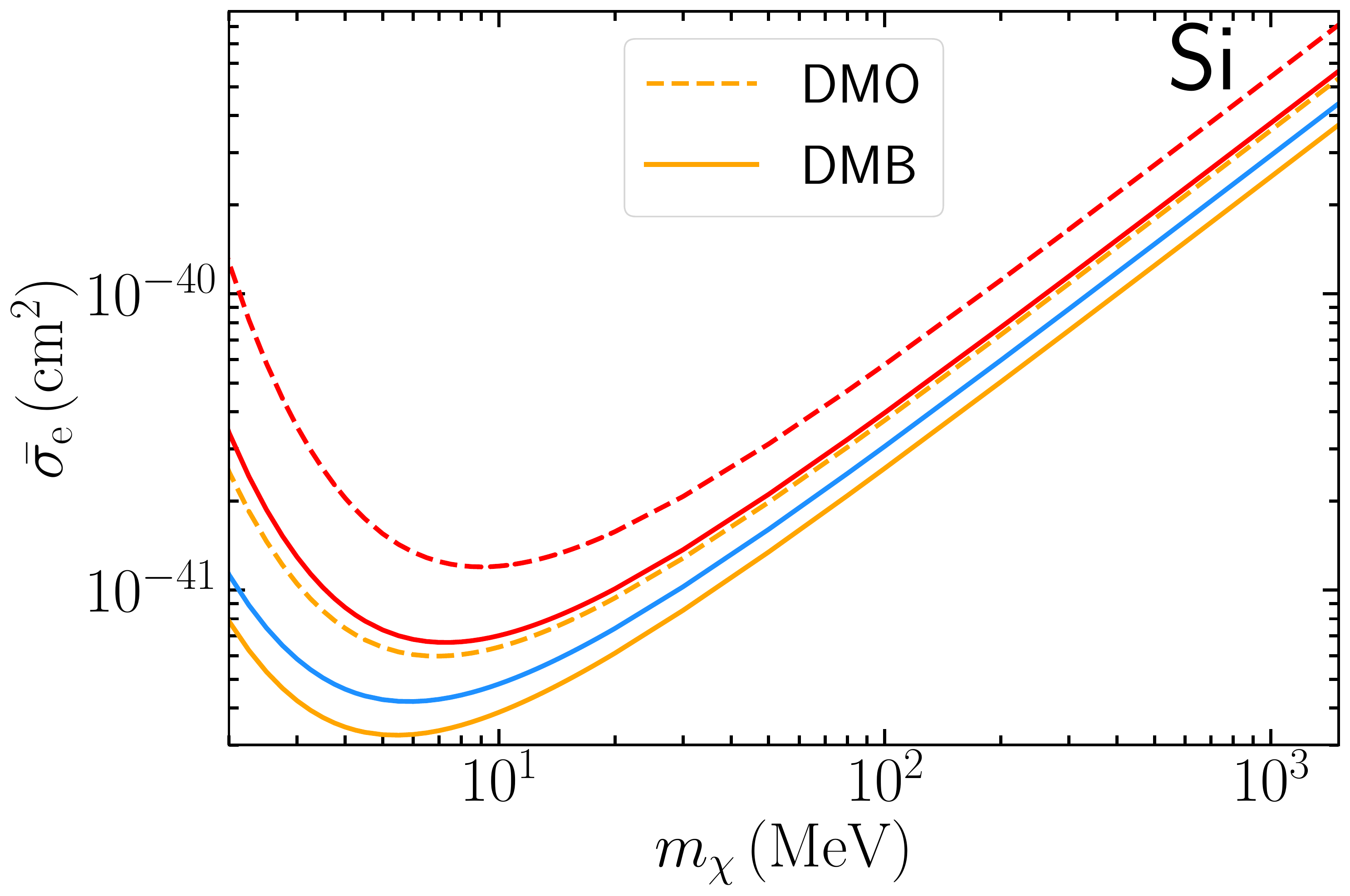}}
\newline
\subfloat[\label{sf:XeTsaq2fa}]{\includegraphics[scale=0.18]{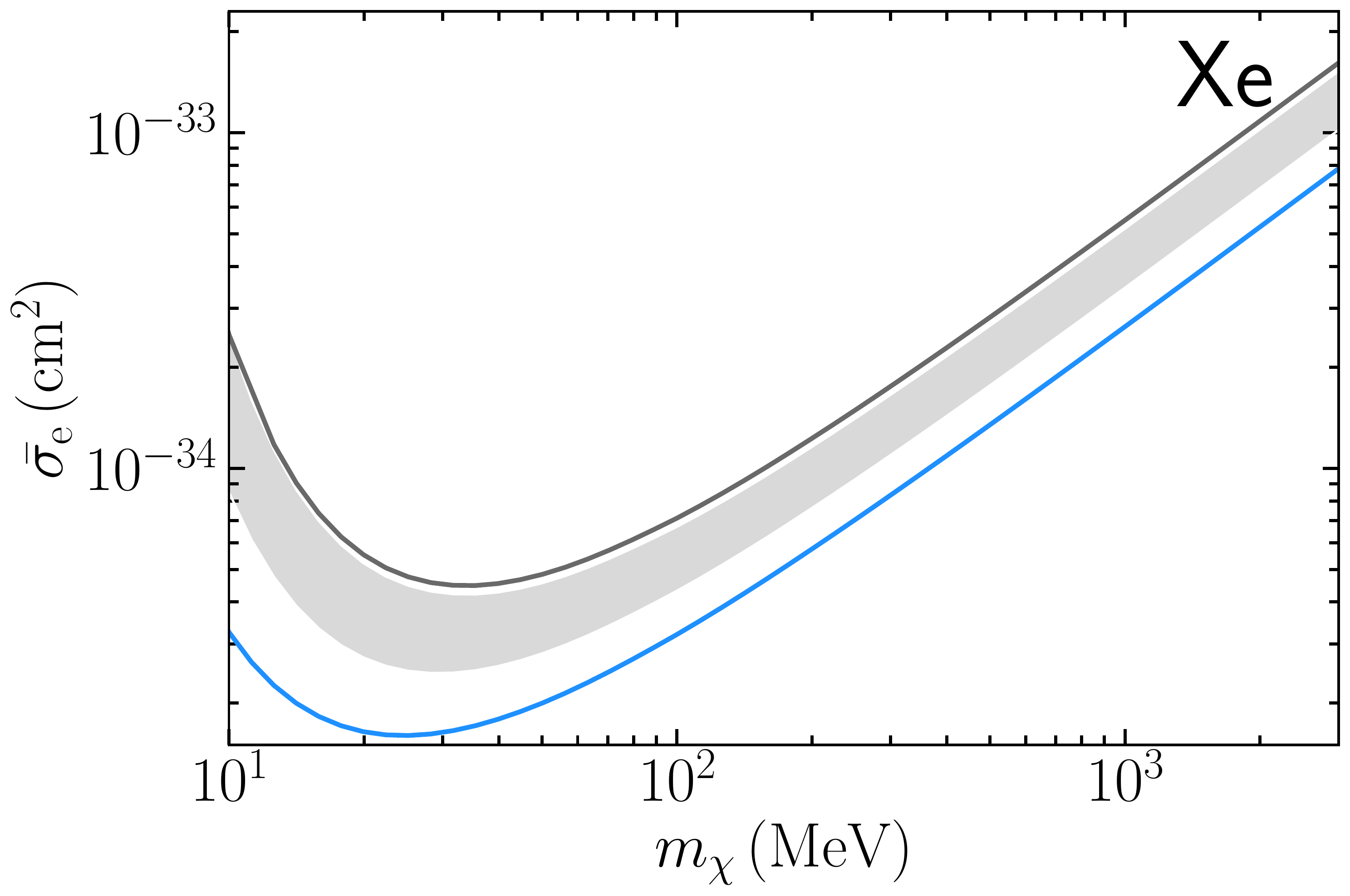}}
\subfloat[\label{sf:SiTsaq2fa}]{\includegraphics[scale=0.18]{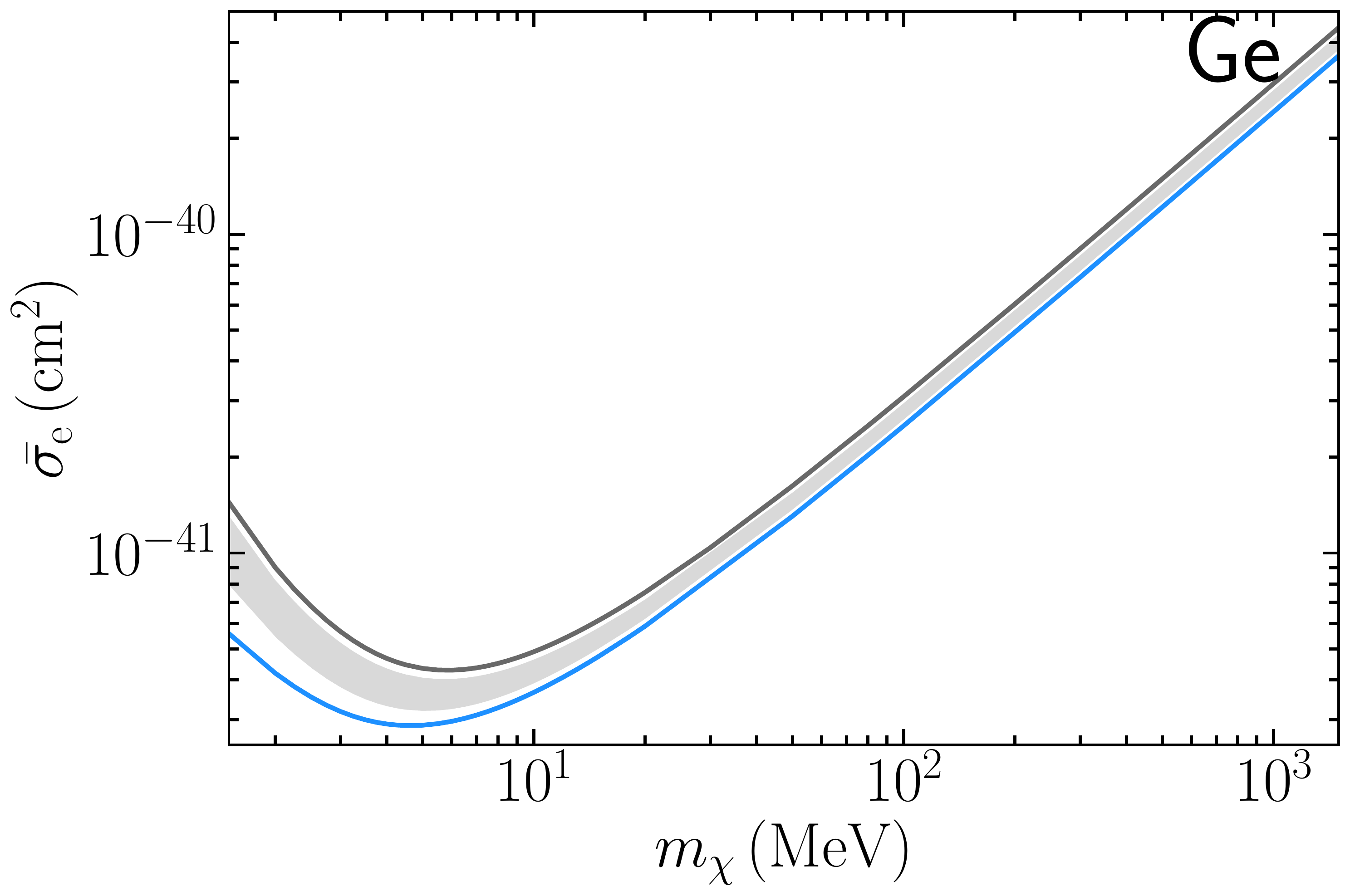}}
\subfloat[\label{sf:GeTsaq2fa}]{\includegraphics[scale=0.18]{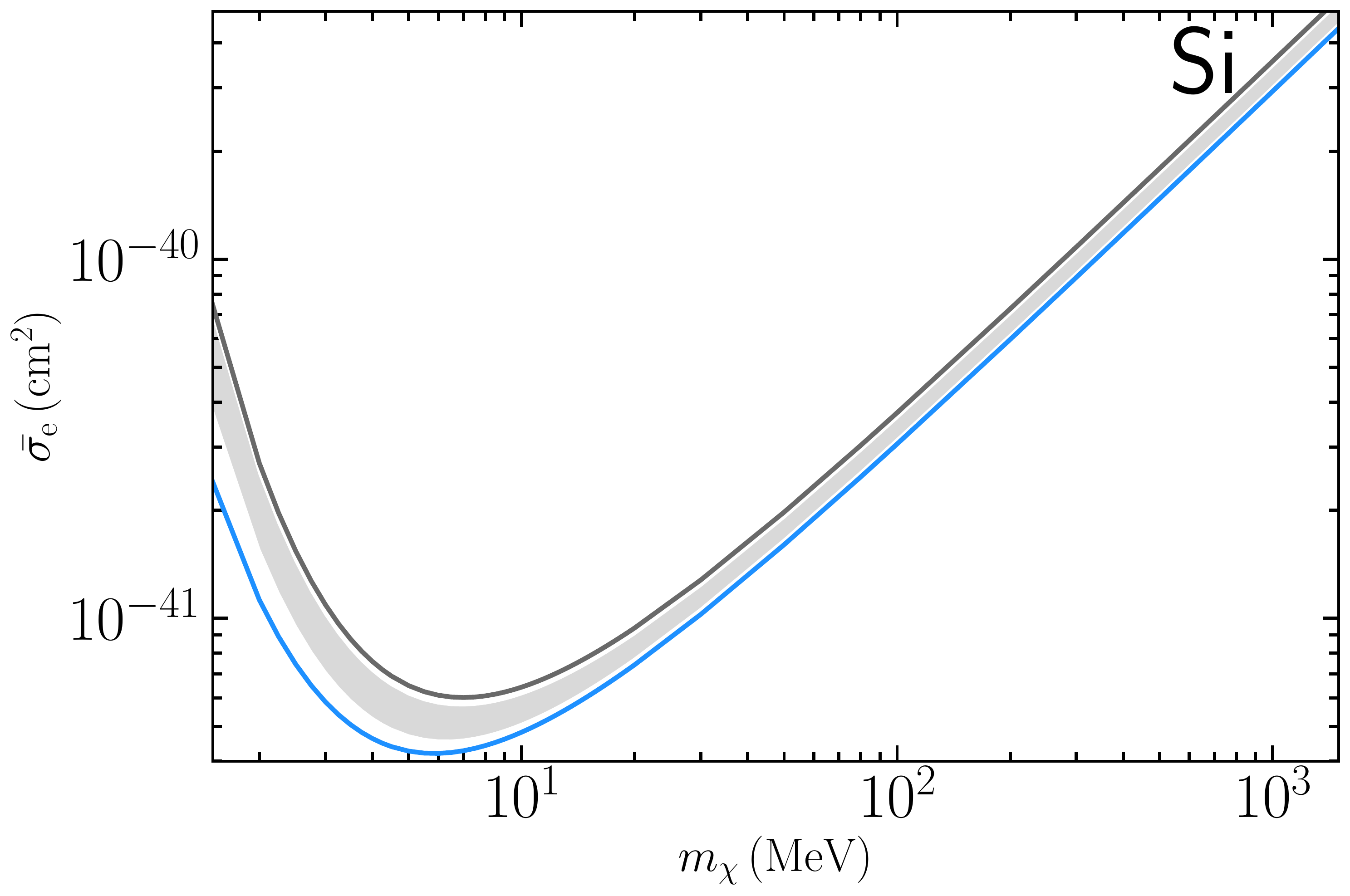}}
\caption{Same as figure \ref{fig:Tsaappq} but for $F_{\rm DM}  \propto q^{-2}$.}
\label{fig:Tsaappq2}
\end{center}
\end{figure*}

\begin{figure*}[t!]
\begin{center}
\subfloat[\label{sf:XeMaoq}]{\includegraphics[scale=0.18]{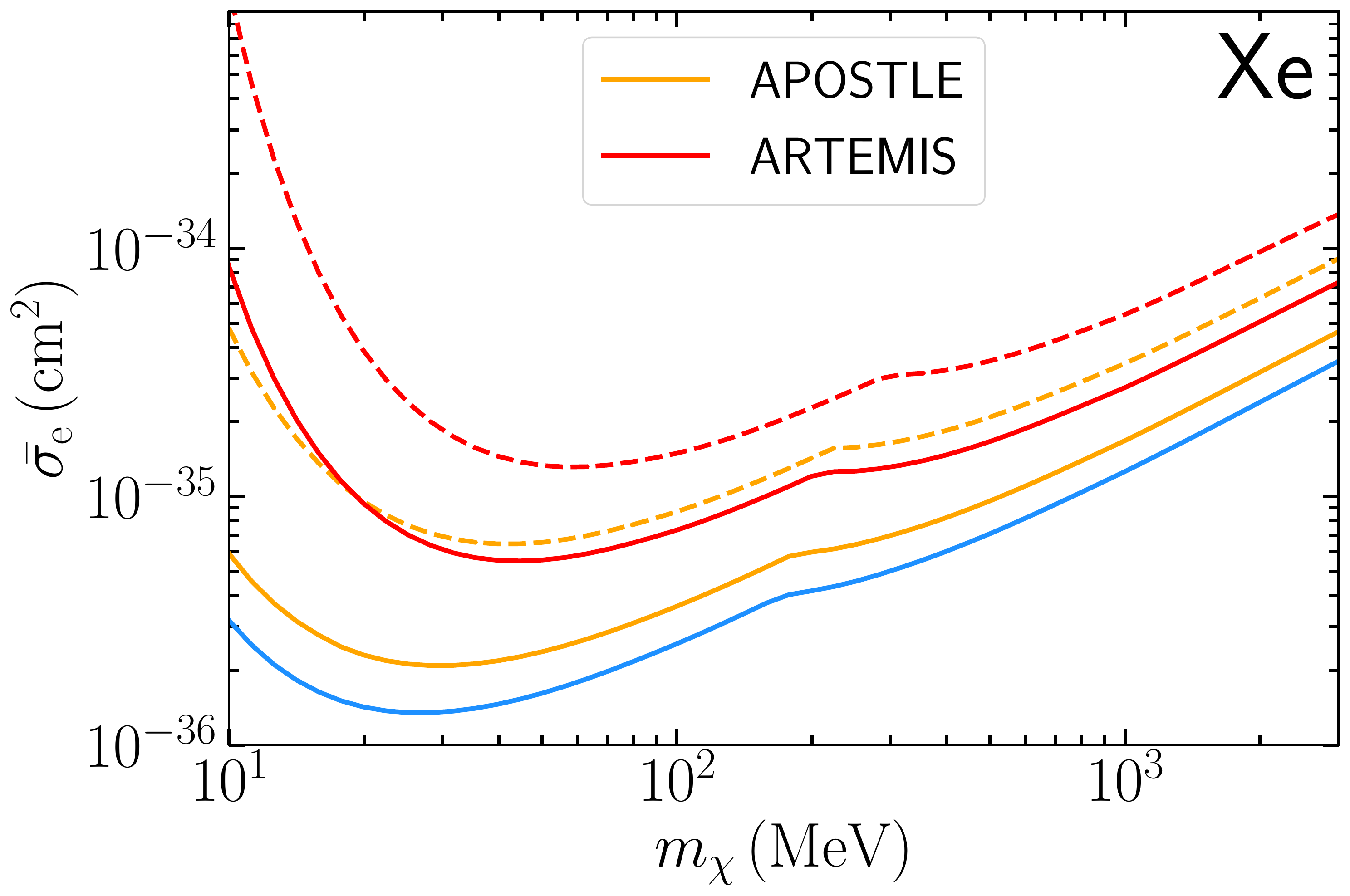}}
\subfloat[\label{sf:SiMaoq}]{\includegraphics[scale=0.18]{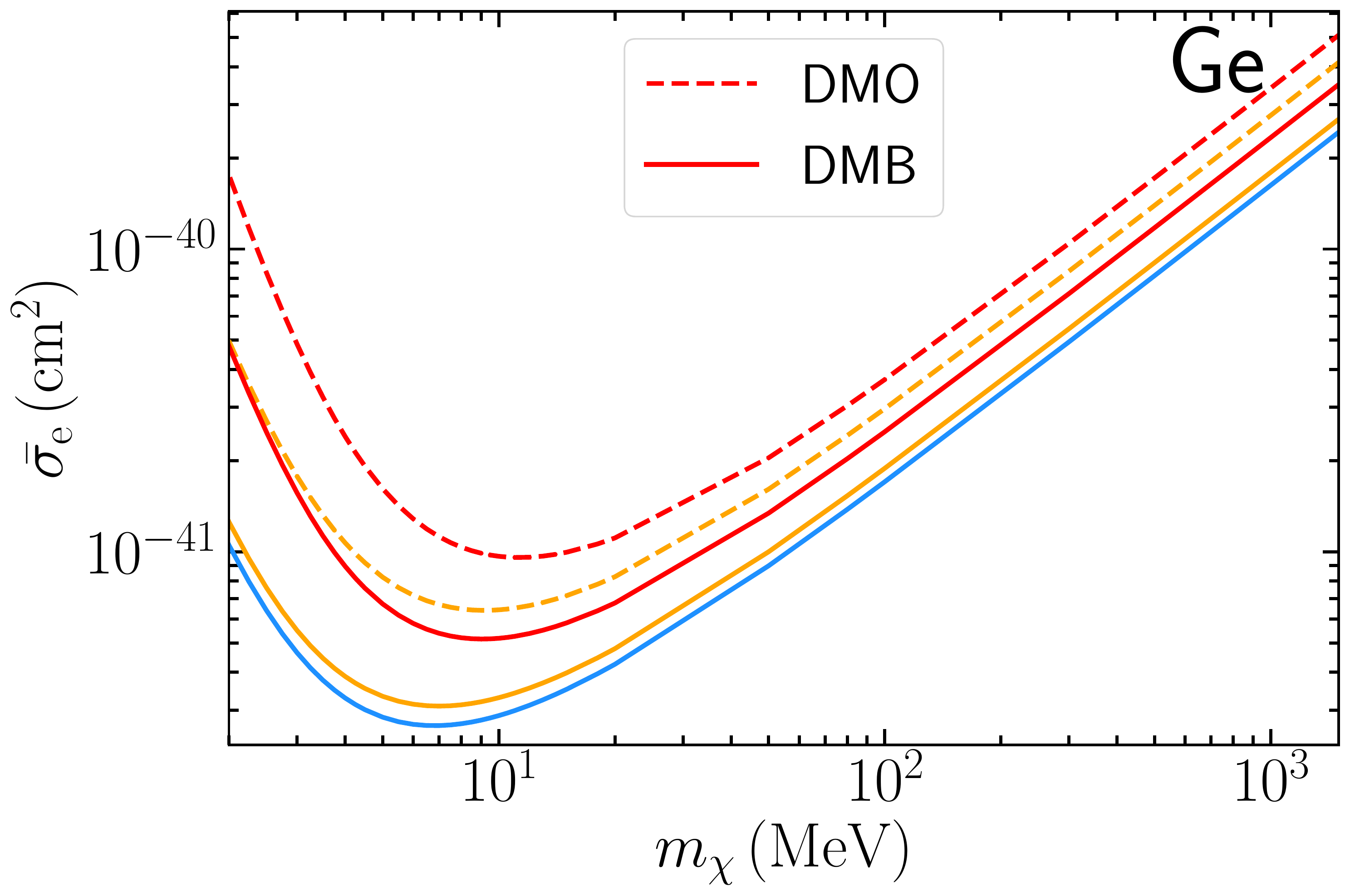}}
\subfloat[\label{sf:GeMaoq}]{\includegraphics[scale=0.18]{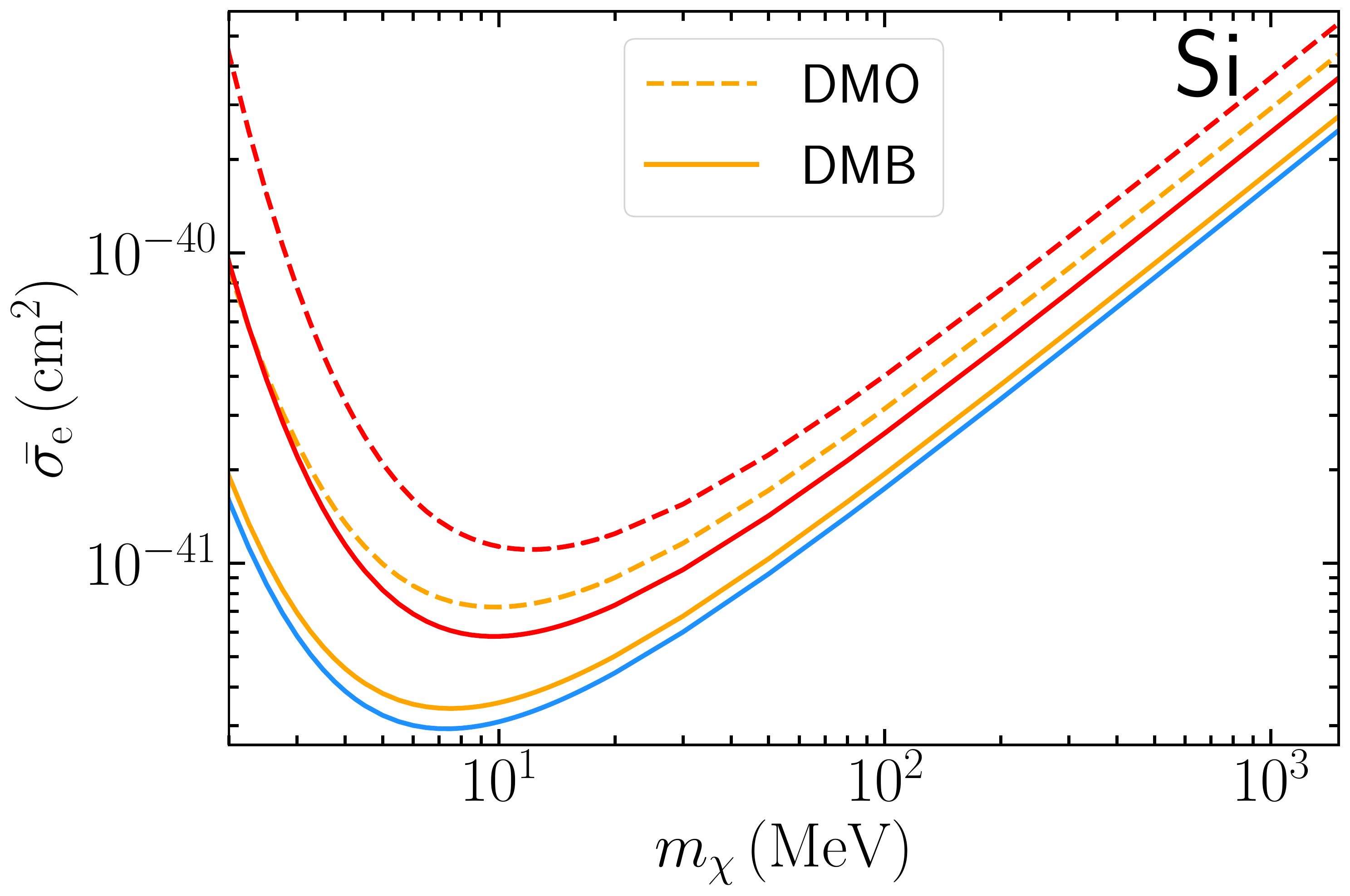}}
\newline
\subfloat[\label{sf:XeMaoqfa}]{\includegraphics[scale=0.18]{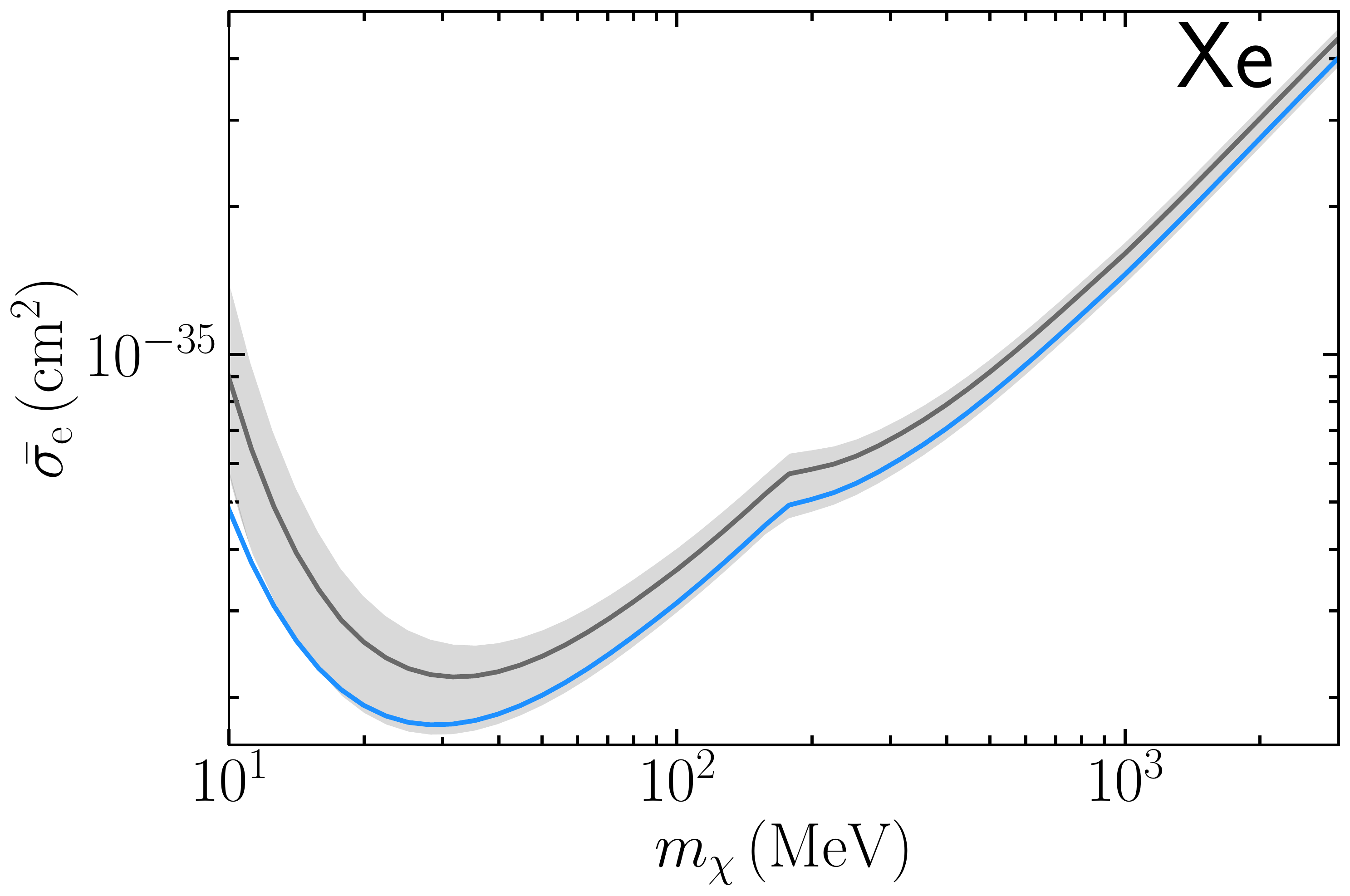}}
\subfloat[\label{sf:SiMaoqfa}]{\includegraphics[scale=0.18]{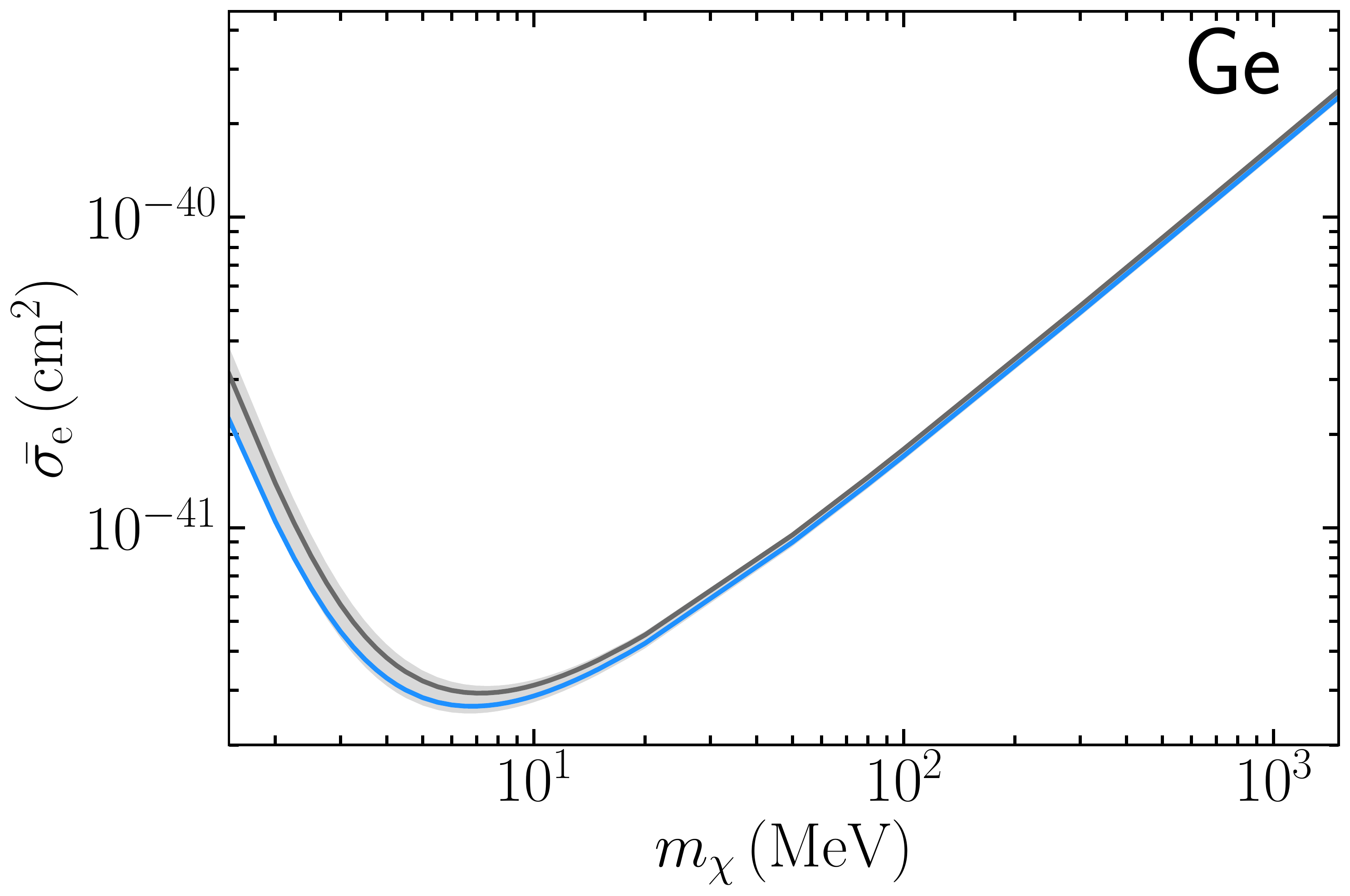}}
\subfloat[\label{sf:GeMaoqfa}]{\includegraphics[scale=0.18]{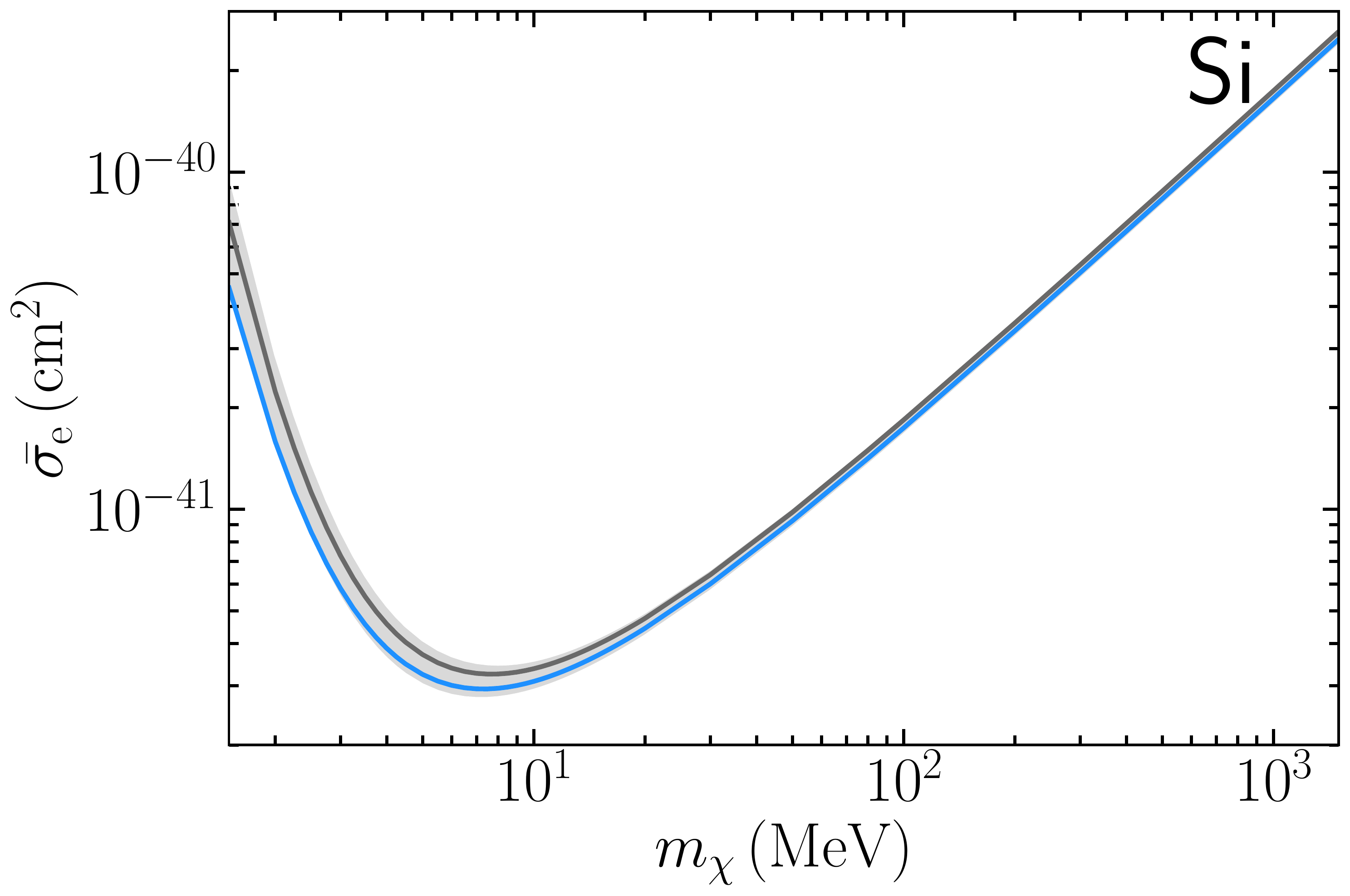}}
\caption{Shifts in the exclusion limits for Mao et. al. distribution with $F_{\rm DM}  \propto q^{-1}$. In the upper panel we show the variations for the best fit values of different  cosmological simulations. The bands in the lower panel represents uncertainties in the recent astrophysical measurements.  The other relevant details are same as of figures \ref{fig:v0-astro} and \ref{fig:Mao}.}
\label{fig:Maoappq}
\end{center}
\end{figure*}

\begin{figure*}[t]
\begin{center}
\subfloat[\label{sf:XeMaoq2}]{\includegraphics[scale=0.18]{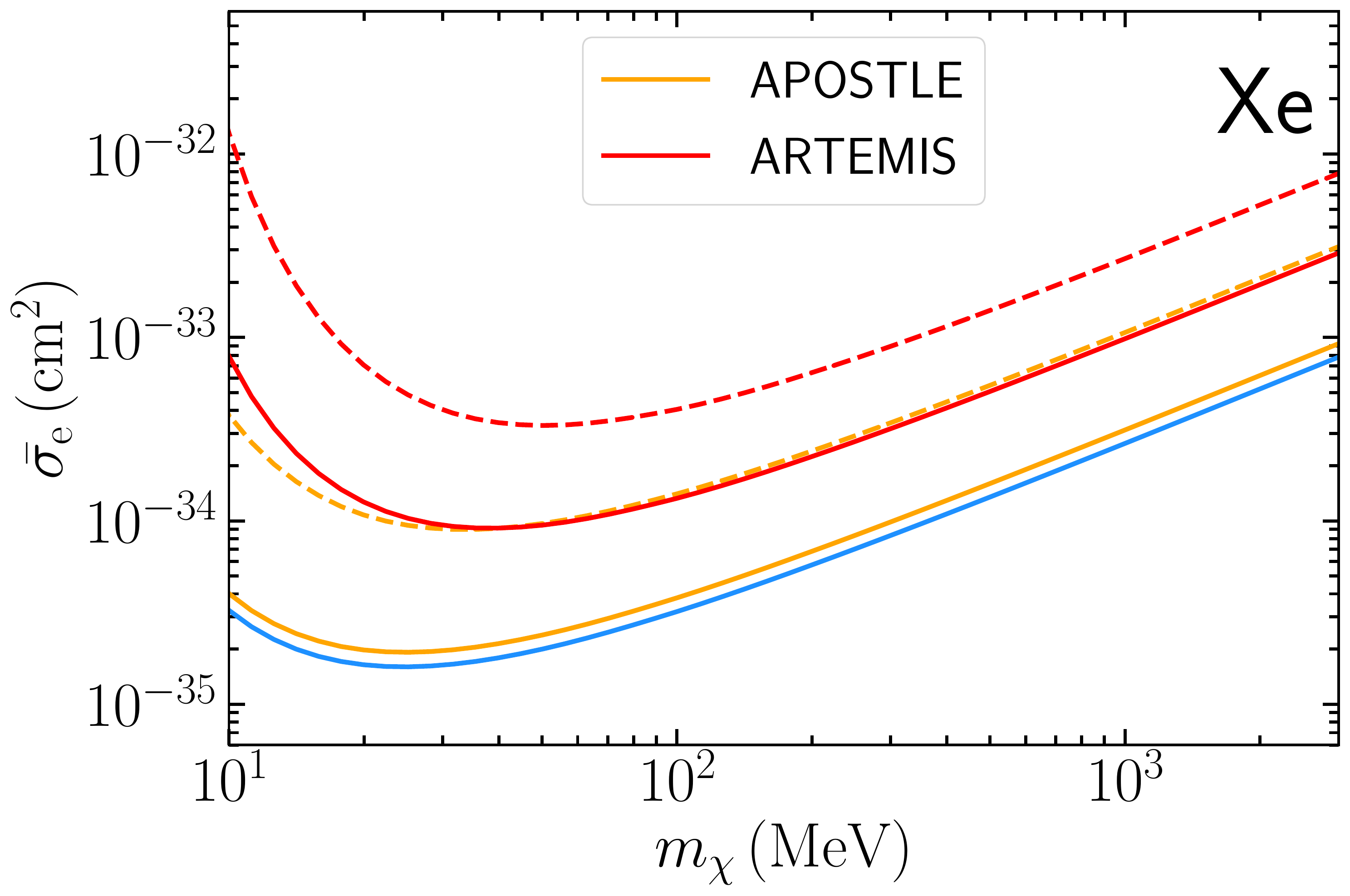}}
\subfloat[\label{sf:SiMaoq2}]{\includegraphics[scale=0.18]{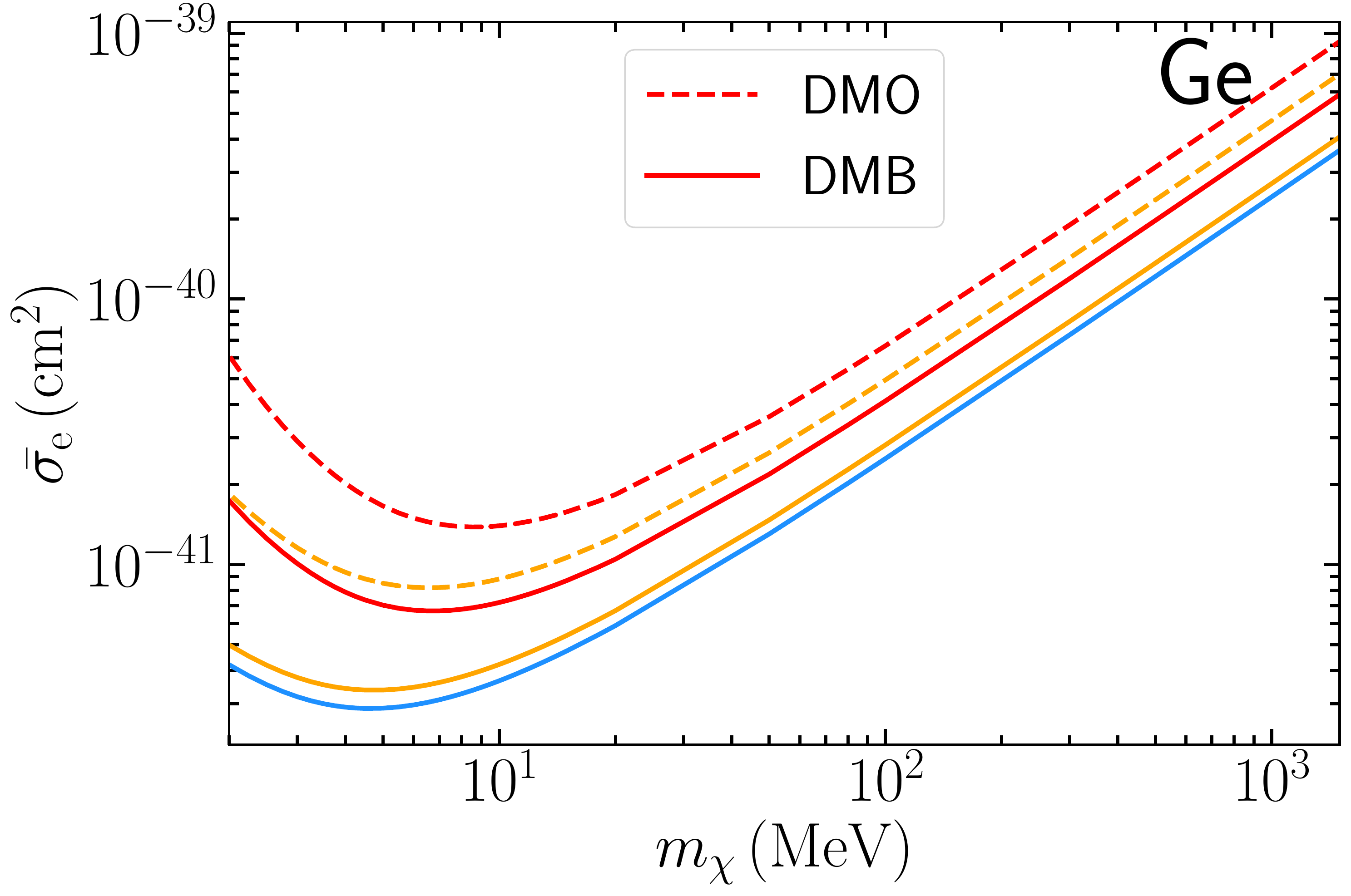}}
\subfloat[\label{sf:GeMaoq2}]{\includegraphics[scale=0.18]{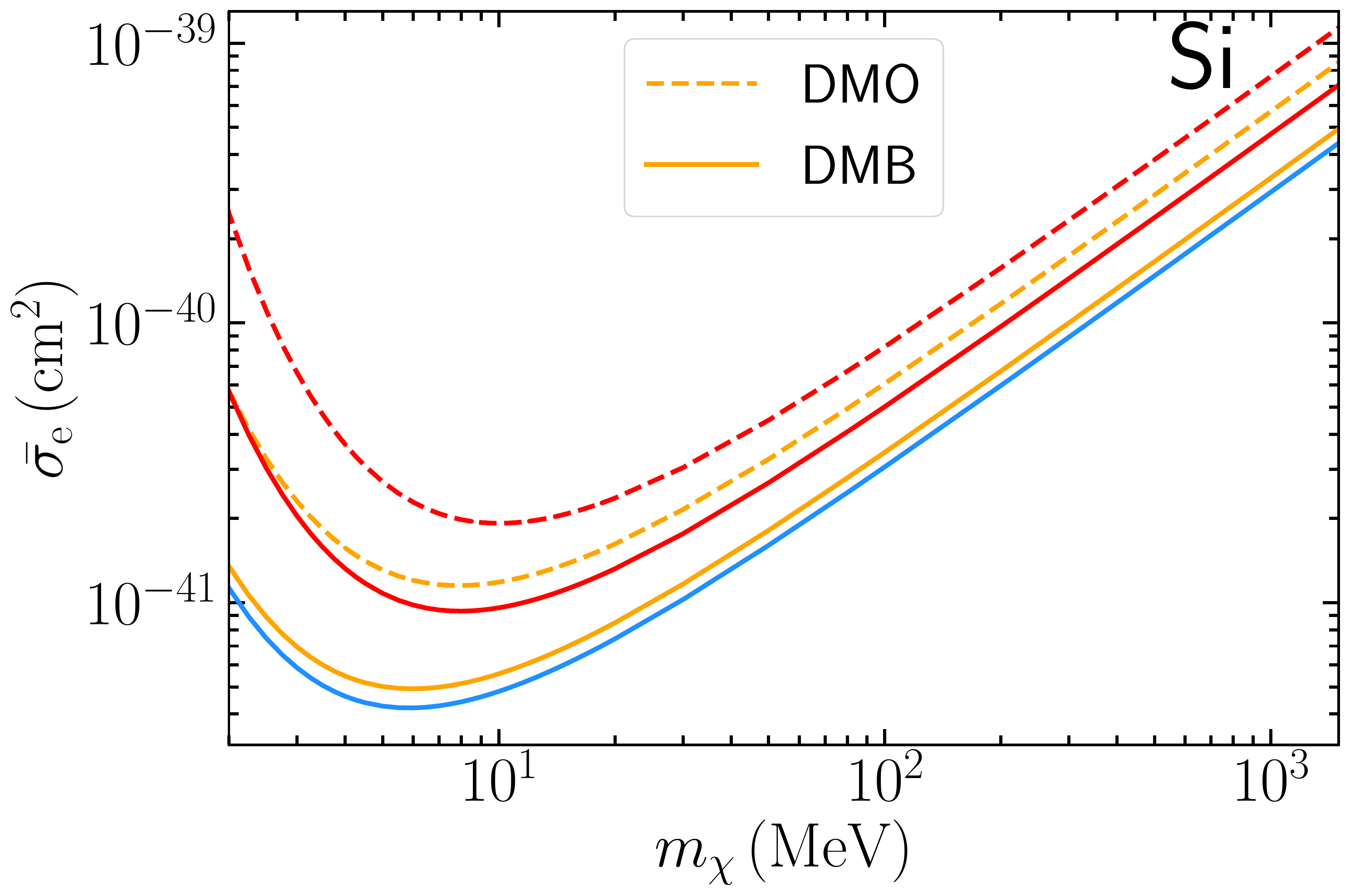}}
\newline
\subfloat[\label{sf:XeMaoq2fa}]{\includegraphics[scale=0.18]{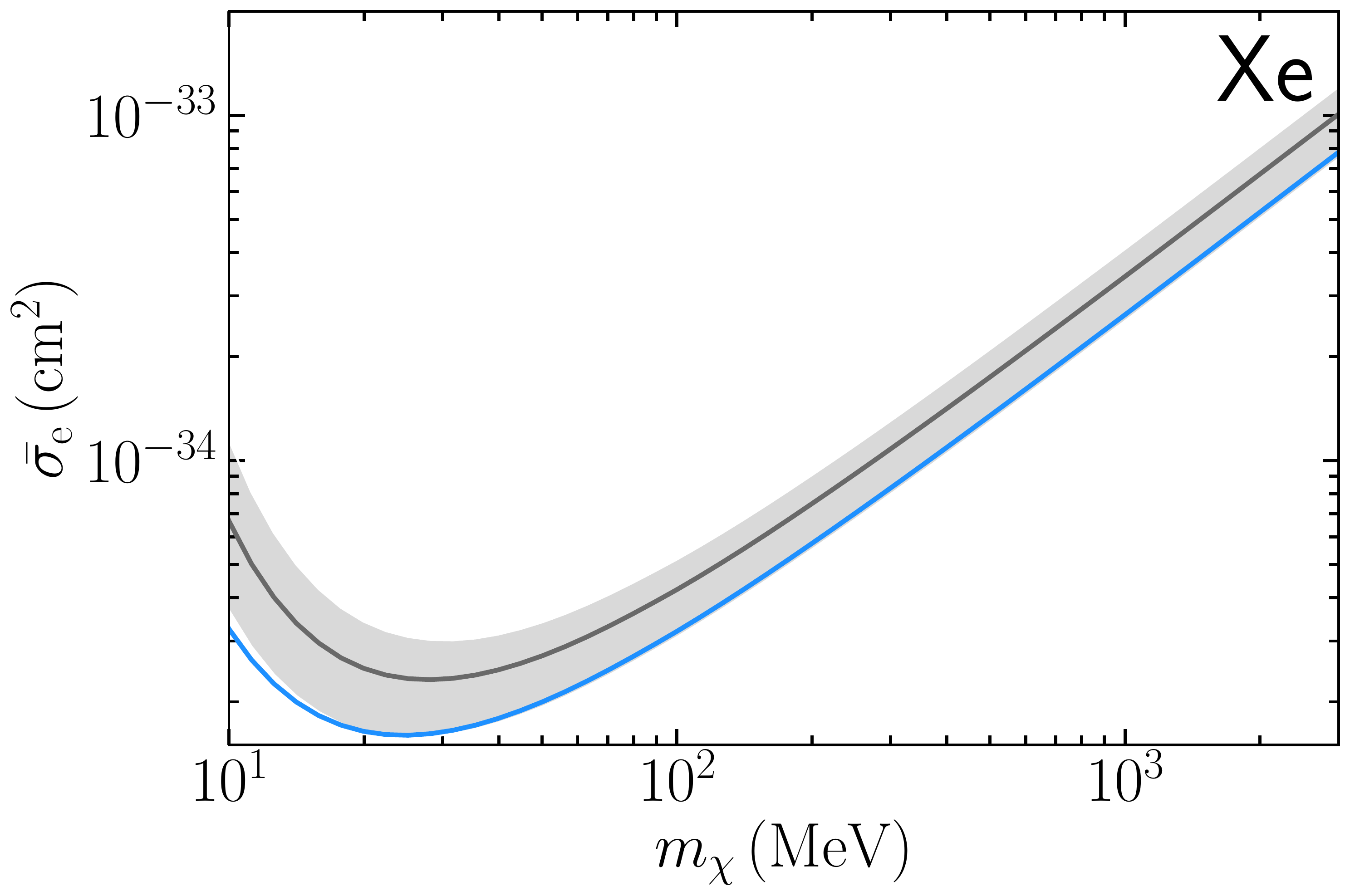}}
\subfloat[\label{sf:SiMaoq2fa}]{\includegraphics[scale=0.18]{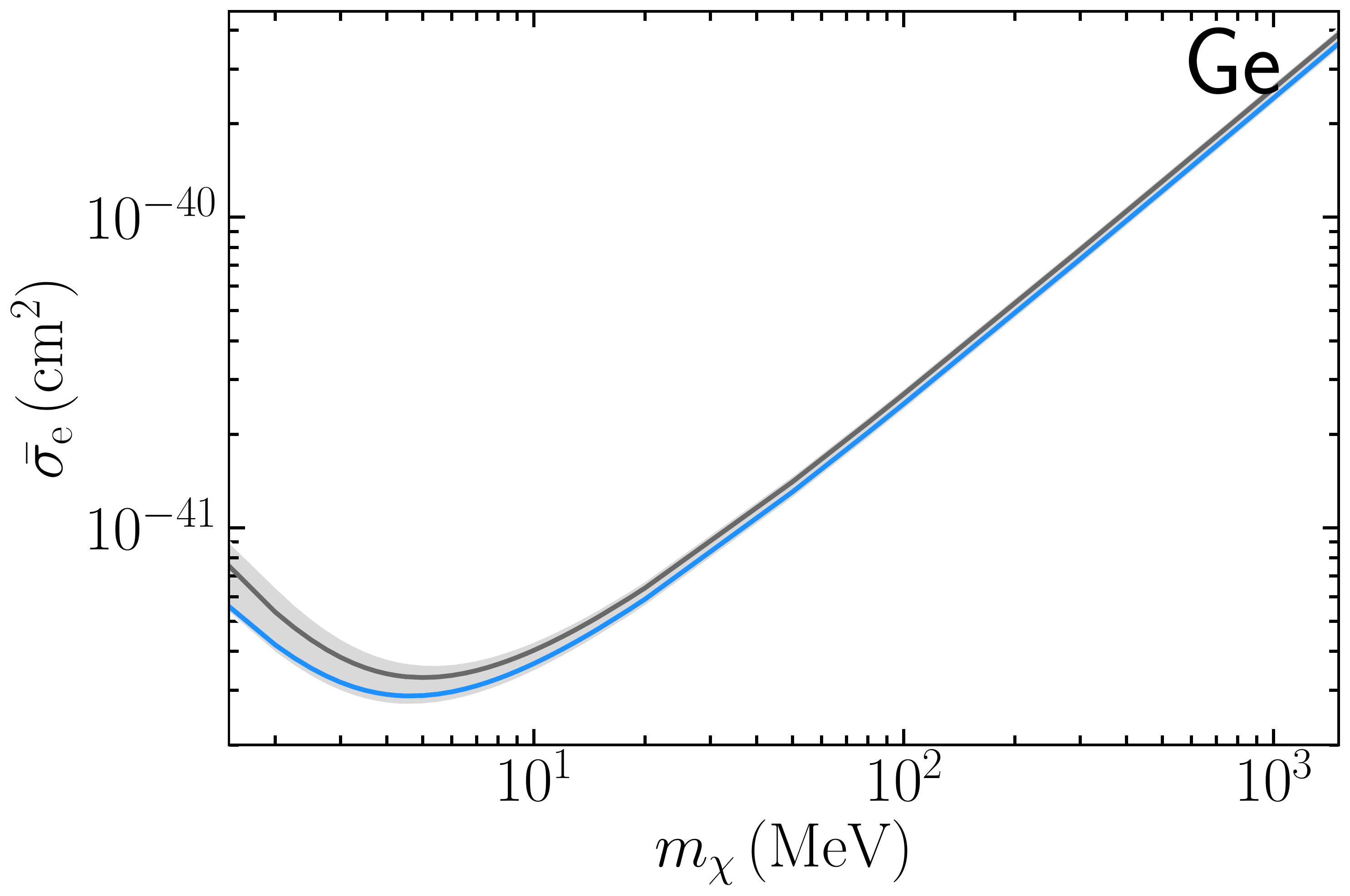}}
\subfloat[\label{sf:GeMaoq2fa}]{\includegraphics[scale=0.18]{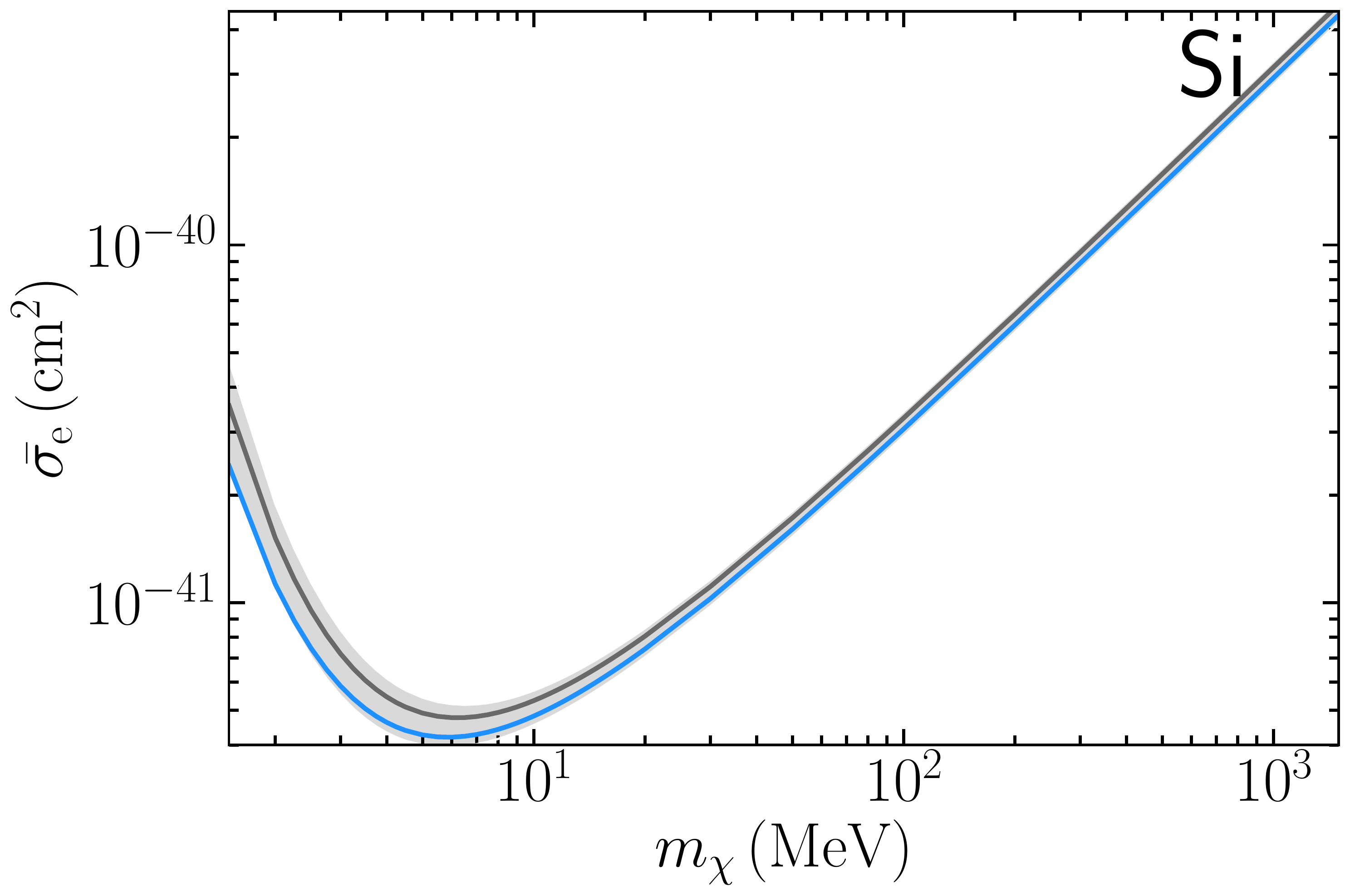}}
\caption{Same as figure \ref{fig:Maoappq} but for $F_{\rm DM}  \propto q^{-2}$.}
\label{fig:Maoappq2}
\end{center}
\end{figure*}

In this appendix we have presented the shifts in the exclusion limits for the other two choices of interactions between DM and electron, namely $F_{\rm DM}\propto q^{-1}$ and $F_{\rm DM}\propto q^{-2}$. The aforementioned variations within the standard halo model have been shown in figures \ref{fig:SHMappq} and \ref{fig:SHMappq2} only for the simulations that have been considered in this work. In figures \ref{fig:Kingappq} and \ref{fig:Kingappq2}  we have provided variations for the King whereas the same for DPL models are depicted in figures \ref{fig:DPLappq} and \ref{fig:DPLappq2}. The bounds from the rest of the non-SHM distributions considered here are shown in  figures \ref{fig:Tsaappq}, \ref{fig:Tsaappq2}, \ref{fig:Maoappq}, and \ref{fig:Maoappq2}.  For each of the figures (except figures \ref{fig:SHMappq} and \ref{fig:SHMappq2}) the upper panels correspond to the variations for the best fit values of the  cosmological simulation and the lower panels correspond to the same considering the updated observational values and their uncertainties. For Xe targets the momentum suppression in the momentum dependent interactions shift the special feature (discussed in section \ref{subsec:king}) towards higher DM masses. For $F_{\rm DM} \propto q^{-1}$ this appears around a DM mass of 200 MeV, however for $F_{\rm DM} \propto q^{-2}$ this crosses the boundary of the figures, as depicted in the left panels of the aforementioned figures.

%

%
\FloatBarrier

\FloatBarrier

\FloatBarrier

\bibliographystyle{h-physrev}
\bibliography{AstroUncertLDM.bib}

\end{document}